\documentstyle[12pt,suthesis_aas,longtable]{report}

\setlongtables
\submitdate{July 1996}
\copyrightyear{1996}

\begin{document}

\newcommand{\cosb}{{\em COS~B}\/}
\newcommand{\sas}{{\em SAS~2}\/}
\newcommand{\cgro}{{\em CGRO}\/}
\newcommand{\egret}{{\em EGRET}\/}
\newcommand{\batse}{{\em BATSE}\/}
\newcommand{\comptel}{{\em COMPTEL}\/}
\newcommand{\osse}{{\em OSSE}\/}
\newcommand{\rosat}{{\em ROSAT}\/}

\newcommand{\sage}{{\tt SAGE}\/}
\newcommand{\like}{{\tt LIKE}\/}
\newcommand{\pulsar}{{\tt PULSAR}\/}
\newcommand{\spectral}{{\tt SPECTRAL}\/}

\newcommand{\gammaray}{$\gamma$-ray}
\newcommand{\gammarays}{$\gamma$-rays}
\newcommand{\Gammaray}{$\gamma$-Ray}
\newcommand{\Gammarays}{$\gamma$-Rays}
\newcommand{\perareasec}{~cm$^{-2}$~s$^{-1}$}
\newcommand{\perareasecmev}{~cm$^{-2}$~s$^{-1}$~MeV$^{-1}$}
\newcommand{\perareasecsr}{~cm$^{-2}$~s$^{-1}$~sr$^{-1}$}
\newcommand{\epm}{$e^\pm$}
\newcommand{\etal}{et~al.}
\newcommand{\sub}[1]{_{\rm #1}}
\newcommand{\expct}[1]{\left\langle #1 \right\rangle}
\newcommand{\idgb}{$I_{{\rm IDGB}}$}

\def\accang{\hbox{$\theta_{67}$}}
\def\chisq{\hbox{$\chi^2$}}
\def\expsr{\hbox{$\cal E$}}
\def\htest{\hbox{$H$-test}}
\def\ipred{\hbox{$\dot E/D^2$}}
\def\ij{\hbox{$_{ij}$}}
\def\kl{\hbox{$_{kl}$}}
\def\psf{\hbox{\rm PSF}}
\def\TS{\hbox{\rm TS}}
\def\sn{\hbox{\rm S/N}}
\def\sqrtTS{\hbox{${\rm TS}^{1/2}$}}
\def\zsqr{\hbox{$Z_m^2$}}

\newcommand{\chapt}[1]{Chapter~\ref{#1}}
\newcommand{\eq}[1]{equation~(\ref{#1})}
\newcommand{\fig}[1]{Figure~\ref{#1}}
\newcommand{\sect}[1]{\S\ref{#1}}
\newcommand{\tbl}[1]{Table~\ref{#1}}

\newcounter{subtable}

\def\aj{AJ}			
\def\araa{ARA\&A}		
\def\apj{ApJ}			
\def\apjl{ApJ}		
\def\apjs{ApJS}		
\def\ao{Appl.Optics}		
\def\apss{Ap\&SS}		
\def\aap{A\&A}		
\def\aapr{A\&A~Rev.}		
\def\aaps{A\&AS}		
\def\azh{AZh}			
\def\baas{BAAS}		
\def\jrasc{JRASC}		
\def\memras{MmRAS}		
\def\mnras{MNRAS}		
\def\nature{Nature}		
\def\nim{Nucl. Inst. and Meth.}		
\def\physrev{Phys.Rev.}		
\def\pra{Phys.Rev.A}		
\def\prb{Phys.Rev.B}		
\def\prc{Phys.Rev.C}		
\def\prd{Phys.Rev.D}		
\def\prl{Phys.Rev.Lett}	
\def\pasp{PASP}		
\def\pasj{PASJ}		
\def\qjras{QJRAS}		
\def\science{Science}		
\def\skytel{S\&T}		
\def\solphys{Solar~Phys.}	
\def\sovast{Soviet~Ast.}	
\def\ssr{Space~Sci.Rev.}
\def\thesis{Ph.D. Thesis}	
\def\zap{ZAp}			
\def\GROworkshop{in Proc. of the Gamma-Ray Observatory Science Workshop,
ed. W. N. Johnson (Greenbelt, MD: NASA)}
\def\EGRETsymp{in The Energetic Gamma-Ray Experiment Telescope (EGRET)
Science Symposium, ed. C. Fichtel \etal\ (Greenbelt, MD: NASA)}

\def\gt{\hbox{$>$}}
\def\lt{\hbox{$<$}}
\newcommand{\ga}{\stackrel{>}{_\sim}}
\newcommand{\la}{\stackrel{<}{_\sim}}
\newcommand{\cross}{{\boldmath \times}}
\newcommand{\vdot}{{\boldmath \cdot}}

\def\deg{\hbox{$^\circ$}}
\def\hr{\hbox{$^{\rm h}$}}
\def\mn{\hbox{$^{\rm m}$}}
\def\sun{\hbox{$\odot$}}
\def\earth{\hbox{$\oplus$}}
\def\sq{\hbox{\rlap{$\sqcap$}$\sqcup$}}
\def\arcmin{\hbox{$^\prime$}}
\def\arcsec{\hbox{$^{\prime\prime}$}}
\def\fd{\hbox{$.\!\!^{\rm d}$}}
\def\fh{\hbox{$.\!\!^{\rm h}$}}
\def\fm{\hbox{$.\!\!^{\rm m}$}}
\def\fs{\hbox{$.\!\!^{\rm s}$}}
\def\fdg{\hbox{$.\!\!^\circ$}}
\def\farcm{\hbox{$.\mkern-4mu^\prime$}}
\def\farcs{\hbox{$.\!\!^{\prime\prime}$}}
\def\fp{\hbox{$.\!\!^{\scriptscriptstyle\rm p}$}}
\def\micron{\hbox{$\mu$m}}
\def\us{\hbox{$\mu$s}}

\def\onehalf{\slantfrac{1}{2}}
\def\onethird{\slantfrac{1}{3}}
\def\twothirds{\slantfrac{2}{3}}
\def\onequarter{\slantfrac{1}{4}}
\def\threequarters{\slantfrac{3}{4}}
\def\ubvr{\hbox{$U\!BV\!R$}}		
\def\ub{\hbox{$U\!-\!B$}}		
\def\bv{\hbox{$B\!-\!V$}}		
\def\vr{\hbox{$V\!-\!R$}}		
\def\ur{\hbox{$U\!-\!R$}}

\input{epsf.sty}

\title{Observations of the Isotropic Diffuse Gamma-ray Background\\
with the EGRET Telescope}

\author{Thomas Daniel Willis}
\dept{Physics}
\principaladviser{Peter F. Michelson}
\firstreader{A.B.C.Walker, Jr.}
\secondreader{V. Petrosian}
 
\beforepreface
\clearpage
\null\vskip 3in
\centerline{To Dad}
\clearpage

\prefacesection{Abstract}
An Isotropic Diffuse Gamma-Ray Background (IDGRB) in the spectral range 30-10,000 MeV was first reported in the early 1970's using measurements made by the \sas\ instrument.  Data recorded by the Energetic Gamma Ray Experiment Telescope (\egret) on board the Compton Gam\-ma Ray Observatory (\cgro) over the last 4 years are analysed in order to extract the best measurement yet made of the IDGRB.  Extensive analysis of the \egret\ instrumental background is presented in order to demonstrate that an uncontaminated data set can be extracted from the \egret\ data.

A model of the high latitude galactic diffuse foreground emission is presented and the existence of an IDGRB is confirmed.  Spatial and spectral analysis of this background is presented.  In addition, point source analysis at high galactic latitudes is performed to reveal the existence of a population of extragalactic sources.  The characteristics of this population are examined and models of its flux distribution are reported. The question of whether the IDGRB is composed of unresolved point sources is addressed using fluctuation analysis. 

Finally, possible future directions for \gammaray\ astronomy are examined through simulations of a future \gammaray\ telescope: the Gamma-ray Large Area Space Telescope (GLAST).  The GLAST baseline design is described and its scientific performance is evaluated.  The ability of this telescope to detect 1,000-10,000 new extragalactic sources is demonstrated and the likely impact on the study of the IDGRB is considered.
\clearpage

\prefacesection{Acknowledgments}
By my count, this constitutes the 6th Stanford PhD thesis generated by the \egret\ instrument. All of us have been the benificiaries not only of the efforts of our predecessors, but also of the entire \egret\ collaboration  as it has evolved through the years.  As such, the list of persons in whose debt we will always remain, grows longer with each dissertation.  I was the last in this line to have known, however briefly, two of the creators of this remarkable telescope and leaders of the Stanford group: Bob Hofstadter and Barrie Hughes.  I am particularly grateful to them for leading this effort which they were unable to see come to fruition.  
\begin{center}
\small
J.~Chiang, R.~Hofstadter$^\ast$, E.~B.~Hughes$^\ast$, B.~B.~Jones, Y.~C.~Lin, 
P.~F.~Michelson, P.~L.~Nolan, S.~K.~Shriver, W.~F.~Tompkins, J.~M.~Fierro \\
{\sc Stanford University} \\[0.07in]

D.~L.~Bertsch, B.~L.~Dingus, J.~A.~Esposito, A.~Etienne, C.~E.~Fichtel,
D.~P.~Friedlander, R.~C.~Hartman, S.~D.~Hunter, D.~J.~Kendig, N.~A.~Laubenthal,
J.~R.~Mattox, L.~M.~McDonald, C.~von~Montigny, R.~Mukherjee, P.~Sreekumar,
D.~J.~Thompson \\
{\sc NASA Goddard Space Flight Center}\\[0.07in]

K.~T.~S.~Brazier, G.~Kanbach, H.~A.~Mayer-Hasselwander, M.~Merck \\
{\sc Max-Planck-Institut F\"ur Extraterrestrische Physik}\\[0.07in]

E.~J.~Schneid\\
{\sc Grumman Aerospace Corporation}\\[0.07in]

D.~A.~Kniffen \\
{\sc Hampden-Sydney College}\\[0.07in]

$^\ast${\em Deceased}
\end{center}

Several of the names mentioned above merit additional mention.  Peter Michelson has been an ideal thesis advisor (when I can find him that is), giving me just enough rope to learn independent research skills and not quite enough to let me hang myself.  Joe Fierro, in his own quirky way, was an enormously helpful person to bounce ideas off of as well as being an infallible font of programming knowledge.  Brian Jones and Bill Tompkins were always good for a fresh perspective when they weren't stealing my egret8 cycles.  At Goddard, I am very much indebted to Dave Bertsch for his work on the diffuse model and to Sreekumar for his consistent help working out the high latitude emission.  John Mattox deserves special mention for his work on \egret\ point source analysis without which we would all have to start from scratch.  The entire collaboration at both Max Planck and GSFC were a constant source of encouragement and ideas.  Finally, none of us could have completed an \egret\ thesis without the help of our resident instrument expert, \egret\ historian and recaller of arcana, statistician, system administrator, Spider pro and general ombudsman for all conceivable problems: Pat Nolan.

This dissertation only represents the last three years of my graduate career at Stanford.  I would also like to acknowledge my other advisor A.B.C.Walker, Jr. for teaching me to respect the challenge of designing and building astrophysical instrumentation and to his research group for fond memories of interminable SSRL shifts. JFL,CCK,CED,RHO,DM,J(H)P,MJA  you know who you are. 

I have taken a long and winding route through graduate school and I never would have made it through without the friendship of a great group of people I met here.  Matteo, Ken, Andre, Paul R., Paul H.; maybe we've got one more problem set in us...or at least one more late night frisbee golf game and good argument.

I gratefully acknowledge the substantial financial support I received  from the NASA grant NAG 5-1605 provided to the
Stanford \egret\ team.

\afterpreface

\chapter{Introduction}

\section{The Isotropic Diffuse Gamma-Ray Background (IDGRB)}

	If one were to be able to look up at the high energy ($E>100{\rm MeV}$) \gammaray\ sky, it would not be dominated by thousands of point sources as is the familiar night sky, but rather it would be seen to be dominated by a bright diffuse swath cut across the sky by the plane of our own Milky Way galaxy.  Diffuse \gammarays\ from our galaxy constitute the single brightest source of celestial gamma radiation.  But even the regions well away from this bright galactic feature would be seen to glow faintly.  Since the 1970's and the results of \sas\ , there has been evidence that the faint \gammaray\ glow at high galactic latitudes is brighter than can be accounted for by the small amount of galactic gas observed looking through the plane of our galaxy.  This glow appears to be isotropic leading to the conclusion that it is likely to be of extra-galactic origin.  It is the nature of this Isotropic Diffuse Gamma-Ray Background (IDGRB) that is the subject of this discussion.  

	  Of course even if our eyes were sensitive to gamma radiation the
spectacular features of the \gammaray\ sky would remain hidden from us because
of the opacity of the earth's atmosphere to \gammarays\ .  The emergence of the
field of \gammaray\ astronomy in the last half century has been driven by the
development of satellite borne instrumentation.  The most sensitive high energy
\gammaray\ telescope yet developed, the Energetic Gamma-ray Experiment Telescope (\egret\ ), was launched in 1991 aboard the Compton Gamma Ray Observatory (CGRO).  This instrument has yielded the best quality data to date on high energy \gammaray\ phenomena.  Traditionally, \gammaray\ astronomy has been limited by poor counting statistics.  Even objects that release comparatively large energies produce relatively few \gammaray\ photons because of the large energy carried by each such event.  \egret 's large collecting area has allowed unprecedented numbers of photons to be collected (over a million photons of celestial origin since launch) which has enabled the investigation of the high energy phenomena of the universe to be conducted with unprecedented power.

	\egret\ is particularly well suited to the study of diffuse radiation.  Its relatively low altitude and low inclination orbit minimize the instrument's exposure to the background cosmic rays and  multiple systems on the instrument reject the cosmic ray induced events with high selectivity.  As will be discussed in Chapter 3, this results in a low instrumental background which is critical to measurements of faint diffuse features.  Furthermore, the existence of a good instrument calibration allows one to do accurate absolute photometry.

\section{Physical Models}

	The details of the measurement of the IDGRB will be discussed in subsequent chapters.  This section is intended to provide the physical motivation for the study of the IDGRB as well as a brief history of the physical models invoked in an attempt to understand this phenomenon.

\begin{figure}[h]
\epsfysize=6.0in
\centerline{\epsfbox{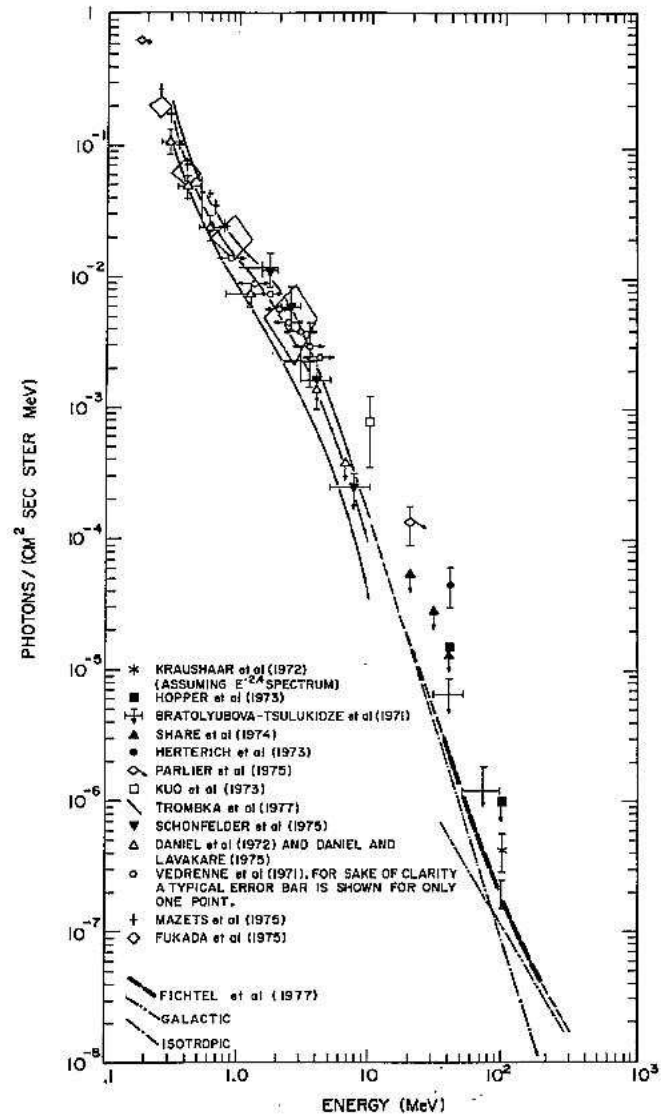}}
\caption{Figure from (Fichtel et al. 1978) showing the \sas\ measurement of the diffuse \gammaray\ background as well as other lower energy balloon measurements.}
\label{Fichtel}
\end{figure}

	\fig{Fichtel} shows a reproduction of the diffuse spectrum reported by Fichtel \etal\ (\cite{Fichtel78}) using \sas\ data. The background was measured to be a power law with a soft spectral index of $2.7\pm0.4$. The integrated intensity above 35 MeV was reported to be  $5.7\pm1.3\times10^{-5}{\rm ph}\;{\rm cm}^{-2}{\rm s}^{-1} {\rm sr}^{-1}$.  Also shown in this figure are balloon measurements of the soft \gammaray\ background.  These early measurements indicated an enhancement above a continuous power law in the energy range from 1-10 MeV (\cite{Trombka77}).  These two measurements motivated much of the early work on the IDGRB.

\subsection{Baryonic Halo Models}

	Interactions between cosmic rays and atomic and molecular gas in the galactic plane are responsible for the copious \gammaray\ production in the plane of our galaxy.  This process will be discussed in detail in Chapter 4.  In addition to the observed matter in our galaxy, there is widely believed to be a dark halo which is invoked in order to explain the measured galactic rotation curves.  If this halo were baryonic in nature it is possible that the same sorts of interactions that are responsible for the diffuse galactic \gammaray\ emission are occurring at large galactocentric radii leading to a comparatively isotropic diffuse source of $\gamma$-radiation. 

	The proposed halo must meet several requirements for this theory to be applicable.  Firstly, the halo must be composed of baryonic matter in order to generate significant targets for the cosmic rays.  Recent measurements by the EROS (\cite{Aubourg93}) and MACHO (\cite{Alcock93}) collaborations have provided evidence in support of this idea.  Monitoring of millions of stars in the LMC has resulted in the detection of a handful of microlensing events indicating halo objects with masses $\simeq 0.1 M_{\odot}$.  Because it is unlikely that there is a mechanism that transforms gas into massive halo objects with perfect efficiency, it is to be expected that some gaseous clouds are also present.  For these clouds to be invisible to terrestrial radio telescopes, they must be composed of molecular hydrogen, ${\rm H}_2$.

	Lastly, this halo must be exposed to a sufficient flux of cosmic rays for this matter to be illuminated.  The cosmic ray flux at large galactocentric distances can be estimated from galactic mass loss (see e.g. \cite{Breitschwerdt91}) but is quite uncertain.

	De Paolis \etal\ (\cite{DePaolis95}) have suggested a scenario in which all the above requirements are met.  In this scenario, the same sorts of proto-galactic clouds which result in globular clusters can result in molecular clouds at larger galactocentric radii.  The factor that determines the fate of such a proto-galactic cloud is the intensity of UV radiation it is exposed to in the early universe.  This emission (due to a population of massive young stars in the galactic center) imprints a characteristic temperature on clouds at distances less than $\sim10$ kpc resulting in their evolution into globular cluster.  Clouds at larger distances cool and collapse simultaneously and consequently are able to fragment into smaller clouds.  These clouds would constitute the raw material from which to form massive halo objects as well as targets for cosmic rays interactions.

	The same authors have predicted the resultant \gammaray\ flux that would emerge from such a scenario.  Assuming that the dark matter halo is entirely baryonic in nature and that the cosmic ray spectrum and composition is similar to that measured locally, they show,
\begin{equation}
I_\gamma \simeq \epsilon \times 1.7\times10^{-6} {\rm ph}\;{\rm cm}^{-2}\;{\rm s}^{-1}\; {\rm sr}^{-1} \; , E> 1 \;{\rm GeV}
\end{equation}
where $\epsilon$ reflects the anisotropy which causes most of the resultant \gammaray\ flux to be directed outward from the galaxy because of the net radial motion of the cosmic rays.  Because this factor is not well known and could easily be smaller than 0.1, these models can be made consistent with the \sas\ measurement.

	The best way to unambiguously test this source of the IDGRB would be to detect an anisotropy resulting from the sun's offset within this halo.  The effect of this offset is to make the emission more intense in directions toward the galactic center.  The \sas\ measurement revealed no such anisotropy and resulted in lower limits on the halo size of $>45$ kpc.  This is somewhat larger than predicted by the above authors but there are large uncertainties associated with measuring the IDGRB flux toward the galactic center because the inverse Compton emission is so pronounced in this direction.

\subsection{Baryon Symmetric Big Bang Theories}

	 Baryon Symmetric Big Bang (BSBB) Cosmologies are cosmological models in which equal amounts of matter and anti-matter are formed in the big bang.   Subsequently, there was a separation of the two forms of matter into zones which then evolved into matter or anti-matter zones that are the size of galactic clusters.  Such cosmologies were first introduced in order to explain how a matter symmetric big bang (which is aesthetically preferable) could result in the dominance of matter of anti-matter in the local universe. While the explanation of this phenomenon now centers on weak CP violation in the early universe, experimental constraints on CP violating interactions are not sufficient to definitively identify them as the cause of the matter asymmetry in the universe.  
	
	BSBB models are summarized in \cite{Stecker78}.  In these models, bubbles of matter and antimatter form in the early universe due to the phase separation between matter and antimatter (\cite{Cisneros73}).  The free energy of the system scales as the ratio of the surface area of such a bubble to the volume and as a result the domains tend to coalesce to form large matter or anti-matter domains (\cite{Omnes72}).  Models to explain galaxy formation in such domains have been proposed (\cite{Stecker72}).

\begin{figure}[h]
\epsfysize=4.0in
\centerline{\epsfbox{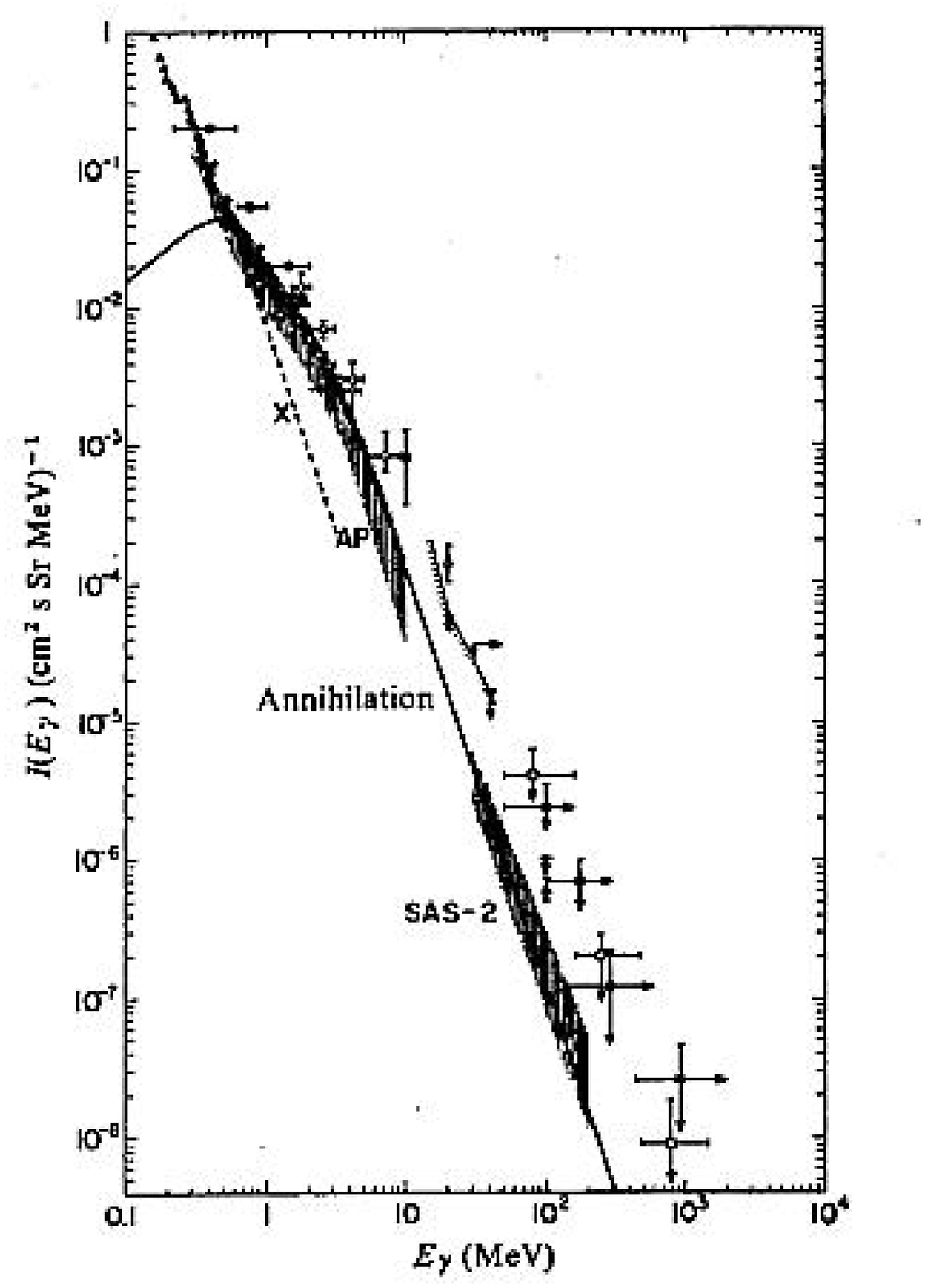}}
\caption{Figure from Stecker (1978) showing the theoretical matter-antimatter annihilation spectrum fit to the IDGRB.  This model reproduces the `MeV bump' as well as the steep spectrum at high energy}
\label{BBSB}
\end{figure}

	The main observational consequence of BSBB models is that the interfaces between matter/antimatter domains would be the site of copious nucleon-antinucleon annihilations.  This would generate a diffuse \gammaray\ flux through the subsequent decay of $\pi^0$'s.  Because these processes are well understood, it is possible to calculate the expected spectrum of such annihilation subject only to cosmological free parameters.  \fig{BBSB} shows a reproduction of a figure from Stecker (\cite{Stecker78}) in which the nucleon-antinucleon annihilation spectrum is fit to the observed IDGRB.  The main features of the IDGRB spectrum are well reproduced by this model.  Both the `MeV-bump' and the steep high energy spectrum match the theoretical spectrum.

	While this early agreement between BSBB models and the IDGRB lends credibility to these theories there are other observational problems.  The most important additional constraint on BSBB theories is the distortion of the microwave background caused by interactions between the supra-thermal electrons that would result from the annihilations and the background photons.  Electron bremsstrahlung would distort the low energy (Rayleigh-Jeans) tail of the microwave background while Thompson scattering would distort the high energy (Wien) tail (\cite{Ramani76}).  Measurements by the Cosmic Background Explorer (COBE) have shown no evidence of such distortion.

	A further challenge to this theory results from big bang nucleosynthesis.  Because the matter/antimatter domains are smaller than the neutron diffusion length at the time of nucleosynthesis, the neutrons are annihilated and cannot participate in nucleosynthesis (\cite{Combes75}).  The He abundance in the universe must instead be generated in galaxies in so-called `little bangs' (\cite{Wagoner74}).  The success of big bang nucleosynthesis at predicting He abundances makes this an unlikely explanation.

	Recent measurements by the COMPTEL instrument aboard \cgro\ (\cite{Kappadath95}) removed one of the strongest pieces of evidence in favor of this theory.  The modulation of the diffuse spectrum with orbital rigidity (see chapter 3) has allowed the unambiguous separation of instrumental and celestial backgrounds for the first time in this energy range.  The most recent reports of the soft \gammaray\ diffuse spectrum suggest that the `MeV-bump' is an instrumental artifact and that the \gammaray\ background in this energy range is instead a power law consistent with the spectrum at higher energies.

\subsection{Exotic Particle Decays}

	The discovery of the IDGRB led in part to interest in the field of particle astrophysics.  Very high energy phenomena which are out of the reach of terrestrial accelerators could be important processes in the early universe.  Observational constraints of the early universe provide the best insight available into physics beyond the standard model.  Many particle theories predict the existence of new massive particles.  The observable end products of the eventual decay of such particles are \gammarays.  These \gammarays\ are produced either directly as products of the particle decay or as the end products of a baryonic shower with neutral pion production and decay.  Attempts have been made to explain the IDGRB using light super-symmetric particle decay (\cite{Ellis84}), and cosmic string interactions (\cite{MacGibbon93}). 

	These theories have the common theme that that some of the IDGRB is due to the decay of Weakly Interacting Massive Particles (WIMPs).  If these WIMPs are long lived compared to the age of the  universe it is possible that they are significant contributors to the dark matter of the universe. The \gammaray\ spectrum expected from a WIMP particle $X$ has been calculated by Kamionkowski (\cite{Kamionkowski95}). The calculation is outlined here.

	If we assume that in the early universe, the particle $X$ was in thermal equilibrium and that $T>>m_{X}$, there would consequently be as many $X$ particles as photons. In this state the $X$ particle is constantly undergoing equilibrated conversions to quark anti-quark pairs or lepton anti lepton pairs: $X\overline{X}\leftrightarrow q\overline{q}$. As the universe cools below $T=m_{X}$ the density of $X$ is exponentially suppressed by the Boltzmann factor.  Consequently, if the universe were to remain in thermal equilibrium as it cools there would be no remaining massive WIMPs.  However, the non-thermal expansion of the universe causes the $X$ to `freeze out'.  This happens when the annihilation rate $\Gamma=\langle\sigma_Av\rangle n_X$ drops below the expansion rate $H$. Kamionkowski shows that the parameter $\langle\sigma_Av\rangle$, the thermally averaged cross-section for annihilation times the relative velocity, determines the relic density that survives the early universe.
\begin{equation}
	\Omega_{X}h^2=(\frac{3\times10^{-27} {\rm cm}^{3}{\rm s}^{-1}}{\langle\sigma_Av\rangle} )\;,
\end{equation}
where $\Omega_X$ is the $X$ density in units of the critical density and $h$ is the Hubble constant in units of $100 \;{\rm km}\;{\rm s}^{-1}{\rm Mpc}^{-1}$.  Note that this result is independent of $m_X$ and that the density increases with decreasing cross section.

	If this relic distribution of WIMPs decays at a characteristic red-shift $z_D$, the resultant \gammaray\ flux can be calculated to be,
\begin{equation}
E\frac{dF}{d\Omega dE}\simeq 5\times10^6 \frac{\Omega_Xh^2}{1+z_D}\frac{1}{E/{\rm MeV}}(\frac{E}{E_0})^{5/2}{\rm exp}\{-(E/E_0)^{3/2}\} {\rm cm}^{-2}{\rm s}^{-1}{\rm sr}^{-1} \;,
\end{equation}
where,
\begin{equation}
E_0=\frac{1}{2}\frac{m_X}{1+z_D}=\frac{m_X}{2}(\frac{\tau}{t_0})^{2/3} \;,
\end{equation}
and $\tau$ is the WIMP lifetime and $t_0$ is the age of the universe.

	If instead the WIMP lifetime $\tau$ is ${\cal{O}}(t_0)$ and the decay happens today,
\begin{equation}
E\frac{dF}{d\Omega dE}\simeq2.4\times10^{4}\frac{\Omega_Xh}{(\tau/t_0)(m_X/GeV)}(\frac{E}{E_0})^{3/2} {\rm cm}^{-2}{\rm s}^{-1}{\rm sr}^{-1} ;,
\end{equation}
for $E<m_X/2$.

\begin{figure}[h]
\epsfysize=4.0in
\centerline{\epsfbox{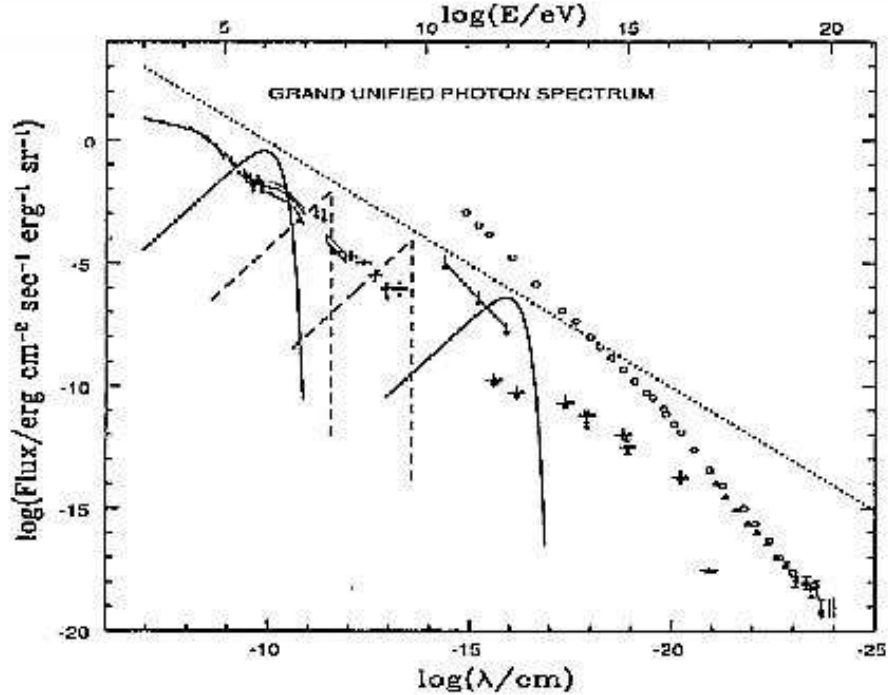}}
\caption{Figure from Kamionkowski (1995) showing the diffuse \gammaray\ background spectrum enveloped by a conservative upper limit (dotted line).  Theoretical calculation of the diffuse \gammaray\ spectrum due to primordial WIMP decays in the early universe (solid lines) and in today's galactic halo (dashed lines) are shown superimposed. }
\label{WIMP}
\end{figure}

	In either case the WIMP annihilation can produce observable levels of background with values of $\Omega_X$ as low as $10^{-7}$.  \fig{WIMP} shows a figure reproduced from Kamionlowski (\cite{Kamionkowski95}) showing these calculated spectra superimposed on the IDGRB spectrum.  While it is evident that a single WIMP line cannot account for the entire IDGRB, it is possible that a spectral `bump' in the IDGRB signals the presence of a WIMP decay.

	The above discussion describes the scenario in which the WIMP particles are unstable on the time scale of the age of the universe.  It is perhaps  more intriguing to consider the case in which the WIMP particles have lifetimes greater than the age of the universe.  In this case WIMP particles could still be abundant today and thus could constitute an important fraction of the dark matter.  These stable WIMPs would cluster along with luminous matter into galaxies.  They are particularly good candidates for massive halos.  Kamionkowski has noted that the expected order of magnitude annihilation cross-section for a particle with weak-scale interactions is, ${\langle\sigma_Av\rangle}\sim\alpha^2/(100 {\rm GeV})^2\sim10^{-25} {\rm cm}^3{\rm s}^{-1}$.  Substitution into equation 1.2 yields the suggestive result that the resulting WIMP density is remarkably close to the critical density.

	While stable WIMPs would not decay into \gammarays\ as described above, there is a finite cross section that couples the WIMPs and anti-WIMPs to ordinary matter.  Thus WIMPs in our galactic halo can annihilate just as they did in the early universe to produce \gammarays.  There are two distinct channels that are of interest.  If the products of the WIMP annihilation are quarks and leptons, the subsequent shower from hadronization of the quarks produce a \gammaray\ spectrum with a broad peak at $\sim 1/10$ of the WIMP rest mass (\cite{Silk84}).  This background could be a significant contribution to the IDGRB even though if the halo is large enough to obscure any quadrupole moment caused by the earth's position within the halo.  

	While by definition there is not direct coupling between WIMPs and \gammarays, it is quite likely that a pair of WIMPs may couple to a pair of photons through Feynmann loop diagrams (\cite{Srednicki86}).  If this is the case, the low mean velocity of a halo WIMP ($\sim 300 {\rm km}\;{\rm s}^{-1}$) leads to the production of essentially mono-energetic \gammarays.  The observation of a \gammaray\ line at several tens of GeV would be a powerful signature of the presence of WIMPs in the galactic halo.

\subsection{Active Galaxies}

	Prior to the launch of \egret, \cosb\ had detected one active galaxy in the Virgo region: 3C 273 (\cite{Bignami81}).  While this detection did not allow much more than speculation as to the mechanisms of particle acceleration in AGN, it was immediately suggested that active galaxies could contribute significantly to the IDGRB (\cite{Schoenfelder78}, \cite{Fichtel81}).  Predictions as to the angular distribution that would result from such a collection of point sources were also put forward (\cite{Gao90}).  As will be discussed at some length in subsequent chapters, the \egret\ detection of a large population of these sources have brought these theories to the forefront.  A large part of this discussion will be devoted to the extraction of the amount of background that can be attributed to these sources.

\section{Summary}

	The relative transparency of the universe to \gammarays\ as well as their ubiquity in high energy processes in the universe give \gammaray\ astronomers a unique insight into otherwise unmeasurable phenomena. The study of the IDGRB which is most likely cosmological in nature, can inform astrophysical models as well as fundamental particle physics theories.  In the following chapters, the \egret\ data will be presented and analyzed to extract the extragalactic signal.  A detailed discussion of the contribution from AGN will be presented and prospects for the future study of this phenomenon will be discussed.

	Throughout this discussion, unless otherwise indicated, ``flux'' will refer to the integrated photon flux from a point source above 100 MeV in units of ${\rm photons}\;{\rm cm}^{-2}\;{\rm s}^{-1}$.  ``Intensity'' will refer to to the integrated diffuse flux above 100 MeV in units of ${\rm photons}\;{\rm cm}^{-2}\;{\rm s}^{-1}\;{\rm sr}^{-1}$.

\chapter{The \egret\ Instrument}
\section{Introduction}
High-energy \gammaray\ observations provide one of the most direct views of
energetic processes occurring in the universe.  Unlike charged particles,
\gammarays\ will pass undeflected through the magnetic fields of interstellar
space,  retaining their directional information.  In addition, the small
interaction cross-section of high-energy \gammarays\ allows them to travel
essentially unattenuated through space.  A high-energy \gammaray\ can pass
through the central plane of the galactic disk with only a one percent chance
of being absorbed.  In contrast, an optical photon can only penetrate about
one-tenth of the distance from the galactic center to the Earth as it travels
through the galactic disk.  Unfortunately, \gammarays\ cannot penetrate very
deeply into the Earth's atmosphere.  Even using high-altitude balloon
observations, the interaction of cosmic rays with the Earth's atmosphere
produces such a large number of background \gammarays\ that only the
strongest point sources can be detected without leaving the Earth's
atmosphere.  Thus, the progress of observational high-energy
\gammaray\ astrophysics was slow until the 1970's when detectors were first
placed on orbiting satellites.

Since the flux of celestial \gammarays\ is relatively low, detectors must
have a large sensitive area, a high detection efficiency, and they must be
able to distinguish a \gammaray\ from charged particle cosmic radiation,
whose flux can be a factor $10^4$ more than that of \gammarays.  In
addition, the detector must measure the incident direction, arrival time, and
energy of the incoming photon.  Using a spark chamber assembly surrounded
by an anti-coincidence dome, \sas\ was able to efficiently reject charged
particle radiation and achieve an angular resolution of $\sim 2\deg$
(\cite{Fichtel75}).  \cosb\ was similar in area to \sas, but it also
incorporated a crystal scintillator to aid in the measurement of the
\gammaray\ energy (\cite{Bignami75}).  The results from these two experiments
moved high-energy \gammaray\ astronomy past the discovery phase and into an
exploration phase.

In the mid 1970's, NASA began the {\em Great Observatories for Space
Astrophysics} program for the express purpose of mapping the electromagnetic
spectrum from infrared to \gammarays\ with unparalleled detail.  As part of
this program, the Compton Gamma Ray Observatory (\cgro) was approved in
1977 to explore the \gammaray\ window from less than 0.1 MeV to more than 10
GeV\@.  Since substantially different detection methods are required to
detect photons in different parts of the \gammaray\ spectrum, four
instruments were selected to be placed on board \cgro.  The Burst and
Transient Source Experiment (\batse) is designed to detect \gammaray\ bursts
over the energy range 25 keV--2 MeV and consists of eight uncollimated
detector modules placed on the eight corners of \cgro, providing nearly
uniform coverage of the sky (\cite{Fishman89}).  The Oriented Scintillation
Spectrometer Experiment (\osse) uses four shielded NaI(Tl)-CsI(Na) phoswich
detectors to study astrophysical sources in the 0.05--10 MeV energy range,
with an effective field-of-view of $3\fdg8 \times 11\fdg4$ full width at
half-maximum (FWHM) (\cite{Johnson93}).  The imaging Compton Telescope
(COMPTEL) explores the 1--30 MeV energy range with a field of view of $\sim$
1~sr, relying on a Compton scattering in one detector array, and a second
interaction in a lower detector array (\cite{Schonfelder93}).  The
\egret\ instrument is a pair-conversion telescope, sensitive to photons in
the energy range 20 MeV to 30 GeV, with a field of view of $\sim$ 1~sr.

\cgro\ was launched aboard the Space Shuttle At\-lan\-tis (STS-37) from the
Ken\-ne\-dy Space Center in Florida on 1991 April 5 and was deployed two days
later.  \egret\ was activated on 1991 April 15 and began taking data on
April 20.  The \egret\ instrument and its calibration have been described
extensively in Hughes \etal\ (1980), Kanbach \etal\ (1988, 1989), Nolan
\etal\ (1992), and Thompson \etal\ (1993).  The \egret\ instrument
has been designed, calibrated, and maintained in a collaborative effort
by scientists from the following institutions:
\begin{itemize}
\item NASA/Goddard Space Flight Center, Greenbelt MD, U.S.A.
\item Stanford University, Stanford CA, U.S.A.
\item Max-Planck Institut f\"ur Extraterrestrische Physik, Garching, FRG.
\item Grumman Aerospace Corporation, Bethpage NY, U.S.A.
\end{itemize}

\section{Design of the Detector}
\label{design}
\begin{figure}[t]
\epsfxsize=\hsize
\centerline{\epsfbox{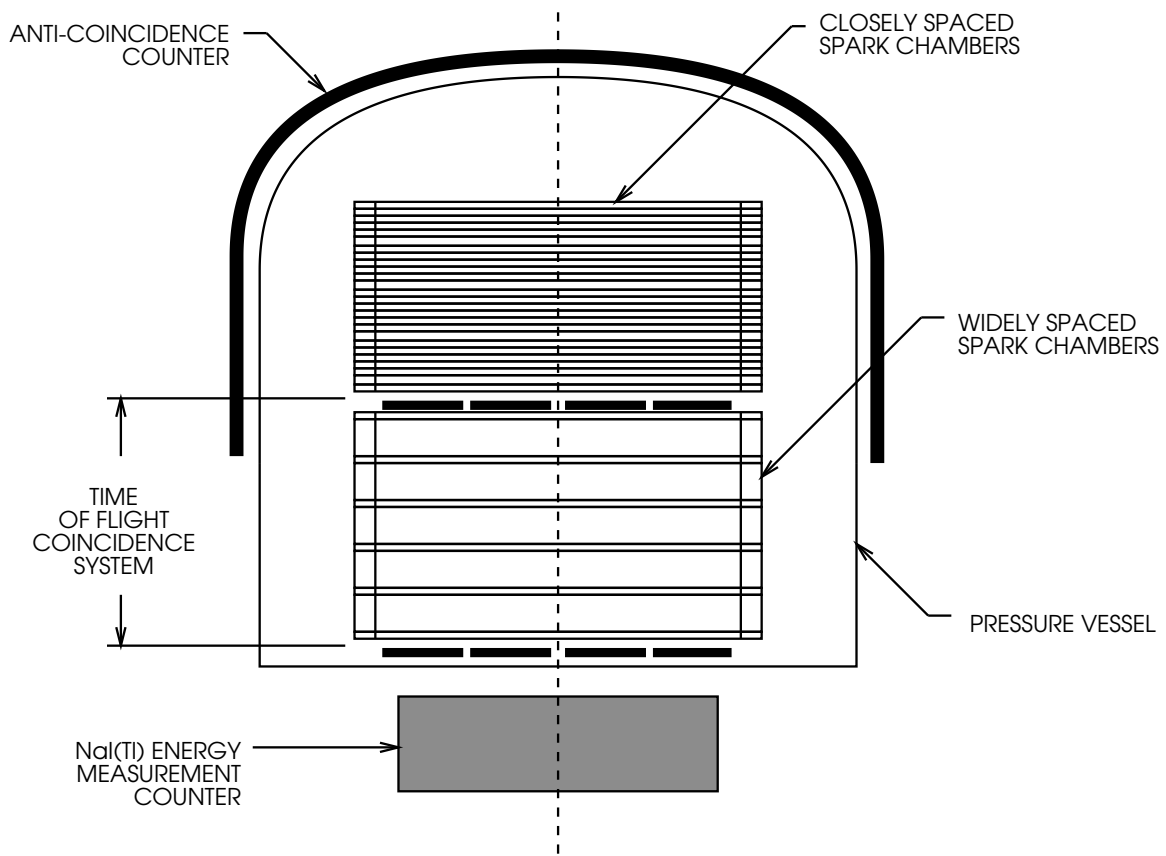}}
\caption[Schematic of the \egret\ Instrument]
{Schematic of the \egret\ instrument.}
\label{egret_schematic}
\end{figure}

\egret\ is multilevel spark chamber triggered by a scintillator coincidence
system, with a large NaI(Tl) crystal spectrometer to measure the energy.  The
instrument is shown schematically in \fig{egret_schematic}.  It is similar in
design to \sas\ and \cosb, but it employs a much larger spark chamber with
improved resolution.  Above the {\em critical energy} of a given material,
generally on the order of a few tens of MeV, pair production becomes the
dominant process by which \gammarays\ interact with matter.  Pair production
is an attractive interaction to observers because the resulting
electron-positron pair retains almost all of the information about the
incident direction and energy of the original photon.  The \egret\ telescope
is designed so that the electron-positron pairs produced by high-energy
\gammarays\ can be clearly separated from other events that might trigger the
detector.

A \gammaray\ which enters the top of the \egret\ instrument will pass
undetected through the large anticoincidence scintillator surrounding the
spark chamber and has a probability $\sim 33\%$ of converting into an
electron-positron pair in one of the thin tantalum (Ta) sheets interleaved
between the 28 closely spaced spark chambers in the upper portion of the
instrument.  The effective critical energy of Ta is 10.7 MeV\@.  The high
atomic number $Z$ of Ta increases the probability of pair production relative
to the Compton process and reduces the energy loss per unit length by the
converted \epm, while the thickness of the sheets is kept to a minimum to
reduce multiple scattering.  If pair production occurs, the secondary charged
particles will ionize the spark chamber gas along their trajectories.

Below the conversion stack are two $4 \times 4$ arrays of plastic
scintillation detector tiles spaced 60 cm apart which register the passage of
charged particles.  The general direction of a charged particle passing
through both scintillator arrays is determined by measuring the
time-of-flight delay with a resolution of $\sim$ 1.5 ns, and by the
combination of individual upper and lower scintillator tiles that detected
the charged particle.  In between the upper and lower arrays of scintillator
tiles are six widely spaced spark chamber modules with very little scattering
material.  These provide additional information on the path of a charged
particle as it travels between the upper and lower scintillator arrays.  If
the time-of-flight delay indicates a downward moving particle which passed
through a valid combination of upper and lower scintillator tiles, and the
anticoincidence system has not been triggered by a charged particle, a high
voltage pulse is applied to orthogonal wire arrays in the spark chamber
modules, and the track information is recorded digitally in ferrite cores,
which are then read out and included in the instrument telemetry.  In this
manner, a three-dimensional picture of the path of the electron-positron pair
is measured, allowing it to be identified with certainty and its basic
properties to be determined.  The high-voltage trigger and path readout will
occur even if only one charged particle is detected by the time-of-flight
coincidence system.

The energy deposition in the NaI(Tl) Total Absorption Shower Counter (TASC)
located directly below the lower array of plastic scintillators is used to
estimate the photon energy.  The TASC has a thickness of about eight
radiation lengths and can determine photon energy from a few tens of MeV up
to several tens of GeV with an energy resolution of $\sim$ 20\% FWHM over most
of that range.  TASC energy measurements must be corrected for the
energy lost by the \epm\ while traveling through the spark chambers and
scintillators.  This correction is estimated from the path length in each
material and is typically in the range 15--30 MeV\@.  Another energy
correction is made to those photons for which one of the electron-positron
pair was not intercepted by the TASC\@.  Once an event has been registered,
it is tagged with an arrival time by the on-board clock which measures
Coordinated Universal Time (UTC) with an absolute accuracy 100 \us, and has a
relative accuracy of 8~\us.  This clock also serves the other three
instruments aboard \cgro\@.  The spark-chamber tracks, energy measurement,
and \gammaray\ arrival time thus comprise the information recorded for a
single event.  There is readout dead time of $\sim$ 100 ms after each
triggered event.

The telescope assembly is surrounded by a thermal blanket, which is necessary
to maintain sufficient temperature control for the anti-coincidence dome.
The thermal blanket also serves to shield the anti-coincidence dome from
micro-meteors which might hit the dome and create a light leak, disabling the
charged particle anti-coincidence system.  Micro-meteors are abundant
in a low Earth orbit, consisting largely of paint flakes from spacecraft.  A
thermal blanket thickness of 0.17 g~cm$^{-2}$ reduces the chance of a
micro-meteor penetration to less than 1\% over the course of the mission and
at the same time does not introduce a significant amount of background
\gammarays\ generated by cosmic-ray interactions in the thermal blanket.

\begin{table}[t]
\centering
\caption[Comparison of High-Energy \Gammaray\ Detectors]
{Comparison of the On-Axis Performance of the High-energy \Gammaray\ 
Detectors}
\bigskip
\begin{tabular}{lccc}\hline\hline
& \sas\ & \cosb\ & \egret\ \\
\hline
Field of View            & 0.25 sr      & 0.25 sr   & 1.0 sr      \\[0.1in]
Effective Area           & 100 cm$^2$   & 70 cm$^2$ & 1200 cm$^2$ \\
~$>$ 100 MeV & & & \\[0.1in]
Angular Resolution       & 1\fdg5       & 1\fdg5    & 0\fdg6      \\
~RMS @ 500 MeV & & & \\[0.1in]
Energy Resolution        & $\sim$ 100\% & 42\%      & 18\%        \\
~FWHM @ 100 MeV & & & \\[0.1in]
Point Source Sensitivity & $10^{-6}$    & $10^{-6}$ & $10^{-7}$   \\
~(photons~cm$^{-2}$~s$^{-1}$) & & & \\
~$>$ 100 MeV & & & \\
~$10^6$~s exposure& & & \\
\hline
\hline
\end{tabular}

\label{detectors}
\end{table}

With its much larger effective area and better angular resolution,
\egret\ has more than an order of magnitude greater sensitivity than
\sas\ and \cosb\@.  Some of the basic properties of the detectors are
summarized in \tbl{detectors}.  The improved angular resolution of
\egret\ comes from the use of thinner conversion plates as well as better
track recognition.  In addition, \sas\ and \cosb\ used less sensitive
\v{C}erenkov detectors to determine the direction of the secondary charged
particles.

\section{Instrument Calibration}
\label{calibration}
Much of the \egret\ calibration was performed in 1986 at the Stanford
Linear Accelerator Center (SLAC).  A beam of electrons with an adjustable
energy between 650 MeV and 30 GeV was back-scattered off pulsed laser
photons to produce a beam of inverse-Compton scattered \gammarays\ from 15
MeV to 10 GeV (\cite{Mattox87}).  This back-scattered beam was collimated
to a cross-section of 1~cm$^2$ at a distance of 170 cm from the interaction
region in order to minimize the angular dispersion.  The beam was then
aimed at the \egret\ instrument to determine the effective sensitive area,
as well as the angular and energy dispersion.

Calibration runs were performed for ten discrete energies (15, 20, 35, 60,
100, 200, 500, 1000, 3000, and 10,000 MeV) with the beam incident at each of
five inclination angles (0\deg, 10\deg, 20\deg, 30\deg, and 40\deg) and three
azimuth angles (0\deg, 22\fdg5, and 45\deg).  Over the course of the
calibration, more than 500,000 events were recorded.  It was found that there
is very little dependence on azimuthal angle.  A smooth instrument response
matrix was formed by applying a two-dimensional fourth-order Chebyshev fit
with respect to energy and second-order with respect to the inclination angle
to the results of the calibration runs.  This fit was stored in three sets of
calibration files, described below, which form the basis for all analysis of
\egret\ data.  Thompson \etal\ (1993) have performed in-flight verification
tests to ensure that the quantities determined during the calibration runs
are still valid.

The sensitive area files contain information on how the instrument effective
area varies with incident energy and inclination angle.  For an on-axis
source, the effective area is greater than 1000 cm$^2$ from 100 MeV out to 3
GeV\@.  Not surprisingly, the effective sensitive area decreases as the
aspect angle increases, such that at an inclination of 30\deg, the sensitive
area is less than 15\% of the on-axis effective area.

The angular dispersion files describe the instrument point spread function
(PSF) at a particular aspect angle and energy.  Although the PSF does not
vary drastically with inclination angle, it is a strong function of photon
energy.  The energy-dependent half-angle \accang\ of a cone containing $\sim
67\%$ of the \gammarays\ from a source is well fit by the function
\begin{equation}
\accang = 5\fdg85 \, (E/{\rm 100~MeV})^{-0.534} \; ,
\label{accang}
\end{equation}
where $E$ is the energy in MeV (\cite{Thompson93}).

The energy dispersion files describe the energy resolution of \egret\@.  For
inclination angles less than 20\deg\ and energies greater than 100 MeV, the
energy resolution is $\sim$ 20\%.  For larger inclination angles or lower
energies, \gammarays\ are much more likely to produce pairs for which one of
the charged particles misses the TASC, leading to a greater degree of
uncertainty associated with the energy estimate.

\section{Instrument Operation and Observations}
\label{corrections}
Upon deployment on 1991 April 7, \cgro\ was put into an nearly circular orbit
of 455 km, with an orbital period of $\sim$ 93 minutes.  The optimal
functioning altitude is between 350 and 450 km, which avoids excessive drag
while remaining below most of particle radiation in the South Atlantic
Anomaly.  Naturally, a gradual deterioration of the orbital radius will
occur, so \cgro\ is equipped with an on-board propulsion system and carries
sufficient fuel to reboost the spacecraft altitude.  By 1993 October,
\cgro\ had declined to a orbital radius of 345 km and the reboost procedure
was initiated.  The reboost was successfully completed in 1993 December,
returning the spacecraft to the original orbital radius of 450 km.  This
procedure can be repeated as necessary in the future.

Due to the low flux of high-energy \gammarays, a typical \egret\ observation
will last from one to three weeks.  During this time, the pointing axis of
the instrument is stable to within 0\fdg5.  The \cgro\ mission has been
divided into phases which focus on different mission objectives.  The
observation dates, pointing directions, and numerical designations for the
first three phases of \cgro\ operation are listed in Tables~\ref{phaseI}--c.
The first month of the mission was dedicated to instrument testing, including
pointings to well-studied objects such as Crab and Vela.  Six observations
from this test phase are of sufficient quality that they are now grouped with
the Phase I observations.  The initial fifteen months of observations that
comprise Phase I were devoted to carrying out the first complete survey of
the \gammaray\ sky (\cite{Fichtel94}).  Phase II lasted from 1992 November
until 1993 September, revisiting some of the more interesting regions found
in Phase I, as well as observing some targets selected through the Compton
Guest Investigator Program (\cite{Bunner89}).  During Phase III, which began
on 1993 August 17 and ended on 1994 October 4, more than 50\% of the
observing time was allocated to Guest Investigators.  Due to scheduling
constraints, the first observation in Phase III occurred between the final
two observations of Phase II, explaining the gap between the initial
observations of Phase III.  A record of all of the \egret\ observations, as
well as periods when the instrument was inactive, is kept in the instrument
{\em timeline} file.

\clearpage
\setcounter{subtable}{1}
\renewcommand{\thetable}{\arabic{chapter}.\arabic{table}\alph{subtable}}
\begin{table}[p]
\scriptsize
\centering
\caption{\cgro\ Phase I Observations}
\bigskip
\begin{tabular}{crrrrrr}
\hline\hline
Viewing &   &   & \multicolumn{2}{c}{Celestial (J2000)} & \multicolumn{2}{c}{Galactic} \\
Period  & \multicolumn{1}{c}{Start Date} & \multicolumn{1}{c}{End Date} & 
\multicolumn{1}{c}{RA} & \multicolumn{1}{c}{DEC} & 
\multicolumn{1}{c}{l$^{\rm II}$} & \multicolumn{1}{c}{b$^{\rm II}$} \\
\hline
0002 & 1991 Apr 22 21:09:02 & 1991 Apr 28 15:12:00 &  86\fdg76 &  22\fdg09 & 186\fdg02 &  -3\fdg28 \\
0003 &      Apr 28 16:02:00 &      May 01 16:37:00 &  89.80 &  15.25 & 193.39 &  -4.25 \\
0004 &      May 01 17:19:00 &      May 04 16:16:00 &  89.77 &  15.24 & 193.39 &  -4.28 \\
0005 &      May 04 16:50:00 &      May 07 15:53:00 &  83.52 &  22.02 & 184.50 &  -5.87 \\
0006 &      May 07 16:32:00 &      May 10 19:40:00 & 162.44 &  57.26 & 150.00 &  53.00 \\
0007 &      May 10 20:15:00 &      May 16 16:39:00 & 135.19 & -45.11 & 266.32 &   0.74 \\[0.1in]
0010 &      May 16 17:19:00 &      May 30 18:51:00 &  88.07 &  17.14 & 190.92 &  -4.74 \\
0020 &      May 30 20:01:00 &      Jun 08 00:08:30 & 301.39 &  36.58 &  73.28 &   2.56 \\
0021 &      Jun 08 01:24:30 &      Jun 15 18:44:00 &  87.83 &  12.47 & 194.86 &  -7.29 \\
0030 &      Jun 15 19:38:00 &      Jun 28 19:30:57 & 191.54 &   2.62 & 299.76 &  65.46 \\
0040 &      Jun 28 20:14:00 &      Jul 12 17:56:16 & 179.84 &  41.52 & 156.18 &  72.08 \\
0050 &      Jul 12 18:48:32 &      Jul 26 19:25:00 & 270.39 & -30.96 &   0.00 &  -4.00 \\
0060 &      Jul 26 20:25:00 &      Aug 08 15:36:00 &  91.28 & -67.96 & 277.99 & -29.32 \\
0071 &      Aug 08 17:00:00 &      Aug 15 17:29:06 & 310.05 &  28.06 &  70.44 &  -8.30 \\
0072 &      Aug 15 18:23:15 &      Aug 22 14:05:00 & 291.98 & -13.27 &  25.00 & -14.00 \\
0080 &      Aug 22 15:01:00 &      Sep 05 14:01:00 & 124.96 & -46.35 & 262.94 &  -5.67 \\
0091 &      Sep 05 15:03:00 &      Sep 12 13:24:10 &   8.34 & -32.31 & 338.98 & -83.50 \\
0092 &      Sep 12 14:34:05 &      Sep 19 13:29:00 & 251.27 &  36.89 &  59.67 &  40.28 \\
0100 &      Sep 19 14:36:00 &      Oct 03 13:11:00 &  30.91 & -60.66 & 287.85 & -54.30 \\
0110 &      Oct 03 14:10:00 &      Oct 17 13:57:40 & 189.02 &   1.06 & 294.25 &  63.67 \\
0120 &      Oct 17 15:15:00 &      Oct 31 14:55:06 & 202.29 & -40.09 & 310.71 &  22.21 \\
0131 &      Oct 31 15:42:00 &      Nov 07 14:34:50 & 291.98 & -13.27 &  25.00 & -14.00 \\
0132 &      Nov 07 15:30:00 &      Nov 14 15:45:50 &   8.34 & -32.31 & 338.98 & -83.50 \\
0140 &      Nov 14 16:50:00 &      Nov 28 11:30:00 & 156.83 & -58.51 & 285.04 &  -0.74 \\
0150 &      Nov 28 12:50:00 &      Dec 12 16:42:00 &  52.00 &  40.24 & 152.63 & -13.44 \\
0160 &      Dec 12 18:00:00 &      Dec 27 16:02:00 & 248.35 & -17.20 & 360.00 &  20.29 \\
0170 &      Dec 27 17:07:00 & 1992 Jan 10 16:15:55 &  83.48 & -72.26 & 283.20 & -31.62 \\
0180 & 1992 Jan 10 18:12:26 &      Jan 23 13:42:00 & 154.60 &  72.04 & 137.47 &  40.49 \\
0190 &      Jan 23 15:08:00 &      Feb 06 15:15:00 & 331.40 &  -1.93 &  58.14 & -43.00 \\
0200 &      Feb 06 16:45:00 &      Feb 20 15:03:00 & 285.28 &   6.37 &  39.70 &   0.76 \\
0210 &      Feb 20 16:05:00 &      Mar 05 15:49:00 &  39.09 &  -1.24 & 171.52 & -53.90 \\
0220 &      Mar 05 16:45:00 &      Mar 19 13:19:39 & 216.00 &  70.74 & 112.47 &  44.46 \\
0230 &      Mar 19 14:15:17 &      Apr 02 12:49:00 & 227.43 & -54.62 & 322.14 &   3.01 \\
0240 &      Apr 02 14:07:00 &      Apr 09 13:02:00 & 223.34 &  11.03 &   9.53 &  57.15 \\
0245 &      Apr 09 13:26:00 &      Apr 16 12:29:28 & 223.34 &  11.03 &   9.53 &  57.15 \\
0250 &      Apr 16 13:17:40 &      Apr 23 12:27:01 & 229.85 &   4.47 &   6.84 &  48.09 \\
0260 &      Apr 23 13:29:22 &      Apr 28 12:44:44 &   1.59 &  20.20 & 108.77 & -41.43 \\
0270 &      Apr 28 13:41:33 &      May 07 14:08:37 & 241.11 & -49.05 & 332.24 &   2.52 \\
0280 &      May 07 14:47:58 &      May 14 14:04:25 &   1.59 &  20.20 & 108.77 & -41.43 \\
0290 &      May 14 14:48:40 &      Jun 04 13:44:28 &  68.97 & -25.09 & 224.01 & -40.00 \\
0300 &      Jun 04 14:50:41 &      Jun 11 14:04:25 & 149.50 & -14.73 & 252.41 &  30.66 \\
0310 &      Jun 11 16:25:00 &      Jun 25 13:17:17 &  88.87 &  49.44 & 163.09 &  11.92 \\
0320 &      Jun 25 14:20:00 &      Jul 02 13:57:08 & 171.17 & -36.81 & 284.20 &  22.89 \\
0330 &      Jul 02 15:06:00 &      Jul 16 15:34:57 & 149.50 & -14.73 & 252.41 &  30.66 \\
0340 &      Jul 16 17:25:00 &      Aug 06 14:38:45 & 345.77 &  57.49 & 108.75 &  -2.37 \\
0350 &      Aug 06 16:35:00 &      Aug 11 00:54:45 & 287.12 & -61.21 & 335.10 & -25.56 \\
0360 &      Aug 11 02:00:00 &      Aug 12 18:22:11 &  68.98 &  30.42 & 169.84 & -11.36 \\
0365 &      Aug 12 19:20:00 &      Aug 20 15:28:00 &  69.39 &  32.90 & 168.17 &  -9.46 \\
0370 &      Aug 20 16:02:00 &      Aug 27 16:49:30 & 358.75 &  18.82 & 104.83 & -42.06 \\
0380 &      Aug 27 18:45:00 &      Sep 01 04:37:00 & 287.12 & -61.21 & 335.10 & -25.56 \\
0390 &      Sep 01 06:18:00 &      Sep 17 15:17:16 &  68.87 &  33.82 & 167.18 &  -9.18 \\
0400 &      Sep 17 16:15:00 &      Oct 08 13:48:00 & 140.88 &  30.40 & 195.90 &  44.71 \\
0410 &      Oct 08 15:11:00 &      Oct 15 16:18:00 & 112.43 & -12.05 & 228.02 &   2.84 \\
0420 &      Oct 15 17:10:00 &      Oct 29 13:31:05 & 319.72 & -41.67 & 359.98 & -44.58 \\
0430 &      Oct 29 15:30:00 &      Nov 03 13:07:47 & 307.83 & -13.95 &  31.13 & -28.33 \\
0440 &      Nov 03 15:10:00 &      Nov 17 14:57:43 & 112.43 & -12.05 & 228.02 &   2.84 \\
\hline
\hline
\end{tabular}

\label{phaseI}
\end{table}
\clearpage
\addtocounter{subtable}{1}
\addtocounter{table}{-1}
\begin{table}[p]
\scriptsize
\centering
\caption{\cgro\ Phase II Observations}
\bigskip
\begin{tabular}{crrrrrr}
\hline\hline
Viewing &   &   & \multicolumn{2}{c}{Celestial (J2000)} & \multicolumn{2}{c}{Galactic} \\
Period  & \multicolumn{1}{c}{Start Date} & \multicolumn{1}{c}{End Date} & 
\multicolumn{1}{c}{RA} & \multicolumn{1}{c}{DEC} & 
\multicolumn{1}{c}{l$^{\rm II}$} & \multicolumn{1}{c}{b$^{\rm II}$} \\
\hline
2010 & 1992 Nov 17 17:10:00 & 1992 Nov 24 16:23:50 & 253\fdg15 &  42\fdg26 &  66\fdg79 &  39\fdg28 \\
2020 &      Nov 24 17:50:00 &      Dec 01 15:29:47 & 251.55 &  45.40 &  70.85 &  40.50 \\
2030 &      Dec 01 17:04:00 &      Dec 22 13:58:00 & 306.59 &  39.34 &  77.85 &   0.69 \\
2040 &      Dec 22 14:44:00 &      Dec 29 15:32:56 & 188.99 &  -0.74 & 294.70 &  61.88 \\
2050 &      Dec 29 16:12:00 & 1993 Jan 05 14:22:45 & 188.84 &  -1.03 & 294.46 &  61.58 \\
2060 & 1993 Jan 05 15:01:26 &      Jan 12 14:54:30 & 188.99 &  -0.74 & 294.70 &  61.88 \\
2070 &      Jan 12 15:42:00 &      Feb 02 14:33:36 & 203.86 & -30.41 & 314.06 &  31.51 \\
2080 &      Feb 02 15:10:00 &      Feb 09 16:12:40 & 198.47 & -41.93 & 307.39 &  20.75 \\
2090 &      Feb 09 18:15:00 &      Feb 22 16:06:50 & 305.69 & -40.81 &   0.24 & -34.01 \\
2100 &      Feb 22 17:15:00 &      Feb 25 14:15:28 & 257.65 & -29.10 & 355.62 &   6.28 \\
2110 &      Feb 25 17:03:00 &      Mar 09 15:08:40 &  18.38 &  58.05 & 125.86 &  -4.70 \\
2120 &      Mar 09 15:56:00 &      Mar 23 16:40:35 & 297.88 &  50.15 &  83.74 &  11.67 \\
2130 &      Mar 23 18:40:00 &      Mar 29 13:02:41 &  80.30 &  22.29 & 182.63 &  -8.22 \\
2140 &      Mar 29 14:57:00 &      Apr 01 15:50:27 & 257.65 & -29.10 & 355.62 &   6.28 \\
2150 &      Apr 01 16:51:13 &      Apr 06 19:27:00 & 203.26 & -39.28 & 311.66 &  22.89 \\
2160 &      Apr 06 20:50:00 &      Apr 12 12:43:27 & 143.66 &  71.46 & 140.75 &  38.11 \\
2170 &      Apr 12 13:50:10 &      Apr 20 14:18:00 & 203.26 & -39.28 & 311.66 &  22.89 \\
2180 &      Apr 20 15:23:01 &      May 05 13:55:45 & 180.75 &  43.09 & 151.41 &  71.26 \\
2190 &      May 05 16:00:00 &      May 08 13:24:04 & 245.35 & -27.22 & 350.10 &  15.86 \\
2200 &      May 08 14:55:00 &      May 13 15:27:36 &  24.11 & -72.07 & 298.09 & -44.63 \\
2210 &      May 13 16:29:00 &      May 24 15:17:05 &  85.21 &  19.46 & 187.52 &  -5.88 \\
2220 &      May 24 17:10:00 &      May 31 13:33:33 & 178.02 &  42.25 & 157.79 &  70.63 \\
2230 &      May 31 14:35:00 &      Jun 03 14:41:08 & 265.98 & -29.72 & 359.14 &  -0.09 \\
2240 &      Jun 03 16:30:00 &      Jun 19 13:03:08 &  24.11 & -72.07 & 298.09 & -44.63 \\
2260 &      Jun 19 14:35:00 &      Jun 29 14:04:56 & 258.44 & -30.35 & 355.00 &   5.00 \\
2270 &      Jun 29 15:00:00 &      Jul 13 13:27:50 & 143.64 &  65.00 & 148.11 &  41.22 \\
2280 &      Jul 13 14:07:00 &      Jul 27 12:43:14 & 145.33 &  63.18 & 149.86 &  42.69 \\
2300 &      Jul 27 14:03:00 &      Jul 30 13:46:10 & 143.03 & -54.64 & 276.66 &  -2.27 \\
2305 &      Jul 30 14:23:00 &      Aug 03 13:45:08 & 149.86 & -53.17 & 278.79 &   1.44 \\
2310 &      Aug 03 20:39:00 &      Aug 10 13:56:02 & 289.94 & -15.33 &  22.22 & -13.08 \\
2290 &      Aug 10 14:52:00 &      Aug 11 16:41:24 & 264.60 & -22.06 &   5.00 &   5.00 \\
2295 &      Aug 12 14:52:00 &      Aug 17 14:16:39 & 264.60 & -22.06 &   5.00 &   5.00 \\
2320 &      Aug 24 14:52:00 &      Sep 07 14:09:19 & 258.02 & -39.35 & 347.50 &  -0.00 \\
\hline
\hline
\end{tabular}

\label{phaseII}
\end{table}
\clearpage
\addtocounter{subtable}{1}
\addtocounter{table}{-1}
\begin{table}[p]
\scriptsize
\centering
\caption{\cgro\ Phase III Observations}
\bigskip
\begin{tabular}{crrrrrr}
\hline\hline
Viewing &   &   & \multicolumn{2}{c}{Celestial (J2000)} & \multicolumn{2}{c}{Galactic} \\
Period  & \multicolumn{1}{c}{Start Date} & \multicolumn{1}{c}{End Date} & 
\multicolumn{1}{c}{RA} & \multicolumn{1}{c}{DEC} & 
\multicolumn{1}{c}{l$^{\rm II}$} & \multicolumn{1}{c}{b$^{\rm II}$} \\
\hline
3010 & 1993 Aug 17 23:04:00 & 1993 Aug 24 13:53:10 & 128\fdg92 & -45\fdg18 & 263\fdg59 &  -2\fdg74 \\
3020 &      Sep 07 19:57:00 &      Sep 09 13:22:00 & 307.63 &  52.63 &  89.13 &   7.82 \\
3023 &      Sep 09 14:19:00 &      Sep 21 14:23:30 & 258.63 & -22.70 &   1.40 &   9.26 \\
3030 &      Sep 21 15:22:00 &      Sep 22 14:44:07 & 157.58 & -42.88 & 277.21 &  12.83 \\
3032 &      Sep 22 16:06:43 &      Oct 01 15:00:00 & 307.63 &  52.63 &  89.13 &   7.82 \\
3034 &      Oct 01 16:00:00 &      Oct 04 14:00:00 & 270.79 &  37.87 &  64.32 &  25.27 \\
3037 &      Oct 17 22:34:33 &      Oct 19 15:11:58 & 307.63 &  52.63 &  89.13 &   7.82 \\
3040 &      Oct 19 16:19:29 &      Oct 25 14:31:00 & 183.29 &   5.68 & 278.21 &  66.70 \\
3050 &      Oct 25 15:40:00 &      Nov 02 14:56:25 & 181.58 &   2.06 & 277.71 &  62.70 \\
3060 &      Nov 02 15:32:42 &      Nov 09 13:09:09 & 180.01 &  -1.62 & 277.60 &  58.70 \\
3070 &      Nov 09 13:47:00 &      Nov 16 14:25:09 & 181.19 &   9.53 & 268.69 &  69.24 \\
3080 &      Nov 16 15:11:00 &      Nov 19 10:01:22 & 187.61 &  12.59 & 283.22 &  74.65 \\
3086 &      Nov 23 22:10:00 &      Dec 01 14:42:22 & 187.61 &  12.59 & 283.22 &  74.65 \\
3100 &      Dec 01 15:42:00 &      Dec 13 15:14:42 &  98.48 &  17.77 & 195.14 &   4.27 \\
3110 &      Dec 13 16:10:00 &      Dec 15 11:04:24 & 187.69 &  12.41 & 283.70 &  74.50 \\
3116 &      Dec 17 23:07:30 &      Dec 20 13:32:53 & 187.69 &  12.41 & 283.70 &  74.50 \\
3120 &      Dec 20 14:15:15 &      Dec 27 15:26:02 & 185.52 &   9.12 & 280.50 &  70.70 \\
3130 &      Dec 27 16:06:05 & 1994 Jan 03 15:34:21 & 190.10 &  16.12 & 289.28 &  78.70 \\
3140 & 1994 Jan 03 16:32:30 &      Jan 16 15:17:40 & 195.69 & -63.83 & 304.18 &  -0.99 \\
3150 &      Jan 16 15:55:00 &      Jan 23 15:30:53 & 195.69 & -63.83 & 304.18 &  -0.99 \\
3160 &      Jan 23 16:16:30 &      Feb 01 14:32:00 & 201.37 & -43.02 & 309.52 &  19.42 \\
3181 &      Feb 01 15:30:00 &      Feb 08 14:46:21 & 301.26 &  30.92 &  68.44 &  -0.38 \\
3211 &      Feb 08 15:39:00 &      Feb 15 14:47:10 &  84.73 &  26.32 & 181.44 &  -2.64 \\
3215 &      Feb 15 15:22:00 &      Feb 17 15:16:23 &  84.73 &  26.32 & 181.44 &  -2.64 \\
3170 &      Feb 17 16:00:00 &      Mar 01 13:32:49 &  37.41 &  10.61 & 158.48 & -45.38 \\
3190 &      Mar 01 14:30:00 &      Mar 08 15:28:23 & 110.48 &  71.34 & 143.99 &  28.02 \\
3200 &      Mar 08 16:40:00 &      Mar 15 14:02:24 & 345.81 &   8.87 &  83.09 & -45.47 \\
3195 &      Mar 15 15:15:00 &      Mar 22 14:04:40 & 105.20 &  68.99 & 146.43 &  26.02 \\
3230 &      Mar 22 15:20:00 &      Apr 05 14:24:21 & 276.44 & -37.11 & 356.84 & -11.29 \\
3220 &      Apr 05 15:35:00 &      Apr 19 14:49:25 & 157.07 &  31.09 & 197.01 &  58.62 \\
3240 &      Apr 19 15:50:00 &      Apr 26 13:16:18 & 269.39 & -13.15 &  15.03 &   5.63 \\
3250 &      Apr 26 14:22:00 &      May 10 14:59:32 &  49.07 &  46.94 & 147.04 &  -9.04 \\
3260 &      May 10 15:57:00 &      May 17 13:39:10 & 156.69 &  31.65 & 195.93 &  58.31 \\
3270 &      May 17 14:55:10 &      May 24 13:44:39 & 348.06 &   5.44 &  82.86 & -49.56 \\
3280 &      May 24 14:33:00 &      May 31 13:55:11 & 298.76 &  28.07 &  64.87 &  -0.03 \\
3290 &      May 31 14:55:00 &      Jun 07 14:01:50 &  69.26 & -47.25 & 253.39 & -42.00 \\
3310 &      Jun 07 15:15:00 &      Jun 10 13:33:36 & 298.76 &  28.07 &  64.87 &  -0.03 \\
3300 &      Jun 10 14:10:00 &      Jun 14 14:02:25 & 275.93 & -13.26 &  18.00 &  -0.00 \\
3315 &      Jun 14 14:48:00 &      Jun 18 14:31:14 & 298.76 &  28.07 &  64.87 &  -0.03 \\
3320 &      Jun 18 15:18:00 &      Jul 05 14:28:24 & 275.93 & -13.26 &  18.00 &  -0.00 \\
3330 &      Jul 05 14:49:42 &      Jul 12 14:31:00 & 298.76 &  28.07 &  64.87 &  -0.03 \\
3350 &      Jul 12 15:26:00 &      Jul 18 13:43:29 &  69.26 & -47.25 & 253.39 & -42.00 \\
3340 &      Jul 18 15:07:00 &      Jul 25 13:47:00 & 279.49 & -25.06 &   9.00 &  -8.38 \\
3355 &      Jul 25 14:45:00 &      Aug 01 13:15:01 &  69.26 & -47.25 & 253.39 & -42.00 \\
3360 &      Aug 01 14:58:00 &      Aug 04 13:22:00 & 349.67 &   9.72 &  88.37 & -46.83 \\
3365 &      Aug 04 14:25:15 &      Aug 09 21:12:27 & 249.14 & -43.04 & 340.43 &   2.86 \\
3370 &      Aug 09 22:22:00 &      Aug 29 14:15:36 &  87.70 &   0.98 & 205.00 & -13.00 \\
3385 &      Aug 31 15:24:15 &      Sep 20 13:47:59 & 128.92 & -45.18 & 263.59 &  -2.74 \\
3390 &      Sep 20 14:49:00 &      Oct 04 12:22:17 & 234.71 &  -1.82 &   4.06 &  40.40 \\
\hline
\hline
\end{tabular}

\label{phaseIII}
\end{table}
\clearpage
\renewcommand{\thetable}{\arabic{chapter}.\arabic{table}}

In order to minimize the amount of Earth albedo photons which might trigger
the detector, \egret\ has specific modes of operation to limit its acceptance
angles as the Earth enters its field of view.  Each mode has a
corresponding set of calibration files which are used to calculate instrument
response for that particular mode.  For each observation, an Exposure History
(EXPHST) file is created to keep track of the times when the \egret\
instrument changes mode and how much livetime is spent in that mode.

In the initial stages of operation, the data from all four instruments was
recorded on-board the spacecraft and then telemetered to the Earth via the
Tracking and Data Relay Satellite (TDRS) once every other orbit.
Unfortunately, the tape recorders failed in 1992 March and all subsequent
data had to be telemetered in real time at 32 kilobits per second during
times of TDRS contact, reducing the \egret\ data coverage to $\sim$ 60\%.
With completion of an
Australian TDRS ground station---named the GRO Remote Terminal System
(GRTS)--- starting in 1994 January, the
\egret\ real time data coverage increased to more than 80\%.

The gas in the \egret\ spark chamber modules gradually deteriorates with
time, causing some degradation in the instrument sensitivity.  Additionally,
lower energy photons have a reduced probability of being detected as the
instrument ages, producing a drift in the instrument spectral response.  This
causes sources to be measured with harder spectra over time.  These effects are corrected for when calculating the
instrument response functions by applying energy-dependent scaling factors
determined from the long-term monitoring of the photon detection rate in five
energy ranges: $>$ 100, 30--100, 100--300, 300--1000, and $>$ 1000 MeV\@.
When the instrument performance degrades below a designated level, the gas in
the spark chamber is replaced.  An on-board gas replenishment system is
capable of refilling the spark chamber volume five times over the lifetime of
the instrument.  Gas refills have occurred on 1991 December 2--3, 1992
December 3--4, 1994 February 8--9, and 1994 November 1--2.

Flight data have indicated that below 70 MeV, the instrument response differs
slightly from the behavior determined by the pre-flight calibrations, which
were known to have limitations at lower energies.  In particular, two effects
were noted by Thompson \etal\ (1993):  (1) the back-scattered calibration
beam, assumed to consist of \gammarays\ produced by inverse-Compton
scattering, had a bremsstrahlung component extending out to calibration
energies; and (2) the calibration measurements were smoothed over a broad
energy range.  In turn, the effective \egret\ sensitive area at lower
energies was over-estimated.  Based primarily on the assumption that the Crab
pulsed photon spectrum has an essentially continuous power-law form over many
decades of energy (\cite{Strong93}), it was
determined that the effective area between 30--50 MeV was too high by a
factor of $2.7 \pm 0.7$, and the 50--70 MeV effective area was overestimated
by a factor of $1.4 \pm 0.2$.  These factors have been incorporated into the
calculation of the instrument response functions.

Secondary pairs formed in the spark chamber are more likely to be detected
and accepted for analysis if their paths are closer to the instrument axis as
opposed to those with larger inclination angles.  This introduces a small
distortion in the derived source locations, systematically shifting the
determined position toward the \egret\ pointing axis.  Flight data has shown
this effect is only consequential for aspect angles greater than 20\deg\ from
the instrument pointing axis (\cite{Thompson93}).  Based on an empirical fit
to the deviations as a function of aspect angle, the data have been corrected
to account for this so-called {\em fisheye effect}.

\section{General Data Processing}

\subsection{Standard Data Files}
The events that are determined to be \gammarays\ are stored in a Primary
Database file corresponding to that observation.  These files contain
all of the measured and derived quantities for each event.  Since these files
are too large to be manipulated easily, pertinent information such as photon
energy, time of arrival, and direction of incidence are stored in Summary
Database (SMDB) files.  The SMDB files are distributed to the participating
institutions and Guest Investigators for analysis.

To facilitate the spatial analysis of the data, the photons in the SMDB files
are binned in $0\fdg5 \times 0\fdg5$ pixels for the following broad energy
ranges: 30--100, 100--300, 300--1000, \gt~100, \gt~300, and \gt~1000 MeV\@.
The photons are also binned for the following ten {\em standard} energy
ranges:  30--50 MeV, 50--70 MeV, 70--100 MeV, 100--150 MeV, 150--300 MeV,
300--500 MeV, 500--1000 MeV, 1--2 GeV, 2--4 GeV, and 4--10 GeV\@.  These maps
are stored in standard FITS (Flexible Image and Transport System) format as
described by Wells, Greisen, \& Harten (1981).  The pixel sizes are chosen to
obtain a reasonable count rate in each bin while avoiding excessive loss of
spatial information.

For each of these photon maps, an analogous $0\fdg5 \times 0\fdg5$ map of
\egret\ exposure is created.  If a detector has exposure \expsr\ to a source
with photon flux $F$, then the number of counts $N$ which will be measured is 
\begin{equation}
N = F \expsr \; ,
\label{countsflux}
\end{equation}
with $F$ in units of photons per unit area per unit time.  In practice, a
detector will only detect photons over a certain energy range, so the flux $F$
in \eq{countsflux} should be defined as
\begin{equation}
F(\Delta E) = \int_{\Delta E} I(E) dE \; ,
\label{flux}
\end{equation}
where $\Delta E$ is the energy range being considered and $I(E)$ is the
differential flux as a function of energy in units of photons per unit area
per unit time per unit energy.  The differential number of counts that will
be detected by \egret\ from a source of intensity $I(E)$ is
\begin{equation}
dN = I(E) \, A(E) \, dE \, dt \; ,
\end{equation}
where $A(E)$ is the energy-dependent effective area of the instrument, and
$dt$ is the differential unit of time.  However, due to energy dispersion,
these counts will not necessarily be measured with their {\em true} energy
$E$.  Taking into account the energy dispersion of the instrument, the
correct expression for the number of counts that will be measured in the
energy range $\Delta E$ from a source of differential flux $I(E)$ is
\begin{equation}
N(\Delta E; \theta, \phi, m) = T(\theta, \phi, m) \, \int_{\Delta E} dE' 
\int_0^\infty dE \; I(E) \, A(E; \theta, \phi, m) \, 
R(E \rightarrow E'; \theta, \phi, m) \; ,
\label{counts}
\end{equation}
where $T(\theta, \phi, m)$ is the amount of instrument livetime spent
observing a source at an aspect angle $\theta$ and azimuth $\phi$ while in
viewing mode $m$; $A(E; \theta, \phi, m)$ is the corresponding effective area
at energy $E$; and $R(E \rightarrow E'; \theta, \phi, m)$ is the probability
per unit energy that a photon of true energy $E$ will be measured with an
energy $E'$.  Substituting \eq{flux} and \eq{counts} into
\eq{countsflux} and solving for exposure shows
\begin{equation}
\expsr(\Delta E; \theta, \phi) = \sum_m T(\theta, \phi, m) \, 
\bar A(\Delta E; \theta, \phi, m) \; ,
\label{exposure}
\end{equation}
with
\begin{equation}
\bar{A}(\Delta E; \theta, \phi, m) = \frac{\int_{\Delta E} dE' \int_0^\infty dE
\; I(E) \, A(E; \theta, \phi, m) \, R(E \rightarrow E'; \theta, \phi, m)}
{\int_{\Delta E} dE \; I(E)} \;.
\label{area}
\end{equation}
Here, $\bar{A}(\Delta E; \theta, \phi, m)$ is the average effective area of
\egret\ over energy range $\Delta E$ at an inclination angle $\theta$,
azimuth $\phi$, and viewing mode $m$.  Typically, it is assumed that the
source photon spectrum behaves as a power law, such that
\begin{equation}
I(E) = I_0 \, E^\alpha {\rm ~photons~cm}^{-2}{\rm ~s}^{-1}{\rm ~MeV}^{-1} \; ,
\label{intensity}
\end{equation}
where $\alpha$ is the spectral index, which is typically $\sim -2.0$.

The average exposure in each $0\fdg5 \times 0\fdg5$ bin is evaluated using
\eq{exposure}.  The total observing time spent in each mode $T(\theta, \phi,
m)$ is found by reading the EXPHST file and determining when a source at
$\theta$ and $\phi$ is occulted by the Earth.  The effective area of
\eq{area} is calculated via interpolation of the results in the calibration
files (\cite{Fierro94}).

\begin{figure}[p]
\epsfxsize=0pt \epsfysize=2in
\centerline{\epsfbox{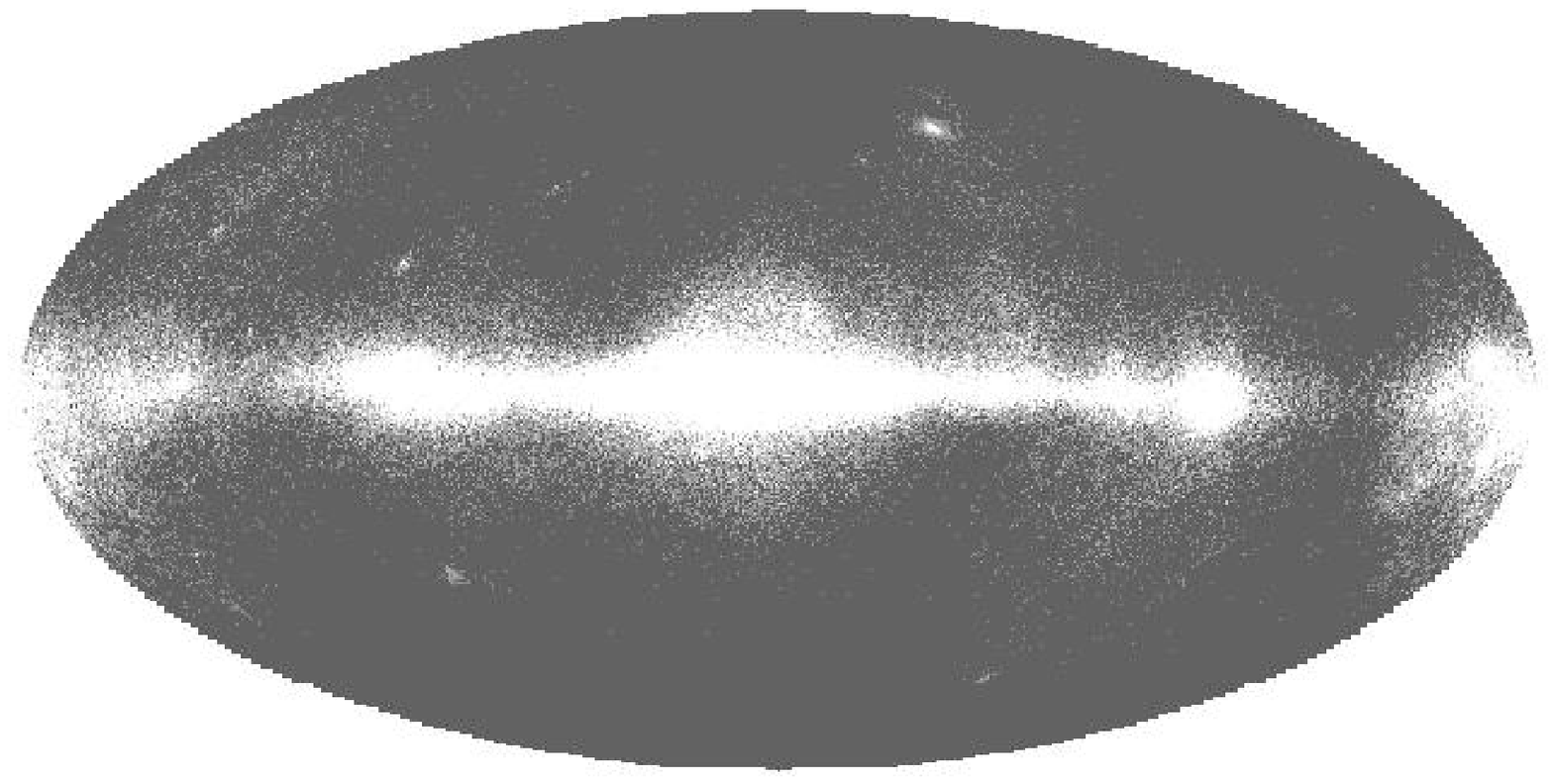}}
\centerline{Photon Counts}
\vspace{0.2in}
\epsfxsize=0pt \epsfysize=2in
\centerline{\epsfbox{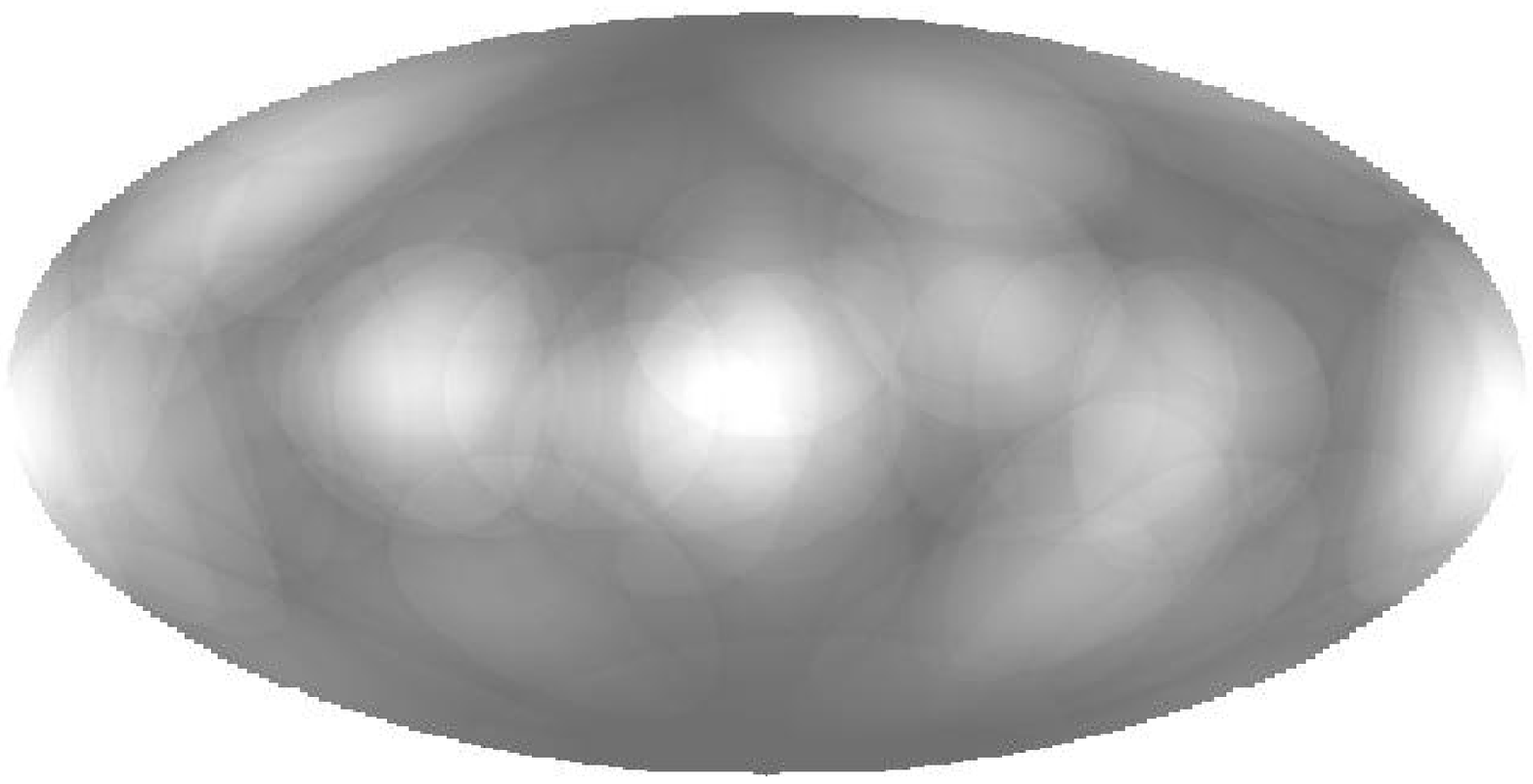}}
\centerline{Instrument Exposure}
\vspace{0.2in}
\epsfxsize=0pt \epsfysize=2in
\centerline{\epsfbox{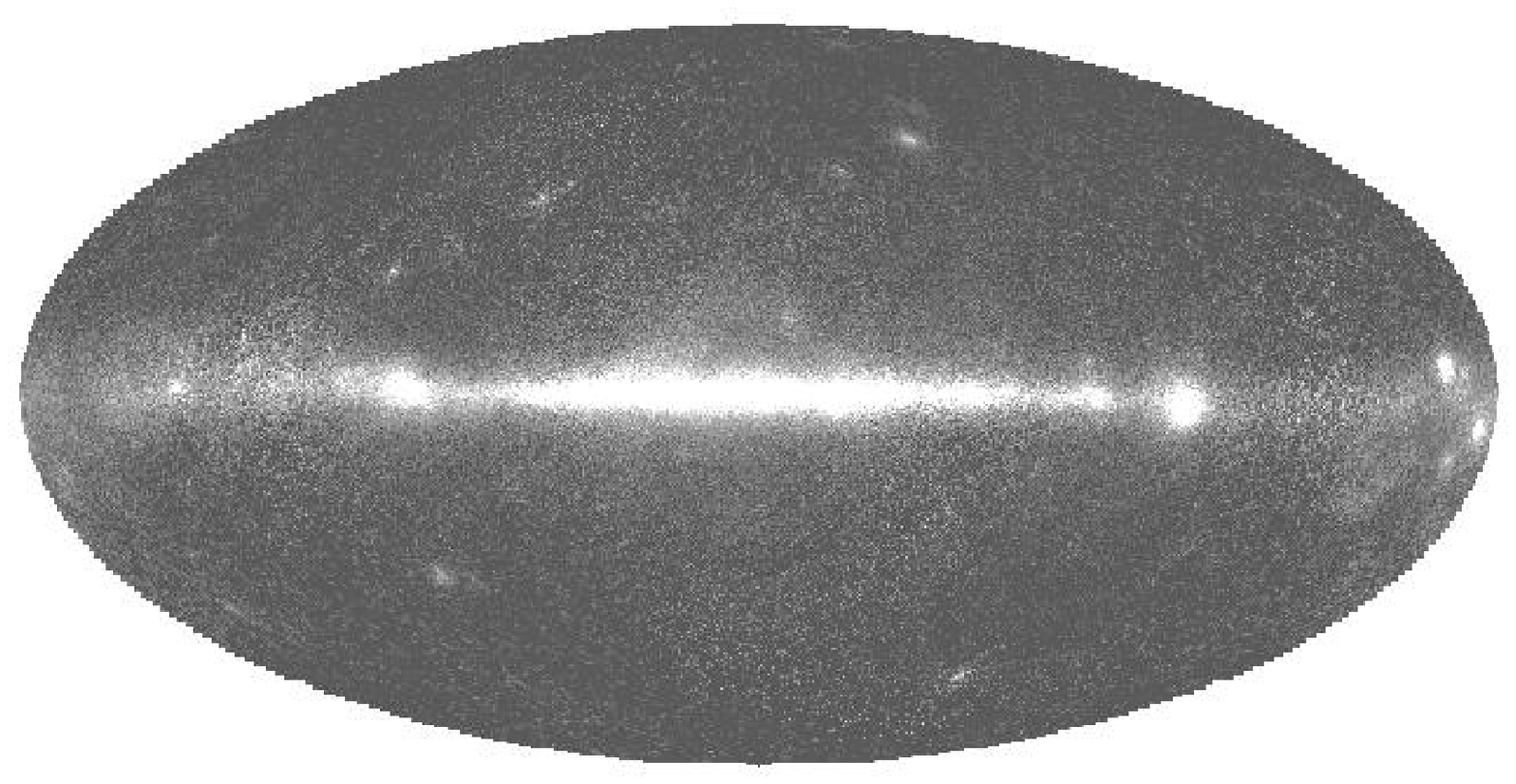}}
\centerline{\gammaray\ Intensity}
\caption[\egret\ All-Sky Maps from Phases I, II, and III Combined]
{\egret\ all-sky maps of photon counts, instrument exposure, and intensity.
These galactic coordinate Aitoff projections were formed from all
\egret\ data above 100~MeV accumulated during Phases I, II, and III.  The
intensity plot in the lower panel is obtained by dividing the \gammaray\ 
counts in the upper panel by the \egret\ exposure in the center panel.}
\label{allsky}
\end{figure}

Once the spatial maps of photon counts and exposure have been made for an
observation, they can be added directly to results from other observations to
produce cumulative counts and exposure maps.  In this way, the photon
intensity maps can be obtained for any set of observations by summing their
corresponding counts and exposure maps, and then dividing the total number of
photons in a bin by the total average exposure for that bin.  The counts,
exposure, and intensity maps above 100 MeV have been summed for all of the
observations in Phase I, II, and III, and the galactic coordinate Aitoff
projections are shown in \fig{allsky}.  The three brightest sources along the
galactic plane are Vela, Geminga, and Crab at galactic longitudes of
263\fdg6, 195\fdg1, and 184\fdg6, respectively.  It should be noted from the
exposure map that instrument exposure is highly non-uniform, with particular
concentration on the galactic center, the anti-center, and the Virgo region.

\chapter{Instrumental Background}

	Absolute measurements of isotropic backgrounds of any kind are always very difficult for the simple reason that there is often no handle with which to distinguish backgrounds which constitute real signal from other instrumental noise.  The revolutionary discovery of the Cosmic Microwave Background (CMB) is an example of a measurement that was possible only because the experimenters had great confidence that they knew the noise characteristics of their detector; in that case a microwave horn.  The measurement of the IDGRB in the MeV range made using balloon measurements (\cite{Trombka77}) showed evidence of an `MeV bump'.  However, measurements in this energy range are complicated by the prevalence of nuclear activity.  A detailed discussion of backgrounds in this energy range are discussed in (\cite{Gehrels92}).  More recent measurements made aboard \cgro\ have shown that these early measurements contained significant instrumental backgrounds which are likely responsible for the `bump' spectral feature (\cite{Kappadath95}).  The modulation of this background in the earth's magnetic field was of critical importance in rejecting instrumental backgrounds.  This and other techniques described in the following discussion are used to evaluate \egret's instrumental background.

\section{Sources of Instrumental Background}

\subsection{Charged Particles}

	The earth's atmosphere at sea level provides $\sim 20$ attenuation lengths of shielding to photons with $E>100$MeV.  For this simple reason it has always been necessary to get above the atmosphere in order to do \gammaray\ astronomy.  The challenge of this approach is that the atmosphere also attenuates cosmic rays ($\sim 27$ radiation lengths at sea level) so operating a \gammaray\ telescope in orbit necessitates the rejection of this cosmic ray flux which can create showers in a pair conversion telescope such as \egret.

	The integrated flux of charged particles in the terrestrial vicinity is well approximated by,
\begin{equation}
	j(>E) =10^{24} E^{-1.74}  {\rm cm}^{-2} {\rm s}^{-1} {\rm sr}^{-1}\;.
\end{equation}
For typical points in the \egret\ orbit, the relevant cosmic ray fluxes are on average $\sim 10^4$ times higher than the corresponding \gammaray\ fluxes.  One must carefully ensure that these particles do not contaminate the data.

{\em a) Downward Moving Cosmic Rays}

	As outlined in the previous chapter, the principal subsystem for the rejection of downward moving charged particles is the Anti-Coincidence Dome (AC dome). Any particle traveling through the spark chamber to the calorimeter must pass first through a 2 cm layer of plastic scintillator.  The majority of the charged particles will pass through as minimum ionizing particles.  The energy deposited in this layer is very efficiently converted to light which is collected by an array of photo-multiplier tubes arranged at the base of the dome.  The probability that a charged particle passing through the AC dome would escape detection was measured before launch to be $< 10^{-6}$.(\cite{Kanbach89})

	A more difficult background to reject is caused by the material in front of the AC dome.  Such material can initiate hadronic showers when bombarded by charged particles.  A common product in such showers are neutral $\pi^0$'s which can then decay into \gammarays\ which yield triggers indistinguishable from those caused by cosmic \gammarays.  For this reason, care was taken to minimize the amount of material in front of the AC dome in the design of \egret.  It was necessary, however, to cover the AC dome with a thermal blanket, which served to stabilize the AC dome temperature, provide a light shield, as well as to shield
against micro-meteoroids.  The resultant blanket has a thickness of $\sim 3 \;{\rm cm}$ and a mass per unit area of $ 0.17 {\rm g}\; cm^{-2}$.

\begin{figure}[h]
\epsfysize=5.0in
\centerline{\epsfbox{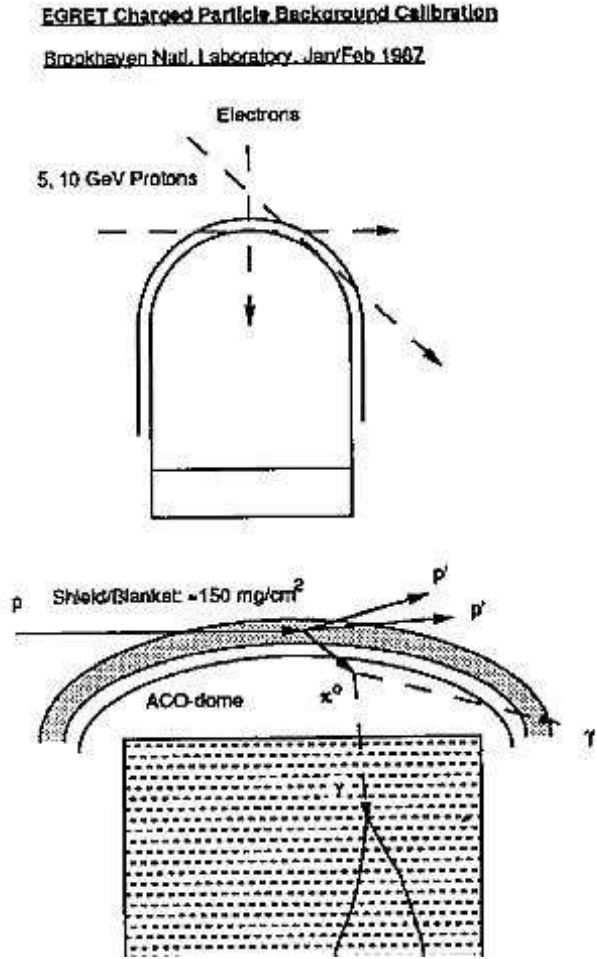}}
\caption{Example of a proton event interacting in the thermal blanket to yield a spurious \gammaray\ event.}
\label{proton_back}
\end{figure}

	The background caused by this unavoidable foreground material was measured at Brookhaven National Laboratory in 1987.  A proton beam was directed
at the instrument at various incident angles and the resultant count rate
was measured.  These measurements were hampered by the high ambient radiation levels in the test beam but an upper limit on this background was estimated to be 
\begin{equation}
 j_{proton}(E> 100 MeV) < 1.3 \times 10^{-6} {\rm cm}^{-2}{\rm s}^{-1}{\rm sr}^{-1}\;.
\end{equation}

{\em b) Upward Moving Cosmic Rays}

	There are additional pathways for cosmic rays to enter the spark chamber that must be guarded against.  The AC dome shields the spark chamber only against downward moving cosmic rays.  It is quite likely that a cosmic ray can enter the telescope from the calorimeter side and travel upward through the detector.  If such a particle were to continue out the front side of the detector it would most likely trigger the AC dome and thus not be problematic.  However, it is also possible that such a particle could undergo a hadronic shower and have the products range out in the detector.  Such an event could conceivably yield a pattern of tracks that could be mistaken for a downward traveling \gammaray\ event.

\begin{figure}[h]
\epsfysize=3.0in
\centerline{\epsfbox{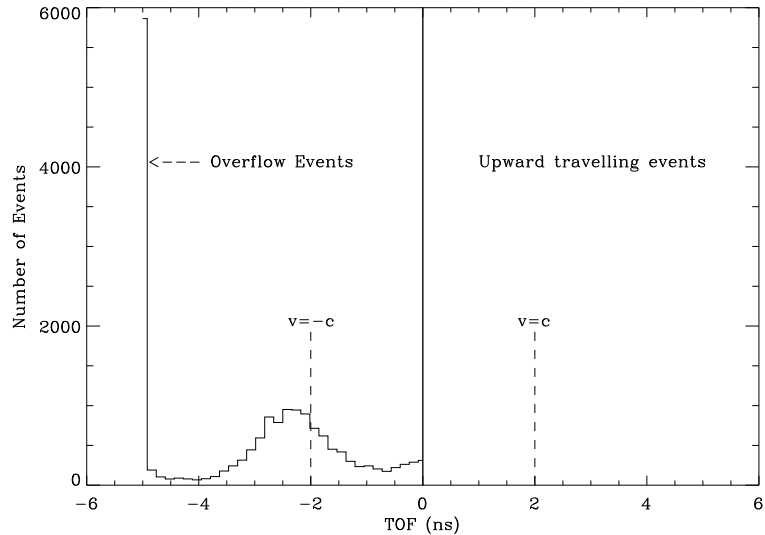}}
\caption{The measured TOF for 17,000 flight events.  Triggers are not generated for upward moving events with TOF$>$0.  In the limit of infinite time resolution there should be no events with -2 ns$<$TOF$<$2 ns because these events measure speeds greater than the velocity of light.  The events in this range are indications that the TOF measurement time resolution is worse than predicted.  The spike at TOF=-5 represents the overflow events which are poorly understood.}
\label{TOF}
\end{figure}

	This source of background has been eliminated by the inclusion of a time of flight scintillator system between the upper spark chamber and the calorimeter.  This system is comprised of two arrays of plastic scintillator separated by 60 cm.  The TOF of any charged particles traveling in the detector can thus be determined allowing the instrument to distinguish between upward and downward moving showers.  The drawback of this approach is that this system severely limits the field of view because it requires a sufficient longitudinal length to allow a TOF measurement of relativistic particles.  The resultant geometric aspect ratio is correspondingly elongated leading to a reduced field of view.

	\fig{TOF} shows the distribution of TOF values for a set of 17,000 flight triggers.  The TOF value is measured and stored as a 6 bit word which is telemetered to the ground along with the event record.  Because events with TOF$>$0 do not generate flight triggers, the distribution is truncated at this point.  The distribution shows a peak at TOF=-2.5 ns. The separation of 60 cm between the two scintillating layers results in a TOF=2 ns for a relativistic particle traveling vertically through the chamber.  Slightly larger TOF values correspond to relativistic particles traveling at an angle through the TOF system. The events in this peak thus constitute downward traveling \gammaray\ events.  There should be a sharp cutoff in the distribution at TOF=-2ns.  The spread of the observed distribution to smaller TOF values is a measure of the time resolution of the TOF system.  The flight data indicate a time resolution of $\sim 1{\rm ns}$.  The pre-flight optimization of the TOF system (\cite{Hunter91}) indicated that the time resolution of this system was $\sim0.45\;{\rm ns}$.  This value was obtained  after careful optimization of cable lengths.  It is possible that the light travel times have slightly altered over time which has slightly degraded this resolution.  This degradation can lead to contamination by allowing the tail of the upwardly moving events to trigger the telescope with TOF values smaller than 0.  There is some direct indication that this is happening in that the distribution is seen to be increasing near TOF=0.  These stray triggers could constitute a significant background and thus must be rejected on the basis of their shower patterns in the spark chamber as will be discussed later.

	It should also be pointed out that there are a large number of \egret\ triggers for which the TOF measurement has overflowed.  These triggers are generated by hits in the two layers separated by more than $\pm 5\;{\rm ns}$.  It is not well understood what causes these triggers that are rejected on the basis of these anomalous TOF values.  A large fraction of them could be due to accidental coincidences between the two scintillator planes.  These planes have a trigger rate in flight of $\sim2-3\;{\rm kHz}$.  Many of these hits will be in coincidence with a veto layer hit however and as such will not generate a trigger.

\subsection{ Neutron Background}

	Cosmic ray spallation in the earth's atmosphere produces a background of secondary particles.  Among the products of such interactions are neutrons.  Furthermore, the active sun can serve as an additional source of neutrons in low earth orbit.  These neutrons are problematic in that they are not vetoed by the AC dome.  They may, however, interact in the spark chamber and produce downward traveling showers.

	The fluxes of these secondary particles are not well measured.  Gehrels (\cite{Gehrels92}) presents a calculated spectrum in the \egret\ energy range.  We can use this spectrum to get a rough estimate of the number of such events to be expected.  His spectrum must be viewed as a very conservative upper limit on the neutron flux because his calculation includes neutrons produced as secondaries from charged particles interacting with a particular spacecraft.  These events will be vetoed by \egret.  The spectrum given yields .3 ${\rm neutrons}\;{\rm cm}^{-2}{\rm s}^{-1}$.  If we assume the tantalum nuclei in the tracker conversion foils convert these neutrons inelastically causing hadronic showers, then the probability of conversion of a neutron into a shower passing through a half radiation length of tantalum is,
\begin{equation}
P=\frac{N_A}{A}l\rho A^{2/3}r_0^2=(\frac{6\times10^{23}}{180} {\rm g}^{-1} )\times (.2\; {\rm cm}) \times (16.6 \;{\rm g}\;{\rm cm}^{-3}) (180)^{2/3} 10^{-26} {\rm cm}^{2} =.02\;,
\end{equation}
where we have used the atomic weight of tantalum ($A$) and its density ($\rho$).  If we use the physical area of the detector as well and assume a live-time fraction of 50\% and a solid angle of $40^\circ$ this results in a trigger rate of 4 Hz.  The rate seen by \egret\ will be much smaller than this because the AC dome will veto any cosmic ray shower that produces neutrons as well as charged particles when interacting with the spacecraft. Nevertheless, there may be some neutron induced hadronic cascades produced in the tracker.  These events must be rejected on the basis of their shower patterns in the spark chamber.

\subsection{ Albedo \Gammarays}

	The \egret\ orbital altitude of 350-450 km ensures that downward traveling \gammarays\ pass unattenuated through the residual atmosphere above the telescope.  Furthermore, downward traveling cosmic rays have a similarly small chance of producing hadronic showers which could produce a spurious \gammaray\ signal.  However, the amount of matter in front of the telescope increases dramatically when the telescope is directed near to the earth's limb.  In these direction, cosmic ray interactions with the earth's atmosphere generate copious \gammarays.

	While these albedo \gammarays\ cause triggers identical to their celestial counterparts, they are also imaged in the same way as the signal \gammarays.  These spurious events can then be rejected simply on the basis of their proximity to the earth's limb.  In fact, \egret\ is capable of operating in several directional modes in which combinations of the trigger scintillator array tiles are disabled and as a result events from near the earth's limb are not permitted to trigger the telescope.  In this way, excessive triggers caused by albedo \gammarays\ are prevented as these events would be eventually rejected based on their origin near the earth's limb.

\section{ Triggering Modes}

	The \egret\ instrument triggers and generates an event record when the following criteria are met:
\begin{itemize}
\item A signal is detected in both the upper and lower arrays of scintillator in the TOF system described above.
\item The TOF between these two signals is within an acceptable range.
\item The energy deposited in the TASC exceeds a defined threshold value (nominally $6 MeV$).
\item No signal from the AC dome is detected.
\end{itemize}
The trigger rate in this mode of operation in orbit is $\sim 0.7 \; {\rm Hz}$.   If one disregards the problematic TOF overflow events this rate drops to $\sim 0.4 \; {\rm Hz}$.  We can estimate the trigger rate caused by \gammarays\ (both celestial and the more dominant albedo) as follows.
\begin{equation}
{\rm Rate}= I_\gamma \epsilon A \Omega (\frac{T_{\rm live}}{T}) \simeq 20\times10^{-5}\times  \frac{1}{3}  \times (80{\rm cm})^2 \times (2\pi(1-cos(40^\circ)) \times \frac{1}{2} \sim .4\; {\rm Hz}\;,
\end{equation}
for $E>30\;{\rm MeV}$ and we have assumed a photon conversion rate of 33\% and used the approximate physical area of the tracker.  While this estimate is not accurate to within a factor 2 it does indicate that the  hardware trigger has been successful in reducing the trigger rate from the several kHz cosmic ray flux (typical TOF deck rate = 2 kHz) down to values less than 1 Hz of which \gammarays\ constitute a large fraction.  This analysis is not sufficient to determine whether there remains any cosmic ray signal.  In fact the distribution of TOF values indicates that there is still a substantial cosmic ray contamination in the hardware triggered events.  Furthermore there may be a significant trigger rate caused by secondary neutrons.  These must be rejected by the characterization of the track patterns in software.

\subsection{Search and Analysis of Gammaray Events (\sage)}
\label{sage}
\sas\ and \cosb\ relied on a tedious event-by-event interactive analysis to select desired events and
determine their structure.  With the significant increase in data recorded by
\egret, this type of manual analysis becomes impractical, and an automated
procedure must be implemented.

For \egret, analysis is done using a computer program called \sage\ (Search
and Analysis of Gammaray Events).  For every event in the \egret\ telemetry,
the spark chamber tracks are analyzed by \sage\ in a nine step process.
First, the event must have generated a given minimum number of sparks in both
orthogonal views.  Second, three sparks must be found within a specified
distance of each other, and must define a reasonable path.  These three spark
lines are known as {\em triplets}.  In the third step, triplets which start
in the top conversion plate or trace back to the instrument walls are
rejected.  In addition, at least one triplet must form before the upper
scintillator array.  In the fourth step, all possible tracks from the
remaining triplets are constructed, with duplicate tracks eliminated.  The
fifth step selects the electron and positron tracks from the candidates
calculated in step four.  Step six checks to see if there is a better choice of
vertex between the two paths.  In the seventh step, the two sets of tracks in
the two orthogonal views are correlated.  The eighth step determines event
descriptors such as the number of spurious sparks and the curvature of the
track.  In the ninth step, the events are stored.

\begin{table}[ht]
\centering
\caption
{Summary of SAGE Rejected Events}
\bigskip
\begin{tabular}{l|r}\hline\hline
Total Triggers (06/03/91-06/04/91)& 55439 \\ \hline
Rejection Condition: &  \\  
Too few sparks in given projection & 7022 \\ 
TOF=0 & 20857 \\ 
No triplets in given projection & 1666 \\ 
Conversion in first plane & 2150 \\ 
Conversion below scintillator plane & 4557 \\ 
Wall event & 7827 \\ 
Angle exceeds $45^\circ$ & 0 \\ 
Track pattern bad & 0 \\ 
Fails to meet minimum point requirement & 5800 \\
Single rejected events & 760 \\ \hline
Total Rejected Events& 50639 \\ \hline
Total Accepted Events& 4800 \\ \hline \hline 
\end{tabular}

\end{table}

The surviving candidate events are divided into two groups, definite
\gammaray\ events and questionable events.  Extensive studies have shown that
the events which are rejected by \sage\ are typically either not
\gammaray\ events or are unsuitable for analysis.  Roughly one good
\gammaray\ in 700 is rejected by \sage\@.  Subsets of the events which are
classified as definite \gammaray\ events are routinely monitored by trained analysts whose determinations agree with the software to better than one part in a thousand.  About 12--15\%
of all events are initially identified as questionable.  The
analysts manually review these events by examining the event tracks on a
graphical display.  These events are usually either accepted or rejected as
structured by \sage\@.  Approximately 0.5\% of the events are interactively
restructured.  As a continuing check of the \sage\ program, 1\% of the
\egret\ data is randomly selected for graphical examination by the analysts.

	Table 1 shows a summary of the cuts made by \sage\ on one days worth of data.  An overall rejection factor of 0.08 results. While many of these rejections represent \gammarays\ whose hit patterns were simply unsuitable for analysis (e.g. rejections for too few sparks) others indicate likely hadronic shower events (e.g. fails to meet minimum track requirement).  The selectivity of this process ultimately determines the residual cosmic ray induced background.

\section{ In Flight Determination of Background}

	The various sources that contribute to \egret's instrumental background have been outlined above as have the systems designed to reject them.  We now turn to the task of using the flight data to determine the efficiency with which backgrounds are being suppressed.

\subsection{Albedo Rejection}

\begin{figure}[h]
\epsfysize=3.0in
\centerline{\epsfbox{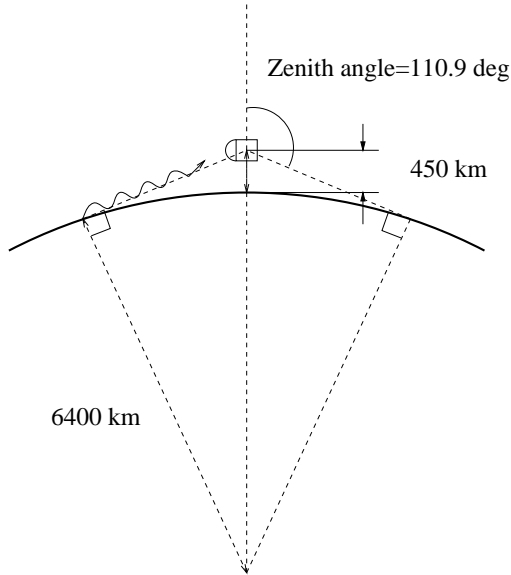}}
\caption{Orbital geometry showing the zenith angle to the earth's limb.}
\label{zengeo}
\end{figure}

	As mentioned previously, albedo \gammarays\ from the earth's limb are rejected on the basis of the proximity of their reconstructed direction to the earth's limb.  This is achieved by making a zenith angle cut on the reconstructed \gammaray\ arrival angle.  As shown in \fig{zengeo}, the zenith angle ($\alpha$) is the angle between a photon arrival direction and the vector from the earth's center to the spacecraft. The earth's limb appears at a zenith angle of $\alpha\sim 111^\circ$ in the \egret\ orbit. For all zenith angles greater than this value, the earth is directly within the field of view.  Because the earth's atmosphere is illuminated from above by cosmic rays, the limb produces the greatest \gammaray\ intensity. Cosmic rays that interact on the far side of the earth's limb produce showers that are directed into the telescopes field of view producing bright hard \gammarays\ (\cite{Thompson81}).  The limb spectrum is therefore brighter and harder than the radiation from the earth's disk. 

	In the limit of infinitely sharp point spread functions, all photons arriving from zenith angles less than $\alpha=111^\circ$ should be celestial in origin.  However, the \egret\ point spread function is several degrees wide.  Moreover, the point spread functions are known to have tails which result in a finite probability of measuring a photon arrival tens of degrees away from the true origin.  This can occur if only one of the $e^+/e^-$ pair tracks are reconstructed.  It is this phenomenon which leads to the so called `fish-eye' effect described in Chapter 2.  While the correction for the `fish-eye' effect does lead to correct point source positions, it does not eliminate the possibility of photons from outside the field of view leaking in.  The result of this effect is that bright features near the edge of the field of view can leak into the data stream.  This is particularly true of the earth's limb which is the brightest \gammaray\ feature in the sky. 

	\fig{zencut} below shows the distribution of measured \gammarays\ in zenith angle, summed over many pointings.  For small zenith angles, the photon count rate is roughly constant.  This is because these photons are dominated by celestial sources which are distributed randomly in zenith angle.  As a result these features tend to average out.  Near the earth's limb the count rate drops.  This is because when \egret\ is directed near the earth's limb, restricted field directional modes are incorporated into the hardware trigger so as to reduce the overall trigger rate.  This process is not perfectly efficient and there is a resulting exclusion of photons that arrive from directions near the limb.  The bright feature clearly visible at $\alpha=111^\circ$ is caused by albedo photons from the earth's limb.  Despite the directional modes intended to reduce the albedo trigger, these albedo photons are a factor 3.5 times as common as celestial \gammarays.  This is a strong background that must be carefully eliminated.  The standard energy-range dependent zenith angle cuts used in \egret\ data analysis are shown superimposed on this figure. It is clear that albedo photons contribute to some extent inside the adopted cuts.  While this does not affect analysis of bright sky features it can have a potentially important impact on analysis of the faint high latitude diffuse emission.

\begin{figure}[h]
\epsfysize=3.0in
\centerline{\epsfbox{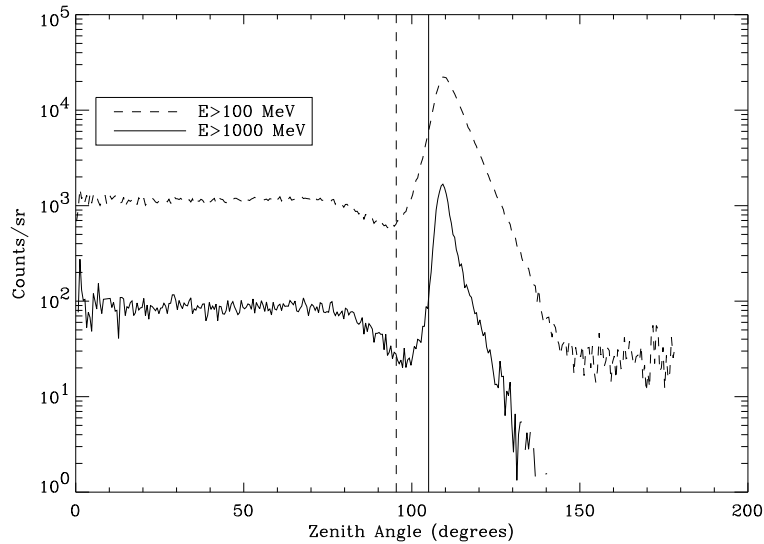}}
\caption{Distribution of photons in zenith angle. The standard acceptance cuts are shown as a vertical line for each energy range.}
\label{zencut}
\end{figure}

	In order to determine the appropriate zenith angle cut, the flight data were analyzed using a variety of zenith angle cuts.  Exposure and counts maps were generated in four independent zenith angle ranges and in two energy ranges  ($E > 100 \; {\rm MeV}$, and $E>1000 \; {\rm MeV}$).  These maps were constructed in celestial coordinates.  The intensity was then calculated in a set of high latitude bins ($\Omega_{ij}=10^{\circ} \times 10^{\circ}$).

	In order to calculate the albedo contamination, several assumptions were made.  First, the intrinsic celestial intensity in each bin is assumed to be constant. This implies that measurements of intensity that are not concurrent should still give the same result.  This is known to be untrue of high latitude point sources which can vary on the time scale of days (\cite{vonMontigny95}). As a result, regions within $10^\circ$ of a strong point source were excluded from this study.  The sources excluded were those found in the combined data from Phase I+II.
	
	Secondly, there was assumed to be negligible albedo contamination in the zenith angle range $\alpha<50^\circ$.  The albedo contamination in each spatial bin, ${I_Z}_{ij}$, is thus calculated to be the difference between the intensity measured in a particular zenith range $Z$ and the intensity calculated using $\alpha <50^\circ$. The data will be shown to be self consistent with this assumption.  The uncertainty in each measurement, $\sigma_{ij}$, is estimated from the counting statistics which allows the weighted mean to be calculated over all spatial bins,
\begin{equation}
	\overline{I_{Z}}=\frac {\sum_{ij}\frac{{I_Z}_{ij}}{\sigma^2_{ij}} }{\sum_{ij} \frac{1}{\sigma^2_{ij}}}\;.
\end{equation}
The uncertainty associated with this value is given by,
\begin{equation}
	\sigma^2_{I_{Z}}=\frac {1}{\sum_{ij}\frac{1}{  \sigma^2_{ij} } }\;.
\end{equation}

\begin{table}[ht]
\centering
\caption
{Summary of Albedo Contamination Results}
\bigskip
\small
\begin{tabular}{l|cccc}\hline\hline
& \multicolumn{4}{c}{$E>100 MeV$}  \\
\cline{2-5}
Zenith Angle range & $\alpha<50.0$      & $50.0<\alpha<65.0$   & $65.0<\alpha<80.0$ & $80.0<\alpha<95.375$        \\
\hline
Fractional Exposure            & 0.38      & 0.24   & 0.24 & 0.14         \\[0.1in]
Differential Albedo            & 0.0      & $0.4 \pm 0.4$   & $1.9 \pm 0.4$ & $6.8 \pm 0.6 $    \\
Intensity ($\times 10^{-6}$)& & & &   \\[0.1in]
Mean Albedo Intensity           & 0.0      & $0.15 \pm 0.15$   & $0.64 \pm 0.2$ & $1.5 \pm 0.2 $    \\
 ($\alpha<\alpha_{max}$)($\times 10^{-6}$)& & & &   \\[0.1in]
Signal:Noise           & N/A      & 66   & 16 & 7    \\[0.1in]
\hline
\hline
& \multicolumn{4}{c}{$E>1000 MeV$}  \\
\cline{2-5}
Zenith Angle range & $\alpha<65.0$      & $65.0<\alpha<80.0$   & $80.0<\alpha<95.0$ & $95.0<\alpha<105$        \\
\hline
Fractional Exposure            &   .60  &  .25  & .13 &    .02      \\[0.1in]
Differential Albedo            &   0    &  $-1.5 \pm 1.0$ & $3.4 \pm 1.5$ &  $48.8 \pm 6.5$   \\[0.1in]
Intensity ($\times 10^{-7}$)& & & &   \\[0.1in]
Mean Albedo Intensity           & 0.0      & $0.0 \pm 0.3$   & $0.45 \pm 0.32$ & $1.41 \pm 0.3 $    \\
 ($\alpha<\alpha_{max}$)($\times 10^{-6}$)& & & &   \\[0.1in]
Signal:Noise           & N/A      & $>33$   & 22 & 7    \\[0.1in]
\hline
\hline
\end{tabular}

\label{zen_results}
\end{table}

\begin{figure}[h]
\epsfysize=3.0in
\centerline{\epsfbox{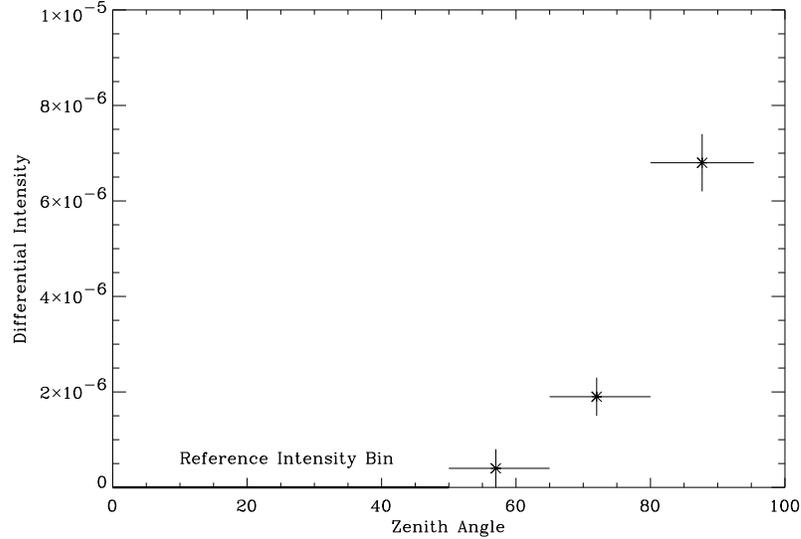}}
\caption{Residual intensity in various zenith angle ranges.  The vertical axis shows the mean additional intensity in a set of high latitude pixels in a given zenith angle range compared to the intensity in the reference zenith range: zenith$<50^\circ$.  The horizontal error bar indicates the extent of the particular zenith ranges and the vertical error bars are statistical uncertainties.}
\label{zenback}
\end{figure}
\begin{figure}[h]
\epsfysize=4.0in
\centerline{\epsfbox{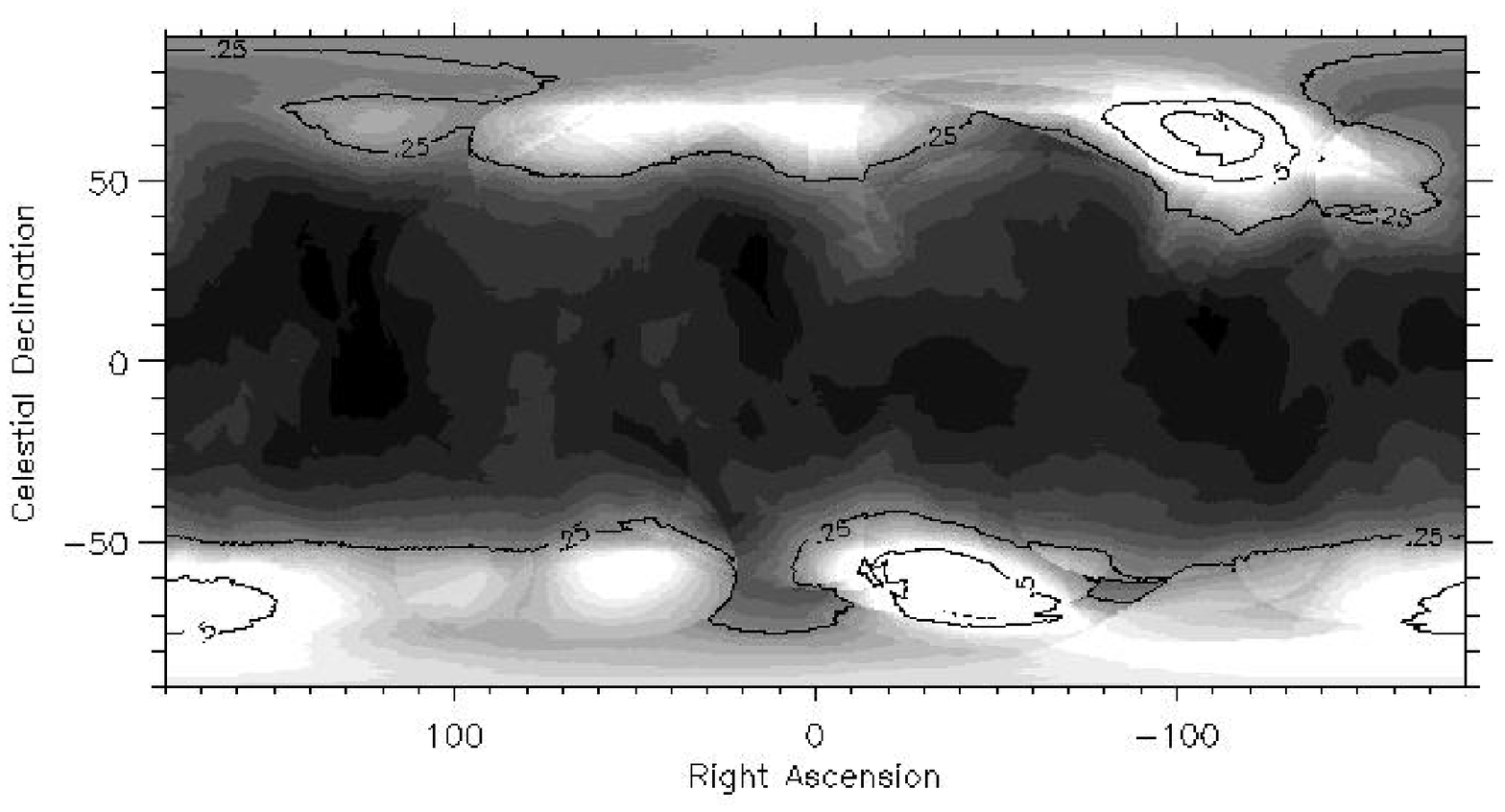}}
\caption{Ratio of exposure in the zenith range $80 < \alpha < 95.375$ to the zenith range $\alpha < 80.0$.  Regions that show up light colored in this plot are dominated by high zenith angle exposure and as a result contain higher than average albedo contamination}
\label{expo_zen}
\end{figure}

	Table 3.2 summarizes the results.  There is a clear signature of albedo contamination outside zenith angles of $65^\circ$.  While one could argue that this should determine the best value of the albedo cut, it should be noted that fully 38\% of the standard exposure occurs in the zenith angle range $65^\circ<\alpha<95.375$.  As a result, such a stringent zenith angle cut eliminates a large fraction of the data.

	A better strategy is to require some minimal signal to noise ratio.  At high latitudes the diffuse intensity for $E>100\; {\rm MeV}$ is $\sim 1\times 10^{-5}$  which sets the standard for measurement accuracy.  This should be compared with the albedo contamination averaged over all included zenith angles.  These values are shown in Table 3.2 along with estimated signal to noise ratios at high latitudes.  The standard \egret\ zenith angle cut in the energy range $E>100\;{\rm MeV}$ of $\alpha_{max}=95.375^\circ$ results in an albedo contamination of $\overline{I_{{\rm albedo}}}=1.5 \times 10^{-6}$ which is  $\sim15\%$ of the high latitude diffuse. A cut of $\alpha_{max}=80.0^\circ$ results in a mean albedo contamination of $\overline{I_{{\rm albedo}}}=0.6 \times 10^{-6}$ while only excluding 14\% of the data.  This is now less than 10\% of the high latitude diffuse which is acceptable.  It is important to note that on the galactic plane, the galactic diffuse emission is at least an order of magnitude greater and as a result the standard zenith angle cuts are acceptable.

	Despite the fact that the earth is not localized in the sky, the albedo contamination it produces  is not isotropic.  The spacecraft orbital pole in a low eccentricity orbit remains at constant zenith angle near to the earth's limb throughout a pointing period.  As a result, pointing near the orbital pole is avoided whenever possible.  However, in order to obtain coverage of the full sky it is necessary to point near the orbital pole on occasion.  The inclination of the \egret\ orbit is $28.5^\circ$ and it precesses about the celestial pole.  As a result regions near the celestial poles and in particular certain regions with declinations near $\delta=\pm 61.5^\circ$ tend to be viewed at higher average zenith angle than regions near the orbital/celestial equator.  \fig{expo_zen} shows the ratio of the exposure in the rigidity range $80 < \alpha < 95.375$ to exposure in the zenith range $\alpha < 80.0$.  Note the coverage of the regions near the celestial pole occurs more frequently at higher zenith angles than regions near the celestial equator.  The result is to distribute the albedo contamination preferentially in these regions of higher average zenith angle.  These regions must be treated carefully to avoid confusion between albedo induced features and real celestial signals.

\subsection{Rigidity Modulation}

	The earth's magnetic field provides the handle that allows us to gain some understanding of the level of charged particle contamination.  If we follow a charged cosmic ray particle in from infinity, the behavior of this particle when it interacts with the earth's magnetic field depends on the particles momentum.  Low energy particles travel in tight orbits around field lines and drift toward the magnetic poles where they are funneled into the trapped radiation belts.  Higher energy particles are able to penetrate all the way to the earth's atmosphere with little deflection.  At a given point in the earth's magnetic field away from areas of trapped radiation, the effect is to introduce a low energy cut-off in the the observed spectrum of charged particles.  This cut-off is traditionally expressed in terms of magnetic rigidity,
\begin{equation}
{\cal R} =(\frac{pc}{ze})=\frac{M}{R^2}\frac{{\cos^4{L}}}{{[{(1+\cos{\theta}{\cos^3{L}})}^{1/2}+1]}^2}\;\;({\rm GV}) \;,
\end{equation}
where $p,z$ are the particles momentum and charge and $M$ is the earth's dipole moment.  Particles entering the earth's magnetic field at an angle of $\theta$ to the tangent  to the latitude circle, $L$ will not penetrate to $R$, the geocentric radius, and are assumed to be trapped if their rigidity is less than that given by the above formula.

\begin{figure}[h]
\epsfysize=4.0in
\centerline{\epsfbox{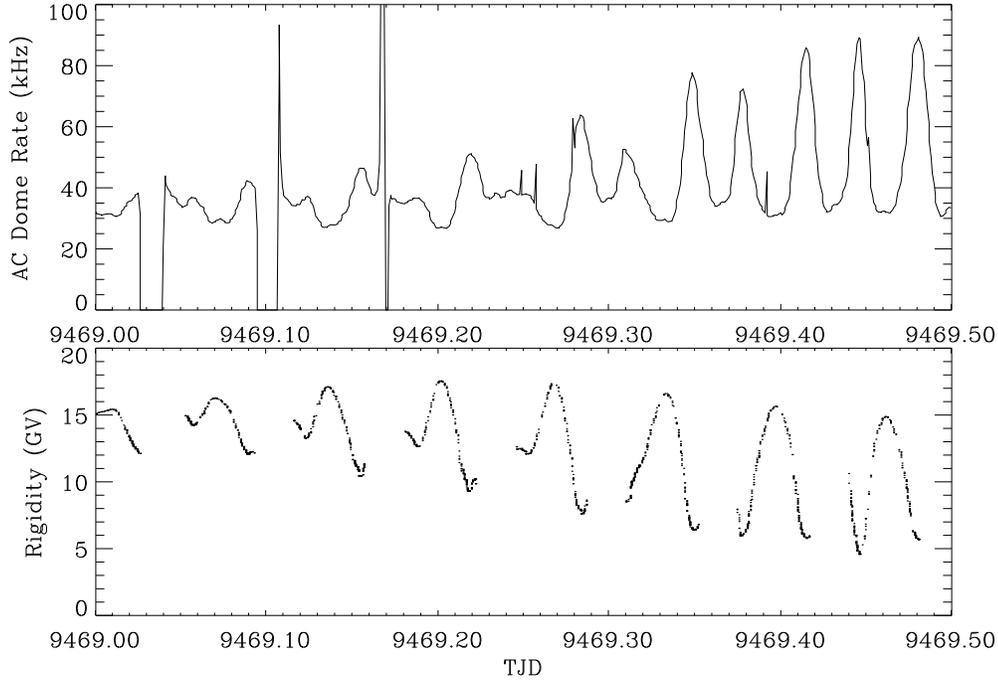}}
\caption{Rigidity history for one day during VP 1.0.  Note the modulation with the satellite orbital period as well with the earth's rotational period.  The gaps represent times when the instrument was off while passing through the South Atlantic Anomaly.}
\label{rigandrate}
\end{figure}

	The end result of this magnetic interaction is that the cosmic ray flux is screened to a greater extent in regions of high magnetic field.  \fig{rigandrate} shows the one day long \egret\ time history showing both the rigidity variation as well as the AC dome rates which are telemetered to the ground as part of the housekeeping data at regular time intervals.  The anti-correlation between rigidity and AC dome rate is clearly visible.  The AC dome rate increases by a factor of four in the orbital regions of lowest magnetic field.  This modulation can be used to check for a modulation in observed gamma ray background rate with orbital rigidity to measure any cosmic ray background leaking into the gamma ray stream.  It is important to note that the AC dome rates do not directly measure cosmic ray rates because they include a significant contribution from hard X-rays.  The well measured hard X-ray background was calculated to result in a AC dome trigger rate of $\sim 20 kHz$ which is entirely consistent with the observed AC dome trigger rates.

	The \egret\ timeline was divided into two sets of intervals: periods when the rigidity was $\gt 14{GV}$ and periods when the rigidity was $\lt14{GV}$.  The resultant intervals are $\sim 40$ minutes long on average.  The two rigidity ranges contain roughly equal numbers of \gammaray\ events. As in the investigation of albedo contamination, exposure and counts maps were constructed in celestial coordinates from the two sets of rigidity intervals.  This results in a measured intensity for each point on the sky in times of high cosmic ray flux and in times of less cosmic ray bombardment.  The difference is a measure of the instrumental background which is not rejected by the various systems. Again it must be assumed that the \gammaray\ signal is not variable on the time scale of a viewing period in regions away from strong point sources.  Because the albedo intensity is also modulated by rigidity, a zenith angle cut of  $\alpha < 50^\circ$ was adopted. 
	
\begin{figure}[h]
\epsfxsize=\hsize
\centerline{\epsfbox{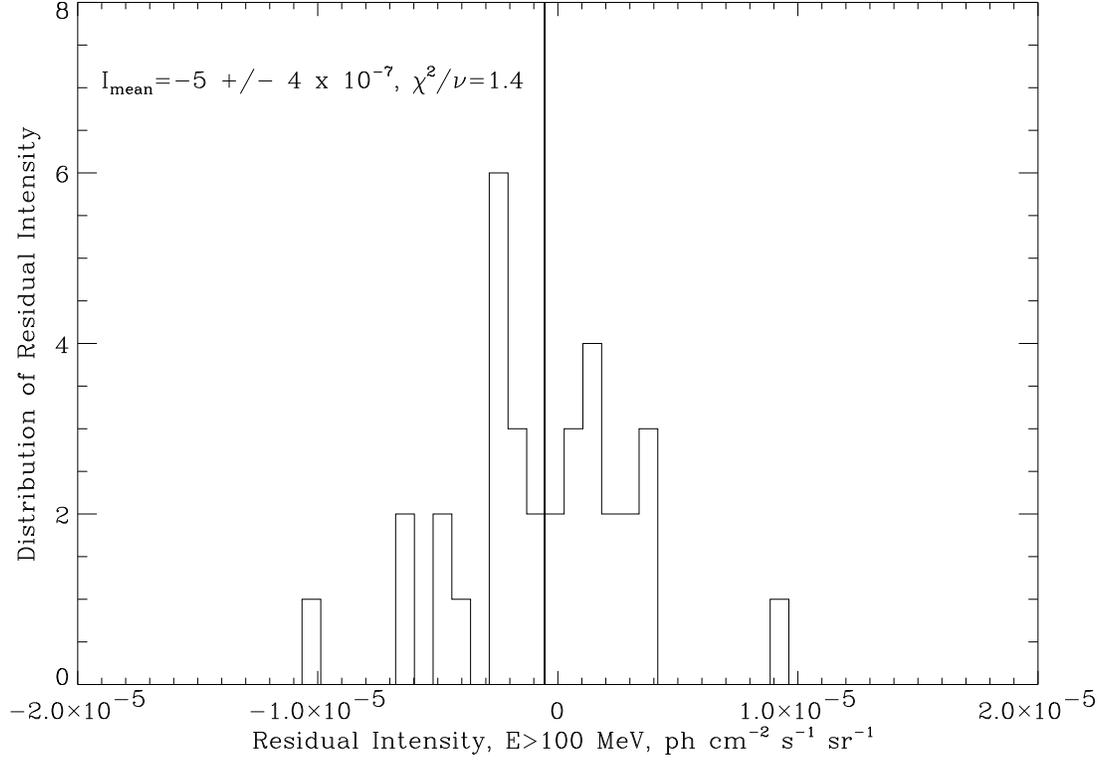}}
\caption{Distribution of instrumental background measurements showing the mean and $\chi^2/\nu$.}
\label{Rigback}
\end{figure}

	\fig{Rigback} shows the distribution of the cosmic ray modulated intensity in a set of $10^\circ \times 10^\circ$ pixels at high latitude.  The weighted mean is calculated in a manner exactly analogous to the technique used in calculating the albedo contamination.  The resultant flux is calculated to be $I_{CR}=-5 \pm 4 \times 10^{-7}$.  This implies that there is no measurable cosmic ray contamination in the data.  The $2\sigma$ upper limit on the residual intensity is $I_{CR}<3 \times 10^{-7}$.	

	In order to extract the upper limit to the mean cosmic ray background from this measurement it is necessary to average the variable instrumental background over the history of the mission.
\begin{equation}
	\overline{I_{CR}}=\frac{\int I_{CR}(t) dt}{T} \;,
\end{equation}
where $I_{CR}$ is the cosmic ray induced instrumental background and T is the total live-time of the mission.  Because we cannot assume that the cosmic ray background is negligible in the high rigidity range, the differential intensity between rigidity ranges does not correspond to a measurement of the absolute cosmic ray background.  Rather we must model the absolute comic ray flux and use the differential measurement to calculate an absolute intensity.  If we assume the X-ray counting rate in the AC dome to be 14 kHz, then we can deduce $I_{CR}(t)$ from the housekeeping data.  Computing $\overline{I_{CR}}$ in each rigidity range we find that,
\begin{equation}
	\overline{I_{CR}}_{<14} \simeq 2\times \overline{I_{CR}}_{>14} \;.
\end{equation}
As a result,
\begin{equation}
	I_{CR_0}={I_{CR}}_{<14}-{I_{CR}}_{>14} = 2\overline{I_{CR}}_{>14} - \overline{I_{CR}}_{>14}=\overline{I_{CR}}_{>14} \;.
\end{equation}
Furthermore, integrating over all viewing periods we find,
\begin{equation}
	T_{<14} \simeq T_{>14} \;.
\end{equation}
This in turn allows us to estimate the mean cosmic ray induced background to be,
\begin{equation}
	\overline{I_{CR}}= \frac{\overline{I_{CR}}_{<14} + \overline{I_{CR}}_{>14}}{2}
		=\frac{2I_{CR_0} + I_{CR_0}}{2}
		=1.5 \times I_{CR_0}	\;,
\end{equation}
where $I_{CR_0}$ is the measured residual intensity between the two rigidity ranges. 

	The resultant upper limit to the average cosmic ray induced instrumental background is $\overline{I_{CR}}<4.5  \times 10^{-7}$.  This value indicates that the cosmic ray contamination is less than an order of magnitude less than the isotropic diffuse background and is negligible compared with the already documented albedo background.  This limit is significantly more stringent than the pre-flight calibration of proton backgrounds.  The ability of \sage\ to reject the cosmic ray events which are not rejected by the hardware trigger has been demonstrated.

	While rigidity directly modulates the charged particle flux it modulates secondary neutron flux more weakly.  A neutron produced at the earth's limb travels through the earth's magnetic field unimpeded until it interacts with the the \egret\ instrument.  As shown in \fig{zengeo}, the interaction point can be $21^\circ$ distant in latitude from the spacecraft latitude.  For the \egret\ orbit this implies that the interaction can occur at a point at which the rigidity is a factor 2 different from the value near the instrument.  The net result is to slightly wash out the modulation of neutrons.  Instead of a factor of 2 difference between the incident flux in the two rigidity ranges used, the averaged flux modulation is closer to 1.2 for typical points in the orbit. Nevertheless,  the lack of a rigidity signal also constrains the neutron induced background.  In light of the fact that charged particles that get through the TOF system are efficiently being rejected on the basis of their hit patterns in the tracker, it is reasonable to assume that the neutron showers are being rejected in a similar manner.  That is, unless the neutron induced trigger rate is significantly larger than the rate of charged particles that escape the hardware trigger, the upper limit calculated above also applies to neutron backgrounds. This is likely to be the case.

\section{Exposure Calibration}

	Certain types of instrumental effects are most easily identified by looking at the data in instrumental coordinates.  If a dead spot in the AC dome were to arise, this would show up very significantly as a bright spot in a particular area of the instrument.  Furthermore, some inaccuracies in the calibration of the sensitivity of the instrument can be identified in flight as a non-uniformity across the instrument.

	Of course the difficulty that must be overcome in looking for this sort of signal is that the \gammaray\ intensity varies across the sky by more than two orders of magnitude from galactic center to galactic pole.  The nonuniform intensity does not necessarily average out to yield a uniform calibration source that can be used to look for any instrumental asymmetries.  Furthermore, there are many bright \gammaray\ sources that move around in instrumental coordinates from viewing period to viewing period.

	In order to be able to attempt an in flight calibration in instrument coordinates, it is necessary construct a rough model of the \gammaray\ sky.  This model consists of three elements: a model of the galactic diffuse emission, a set of bright point sources, and some level of isotropic diffuse background.  The galactic diffuse model will be described in detail in the next chapter.  The point sources are extracted from the data using the Maximum Likelihood technique (see Appendix A), and the isotropic diffuse level is chosen to make the data match the model along the pointing axis.  It is important to note that because a given point source is not localized in instrumental coordinates it is necessary to evaluate the strength of the source during each pointing in order to properly model its impact on the data.  The general strategy is not to create a perfect model of the sky but rather to approximate a model of the sky.  Any inaccuracies in this model should not be localized in instrumental coordinates and as a result should tend to become evenly distributed as one averages over many pointings.  Instrumental effects such as inaccuracies in the exposure calibration are localized in this coordinate system and as a result tend to accumulate.

\begin{figure}[h]
\epsfysize=5.0in
\centerline{\epsfbox{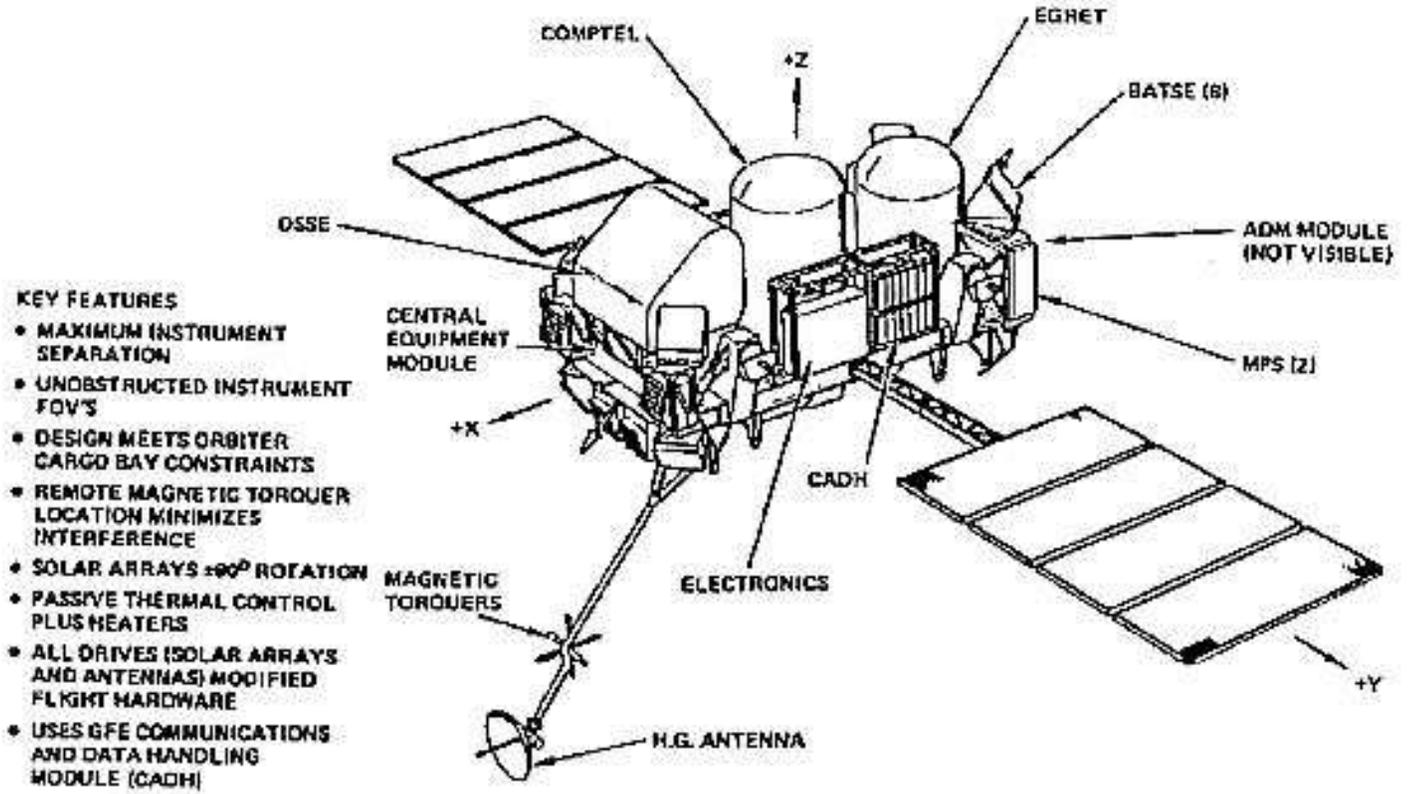}}
\caption{Schematic of the \cgro spacecraft showing the spacecraft axes and the instrument positions.}
\label{cgro}
\end{figure}

\begin{figure}[h]
\epsfysize=3.0in
\centerline{\epsfbox{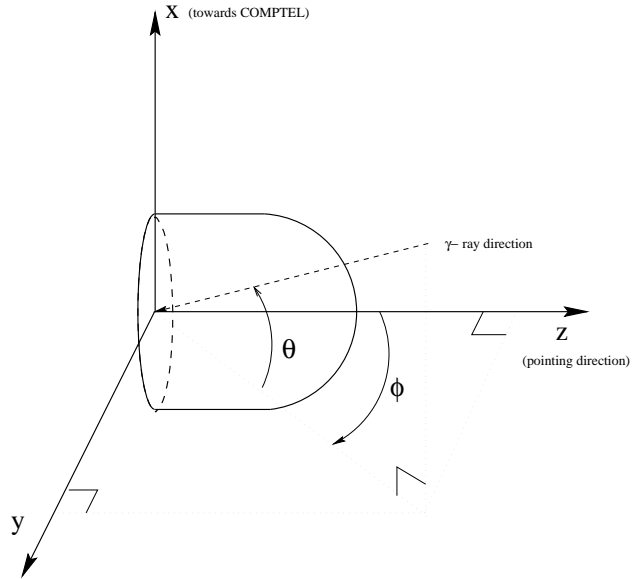}}
\caption{Definition of the instrumental coordinates used in this section.  They are chosen so as not to have a singularity along the pointing direction.  $\theta,\phi$ are analogous to $l,b$ with the $x$-axis representing the polar axis and the $z$-axis representing the galactic center axis.}
\label{instcoord}
\end{figure}

	It is necessary first to define a set of instrumental coordinates.  This choice is dictated by the fact that we do not want a coordinate singularity in the center of the field of view.  As shown in \fig{instcoord} the coordinates are defined by,
\begin{equation}
	\sin(\theta)=\hat{r} \cdot \hat{x} \;,
\end{equation}
and,
\begin{equation}
	\cos(\phi)=\frac{(\hat{r} \times \hat{x})}{\cos(\theta)} \cdot \hat{z}\;,
\end{equation}
where $\hat{x},\hat{y},\hat{z}$ are the spacecraft axes and $\hat{r}$ is the photon arrival direction.  These coordinates are exactly analogous to galactic coordinates with the spacecraft $\hat{x}$-axis corresponding to the galactic pole and the $\hat{z}$-axis corresponding to the galactic center direction.  As a result $\phi,\theta$ will often be referred to as instrumental longitude and instrumental latitude respectively.  It should be noted that the $\hat{z}$-axis is given by $(\phi,\theta)=(0,0)$, the $\hat{x}$-axis is given by $(\phi,\theta)=(0,90)$, and the $\hat{y}$-axis is given by $(\phi,\theta)=(90,0)$.  Also note that the COMPTEL instrument lies along the $+\hat{x}$-axis while the $-\hat{x}$-axis is directed toward the edge of the spacecraft.

	For each pointing, the coordinate transformation matrix is constructed to convert a vector in galactic or celestial coordinates into instrumental coordinates.  This matrix is essentially the product of two rotation matrices. A map of exposure in instrumental coordinates is constructed from the standard map in galactic or celestial coordinates using the following technique.  The center position of each bin in the instrumental map is transformed into galactic coordinates.  The exposure per steradian at that point is calculated from the interpolation between the nearest bins in the celestial exposure map.  The exposure is assumed to vary slowly across a $.5^\circ \times .5^\circ$ bin so that the exposure is simply the bin size times this value.  This same technique is employed in order to construct a map of sky intensity in instrumental coordinates.  The maps are multiplied to produce a map of predicted counts in each instrumental bin which can be compared to the data which can be directly binned in instrumental coordinates.

\begin{figure}[p]
\epsfysize=3.0in
\centerline{\epsfbox{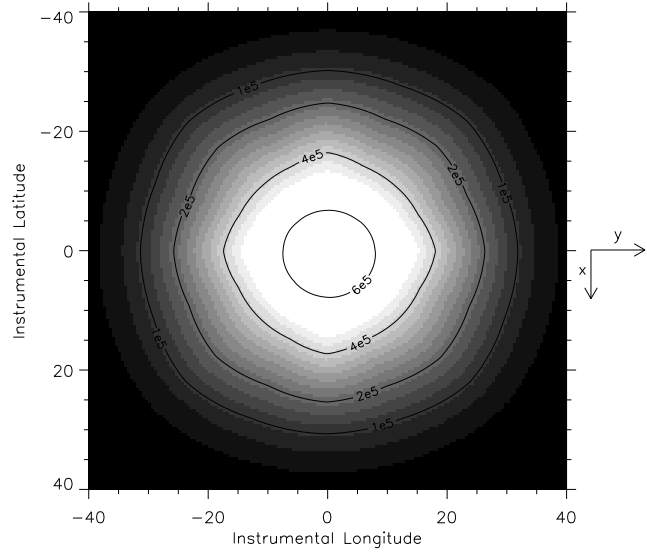}}
\caption{Exposure map in instrumental coordinates.  The energy range is $E>100 MeV$.}
\label{exposr.inst}
\end{figure}

\begin{figure}[p]
\epsfysize=3.0in
\centerline{\epsfbox{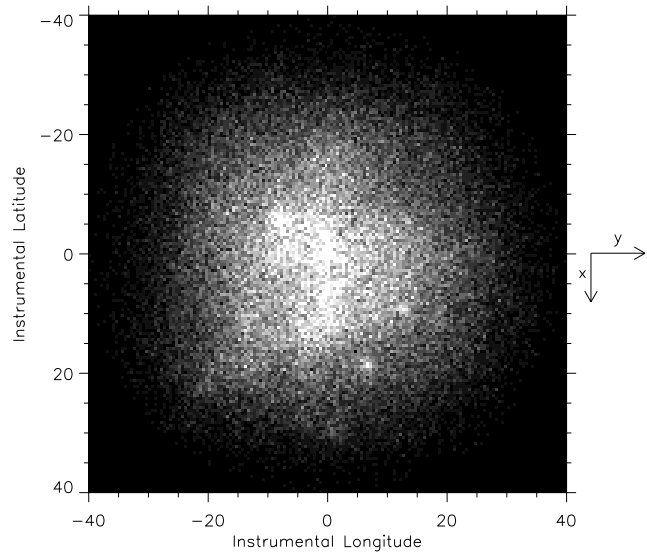}}
\caption{Measured counts in instrumental coordinates.  The energy range is $E>100 MeV$.}
\label{counts.inst}
\end{figure}

	\fig{exposr.inst} and \fig{counts.inst}  show the exposure and counts maps in instrumental coordinates for a set of 62 coadded viewing periods.  The `latitude' and `longitude' angles are those defined in \fig{instcoord}.  The departure of the exposure map from azimuthal symmetry is caused by the embedded square structure of the spark chamber and calorimeter.  The following cuts have been made on the photons in the SMDB:
\begin{itemize}
\item Measured Energy $E>100\;{\rm MeV}$.
\item Energy in TASC greater than threshold of 6 MeV.
\item Photon does not arrive during an excluded interval.
\item Zenith angle $\alpha<80^\circ$.
\end{itemize}
The total number of photons included in this study is 174,000.


\begin{figure}[p]
\epsfysize=3.0in
\centerline{\epsfbox{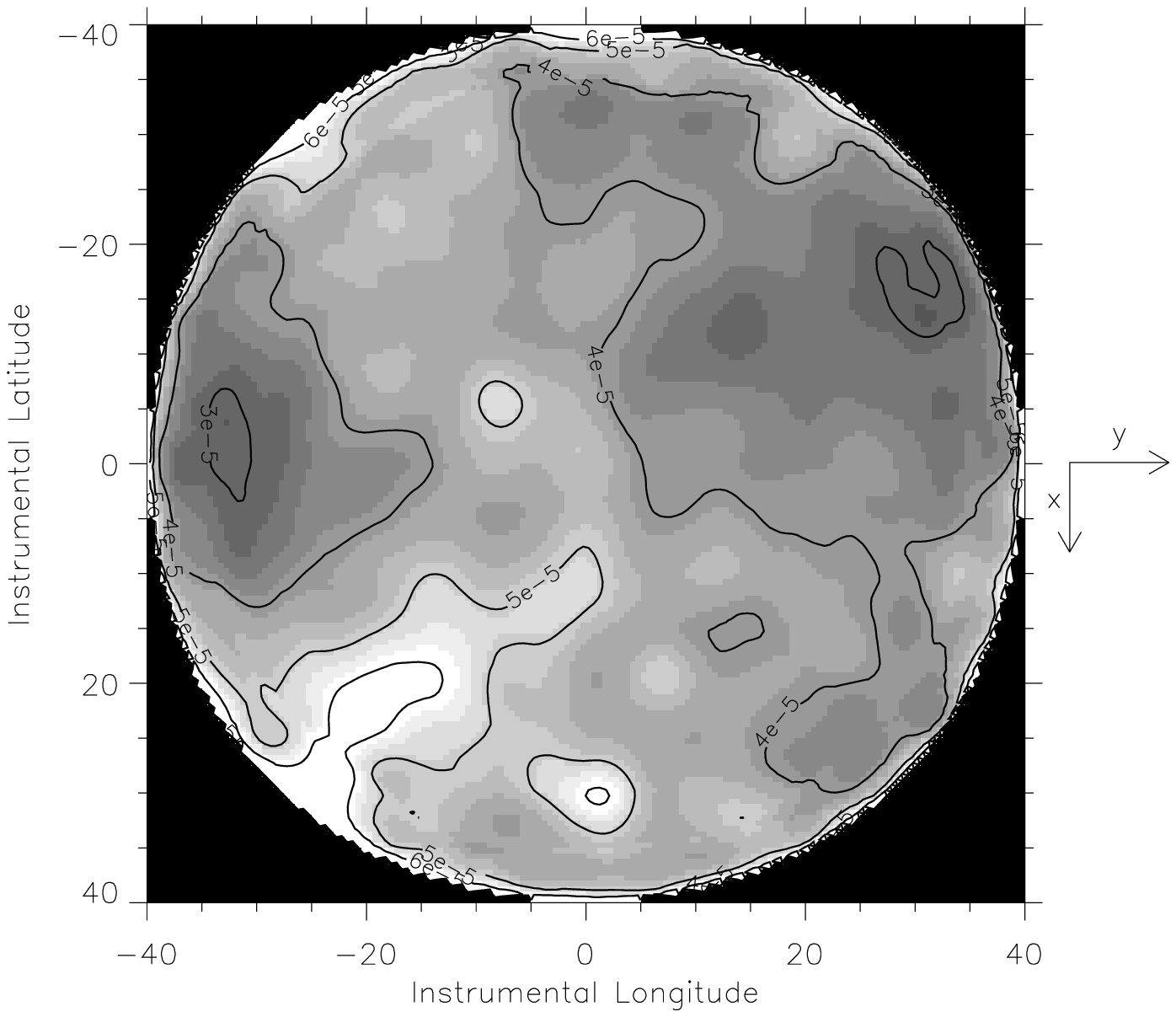}}
\caption{Measured intensity in instrumental coordinates.  The energy range is $E>100 MeV$.}
\label{measured.inst}
\end{figure}

\begin{figure}[p]
\epsfysize=3.0in
\centerline{\epsfbox{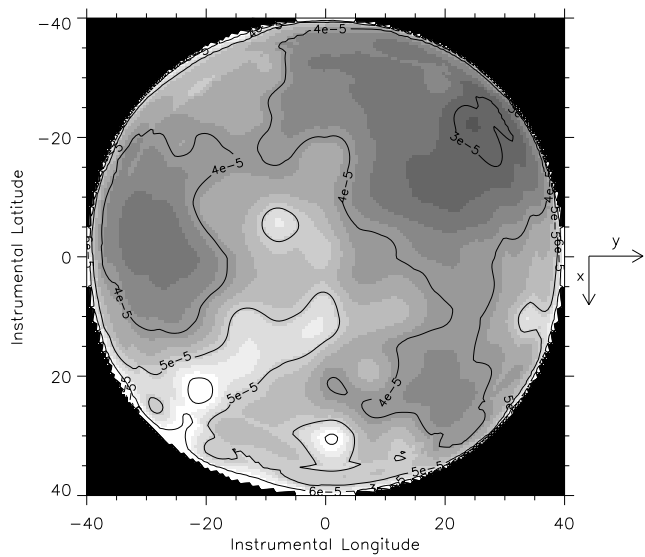}}
\caption{Modelled intensity for the same set of viewing periods shown above.  The model used is the Bertsch et al. model (see Chapter 4). A time variable set of point sources have been included as has an isotropic diffuse background.}
\label{modelled.inst}
\end{figure}

	The measured intensity is shown in \fig{measured.inst}.  The modeled intensity is shown in \fig{modelled.inst}.  In order to minimize the effect of the counting statistics, both maps are convolved with a Gaussian smoothing function with a $2^\circ$ half-width.  The viewing periods for this study were selected based on two criteria.  All viewing periods within $80^\circ$ of the Galactic Center were excluded because the galactic model is known to be inaccurate in this region.  Furthermore, no viewing periods were included in which the galactic ridge lay in the $x-z$ or $y-z$ plane.  This was done so as to decouple any symmetries in the celestial features with the instrumental symmetries.  Unfortunately, this eliminated many galactic viewing periods because due to the constraints of the \osse] detector which points in the $xz$ plane, the galactic ridge is often polarized in one of these two planes.

\begin{figure}[p]
\epsfysize=3.0in
\centerline{\epsfbox{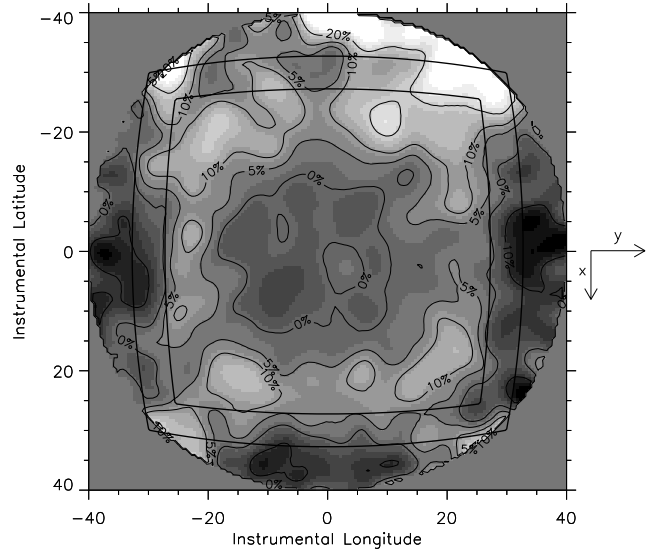}}
\caption{Normalized residual intensity in instrumental coordinates. The values plotted are defined in equation 3.15. A clear instrumental pattern is evident.  The squarish profile shows the projection of the top of the spark chamber into these coordinates to emphasize the instrumental symmetry.}
\label{resid.inst}
\end{figure}

\begin{figure}[p]
\epsfysize=3.0in
\centerline{\epsfbox{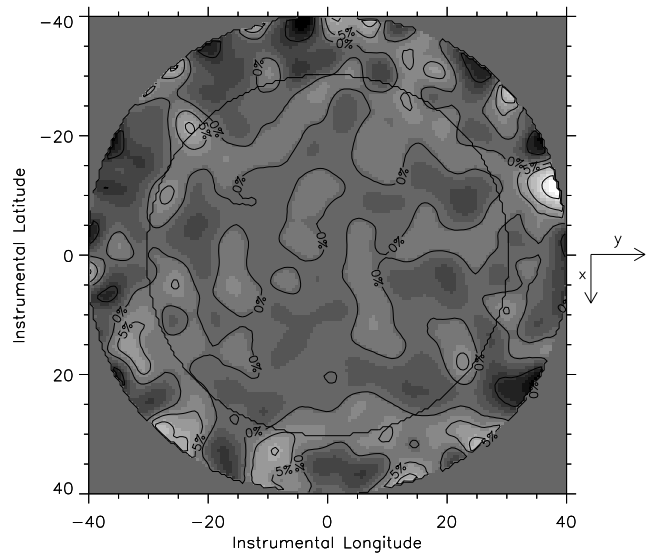}}
\caption{A residual map of a Poisson sample of the background model indicating the level of noise in the above map.  The major features are clearly significant.}
\label{noise.inst}
\end{figure}

	The instrumental effect is measured as the fractional residual at each location in instrumental coordinates,
\begin{equation}
	R_{ij}=\frac{C_{ij}-M_{ij}}{M_{ij}} \;,
\end{equation}
where $C_{ij},M_{ij}$ are the measured and modelled counts in the bin $ij$.  \fig{resid.inst} shows the normalized residual map for the data set shown above.  There is clear evidence of features correlated with the instrumental structure.  \fig{noise.inst} shows the normalized residuals for a Poisson sample of the model in order to indicate the level of statistical fluctuation present in this map. The features seen in the data clearly cannot be statistical fluctuations. 


	There are two possible explanations for these features: background contamination or exposure calibration inaccuracies.  If the former explanation holds true, the excesses seen in the residuals are due to some sort of contamination which is particularly concentrated in certain areas of the field of view.  Because this would imply that the contaminating intensity is uncorrelated to the \gammaray\ signal, the residual intensity should be independent of the \gammaray\ intensity of the sky being observed.  This assumption can be tested by seeing whether the residual intensity at a particular position in the field of view correlates with the residual intensity at the same instrumental position measured using a very different set of pointings which give different intrinsic sky intensities at that point.  On the other hand if the residual intensity is due to a miscalibration of the exposure, the excess intensity will depend on the intrinsic celestial \gammaray\ intensity.  In this case the normalized residual should correlate between different sets of pointings. 
	
\begin{figure}[p]
\epsfysize=2.0in
\centerline{\epsfbox{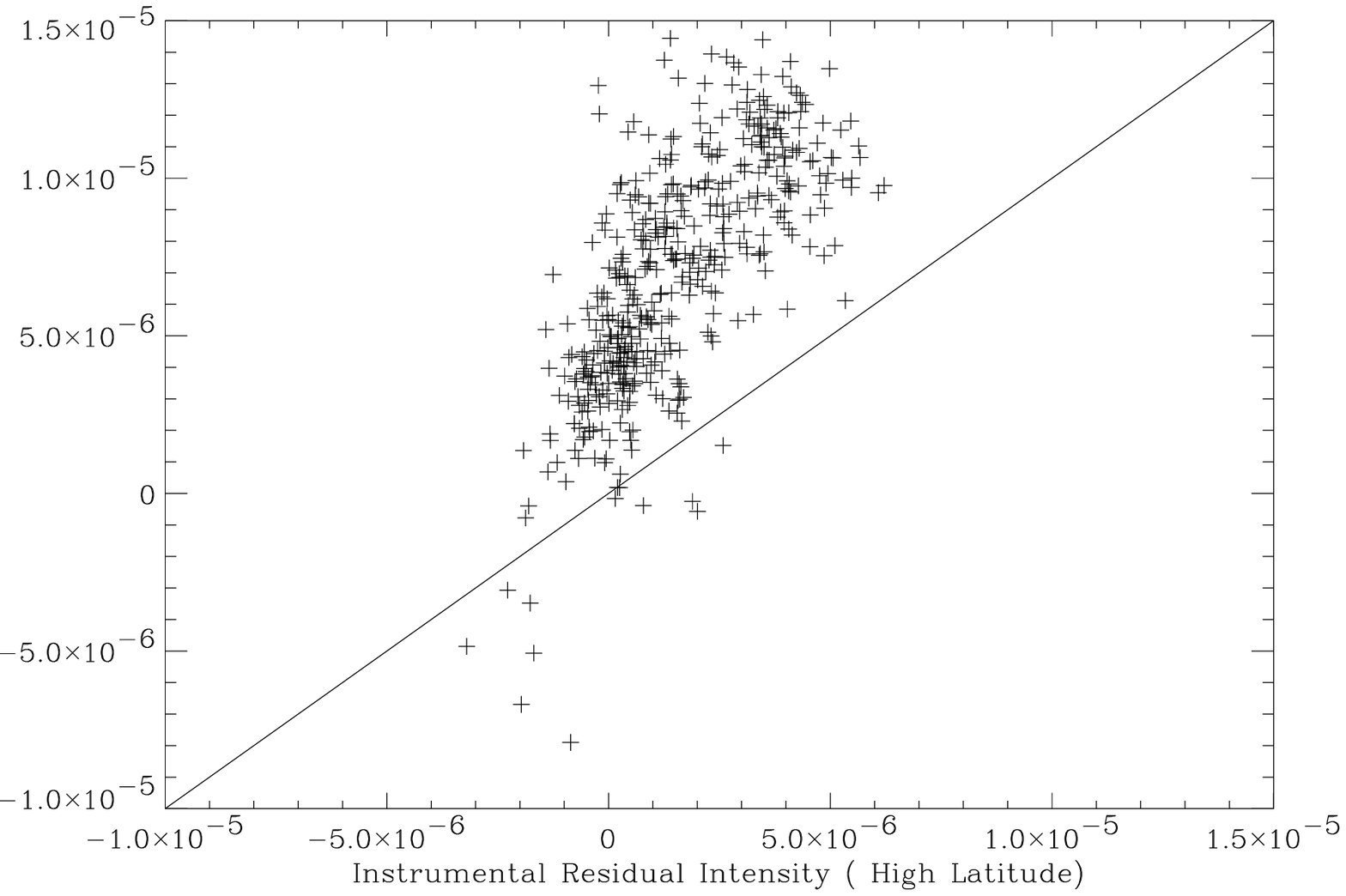}}
\caption{A correlation between the instrumental residual intensity maps for two independent data sets.  The vertical axis represents the measured residual intensity in a set of galactic plane viewing periods.  The horizontal axis represents the same values calculated using high latitude viewing periods.}
\label{intens.corr}
\end{figure}

\begin{figure}[p]
\epsfysize=2.0in
\centerline{\epsfbox{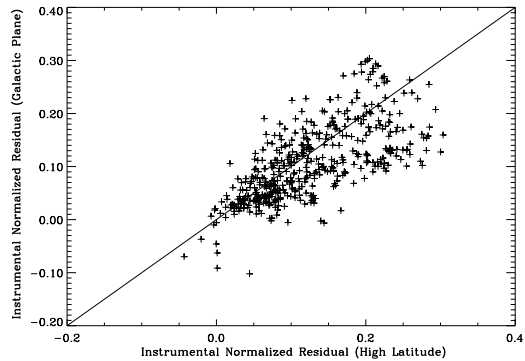}}
\caption{A correlation between normalized instrumental residual maps for two independent data sets.  The vertical axis represents the normalized residual intensity in a set of galactic plane viewing periods.  The horizontal axis represents the same values calculated using high latitude viewing periods.  The correlation is clearly superior to that shown above indicating that the exposure calibration is more likely to be a source of the inaccuracies than contaminating radiation.}
\label{resid.corr}
\end{figure}

	This test was done using a set of high latitude pointings and a set of galactic plane pointings.  The two data sets were completely independent and the modelled intensities were much higher for the galactic plane pointings.  In each case the residual intensities were calculated in instrumental coordinates as were the normalized residuals in instrumental coordinates.  \fig{intens.corr} shows the instrumental residual intensity for the galactic plane pointings plotted against the same quantity from the high latitude map.  The correlation is very poor.  In contrast, \fig{resid.corr}, shows the correlation using the normalized residual maps.  The correlation is much better.  This test provides strong evidence in favor of the source of the residuals being an exposure miscalibration.

	The first feature of note is the pronounced hole along the $\pm\hat{y}$-axis and the $+\hat{x}$-axis at inclination angles of $> 30^\circ$.  In order to understand this, the projection of the top plane of the spark chamber into these coordinates is shown in \fig{resid.inst} as the outer squarish contour.  It is evident that there is a correlation between the symmetry of the detector and the instrumental residuals.  Furthermore, there are residual excesses that form a sqarish feature at inclinations greater than $\sim15\%$ which is most pronounced along the lines which lie at $45^\circ$ to the $\hat{x}$ and $\hat{y}$ axes.  This effect is at worst $\sim20\%$.

	The preflight calibration of \egret\ was intended to be completed to yield accuracies within $\sim 1\%$.  The source of the features that are apparent in \fig{resid.inst} are poorly understood.  Some suggestions follow.

	The instrument was only calibrated in one octant.  The calibration for the remaining 7 octants was obtained through symmetry operations.  There is an apparent variation from corner to corner in the residuals of about 10\%.  This could be due instrument asymmetries that were not detected.

	The most likely source of this calibration inaccuracy is the convolution of the model with the PSF.  The correct calculation of the measured count rate as a function of instrumental position, $C(\theta,\phi)$, is as follows.
\begin{equation}
C(\theta,\phi)=\int_{\theta^{\prime},\phi^{\prime}} I(\theta^{\prime},\phi^{\prime}) \; {\cal E}(\theta^{\prime},\phi^{\prime}) \; PSF(\theta^{\prime},\phi^{\prime};\theta,\phi) \; d\Omega^{\prime}  \;,
\end{equation}
where $I(\theta,\phi)$ represents the intrinsic sky intensity,  ${\cal E}(\theta,\phi)$ represents the instrumental exposure in a given direction, and $PSF(\theta^{\prime},\phi^{\prime};\theta,\phi)$ is the appropriate point spread function.  It is, tedious, however to carry out this convolution for every instrumental pointing and thus the analysis is expedited by calculating one convolved background model,
\begin{equation}
I_{{\rm Conv}}(\theta,\phi)=\int_{\theta^{\prime},\phi^{\prime}} I(\theta^{\prime},\phi^{\prime}) \; PSF(\theta^{\prime},\phi^{\prime};\theta,\phi) \; d\Omega^{\prime} \;, 
\end{equation}
leading to a modelled counts map,
\begin{equation}
M(\theta,\phi)= I_{{\rm Conv}}(\theta,\phi) \; {\cal E}(\theta,\phi) \; .
\end{equation}

	In the limit of a featureless background, the resultant discrepancy between measured counts and modelled counts is given by,
\begin{equation}
	R(\theta,\phi)=\frac{[\int_{\theta^{\prime},\phi^{\prime}} {\cal E}(\theta^{\prime},\phi^{\prime}) \; PSF(\theta^{\prime},\phi^{\prime};\theta,\phi) \; d\Omega^{\prime} - {\cal E}(\theta,\phi)]}{{\cal E}(\theta,\phi)} 
\end{equation}
In other words, the residuals map is given by the difference between the exposure map convolved with the PSF and the unconvolved exposure map.  In the limit of a symmetric PSF, this does not introduce substantial error, however, the \egret\ PSF is not symmetric because of the `fish-eye' effect.  The `fish-eye' correction ensures that point sources are correctly positioned, but it does distort the background.

\begin{figure}[h]
\epsfysize=3.0in
\centerline{\epsfbox{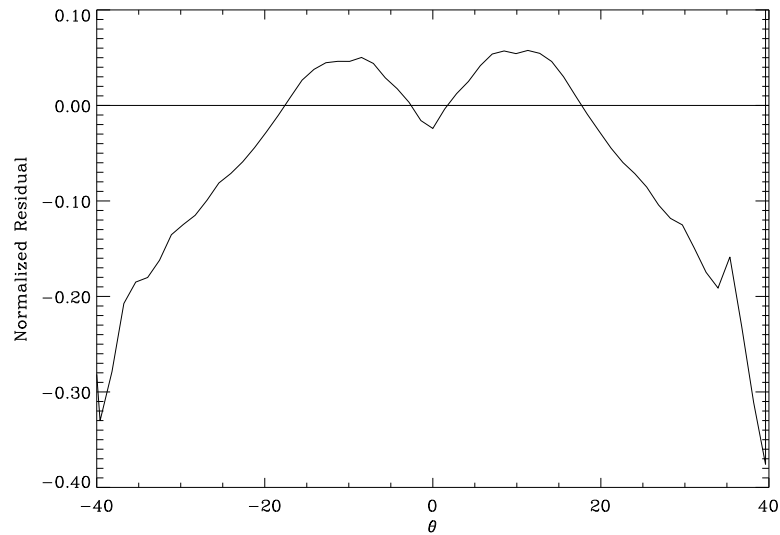}}
\caption{Residual introduced by convolution with a skewed Gaussian PSF.  The plot represents a slice through the residuals map along the y-axis in instrumental coordinates.  The features apparent in the measured data residuals are reproduced.}
\label{Expconv}
\end{figure}

	Figure \fig{Expconv} shows the residuals calculated using equation 3.16.  For the purposes of this demonstration, a skewed Gaussian PSF is used.  The features apparent in the flight data are reproduced using this method.  It should in principal be possible to exactly reproduce these features given knowledge of how the PSF skewness varies across the instrument.  For the purposes of this study, it is sufficient to correct the exposure maps in order to remove this instrumental effect.  This technique is sometimes called `flat-fielding'.  It is important to note that this is only
applicable for high latitude data for which the sky intensity is roughly uniform.

\section{Summary}

	The in-flight instrumental performance of \egret\ has been studied in detail. Instrumental background due to cosmic ray and neutron backgrounds in the energy range $E>100\;{\rm MeV}$ have been shown to be smaller than $4.5\times10^{-7}$.  This amounts to better than an order of magnitude signal to noise ratio for the detection of the IDGRB.  This demonstrates the excellent selectivity with which the various \egret\ hardware and software systems reject the high flux of cosmic rays in low earth orbit.
	
	The albedo \gammarays\ from the earth's limb are more problematic.  Using an albedo cut of $\alpha<80^\circ$ minimizes this problem but still results in a background intensity above 100 MeV of $6\times10^{-7}$.  This background is concentrated in certain regions near the celestial pole requiring particular care when analyzing the diffuse background in these areas.

	Finally, the exposure calibration has been reviewed using the flight data.  Significant discrepancies between the flight data and the preflight calibration have been revealed.  These discrepancies are most likely due to PSF skewness but can be corrected for through the `flat-fielding' of the exposure maps.

\chapter{ Diffuse $\gamma$ Radiation}

\section{Galactic Diffuse Radiation}

	The measurement of a diffuse source of galactic \gammarays\ dates back to the earliest days of \gammaray\ astronomy.  A bright feature in a narrow band along the galactic plane was first seen by $OSO 3$ (\cite{Kraushaar72}) as well as by a high altitude balloon instrument (\cite{Fichtel72}).  The subsequent instruments ($SAS\;2$ and $COS\;B$) were able to verify the correlation between \gammaray\ features and galactic structural features (\cite{Fichtel78}, \cite{Strong88}).

	This radiation was quickly interpreted based on its spatial distribution and spectrum as being associated with cosmic-ray/matter interactions and to a lesser extent cosmic-ray/photon interactions.  The most compelling evidence for this origin of the galactic diffuse radiation was provided by the strong correlation between galactic matter tracers such as 21 cm radio emission and diffuse \gammaray\ emission (\cite{Fichtel78}).  It was also realized \gammarays\ could provide valuable information about the distribution of cosmic-rays in the galaxy because of the relative transparency of the galaxy to \gammarays.

	In this chapter, a model of the diffuse galactic emission will be presented.  This model was developed at Goddard Space Flight Center by a team of scientists led by D.L.Bertsch.  This model will then be used to extract a measurement of the IDGRB.

\section{Cosmic-ray Interactions and \Gammaray\ production}

	As discussed previously, cosmic-rays are relativistic charged particles.  Most evidence points to a galactic source of the cosmic rays at least up to energies of a few TeV. Most theories speculate that supernovae are the predominant sources of these particles (\cite{Biermann95}).

	The interactions between such particles and galactic matter and photons are well understood in the \egret\ energy range from terrestrial accelerator measurements.  The dominant processes which yield \gammarays\ are: 
\begin{itemize}
\item hadronic showers yielding pions which decays to \gammarays.
\item cosmic ray electron bremsstrahlung.
\item inverse Compton interactions between relativistic electrons and soft photons.
\end{itemize}
These three processes are characterized by the three \gammaray\ production functions $q_{pion}(E),q_{brem}(E),q_{IC}(E)$ which measure the differential \gammaray\ energy spectrum produced per target nucleon (or photon in the case of the inverse Compton function).  These functions are determined using an assumed cosmic ray spectrum which is measured locally.

\subsection{Nuclear Interactions with matter}

	The dominant source of \gammarays\ above $\sim 70\;{\rm MeV}$ is the decay of neutral and charged pions.  Pions are produced in hadronic showers with center of mass energies greater than the rest mass of a pion.  Neutral pions decays directly into 2 \gammarays\ whereas charged pions decay into electrons and positrons which in turn annihilate to produce \gammarays.  Several calculations of the \gammaray\ production function from the decay of pions, $q_{pion}$, have been made (\cite{Carvallo71}, \cite{Stecker70}, \cite{Dermer86}). This function has a broad maximum at the rest energy of the pion, 68 MeV. 

\subsection{Electron Bremsstrahlung}

	The electrons also contribute to the production of \gammarays\ through bremsstrahlung.  The bremsstrahlung production function, $q_{brem}$, is difficult to ascertain because this interaction becomes significant at lower energies where the local electron spectrum is difficult to measure due to the solar modulation. Fichtel et al. (\cite{Fichtel91}) have estimated the local electron spectrum using \gammaray\ data matched with synchrotron data.  Their resultant bremsstrahlung production function becomes the dominant source of \gammarays\ at energies below $\sim20 \;{\rm MeV}$.

\subsection{Inverse Compton}

	Cosmic ray electrons can also boost soft photons up to \gammaray\ energies through the inverse Compton process.  \Gammarays\ produced through the inverse Compton process have a mean energy which is a function of the product of the electron energy and the photon energy,
\begin{equation}
	\overline{E_{\gamma}}=\frac{4}{3}\gamma^2\overline{E_{{\rm photon}}}=\frac{4}{3}(\frac{E_e}{m_ec^2})^2 \overline{E_{{\rm photon}}} \;,
\end{equation}
where $E_{\gamma},E_e,E_{{\rm photon}}$ are the energies of the resultant \gammaray, incident electron, and incident photon respectively.  The two factors of $\gamma$ result from the Doppler shift of the target boost of the photon into the electron's rest frame and then the subsequent Doppler boost back into the observer's rest frame.  In order to produce a 100 MeV \gammaray\ using cosmic ray electrons from with energies from 1 GeV to 100 GeV requires target photons with wavelengths from ( $50 \;{\rm nm} -500\; \mu {\rm m}$).  As a result the important photon distributions include the microwave background, Far Infra-Red (FIR), Near Infra-Red (NIR), optical, and UV.  

\section{Galactic Diffuse \Gammaray\ Model}

	The \egret\ team has developed a galactic diffuse \gammaray\ emission model (\cite{Bertsch93}).  This model was revised and analyzed spectrally by (\cite{Hunter96}).  The model will be outlined here.  The galactic diffuse \gammaray\ intensity in a given direction $(l,b)$, is given by the line of sight integral,
\clearpage
\begin{eqnarray}
j(E,l,b)=\frac{1}{4\pi}&\times & \int [c_e(\rho,l,b)q_{brem}(E) + c_n(\rho,l,b)q_{pion}(E)] \nonumber \\
& \times & [n_{HI}(\rho,l,b) + n_{HII}(\rho,l,b) + n_{H_2}(\rho,l,b)] d\rho \nonumber \\
& + & \frac{1}{4\pi}\sum_i  \int  c_e(\rho,l,b)q_{IC}(E,\rho)u_{IC}(\rho,l,b) d\rho \;.
\end{eqnarray}
In the above equation, $q_{pion},q_{brem},q_{IC}$ represent the \gammaray\ production functions for nuclear, bremsstrahlung, and inverse Compton processes discussed above.  These functions measure the \gammaray\ production per target atom when exposed to the local cosmic ray flux. The number densities of atomic hydrogen, ionized hydrogen, and molecular hydrogen are denoted by $n_{HI}$,$n_{HII}$,and $n_{H_2}$.  These densities are the matter densities at the position $(\rho,l,b)$.  The cosmic rays densities are also functions of position in the galaxy.  The functions $c_e(\rho,l,b),c_n(\rho,l,b)$ are the ratios of the electron and nucleon cosmic ray intensities relative to the local intensities.  These functions could also depend on energy if the spectral indices vary with location within the galaxy.  The second line of sight integral accounts for the inverse Compton contributions.  The production functions are calculated for four independent photon energy bands  resulting in a summation over photon energy band.  The functions $u_i(\rho,l,b)$ reflect the photon densities in the $i^{th}$ band.

\subsection{Matter Distribution}

	Equation 4.2 provides a recipe for calculating the \gammaray\ intensity given knowledge of the matter, cosmic ray, and photon distributions everywhere in the galaxy.  In this section the state of knowledge of the matter distributions will be discussed.

	The bulk of the galactic mass is in the form of atomic hydrogen, HI.  The distribution of this component is traced using the 21 cm hyperfine transition of hydrogen.  Galactic rotation induces Doppler shifts which allow the three dimensional structure of this gas to be determined. Knowledge of the galactic rotation curve allows one to place the atomic hydrogen along a given line of sight at a given distance based on the projected velocity of that position along the line of sight.  In the inner galaxy (i.e. within the solar galactocentric radius) there is a near-far ambiguity that must be resolved.  Matter at a given velocity can lie at either of two intersection points between the line of sight and the relevant line of constant galactocentric radius.  In this case the {\em ad hoc} assumption is made to split the gas evenly between each position.

	In the inner galaxy there is a considerable of matter in the form of molecular hydrogen, $H_2$.  This gas is not directly observable.  The best available tracer for galactic molecular matter is the 2.6mm $J=1 \rightarrow 0$ rotational transition in CO molecules (\cite{Kutner85},\cite{Maloney88}).  Observations of this line can be deconvolved in the same way as the 21cm data in order to construct a model of the CO distribution in the galaxy. There remains the task of converting this measurement into a map of the $H_2$ density.  The conversion factor between CO line intensity and $H_2$ density is given by the X-factor, $X$.  This value is a free parameter in the model presented here.

\begin{figure}[p]
\epsfysize=7.0in
\centerline{\epsfbox{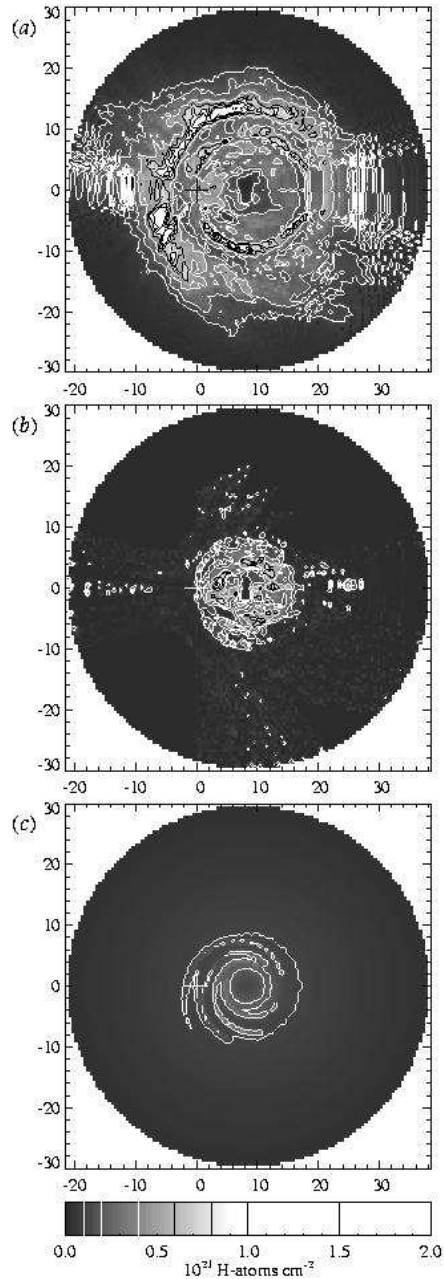}}
\caption{A reproduction of a figure from Hunter et al. (1996) showing the two dimensional plots of the matter distributions as seen looking face on our galaxy.  The scale indicates the distance from the solar position which is marked with a cross.  Figure (a) shows the distribution of HI, figure (b) shows the distribution of $H_2$ and figure (c) shows the distribution of ionized hydrogen, HII.}
\label{2d_sd_plot_1}
\end{figure}

	The radio survey data used to construct the galactic matter distribution are described in (\cite{Bertsch93}).  The deconvolved matter distributions that are constructed using this data are shown in \fig{2d_sd_plot_1} which is reproduced from (\cite{Hunter96}).  The spiral arm structure of the galaxy is clearly visible.  The dominance of the molecular hydrogen in the inner galaxy is also evident.

\subsection{\Gammaray\ Production in the Galactic Plane}

	In order to use this matter distribution to predict the galactic diffuse \gammaray\ emission, the distribution of cosmic rays throughout the galaxy must be known in order to perform the integral in equation 4.2.  Because there is no direct measurement of cosmic ray density available except for the local measurement, some assumption about cosmic ray distribution must be adopted.  The assumption used in generating the diffuse \gammaray\ model is that the cosmic rays are in dynamic balance with the magnetic fields and the gravitational attraction of the galaxy (\cite{Bertsch93}).  The result of this balance is a coupling between cosmic rays and matter in the galaxy.  The cosmic ray densities used to predict the diffuse \gammaray\ emission are thus derived through convolving the matter distribution with a characteristic smoothing function which is the second free parameter in the model.

\begin{figure}[p]
\epsfysize=6in
\centerline{\epsfbox{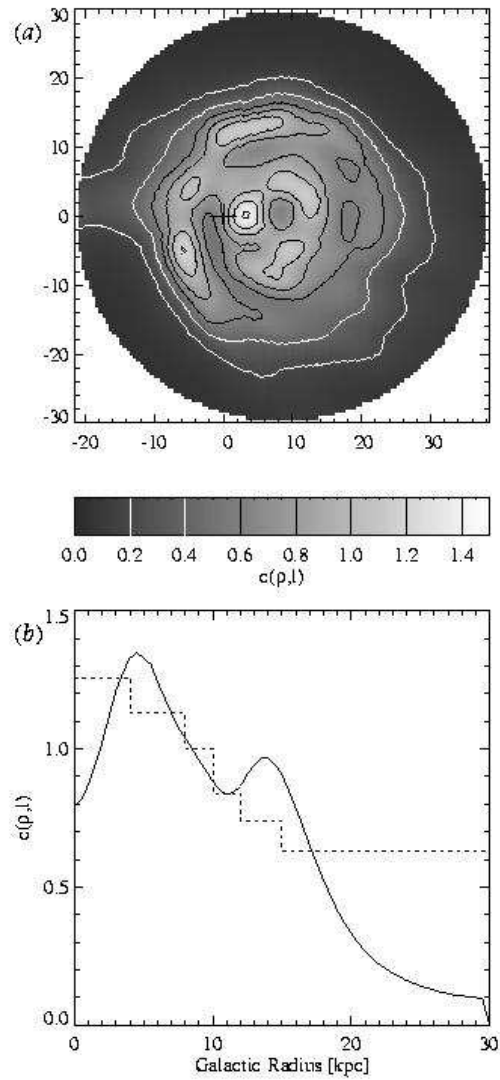}}
\caption{A reproduction of a figure from Hunter et al. 1996 showing two dimensional plots of the cosmic ray distributions as seen looking face on our galaxy.  The scale indicates the distance from the solar position which is marked with a cross.  Figure (a) shows the distribution of cosmic rays using the best fit parameters from the model. Figure (b) shows a comparison between the galactic radial distribution of cosmic rays using matter/cosmic ray coupling (solid line) and assuming azimuthal symmetry to the cosmic rays (dotted line).}
\label{2d_cr_plot}
\end{figure}

	\fig{2d_cr_plot} shows the resultant cosmic ray distribution which is derived fitting the two free parameters in the model to the \egret\ data.  The details of this analysis are described in (\cite{Hunter96}).  The similarity between the map of cosmic rays and the input matter distribution is the direct result of the assumption about a coupling between these two quantities. 

	An alternative approach to the determination of the cosmic ray distribution is to instead assume an azimuthally symmetric cosmic ray distribution in the galactic plane.  The gradient of this distribution can then be determined from the \gammaray\ data.  This approach was used by Strong et al. (\cite{Strong88}) in analyzing the \cosb\ data. \fig{2d_cr_plot} shows a comparison between the radial cosmic ray distribution as determined under the assumption of dynamic balance from the \egret\ data, and the cosmic ray gradient determined by Strong et al. using \cosb\ data.  In general the azimuthally symmetric method yields a smaller cosmic ray gradient than is predicted by dynamic balance.

\section{High Latitude Diffuse Emission}

	The problem of modeling the diffuse intensity at high latitudes presents a slightly different set of challenges.  \cite{Sreekumar96} have extended the galactic diffuse emission model to latitudes $\mid b\mid > 10^\circ$. The density of neutral galactic gas (molecular and atomic) falls off rapidly away from the plane of the galaxy.  As a result, the neutral gas along lines of sight with $b>10^\circ$ is dominated by gas within a few hundred pc of the solar position.  Local cosmic ray conditions are assumed to be valid at this range.  This assumption obviates the need to deconvolve the matter distributions at high latitudes.  The contribution from neutral gas is thus readily calculated using equation 4.2 where we use local electron and nucleon cosmic ray spectra and assume the value for the $X$ parameter to be 1.5 consistent with the  derived value from the galactic plane.  

	The cosmic ray distributions undoubtedly extend to much larger distances from the galactic plane than does the neutral gas.  As a result, the diffuse emission along high latitude lines of sight contains significant contributions from \gammarays\ produced in interactions well above the galactic plane. There are three relevant interactions that need to be modeled: interactions between cosmic ray nucleons and ionized hydrogen (HII), interactions between cosmic ray electrons and HII, and inverse Compton interactions between cosmic ray electrons and the soft galactic radiation field.

	In order to model the contributions from these large scale height components it is necessary to construct a model of the cosmic ray distribution above the galactic plane as well as the distribution of the various targets.  The cosmic rays are assumed to be galactic in origin with most models predicting the source of all but the highest energy cosmic rays to be supernovae.  The diffusion perpendicular to the plane of the galaxy is characterized by a cosmic ray scale height, $z_p,z_e$.  A lower limit on the electron scale height can be extracted from galactic synchrotron radio data (\cite{Osborne95}).  This upper limit has been shown to be $\sim 1\;{\rm kpc}$.  In an attempt to make a minimal set of assumptions in constructing our model, the cosmic ray electron scale height $z_e$ is assumed to be at the lower limit of the possible range: $z_e=1\;{\rm kpc}$.  Furthermore, the cosmic ray nucleon scale height is assumed to be identical: $z_p=z_e$.  Lastly the electron and nucleon cosmic ray spectra are assumed to be identical. A three dimensional model of the cosmic ray distribution is then constructed by projecting the cosmic ray densities, inferred from the galactic plane data, to higher latitudes using the scale height of 1 kpc.  Uncertainties associated with these assumptions will be discussed in more detail below.

	The only available constraints on the distribution of ionized hydrogen come from dispersion measures of radio pulsars.  Taylor \& Cordes (\cite{Taylor93}) have used a population of pulsars to constrain a model for the distribution of ionized matter in the galaxy which is adopted for this work.

	As discussed previously, the important the important photon distributions include the microwave background, Far Infra-Red (FIR), Near Infra-Red (NIR), optical, and UV.  With the exception of the microwave background radiation, the photon distributions are spatially dependent and not very well constrained.  The following input assumptions constitute the assumed photon radiation field,
\begin{itemize}
\item Blackbody radiation: isotropic at a temperature of 2.7 K.
\item FIR: the cold dust emission model of Cox, Krugel \& Metzger (1996) with the total FIR luminosity of the galaxy normalized to $1.5\times10^{10}L_{\odot}$.  The resultant distribution is radially symmetric with a peak at $\sim4\;{\rm kpc}$ and a scale height of 100 pc
\item (NIR+Optical+UV):  These components are stellar in origin and the model of Chi \& Wolfendale (1991) is used.
\end{itemize}

	Given these models for the cosmic ray distribution, matter distribution, and photon distribution, equation 4.2 is again used in order to generate diffuse \gammaray\ intensities along each line of sight.

\section{Model Predictions}

\begin{figure}[p]
\epsfxsize=0pt \epsfysize=2.5in
\centerline{\epsfbox{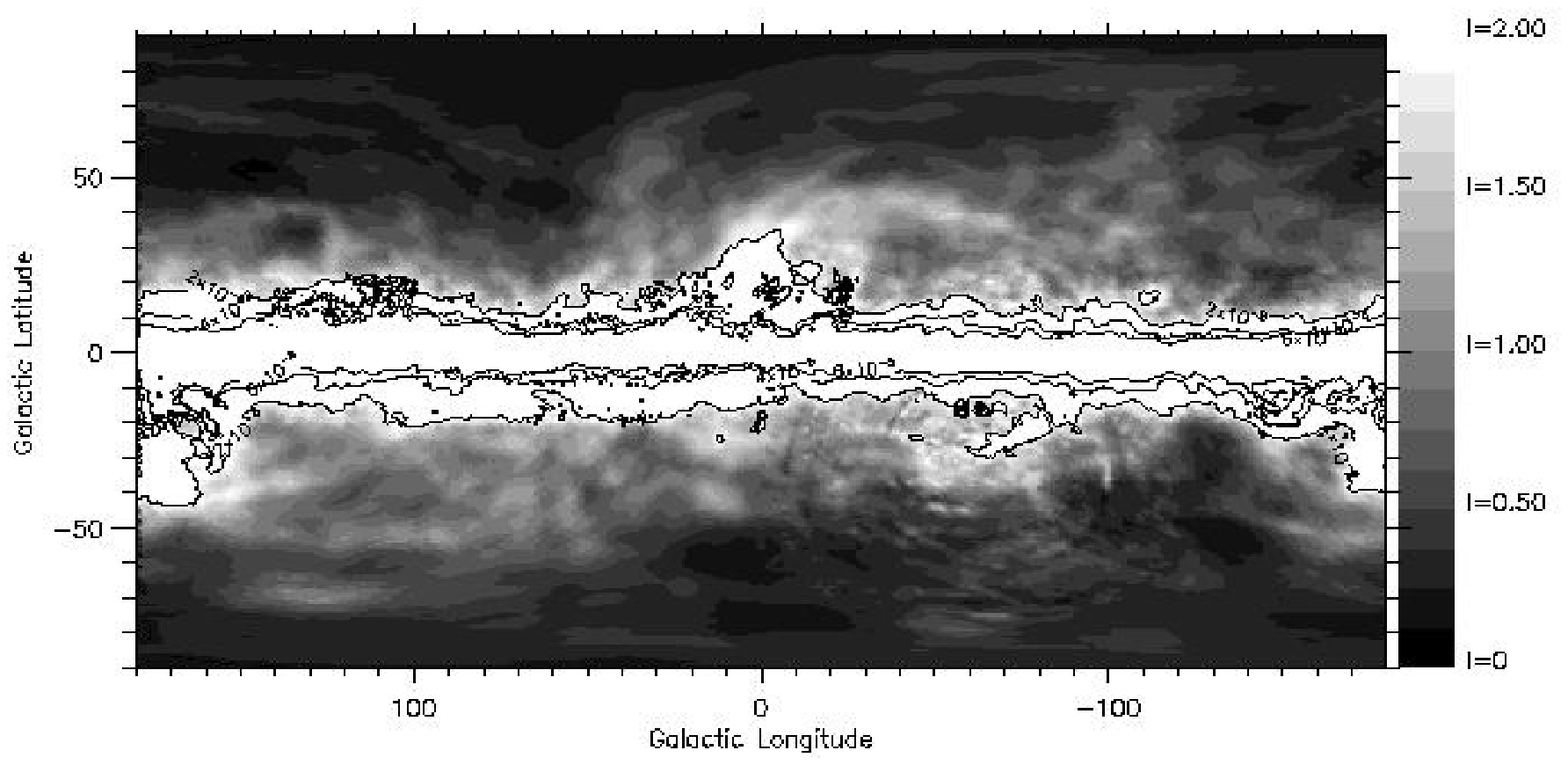}}
\centerline{Diffuse Intensity Due to HI and ${\rm H}_2$}
\epsfxsize=0pt \epsfysize=2.5in
\centerline{\epsfbox{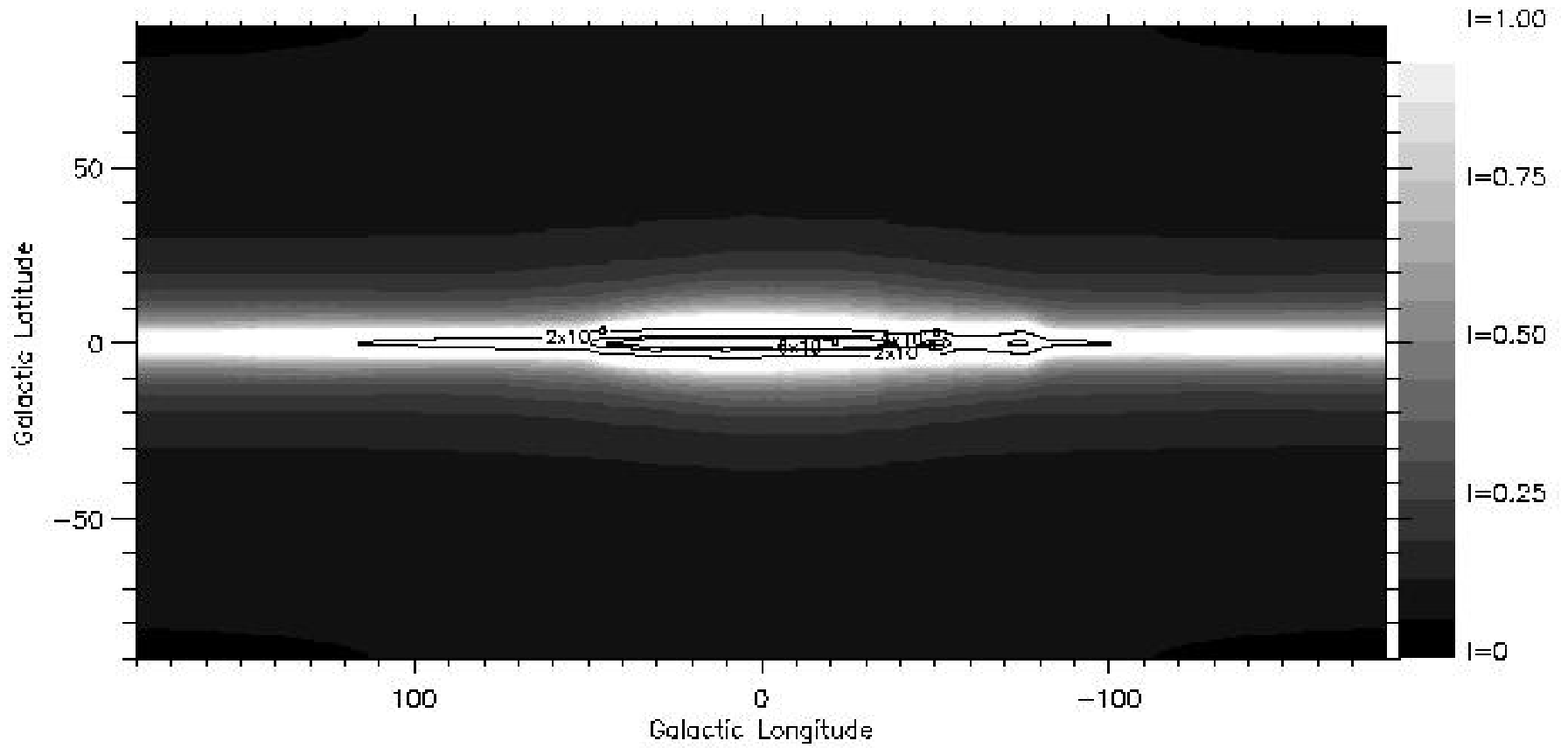}}
\centerline{Diffuse Intensity Due to HII}
\epsfxsize=0pt \epsfysize=2.5in
\centerline{\epsfbox{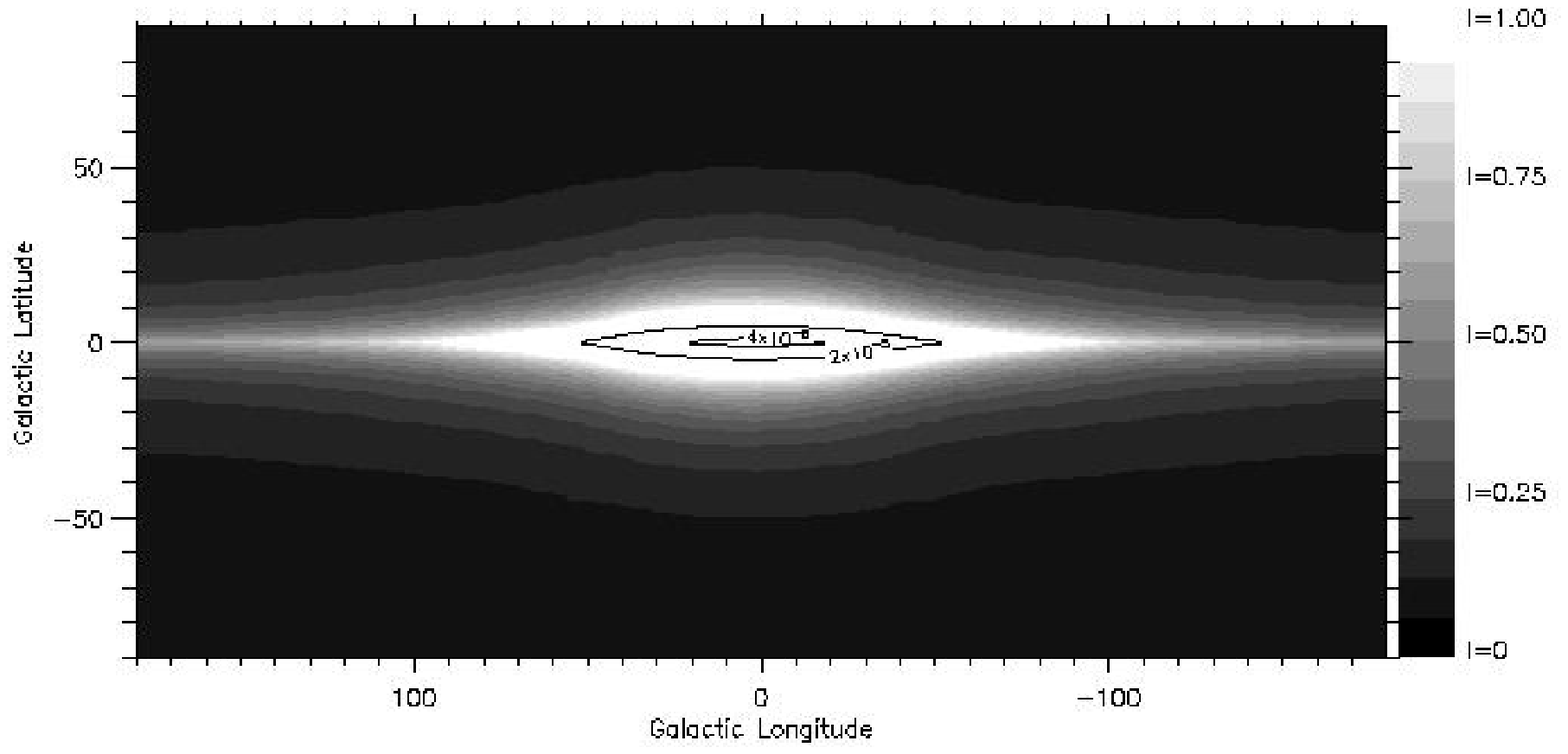}}
\centerline{Diffuse Intensity Due to Inverse Compton Scattering}
\caption{The galactic contributions to the diffuse \gammaray\ emission.  The scales are in units of $10^{-5} {\rm ph}\;{\rm cm}^{-2}{\rm s}^{-1} {\rm sr}^{-1}$}
\label{components}
\end{figure}

 \fig{components} shows the three principal components to the galactic diffuse \gammaray\ emission.  The first panel shows the contribution from neutral atomic and molecular hydrogen.  Because we are only interested in the high latitude diffuse emission the color scale saturates on the galactic plane but reaches values as high as $8\times10^{-4}$ in the galactic center.  The second panel shows the contribution from HII and the third shows the inverse Compton component.  All three panels show the integrated flux with $E>100\; {\rm MeV}$.

\section{Measurement of the IDGRB}

\subsection{Likelihood Analysis}

	The combined \egret\ data set from the first four years of observations was used for the diffuse analysis described below.  The photons used in this study were selected using the same criteria used in the instrumental calibration.  This selection encompasses the standard selection criteria for $E>100\;{\rm MeV}$ except that the more restrictive albedo cut of $\alpha<80^{\circ}$ is used to avoid albedo contamination. Furthermore, only photons within $30^\circ$ of the instrument pointing were included.  A total of 630,416 photons resulted from these cuts.

	These photons were binned in celestial coordinates using $.5^\circ\times.5^\circ$ bins.  Celestial coordinates were chosen so as to avoid coordinate singularities at the galactic poles.  The exposure maps were calculated for the appropriate data cuts.  These maps were combined to generate a full sky exposure map.  Two such maps were generated: one using the preflight calibration and a second using the in-flight correction outlined in the previous chapter.  The diffuse model outlined above was convolved with the \egret\ point spread function to generate an observational background model.

	These three maps were analyzed using the \egret\ mmaximum likelihood software (LIKE v 4.5) (\cite{Mattox96}) (see also Appendix A).  This package searches for point sources using the maximum likelihood method.  To this end, a 3 free parameter fit was made to the data at every point on the sky. The 3 parameters that were estimated using this method were the source flux ($S$) of a putative source at a given location ($l_0,b_0$) as well as 2 diffuse parameters ($I_{{\rm IDGRB}},g_m$) which measure the IDGRB intensity and the galactic diffuse emission model scaling, respectively. 
\begin{equation}
	I(l,b)= S\times PSF(l,b;l_0,b_0) + g_mG(l,b) + I_{{\rm IDGRB}} \;.
\end{equation} 
These 3 parameters were optimized at a given point using data within $15^\circ$ of that point.

	Following the procedures used to generate the 2nd \egret\ source catalog (\cite{Thompson96}), this process was carried out until all sources with significances $\sigma \geq 4$ were found. The number of sources isolated using this method was 147. The point source detections will be discussed in detail next chapter.  Maps of the 2 diffuse parameters were also generated.  

	If the 2 diffuse sources were totally independent, the maps of the diffuse parameters would provide measurements of the IDGRB intensity.  However at high latitudes, the relative featurelessness of the galactic diffuse emission causes the diffuse parameters to be highly correlated.  As a result, it is only possible to extract the sum of the two components.

\subsection{Residual Emission}
\begin{figure}[p]
\epsfxsize=0pt \epsfysize=3.0in
\centerline{\epsfbox{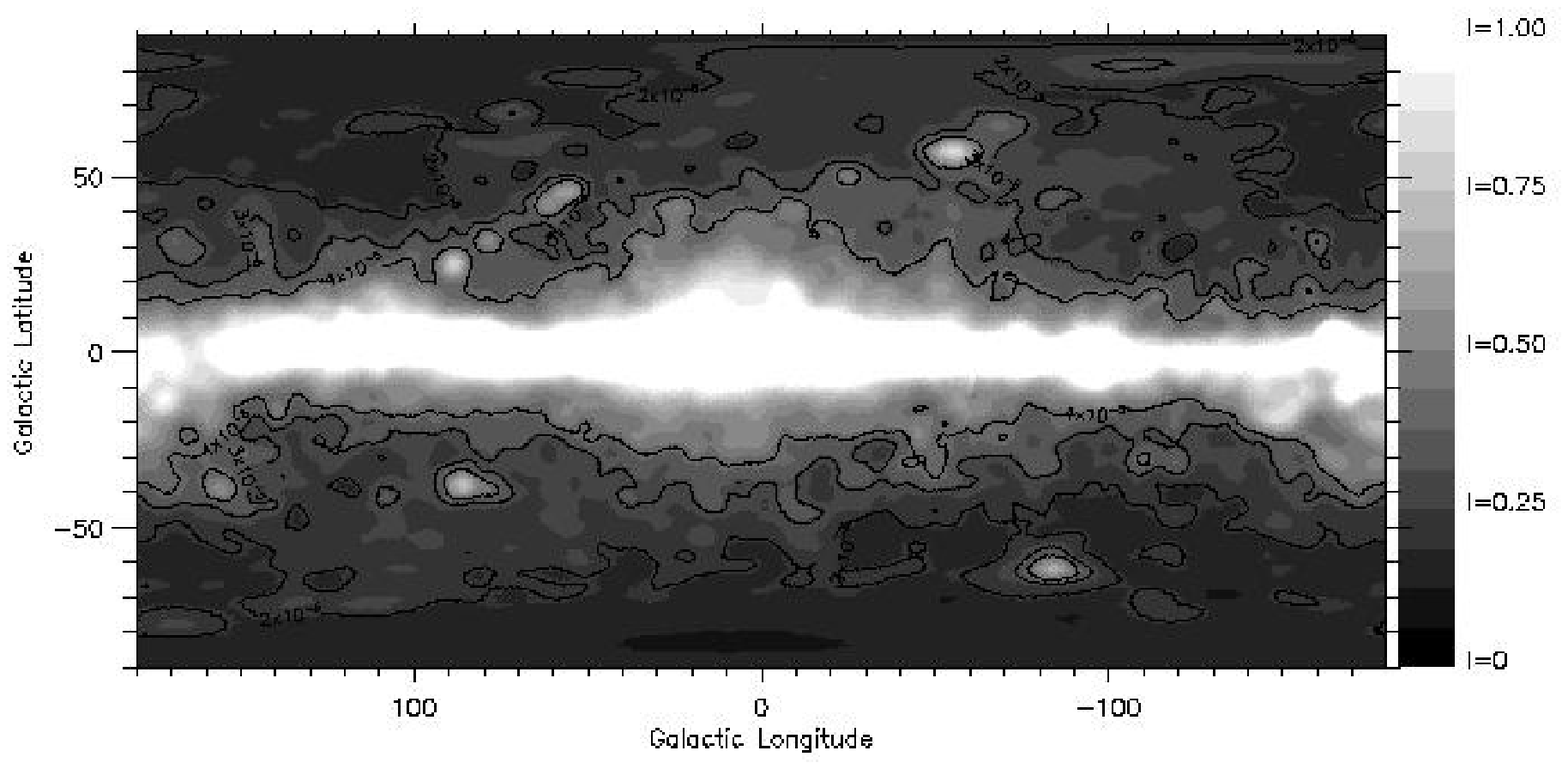}}
\centerline{Measured Intensity}
\epsfxsize=0pt \epsfysize=3.0in
\centerline{\epsfbox{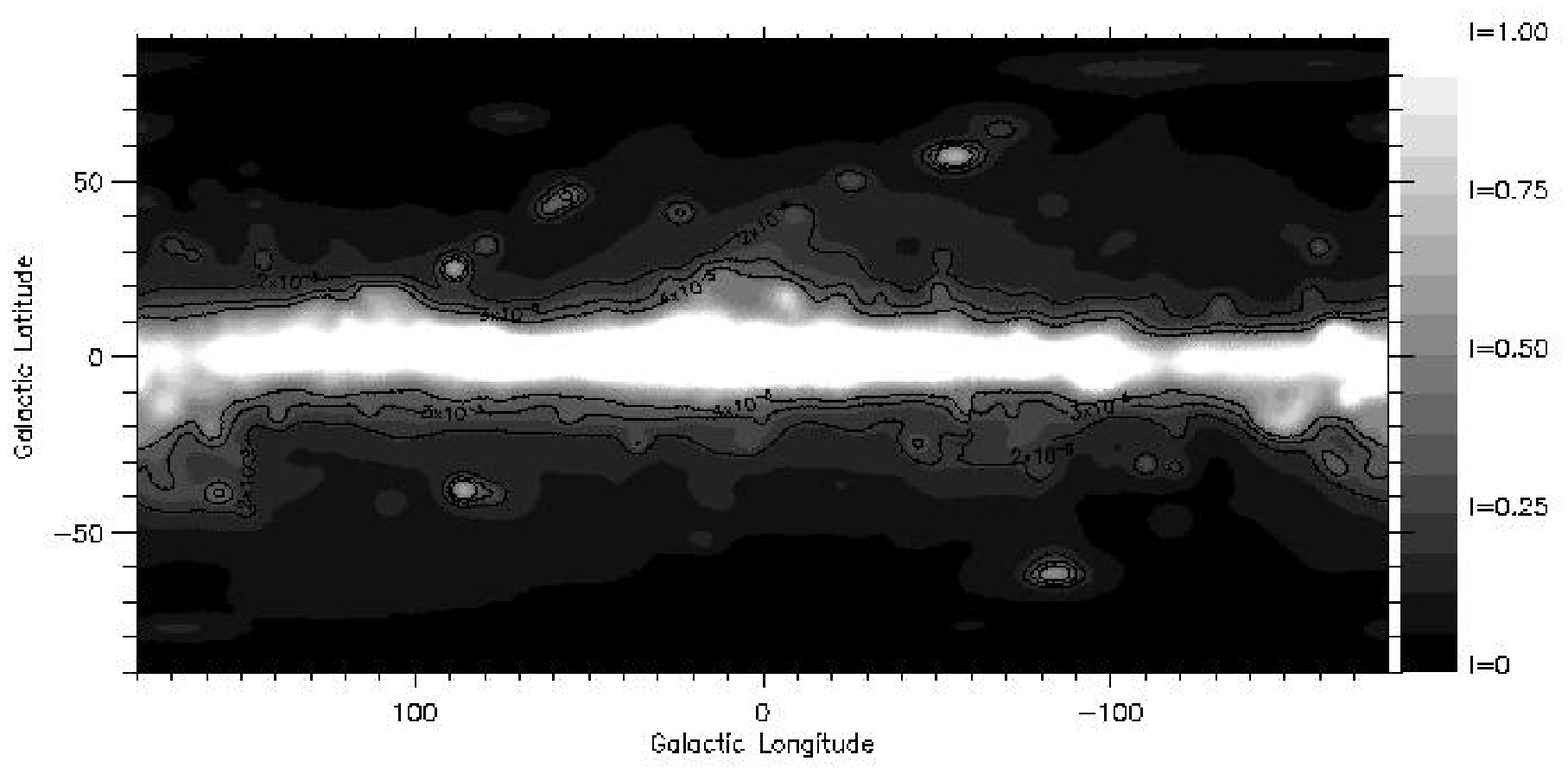}}
\centerline{Modeled Galactic Diffuse Intensity with Resolved Point sources}
\caption{Measured intensity and modeled galactic diffuse intensity.  The likelihood reconstructed point sources with significances $\sigma \geq 4$ are also included.  The scales are in units of $10^{-4} {\rm ph}\;{\rm cm}^{-2}{\rm s}^{-1} {\rm sr}^{-1}$}
\label{raw}
\end{figure}

	In order to find the level of isotropic diffuse emission that is consistent with the \egret\ data, the galactic diffuse model is directly correlated with the measured intensity.  To properly account for the effect of the point sources, the point sources isolated using the likelihood technique are added to the galactic diffuse model using the appropriate point spread function.  \fig{raw} shows the measured intensity as well as the modeled intensity including the point sources reconstructed using maximum likelihood.  The intensities in each case have been smoothed using a Gaussian with a $5^\circ$ width in order to decrease the Poisson noise and the galactic plane has again been saturated in order to reveal the structure at high latitude.

	The first panel in \fig{residuals} shows the difference between these two quantities.  This is a map of all the intensity in the sky that is not associated with resolved point sources and is not accounted for in the galactic diffuse model.  There is clear evidence of excess diffuse emission at high latitudes.

	In order to ascertain the level of the IDGRB a set of $6^{\circ}\times6^{\circ}$ in celestial coordinates are used.  The measured intensity is correlated to the galactic diffuse model intensity in these pixels:
\begin{equation}
	I_{ij}=g_mG_{ij} + I_{{\rm IDGRB}} \;.
\end{equation}
The parameters $g_m,I_{{\rm IDGRB}}$ are estimated by linear regression.  A one parameter fit is also performed using a fixed value of $g_m=1$.

\begin{figure}[h]
\epsfysize=4.0in
\centerline{\epsfbox{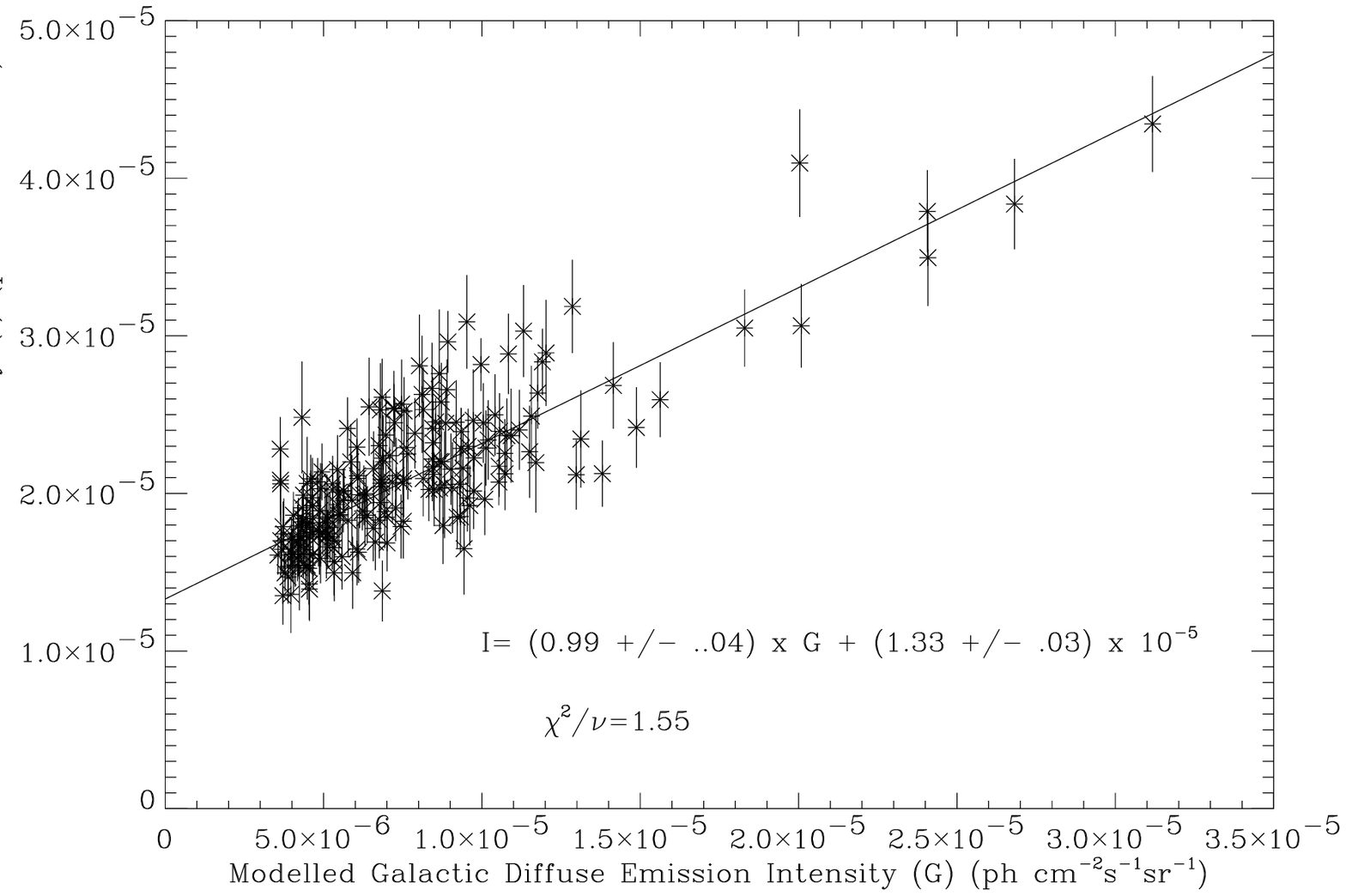}}
\caption{Measured diffuse intensity versus modeled galactic diffuse intensity for a set of high latitude pixels.  Regions within $90^\circ$ of the galactic center are excluded.  The extrapolation to zero galactic diffuse intensity yields a measurement of $I_{{\rm IDGRB}}=1.33 \pm .03 \times 10^{-5}$.  The departure of $\chi^2/\nu$ from 1.0 indicates that local diffuse model inaccuracies remain.}
\label{extrapolate}
\end{figure}

	While in theory, one could use data from the entire sky to contribute to this analysis, this approach is complicated by several factors.  Firstly, the galactic plane emission is very intense compared with the corresponding high latitude intensities.  Small inaccuracies in the galactic model at low latitudes could severely bias the analysis and as a result regions within $30^\circ$ of the galactic plane have been excluded.  Secondly, regions in the direct vicinity of strong point sources are dominated by the emission from such sources and the residual emission is likely to be dominated by small inaccuracies in the fitting of the point source.  As a result, regions within $3^\circ$ of any of the 147 resolved point sources are excluded.

	It is evident from inspection of the residual intensity in \fig{residuals} that there is a prominent feature surrounding the galactic center which is not the result of isotropic emission.  In order to gauge the effect of this feature on the measurement of the IDGRB, regions within $\theta_{\rm GC}$ of the galactic center are excluded from the study.  $\theta_{\rm GC}$ is allowed to vary from $40^\circ$ to $110^\circ$ in order to determine whether the effect of this feature extends to large viewing angles.

\begin{table}[ht]
\centering
\caption
{Residual intensity as a function of sky region.}
\bigskip
\begin{tabular}{l|cc|ccc}\hline\hline
& \multicolumn{2}{c|}{1 Free Parameter} & \multicolumn{3}{c}{2 Free Parameters} \\ \hline
$\theta_{{\rm GC}}$ &  $I_{{\rm IDGRB}}$ &  $\chi^2/\nu$ &  $I_{{\rm IDGRB}}$& $g_b$ &  $\chi^2/\nu$ \\ 
$(^\circ)$ &  $\times10^{-5}$ &     &  $\times10^{-5}$&  &   \\ \hline

110	& $1.23 \pm .01$& 1.6	&	$1.30 \pm .04$	& $.98 \pm .05$	& 1.6 \\
100	& $1.26 \pm .01$& 1.4	&	$1.34 \pm .03$	& $.95 \pm .05$	& 1.4 \\
90	& $1.28 \pm .01$& 1.6	&	$1.33 \pm .03$	& $.99 \pm .04$	& 1.6 \\
80	& $1.29 \pm .01$& 1.9	&	$1.32 \pm .03$	& $1.02 \pm .04$ & 1.8 \\
70	& $1.32 \pm .01$& 2.3	&	$1.30 \pm .03$	& $1.07 \pm .04$ & 2.3 \\
60	& $1.32 \pm .01$& 2.5	&	$1.30 \pm .02$	& $1.10 \pm .03$ & 2.4 \\
50	& $1.35 \pm .01$& 2.6	&	$1.27 \pm .02$	& $1.18 \pm .03$ & 2.5 \\
40	& $1.37 \pm .01$& 2.7	&	$1.24 \pm .02$	& $1.24 \pm .03$ & 2.5 \\ \hline \hline
\end{tabular}

\label{IDGRB}
\end{table}

	Table 4.1 summarizes the results of this study. For values of $\theta_{\rm GC}<90^\circ$ the values of $g_m$ depart significantly from 1, indicating that the galactic diffuse model is incorrectly predicting the \gammaray\ intensity.  However, the fact that the values of $\chi^2/\nu$ depart significantly from 1 in these regions indicates that the data are inconsistent with the additional emission being isotropic.  For values of $\theta_{\rm GC}>90^\circ$ the influence of this additional emission is not important.  In these regions, the gas-map appears to correlate well with the measured intensities as evidenced by values of $g_m$ very close to 1.  Furthermore, values of $\chi^2/\nu$ close to 1 indicate that the additional emission is consistent with the hypothesis that this emission is isotropic in nature.  \fig{extrapolate} shows the explicit correlation for $\theta_{\rm GC}=90^\circ$.  The measured value of $I_{\rm IDGRB}$ is $(1.33 \pm 0.03)\times10^{-5}\;{\rm ph}\;{\rm cm}^{-2}{\rm s}^{-1} {\rm sr}^{-1}$.

\begin{figure}[p]
\epsfxsize=0pt \epsfysize=2.5in
\centerline{\epsfbox{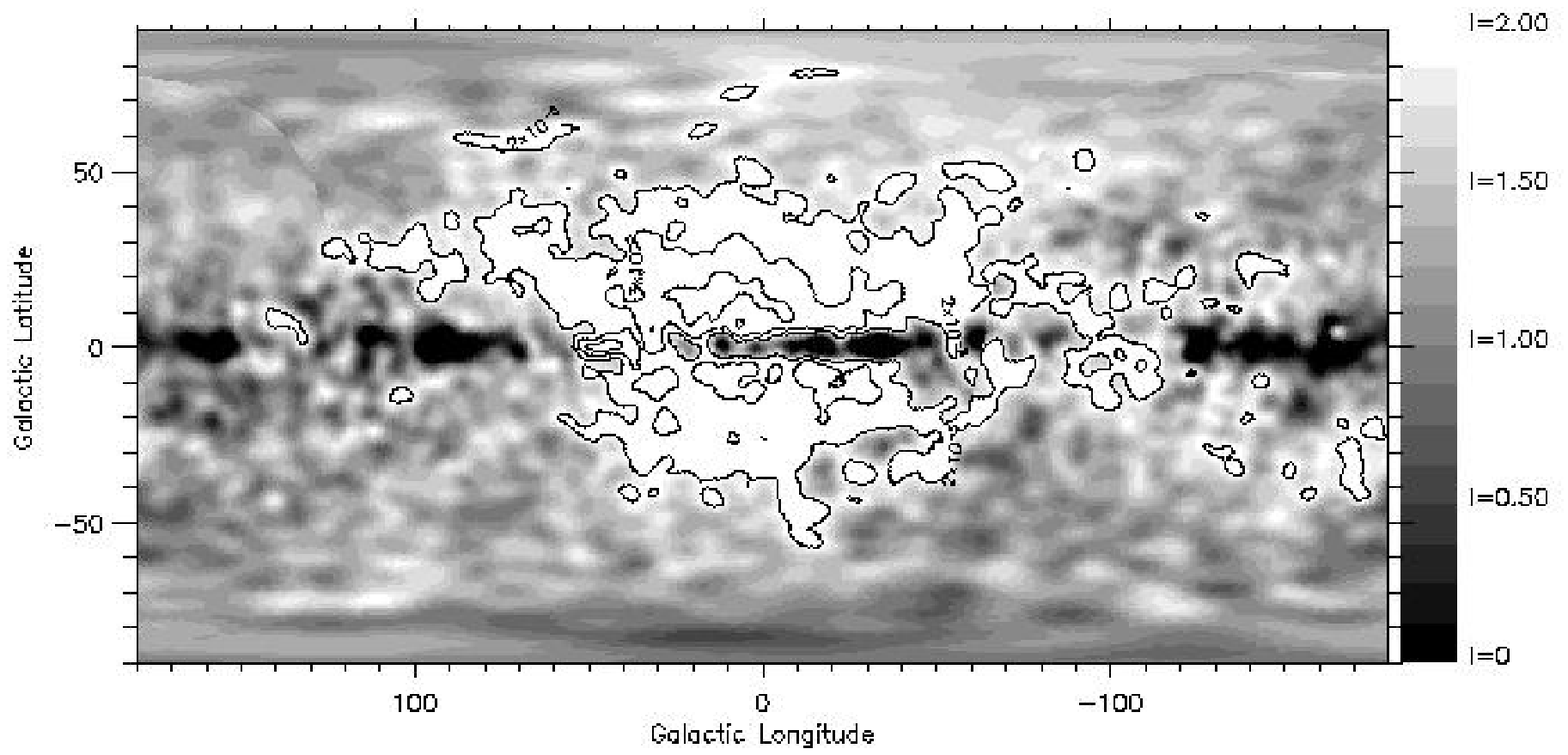}}
\centerline{Residual Intensity using the high latitude galactic diffuse model}
\epsfxsize=0pt \epsfysize=2.5in
\centerline{\epsfbox{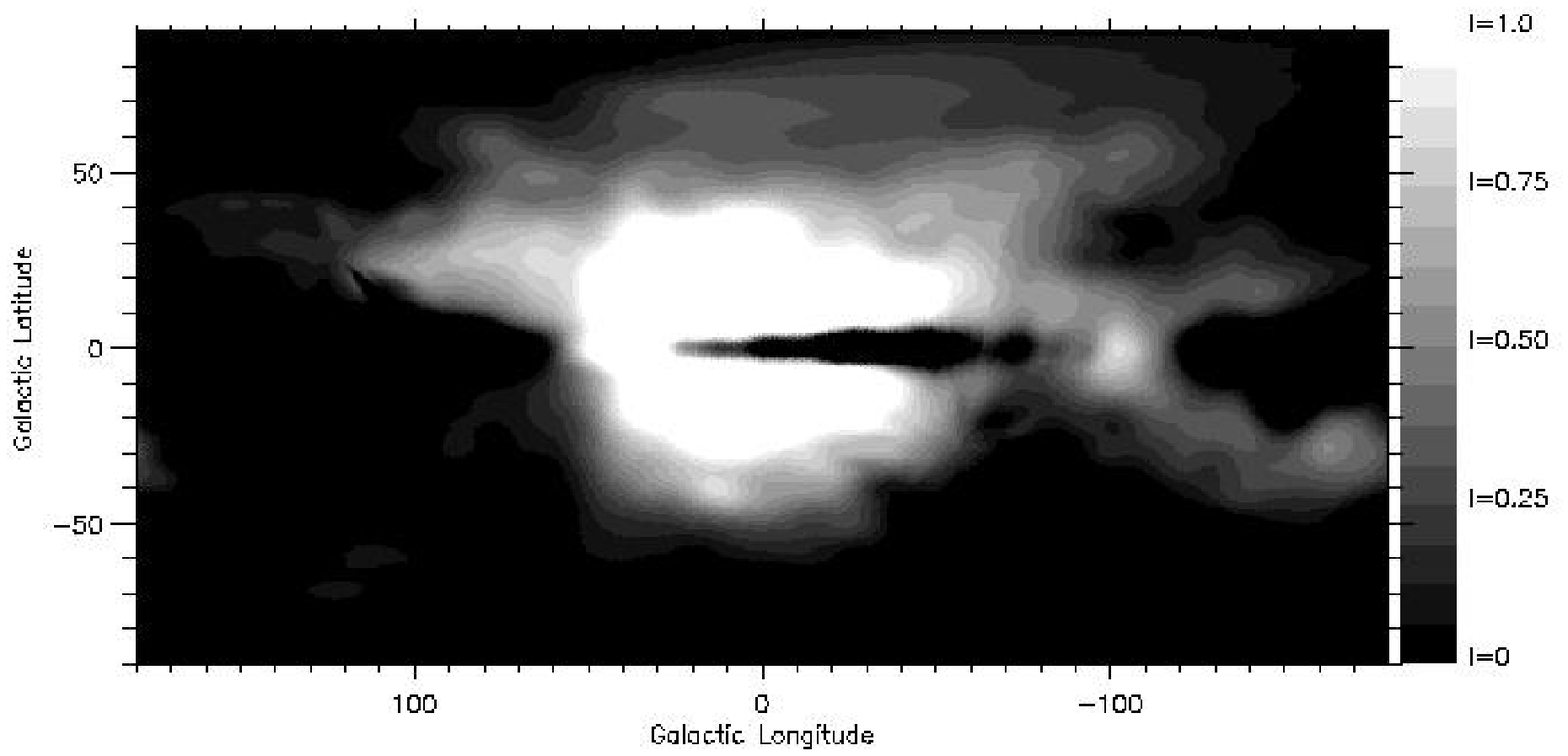}}
\centerline{Large Scale Residual Intensity}
\epsfxsize=0pt \epsfysize=2.5in
\centerline{\epsfbox{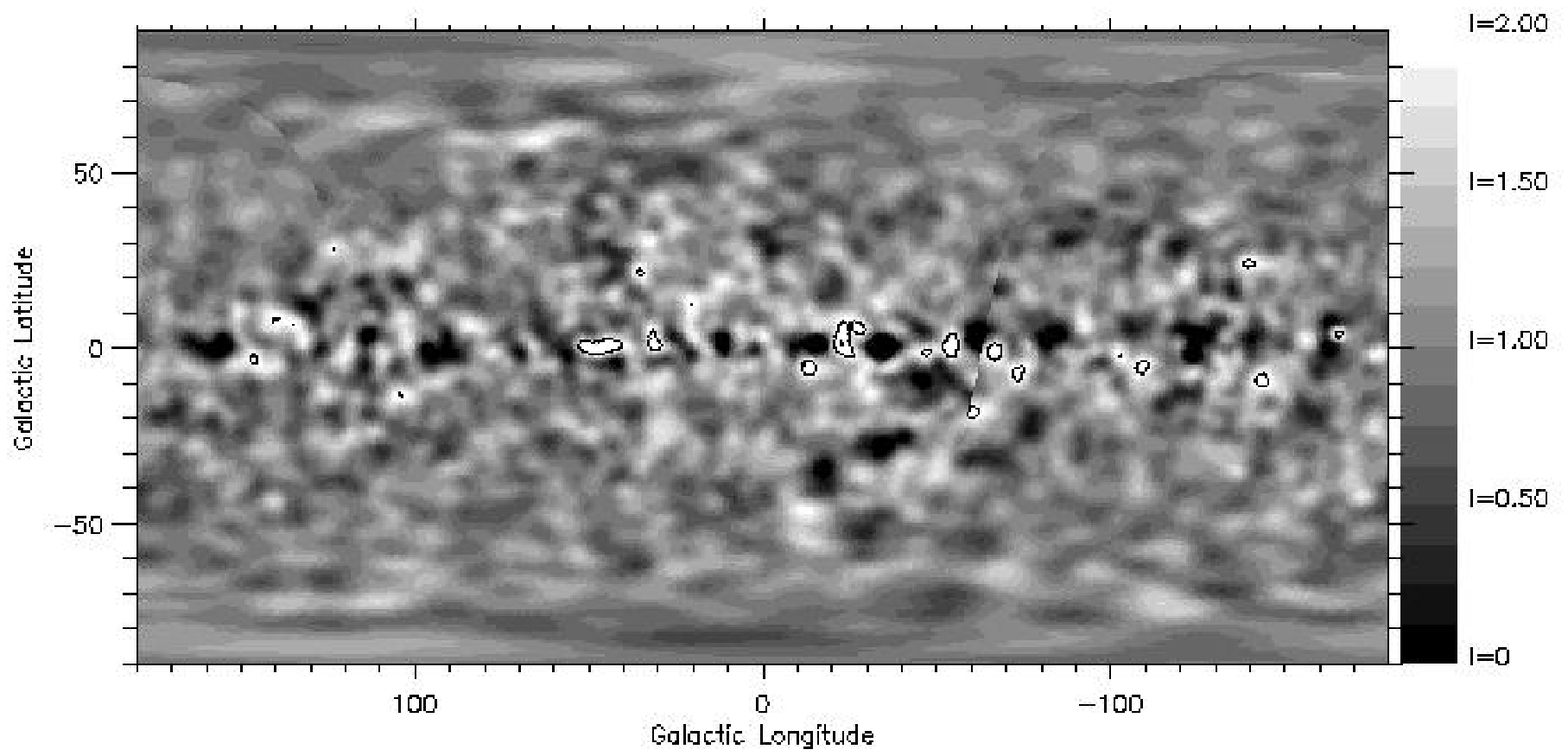}}
\centerline{Small Scale Residual Intensity}
\caption{The residual intensity after subtracting a galactic diffuse model and resolved point sources.  The scales are in units of $10^{-5} {\rm ph}\;{\rm cm}^{-2}{\rm s}^{-1} {\rm sr}^{-1}$}
\label{residuals}
\end{figure}

\subsection{Likelihood Fit to the Diffuse Background}

	In order to extract the galactic bulge feature, the likelihood fit to the diffuse parameters is used.  While these parameters cannot be directly separated due to their high degree of correlation, they do give a measure of the best fit to the diffuse intensity within $15^{\circ}$ of each point on the sky.  These values can then be used to construct a maximum likelihood diffuse map in the following way,
\begin{equation}
I_{ij}={g_m}_{i j}G_{ij}+ {I_{\rm IDGRB}}_{ij} \;.
\end{equation}
This method gives a model of the total diffuse emission which has been fit to the data on large scales. 

	In order to analyze the large scale features, the galactic diffuse model is subtracted from the maximum likelihood diffuse model as is the IDGRB as measured above.  What remains is the residual emission on large scales.  A map of this emission is shown in the second panel of \fig{residuals}.

\subsection{Possible Sources of Large Scale Residual Diffuse Features}

	The most prominent diffuse feature that remains after subtracting our model of the \gammaray\ intensity is a prominent feature around the galactic center.  The position of this feature in the galactic bulge is strongly suggestive that it is the result of imperfectly modeled inverse Compton emission.  A few of the uncertainties in this calculation will now be discussed and their impact on the data assessed.

	There is much uncertainty associated with the photon distributions used in this model.  The FIR photon distribution was taken to be 100 pc (\cite{Sodroski94}).  It is possible that these values are biased by the fact that photons that escape the galactic disk are not observed directly despite the fact that they can serve as target photons for cosmic ray electrons above the galactic plane.  A scale height of 1 kpc for these photons would increase the inverse Compton contribution by $\sim20\%$ at mid-latitudes.

	Furthermore, the absolute normalization of the FIR photon distribution is uncertain. Sodroski et al. (\cite{Sodroski94}) quote a galactic FIR luminosity of $1.6-2.2 \times10^{10}L_{\odot}$ using COBE/DIRBE data whereas Wright et al. (\cite{Wright91}) report a value of $(1.8\pm 0.6)\times10^{10}L_{\odot}$ from an analysis of COBE/FIRAS.  It is clear that the total FIR luminosity could be increased by 20\% without being in conflict with the available FIR data.

	Similar uncertainties exist in the work of \cite{Chi91}.  Both the photon scale heights and normalizations are uncertain to within 30\%.

	Finally, the cosmic ray scale height itself is very uncertain.  The value adopted for this analysis was the lower limit from the deconvolution of the galactic synchrotron data.  Youssefi \& Strong (\cite{Youssefi95}) used the opposite approach in analyzing the \cosb\ diffuse \gammaray\ data.  In this work, the authors use the \gammaray\ data to constrain the inverse Compton emission leaving the cosmic ray scale height as a free parameter.  The reported electron cosmic ray scale height is found to be $3-5\;{\rm kpc}$.  The same authors have proposed a cosmic ray diffusion model (\cite{Strong95}).  Under assumptions about the galactic magnetic field distribution as well as the cosmic ray injection spectrum and distribution, an energy dependent cosmic ray electron scale height is derived.  This scale height peaks at $\sim 1\;{\rm GeV}$ after which inverse Compton losses tend to contain the cosmic rays to a tighter distribution about the galactic plane.  While more data is needed in order to determine the electron scale height accurately, it is not unlikely that much of the residual surrounding the galactic center is the result of additional a inverse Compton emission.

	More puzzling, perhaps, than the bright residual in the galactic bulge is the North-South anisotropy at high latitudes.  In the North Galactic polar region there are several large scale residual excesses.  The region extending from the Virgo region to the North Galactic pole appears to be  significantly brighter than modelled.  Furthermore, the South Galactic pole intensity is significantly underpredicted by the model.  A third region near $(l,b)=(70,60)$ has been  reported as a region of excessive diffuse emission (\cite{Chen95}).  However, the proximity of this region to the orbital pole in the principal observation of this region means that this observation was most likely contaminated by albedo photons.   Reanalysis of this region using the more restrictive zenith angle cut also reveals a point source in this region (see next chapter).  The remaining excess in this region is most likely a statistical fluctuation.

	The asymmetries at such high latitudes argues in favor of a local source of these anisotropies.  A correlation between the northern excesses and the large radio supernova remnant features (Loop 1 and Loop 3) has been suggested (\cite{Osborne95}) but the evidence is not very compelling.  The most likely source of these features is a local variation in the cosmic ray density within a few hundred parsecs.

	The underprediction of the model near the South Galactic pole is potentially due in part to the sensitivity drift of the \egret\ instrument which is caused by gas degradation.  This effect is accounted for by looking at overlapping diffuse regions and fitting the sensitivity trend versus time.  Several of the dominant observations of the South Galactic Pole exhibit apparent sensitivities which are below the trend indicating that there may have been slower than average degradation during these observations. 

\subsection{ Effect on the Measurement of the IDGRB}

	In light of the problems with the galactic diffuse model the impact on the measurement of the IDGRB must be assessed.  In the limit of infinite cosmic ray scale heights, the total isotropically distributed \gammaray\ emission that would result from interactions between cosmic rays and microwave background photons well above the galactic plane amounts to $\sim 20\%$ of the IDGRB (\cite{Sreekumar96}).  Thus it is clearly impossible to attempt to explain the IDGRB as the result of an extended cosmic ray halo. However, it is clear that the systematic uncertainties in the foreground model limit our ability to measure the IDGRB intensity.  These uncertainties are significantly larger than the statistical uncertainties.

\section{ Spectral Measurement of the IDGRB}

	While this study is primarily concerned with extracting the spatial signals from the diffuse background, the spectrum of the IDGRB is also of critical importance.  A measurement of the IDGRB spectrum will be outlined here.  This work follows the basic techniques outlined in (\cite{Kniffen95}).  A more detailed analysis of the IDGRB spectrum is to appear in (\cite{Sreekumar96}).

	Maps of the summed \egret\ counts through Viewing period 404.0 were generated in ten energy ranges: ($E\;$ (MeV): 30-50,50-70,70-100,100-150,150-300,300-500,500-1000,1000-2000,2000-4000,4000-10000).  Corresponding exposure maps were also generated.  To minimize the albedo contamination in making this sensitive measurement, an albedo cut of $\alpha<80^\circ$ was again adopted for all energy ranges.  The sky at latitudes $\mid b\mid >30^\circ$ was divided into equal solid angle ($20^\circ \times 20^\circ$) bins in celestial coordinates.  Regions within $5^\circ$ of a strong point source were excluded from the analysis.  In order to avoid the regions in which the unmodelled inverse Compton emission is important, regions within $90^\circ$ of the galactic center are excluded from the analysis as well.

	The galactic diffuse contribution to the intensity of each pixel was calculated using the extrapolation technique described above.  The scaling of the galactic diffuse model was left as a free parameter in each energy range which was determined by the correlation between measured intensity in each pixel
and the modelled intensity in that pixel.  In each energy range this scaling was consistent with unity except in the highest three ranges in which the galactic model significantly underpredicted the diffuse emission.  For a discussion of this effect see Hunter et al. (\cite{Hunter96}).

	The number of residual non-galactic counts in each bin in each energy range were thus determined.  The spectrum was then deconvolved from this data using the SPECTRAL program (\cite{Nolan93}).  This program performs a maximum likelihood fit to the measured counts amongst a space of power law input spectra which are folded through the \egret\ energy response functions.

\begin{figure}[p]
\epsfxsize=0pt \epsfysize=3.0in
\centerline{\epsfbox{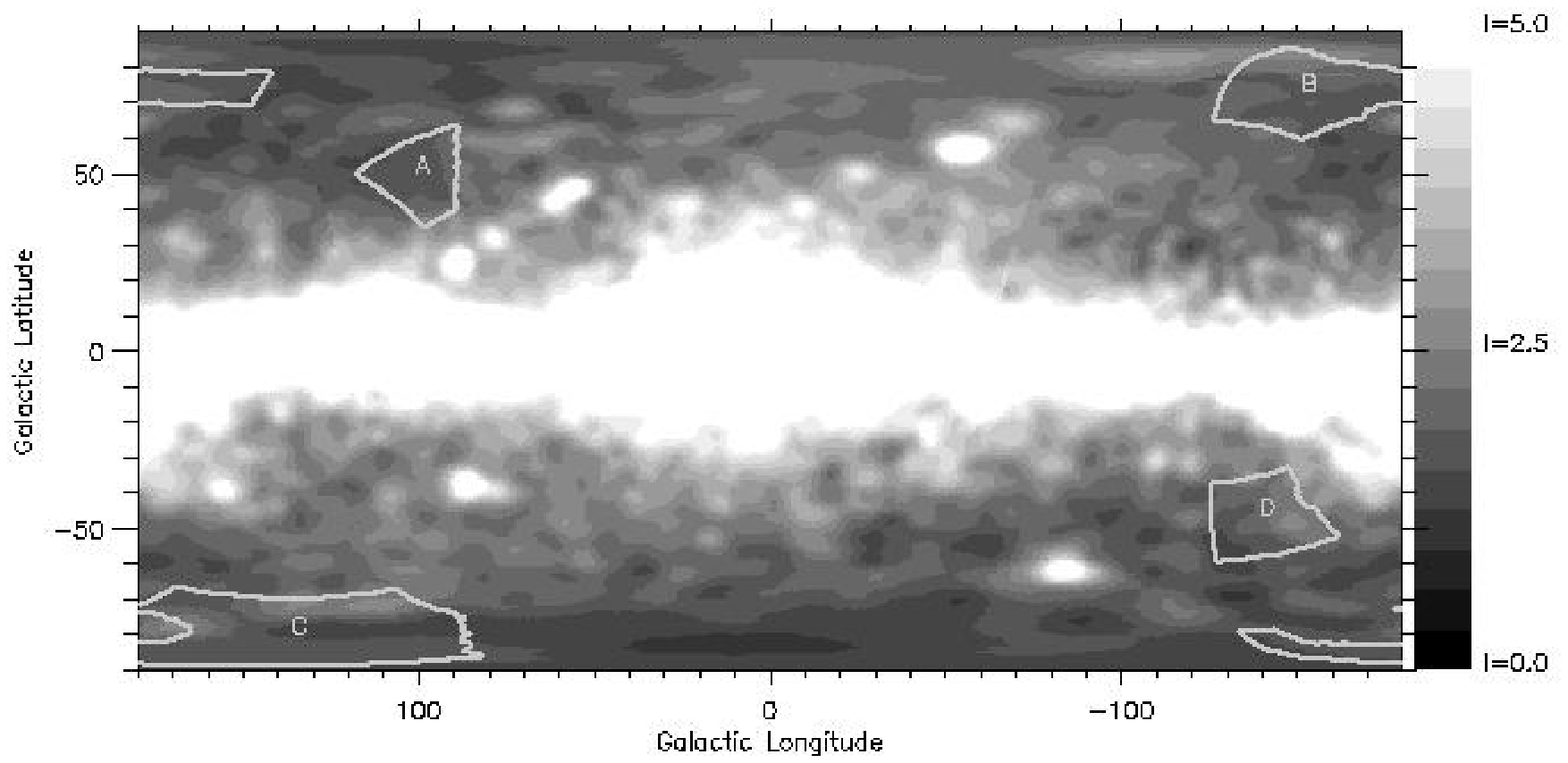}}
\epsfxsize=0pt \epsfysize=4in
\vspace{0.2in}
\centerline{\epsfbox{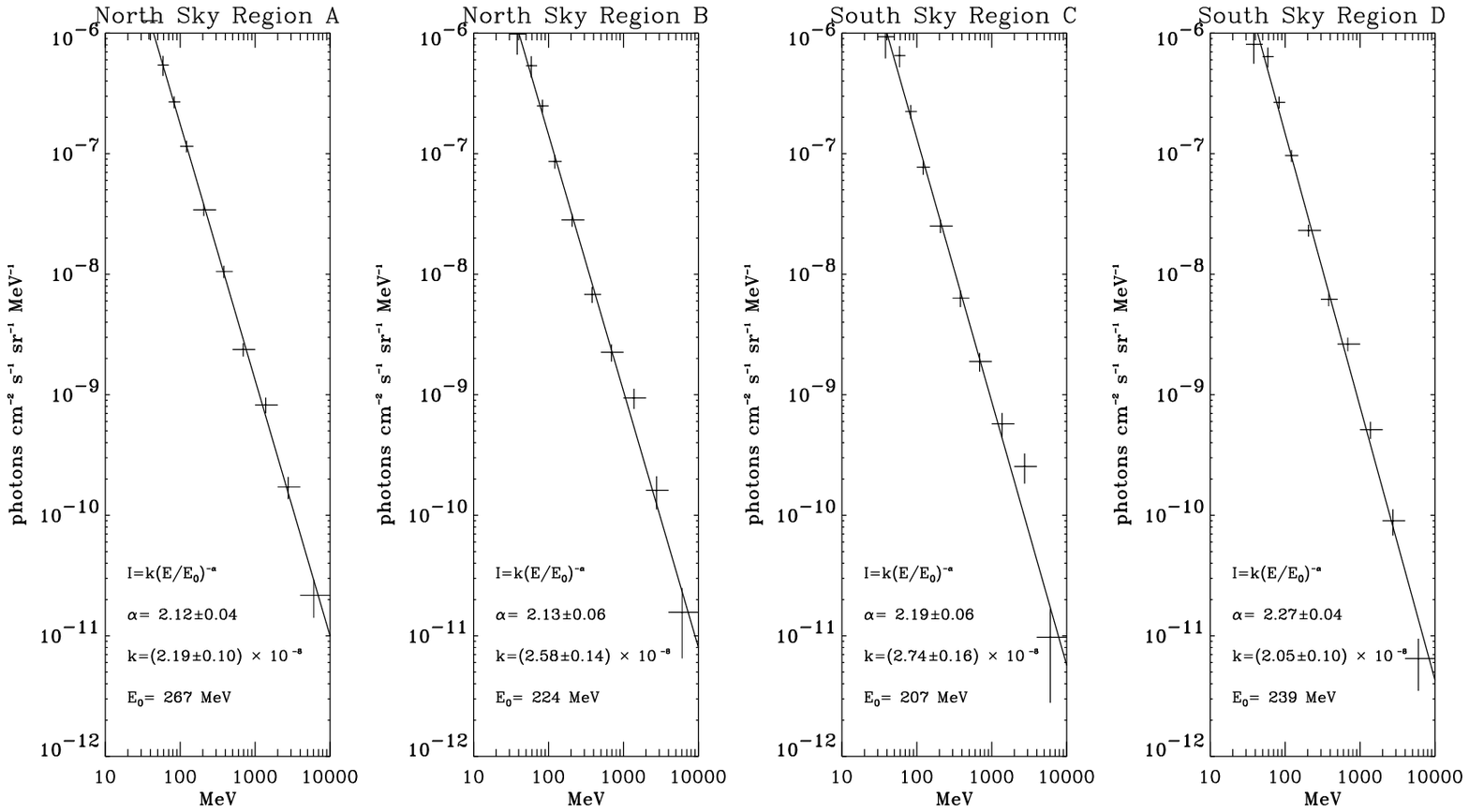}}
\caption{Four independent measurements of the IDGRB spectrum.  The regions used to calculate each spectrum are shown in the top panel.  Each region is $20^\circ\times20^\circ$.  Regions within $5^\circ$ of strong point sources are excluded.  There is good consistency between the best fit spectra.}
\label{Spectra}
\end{figure}

	\fig{Spectra} shows several of the measured spectra across the sky.  The map indicates the regions for which the spectra are shown.  Even with the stringent data cuts that are made in order to obtain clean a clean data set, well constrained spectra are available.  All the spectra are good fits to a power law.  There is a slight indication that there is an excess in the 2000-4000 bin and a deficit in the 4000-10000 bin in all the spectra.  This is due to an inaccuracy in the energy determination of the Class B events in which an insufficient amount of energy was deposited in the TASC  to pass the photo-multiplier threshold.  These events had to have their energies calculated from their track patterns.  The power law indices are all within one or two standard deviations of each other.

\begin{figure}[h]
\epsfysize=3.0in
\centerline{\epsfbox{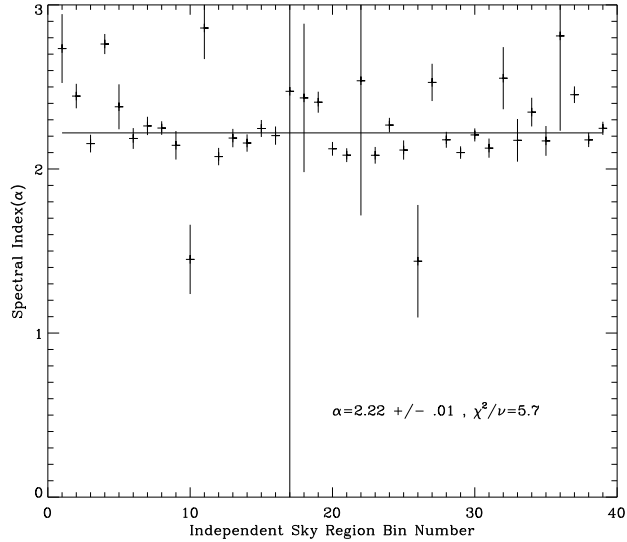}}
\caption{Several independent measurements of the IDGRB spectral index ($\alpha$).  The error bars represent the statistical errors from the deconvolution.  The high value of $\chi^2/\nu$ indicates that the systematic errors in the galactic diffuse model are dominant.}
\label{alphadist}
\end{figure}

	The distribution of the spectral index across the sky is shown in \fig{alphadist}.  The variation in the error bar size indicate the variable statistical errors in each measurement due to exposure variation as well as the truncation of some of the analysis regions by their proximity to point sources.  The weighted mean is also given in this figure.  The best fit value is $\alpha=2.22\pm.01$.  Also shown is the minimum $\chi^2/\nu$ value for the data set.  The elevated value of this number is indicative that systematic errors in the galactic diffuse model dominate over statistical errors.  There is no evidence for a smooth variation of the spectral index across the sky.

\begin{figure}[h]
\epsfysize=6.0in
\centerline{\epsfbox{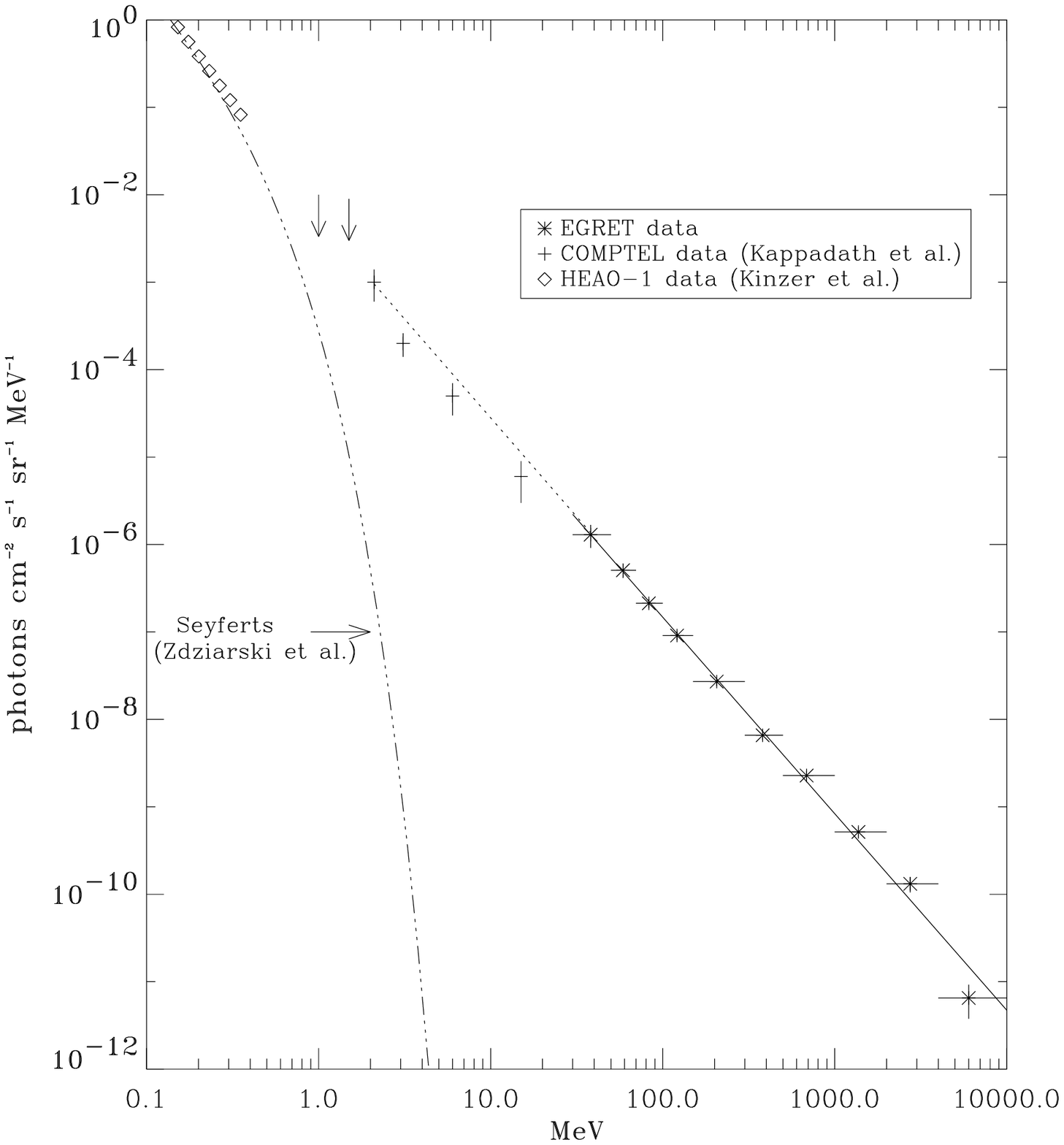}}
\caption{The cosmic background radiation spectrum from hard X-rays through the \egret\ range.  The X-ray data is from recently re-analyzed work by Kinzer et al. 1996 in which new measurements of the instrumental background were obtained.  The MeV data are from the \comptel\ instrument.  The pronounced spectral bump at $\sim 1\;$ MeV is not evident in this data after a careful measurement of the rigidity dependence of the background (Kappadath et al. 1995).  Also shown is the spectrum of calculated Seyfert contribution to the background (Zdziarski et al. 1995).}
\label{Spectralcomp}
\end{figure}

	The cosmic background spectrum can now be placed in its spectral context.  \fig{Spectralcomp} shows the best available measurements of the cosmic background radiation from hard X-rays through the \egret\ range.  The X-ray data is from HEAO-1 as analyzed by Kinzer et al. (\cite{Kinzer96}).  The data in the MeV region is from \comptel\ (\cite{Kappadath95}).  The dramatic difference between this and other measurements of the background in this spectral region results from the fact that instrumental background was able to be subtracted through its rigidity modulation.  Previous balloon measurements were performed at a single rigidity, prohibiting the use of this powerful technique.

	Also shown on in \fig{Spectralcomp} is the contribution of X-ray Seyferts to the background   (\cite{Zdziarski95}).  Essentially all of the hard X-ray background is now believed to be due to unresolved X-ray Seyferts.  However, these spectra are well fit by thermal comptonization spectra which are exponentially suppressed in the \gammaray\ region.  Indeed no X-ray Seyferts have been detected by \egret.  These sources do not contribute substantially to the background above 1 MeV.

	The remaining high energy spectrum is a fairly consistent power law throughout the measured spectral region.  There is no evidence of spectral features that would indicate the presence of a high energy particle decay.  Furthermore, the absence of the `MeV bump' eliminates the strongest evidence in favor of an annihilation feature in that spectral region.

\section{Discussion}

	\egret\ has proved to be a powerful tool in the study of high energy diffuse \gammaray\ emission.  The large effective area, good spatial resolution, and low instrumental background have allowed measurement of both galactic and extra-galactic diffuse emission with unprecedented accuracy.

	A model of the diffuse \gammaray\ emission has been presented.  This model with only two free parameters is an excellent fit to the observed \gammaray\ data.  At intermediate galactic latitudes toward the galactic center there remains a clear deficiency in the model.  The high residuals in this region are almost certainly the result of unaccounted for inverse Compton interaction between galactic cosmic ray electrons and soft background photons.  The poor constraints on the distribution of cosmic ray electrons remains a difficulty in the construction of more accurate models.  The \gammaray\ data can be used to constrain this distribution but such work is difficult because of the many uncertainties involved.

	Despite these difficulties, there is clear evidence of an IDGRB that cannot be explained by the known galactic sources of diffuse emission.  It has been shown that the intensity of this emission above 100 MeV is,
\begin{equation}
 I_{{\rm IDGRB}}=(1.33 \pm .03 \times10^{-5}\;{\rm photons}\;{\rm cm}^{-2}\;{\rm s}^{-1}\;{\rm sr}^{-1} \;.
\end{equation}
The statistical error quoted is small compared to the systematic uncertainties involved in the calculation.  Most importantly the poorly constrained cosmic ray scale height allows for up to 20\% of this emission being galactic in nature.

	Spectral measurements of the IDGRB have shown that this emission is well described by a power law with spectral index of $\alpha=2.22\pm.01$.  This spectrum does not appear to vary across the sky lending credibility to the hypothesized extra-galactic origin.  The spectrum is harder than measured by \sas\ but the measured value is within the uncertainty of that measurement.  This spectrum matches well to the new Comptel spectrum.  The combination of the harder high energy spectrum and the disappearance of the `MeV' bump have most likely ended speculation about Baryon Symmetric Big Bang theories of the origin of this radiation.  The fact that this spectrum is now consistent with measured AGN spectra has fueled speculation that this emission is due to unresolved members of that population.  This notion will be explored in detail in the subsequent chapters.

\chapter{High Latitude Point Sources}

	Prior to the launch of \egret, the only extra-galactic source detected in the high energy \gammaray\ band was the quasar 3C 273 (\cite{Bignami81}).  One of \egret's most dramatic results has been the detection of dozens of sources at latitudes greater than $10^\circ$.  The majority of these sources have been identified as being extra-galactic based on their positional co-incidence with bright Flat Spectrum Radio Quasars (FSRQ's) (\cite{vonMontigny95}).  This discovery has focussed the discussion about the origin of the IDGRB on the contribution from active galactic nuclei (AGN).  In this chapter, the nature of the high latitude point sources detected by \egret\ will be examined and several models to explain their distribution will be discussed.

\section{High Latitude Point Source Analysis}

	The second \egret\ catalog (\cite{Thompson96}) summarizes the point source analysis of the \egret\ data through August 17 1993.  In this study the maximum likelihood technique is used to locate point sources and to determine their fluxes.  This technique is applied to each viewing period individually as well as to the summed all-sky data.  The point source analysis technique is discussed is detail in (\cite{Mattox96}) and is summarized in Appendix A.

	The point source and diffuse model parameters are fit to the data within a $15^\circ$ radius of a given test point.  In this way local inaccuracies in the diffuse model such as the large scale underprediction of the model in the galactic bulge region, do not bias the analysis.  The point source significance is then estimated using the distribution of the likelihood statistic in the null hypothesis (see Appendix A).  This analysis allows the simultaneous determination of the value of the diffuse backgrounds  at a given point as well as the significance and flux of a putative point source at this location.

	In order to simultaneously analyze the many point sources that \egret\ has discovered, it is necessary to account for the coupling between sources that is the result of a finite point spread function.  This is done using an iterative technique.  A likelihood map is constructed and sources above a given significance threshold are extracted.  The three point source parameters ($l,b,S$), for each source are then iteratively adjusted and the effect on the likelihood is determined.  Because simultaneous optimization of all three parameters for each source would be computationally infeasible, the source fluxes are adjusted first and then held constant while the positions of these sources are adjusted.  When a simultaneous maximum has been achieved, these sources are then added to the diffuse map as background features and the data are reanalyzed to see if any additional sources that were previously obscured are thus revealed.  All sources are then simultaneously optimized.  This process is then repeated until no sources above a given significance threshold remain.

	  The above process is reliable so long as the detected point source density is sufficiently low so as not to create false convergences under optimization.  Simulations have shown that sources with significances of greater than or equal to $4\sigma$ can be reliably reconstructed using this technique (see Appendix A).  It is, however, impossible to push this technique to lower significance as the results tend to change with the order in which the sources are added.

	The point source analysis of the summed data carried out for the Phase I+II catalog has been redone with the following modifications:
\begin{itemize}
\item Data through 11 April 1995 (Viewing Period 419.1) are included.
\item Only regions with $|b|\geq20^\circ$ are analyzed.
\item A zenith angle cut of $80^\circ$ is adopted.
\end{itemize}

\section{Point Source Results}

	Table 5.2 summarizes the results of the high latitude point source analysis.  A total of 55 sources were detected with $|b|\geq20^\circ$.  Of these 33 are blazars which were also detected in Phase I+II (P12) data.  Another 4 can be identified with blazars that were undetected in the P12 data.  The only other identified high latitude source is the Large Magellanic cloud (LMC) which is also seen in the P12 data (\cite{Sreekumar92}).  The remaining 17 sources are unidentified.  Of these 6 were
detected in the P12 data.  The remaining 11 sources are new unidentified sources.  The fact that most of these sources are near the significance threshold is indicative of the fact that they are sources rendered visible through increased exposure.  At least one of these sources ( $l=73.09,b=68.18$) is visible in the P12 data if one adopts the more restrictive albedo cut made in this analysis.  This is a result of the fact that this source appeared near the orbital pole during its observation which increases the albedo background as discussed in Chapter 3. This source contributes to the diffuse emission enhancement reported by (\cite{Chen95}).

\begin{table}[ht]
\centering
\caption
{Results of the likelihood analysis performed using all data up to VP 419.1 using an $80^\circ$ zenith angle cut.}
\bigskip
\tiny
\begin{tabular}{l|cccccc}\hline\hline
Category & $l$& $b$ & $S(E>100 {\rm MeV})$ &$\sigma$ & P12? &ID \\ \hline
Blazars&23.52& 41.02& $27.0 \pm  5.1$&   6.7& x&  1606+106       \\
Detected in P12&35.98& -24.58& $22.8 \pm  4.2$&   6.7& x&  2022-077       \\
(=33 sources)&55.81& 46.02& $53.1 \pm  7.7$&   9.9 &x&  1611+343       \\
&61.69& 42.16& $40.4 \pm  6.7$&   8.0&  x&  1633+382       \\
&77.71& -39.09& $20.9 \pm  3.0$&   8.8&  x&  CTA 102        \\
&79.43& 31.60& $28.4 \pm  5.1$&   7.1&  x&  1739+522       \\
&85.98& -37.97& $68.5 \pm  4.4$&  22.2& x&  3C 454.3       \\
&143.54& 34.55& $12.1 \pm  2.1$&   7.0&  x&  0836+710       \\
&143.55& 28.12& $19.0 \pm  2.3$&  10.5&  x&  0716+714       \\
&147.65& -44.57& $10.0 \pm  3.3$&   3.6&  x&  0202+149       \\
&150.17& -28.73& $12.1 \pm  2.9$&   5.0&  x&  0234+285?      \\
&156.29& -38.83& $32.5 \pm  4.4$&   9.4& x&  0235+164       \\
&159.67& 47.54& $9.0 \pm  1.8$&   6.1& x&  0954+556	\\
&169.96& 32.09& $18.0 \pm  3.0$&   7.5& x&  0804+499      \\
&177.00& 44.28& $13.8 \pm  2.2$&   7.9&  x&  GRO J0916+43   \\
&179.89& 64.89& $14.0 \pm  2.0$&   9.0&  x&  Mrk 421        \\
&-172.88& -20.34& $20.2 \pm  3.1$&   7.5&  x&  0446+112       \\
&-165.21& -32.17& $19.8 \pm  3.7$&   6.3&  x&  0420-014       \\
&-160.11& 31.36& $23.7 \pm  4.1$&   7.6&  x&  0827+243       \\
&-158.94& -25.35& $10.4 \pm  2.8$&   4.2&  x&  0458-020      \\
&-118.99& -31.54& $19.3 \pm  3.8$&   6.3&  x&  0521-365       \\
&-109.97& -30.84& $24.4 \pm  3.5$&   9.0&  x&  0537-441       \\
&-83.78& -61.78& $71.2 \pm  6.7$&  16.1&  x&  0208-512       \\
&-83.05& 43.53& $10.2 \pm  2.5$&   4.7&  x&  1127-145       \\
&-69.30& 64.07& $19.5 \pm  1.9$&  12.3&  x&  3C 273         \\
&-55.06& 56.98& $81.2 \pm  3.0$&  39.8&  x&  3C 279         \\
&-51.52& 28.09& $17.2 \pm  2.9$&   7.0&  x&  1313-333      \\
&-25.73& 50.48& $26.6 \pm  2.9$&  11.9&  x& 1406-076        \\
&-9.00& 40.39& $15.8 \pm  4.0$&   4.6&  x& 1510-089        \\
&-105.92& 81.70& $15.2 \pm  2.0$&   9.4&  x&  1222+216       \\
&168.19& -77.32& $13.6 \pm  3.7$&   4.6&  x& 0130-171     \\
&145.92& 44.10& $9.2 \pm  1.8$&   5.8&  x& 0954+658        \\ 
&107.12& -41.72& $11.7 \pm  3.3$&   4.2&  x& 2356+196        \\ \hline
New Blazars&-162.64& -28.84& $20.1 \pm  3.8$&   6.4&   &  2251+158  \\
(=4 sources)&-161.24& 83.43& $11.2 \pm  2.0$&   6.7&   &  1219+285  \\
&27.03& 20.65& $17.2 \pm  4.3$&   4.6&   &  1725+044  \\
&17.76& -52.26& $12.4 \pm  3.4$&   4.5&   &   2155-304 \\ \hline
Other Identified&-80.50& -32.28& $14.4 \pm  2.8$&   6.1& x&  LMC            \\ \hline
Unidentified&88.63& 25.05& $69.8 \pm  5.4$&  18.2&  x&  GRO J1837+59   \\
(=6 sources)&163.14& 29.11& $13.9 \pm  2.8$&   6.1&  x&  GRO J0744+54   \\
&179.64& -22.86& $22.7 \pm  3.3$&   8.0&  x&  GRO J0421+15   \\
&-66.22& 67.09& $10.7 \pm  1.7$&   7.3&  x&  GRO J1237+04   \\
&-44.99& -24.81& $15.6 \pm  5.1$&   3.6& x&  GRO J1731-78   \\
&147.40& 53.31& $6.1 \pm  1.6$&   4.6&  x&  GRO J1047-58   \\ \hline
New Unidentified&15.93& 24.59& $15.1 \pm  3.8$&   4.5&   &    \\
(=11 sources)&73.09& 68.18& $15.6 \pm  4.8$&   4.2&   &    \\
&-176.46& -31.74& $12.8 \pm  3.3$&   4.4&   &    \\
&-119.60& -21.53& $13.2 \pm  3.4$&   4.7&   &    \\
&-63.99& 58.23& $7.5 \pm  1.7$&   5.0&   &    \\
&10.71& 23.62& $14.5 \pm  3.5$&   4.6&   &    \\
&-83.78& -56.59& $12.6 \pm  3.6$&   4.2&   &    \\
&-152.87& 35.39& $11.3 \pm  3.2$&   4.3&   &    \\
&-48.29& 57.30& $7.2 \pm  1.8$&   4.4&  &    \\
& 4.25& -24.03& $11.0 \pm  2.9$&   4.2&   &    \\
&-117.68& -47.12& $9.5 \pm  2.7$&   4.3&   &    \\ \hline \hline
\end{tabular}

\label{Table.big}
\end{table}

\subsection{Time Variability}

	Before it is possible to analyze the source flux distribution, it is necessary to understand the time variability of high latitude sources.  The time variability of blazars is most dramatically exemplified by the \gammaray\ flare that occurred in June 1991 in the source 3C279 (\cite{Kniffen95}).  In this observation the \gammaray\ flux was seen to change on the time scale of several days.  This short time scale variation in such a luminous source is compelling evidence that the source is compact and that a jet is involved in the acceleration mechanism.

\begin{figure}[p]
\epsfxsize=0pt \epsfysize=2.5in
\centerline{\epsfbox{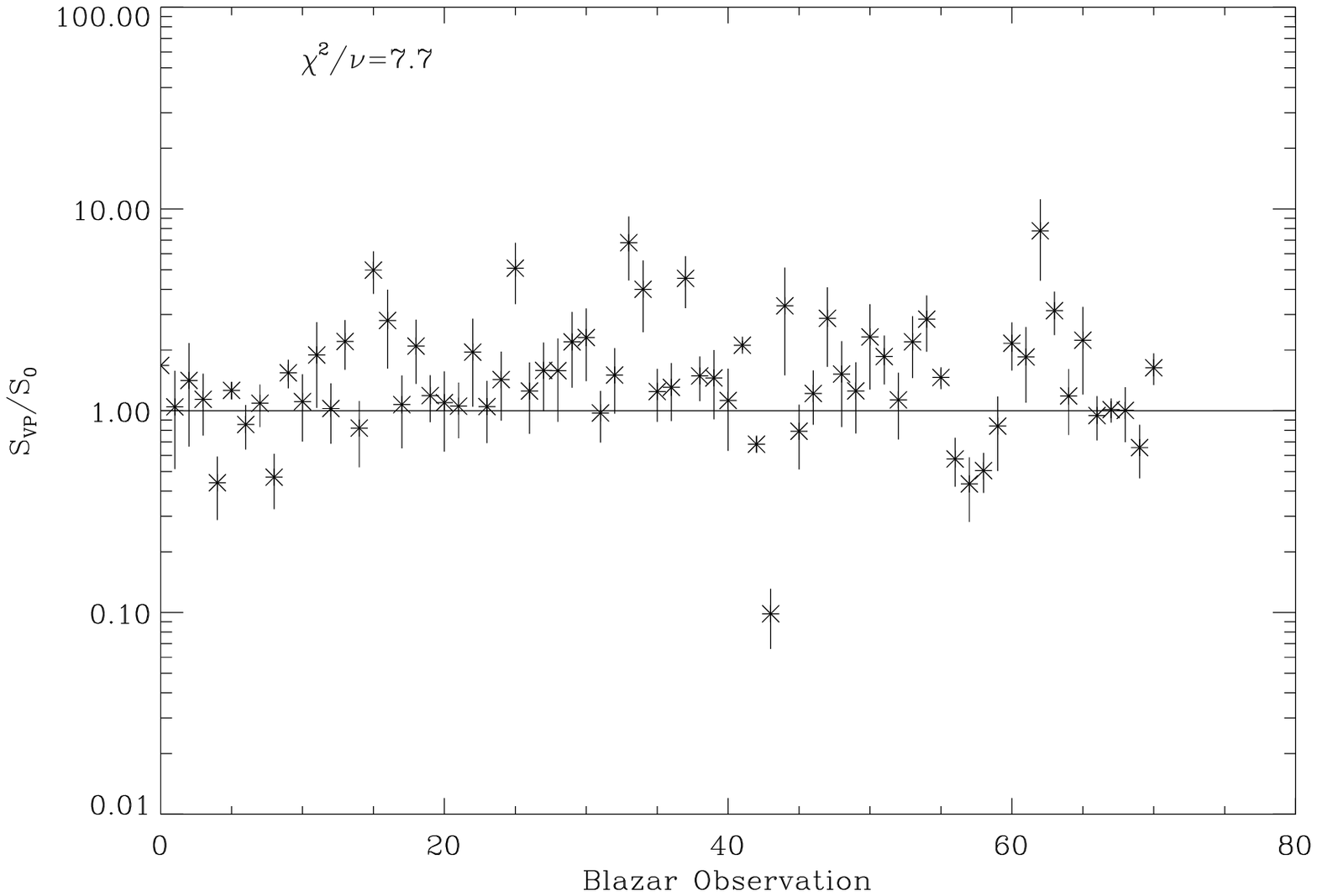}}
\centerline{Blazar flux variations.}
\epsfxsize=0pt \epsfysize=2.5in
\centerline{\epsfbox{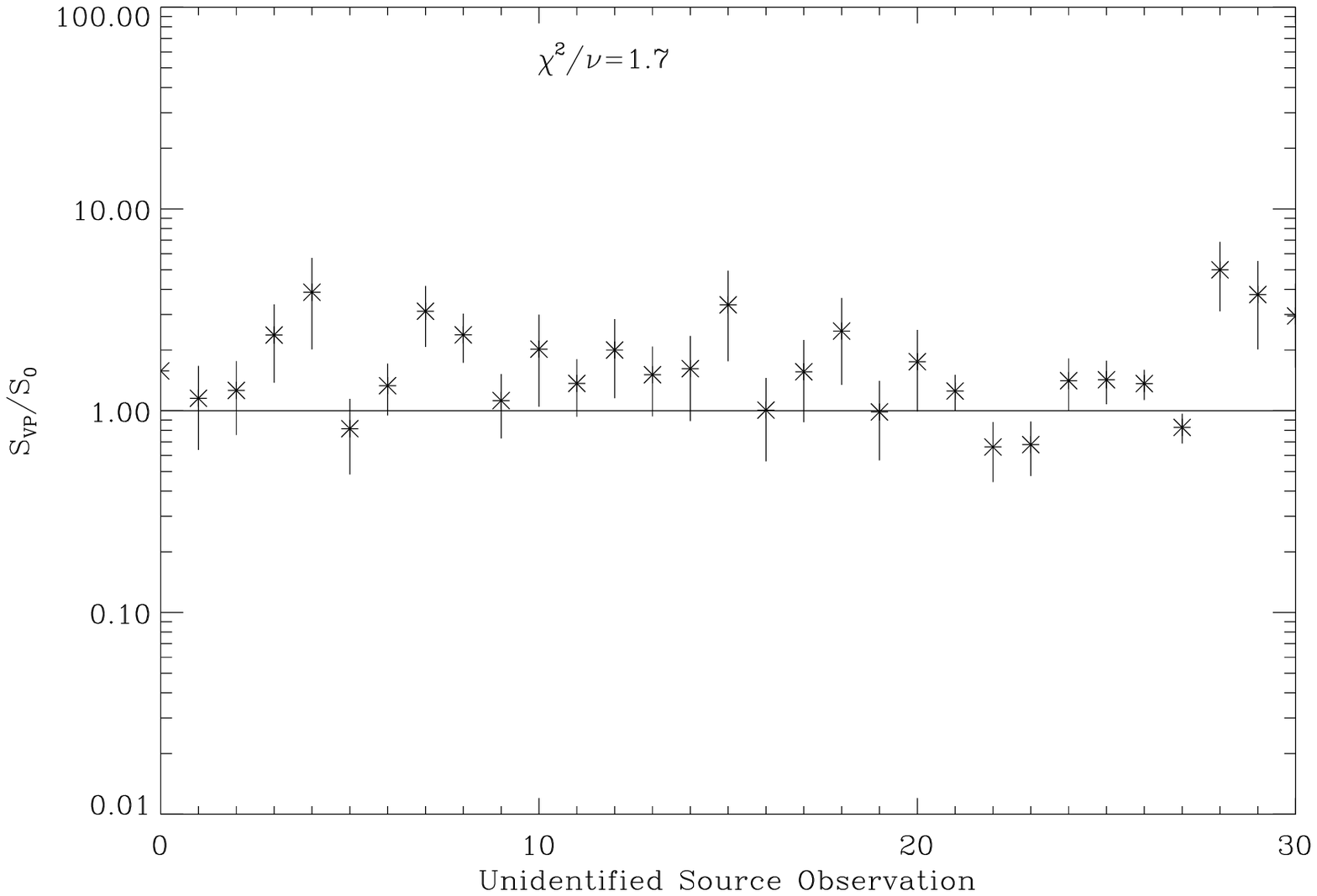}}
\centerline{Unidentified source flux variations.}
\caption{Distributions of flux in a given viewing period normalized by the mean flux of that source.  Only detections of greater than $4\sigma$ are shown.  The asymmetry about 1 is caused by the absence of the upper limits for which a flux measurement is impossible.  The high value of $\chi^2/\nu$ for the blazars demonstrates strong time variability in these sources.  There is weaker evidence for this variation in the unidentified sources although this may be result of truncation at the detection threshold for these comparatively weaker sources. }
\label{var}
\end{figure}

	The data from the Phase I+II catalog are used here to address the question of time variability for the blazars as a class.  \fig{var} shows the distribution of fluxes in each independent observation of a source normalized by the mean flux calculated using the summed data.  Only observations in which the source was detected with $\sigma \geq 4$ are shown.  This results in an asymmetry about unity.  The error bars are calculated by compounding the uncertainty of a given measurement with the uncertainty of the mean.  The distributions are shown for blazars as well as for unidentified sources.

	The blazars show clear evidence of time variation on the time scale of weeks (the average duration of a viewing period).  This is evidenced by high value of $\chi^2/\nu$.  The unidentified sources show weaker evidence of variation.  This may be evidence that some fraction of these sources are not blazars but some foreground population of time independent sources.  However, this conclusion is not straightforward because of the upper limits are not included in this method.  The distributions shown are truncated at the detection threshold.  The effect of this is to make the distributions  of weak sources cluster about the detection threshold.  Because the unidentified sources are weaker on average than the blazars, the effect is to introduce a bias which tends to remove the evidence of time variability for these sources. 

\begin{figure}[h]
\epsfysize=3.0in
\centerline{\epsfbox{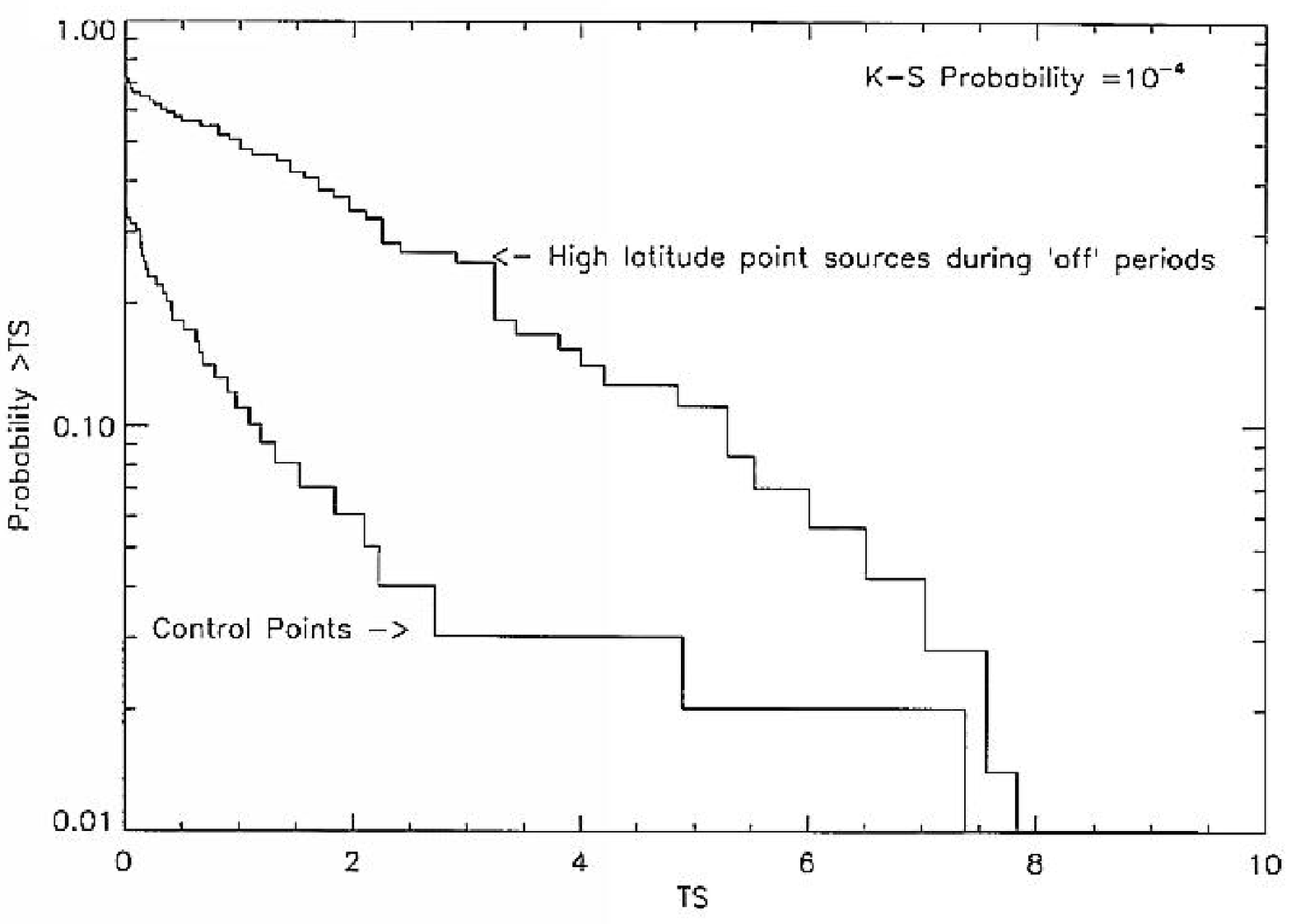}}
\caption{The distribution of the likelihood test statistic (TS) for a set of observations of blazars for which only upper limits were obtained.  This is compared to a set of control points away from any strong sources.  The result clearly indicates that these sources do not completely turn off but rather exhibit some quiescent flux. }
\label{ul.agn}
\end{figure}

	The observations of blazars in which upper limits were obtained can be collectively analyzed to ascertain whether  there is evidence of continued quiescent flux during these periods.  This is done by examining the distribution of the likelihood test statistic at the positions of known sources during periods in which only an upper limit was obtained for their flux.  This distribution is then compared to the a similar distribution of test points which were randomly selected in the sky avoiding regions near known point sources.  The result is shown in \fig{ul.agn}.  There is clear evidence for the continued emission during these `off' periods of the blazars. 

	There are several ways to calculate the source flux distribution in light of this time variability of the sources.  The simplest method which is described above is to sum all the data together and calculate the source flux.  This can be called an exposure weighted mean in the sense that observations near the the instrumental axis and during times of high instrumental efficiency (when the effective area, $A_i$, is high) are more highly weighted in the mean flux as they provide a greater fraction of the total counts $C$, 
\begin{equation}
	\overline{S}_{\rm{Summed}}=\frac{\sum_i C_i}{\sum_i {\cal{E}}_i} = \frac{\sum_i S_i {\cal{E}}_i}{\sum_i{\cal{E}}_i} =\frac{\sum_i S_i A_i t_i}{\sum_i A_i t_i}\;.
\end{equation}
 The result is that some sources are dominated by one or two observations and the flux is not averaged over a very long period.  The more meaningful average flux is the time averaged flux,
\begin{equation}
	\overline{S}_{\rm{t}} =\frac{\sum_i S_i t_i}{\sum_i  t_i}\;.
\end{equation}
The difficulty in calculating this sort of average is the problem of how to treat upper limits.  For the purpose of this study only upper limits that are inconsistent with the exposure averaged flux are considered.  These observations are assigned a flux equal to half the upper limit flux.

\begin{figure}[h]
\epsfysize=3.0in
\centerline{\epsfbox{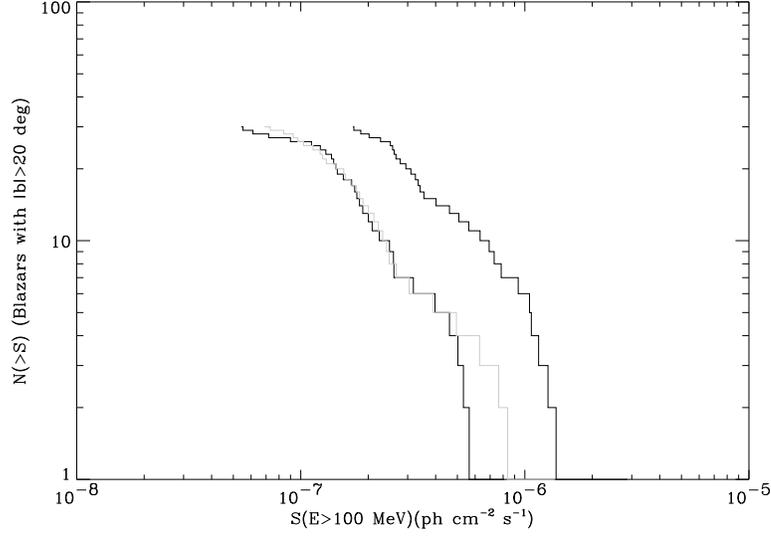}}
\caption{Integral source distributions for the high latitude blazars.  Three different measures of source flux are shown.  The distribution to the right uses the peak flux measured during the mission.  The other two distributions represent two measures of mean flux.  The gray line shows the distribution using the exposure weighted (summed data) flux, while the black line shows the distribution using the time averaged flux.  There is a evidence that the distribution steepens as one averages over longer times.}
\label{srcdist_time}
\end{figure}

	\fig{srcdist_time} shows the integral source distributions ($logN-logS$ distributions) for the high latitude blazars calculated using different flux measures.  The summed flux distribution is shown as a gray line.  The time averaged flux distribution is shown as a black line.  Also shown for reference is the distribution of peak fluxes.  The two flux averages are similar but there is some evidence of a steepening of the distribution as one averages over longer times.

\subsection{Spatial Distribution}

	If the sources are extra-galactic in origin, their spatial distribution should be uniform.  In order to verify this, it is necessary to be able to account for the variations in detection threshold across the sky.  The detection threshold is determined jointly by the exposure and background at a given point.  To a lesser extent, the spatial distribution of the background also affects a source's ability to be detected (e.g. it is hard to detect a point source on top of a spike in the background than a comparably intense flat background).  A simple model of the relationship between flux, $S$, and significance, $\sigma$, of a source is given by the statistical fluctuation of the background,
\begin{equation}
		\sigma=\frac{C_S}{\sqrt{C_S+C_B}} \;,
\end{equation}
where $C_S,C_B$ are the counts from the source and background respectively.  For low significance sources, the significance is dominated by the background fluctuations $C_B>C_S$.  In this assumption, the significance of a source of strength $S$ to which there is exposure $\cal{E}$ and which sits on a diffuse background of strength $B$ is,
\begin{equation}
\sigma=\frac{S\cal{E}}{\sqrt{B{\cal{E}}\Omega}} = kS\sqrt{\frac{\cal{E}}{B}} \;,
\end{equation}
where $k$ is a constant which is determined by the point spread function. In this way source significance is shown to be proportional to its flux near the detection threshold.

\begin{figure}[h]
\epsfysize=3.0in
\centerline{\epsfbox{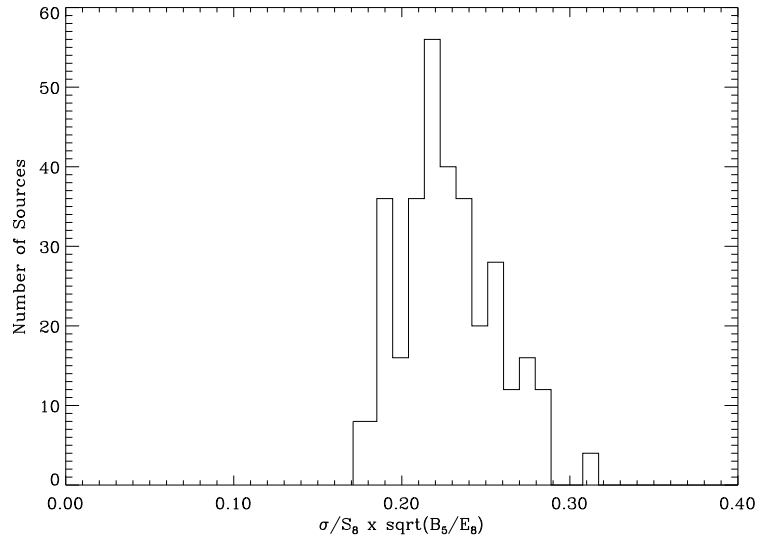}}
\caption{The distribution of the quantity $k$ defined in equation 5.4 for sources detected in Phase I+II. $\sigma,S,B,\cal{E}$ are respectively the source significance,flux,diffuse background at the source location, and exposure.  The tight distribution of this parameter indicates that the several orders of magnitude variation in the input parameters have been successfully scaled out.}
\label{Visdist}
\end{figure}

	This relationship was fit to the data from Phase I+II.  The parameter $k$ was evaluated for all sources with $4.0<\sigma<10.0$.  The exposure was taken from the calculated exposure map and the background was assumed to be represented by the best fit to the diffuse background (galactic and extra-galactic).  The resulting distribution of $k$ is shown in \fig{Visdist}.  The distribution is sharply peaked about $k=0.23$.  The several orders of magnitude variation in the different constituent parameters has been successfully scaled out.

\begin{figure}[h]
\epsfysize=3.0in
\centerline{\epsfbox{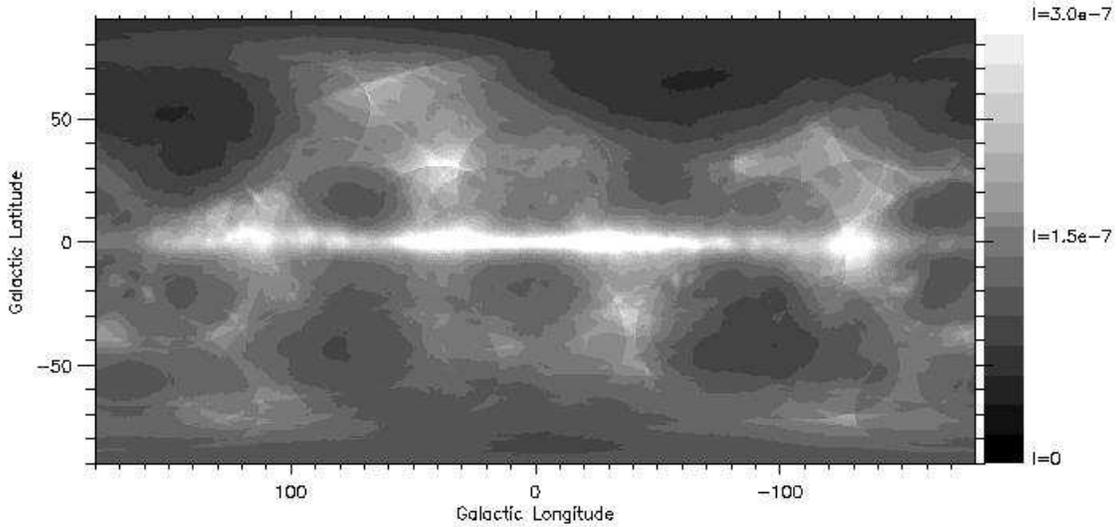}}
\caption{The $4\sigma$ detection threshold map for the combined data through VP 419.1 calculated using the above method.  The values at high latitudes range between $5\times10^{-8}$ and $2\times10^{-7}$.}
\label{tmap}
\end{figure}

	This model allows us to determine a $4\sigma$ detection threshold map for the data set in question.  \fig{tmap} the detection threshold map for the summed data through VP 419.1.  Note that the non-uniform exposure creates almost an order of magnitude variance in the detection threshold at high latitude.

	The source distribution can now be analyzed to check for uniformity.  For a given range of $|b|$ the predicted number of sources can be calculated using some expected source flux distribution.  A Euclidean source distribution with $N(>S)=AS^{-1.5}$ is used.  The source distributions shown above appear to be consistent with this assumption. The total number of sources, $N_b$,  expected in a band of latitude $\delta |b|$ is thus given by,
\begin{equation}
	N_b=\int_{\delta |b|} A S_{{\rm Threshold}}^{-1.5} d\Omega \;.
\end{equation}
The ratio of this number to the number of sources detected in the same latitude range is a measure of the true source density on the sky. 

\begin{figure}[h]
\epsfysize=3.0in
\centerline{\epsfbox{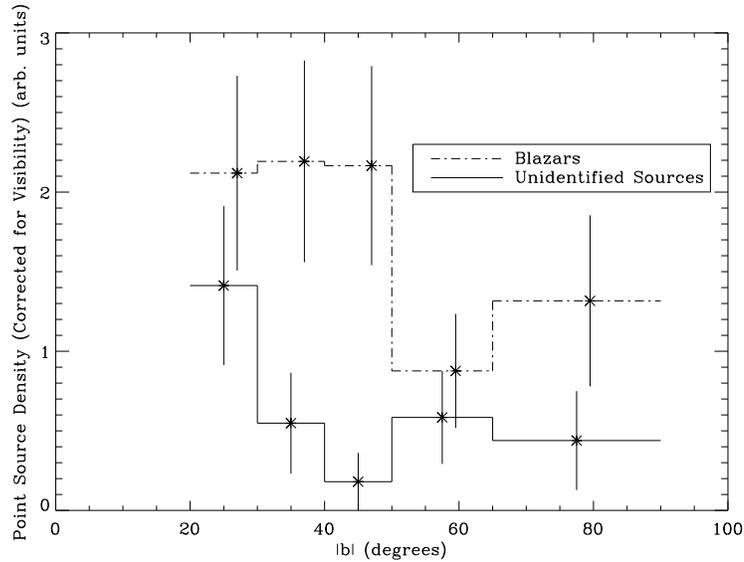}}
\caption{The detected source density as a function of galactic latitude.  The source density has been corrected for the variable exposure and background.  The blazar distribution appears uniform, however, there is slight evidence for an excess of unidentified sources at medium ($20^\circ-30^\circ$) latitudes.}
\label{latdist}
\end{figure}

	\fig{latdist} shows the corrected source density as a function of latitude for the blazars as well as for the unidentified sources.  The distribution of blazars appears to be uniform as expected.  The unidentified sources show a possible excess at medium latitudes between $20^\circ$ and $30^\circ$.  Kanbach (\cite{Kanbach96}) has used similar analyses to suggest that on the basis of the spatial distribution as well as the time variability differences between the two populations of high latitude sources, there is evidence for a local galactic component in the unidentified sources.  The evidence, however, inconclusive.

\section{Source Distribution Models}

	The first step towards understanding the extent to which unresolved members of this population of sources contributes to the IDGRB involves gleaning information from the distribution of detected source fluxes.  As will be seen shortly, the number of sources ($N$) with a flux greater than some value $(S)$ is most often given by a power law and because power laws are most often displayed on $log-log$ plots, the flux distribution $N(>S)$ is often called a $logN-logS$ relation. 

	If an observer sits at the center of an infinite distribution of standard candles of strength $L_0$, the observed flux of these candles will depend on their distance as,
\begin{equation}
	S=\frac{L_0}{4\pi R^2} \;,
\end{equation}
while the number of sources observed with a given flux or greater is given by the volume of the sphere enclosed of radius $R$,
\begin{equation}
	N(>S)=\frac{4\pi}{3}\rho R^3= k S^{-3/2} \;.
\end{equation}
This simple form of the $logN/logS$ relationship is called the Euclidean distribution.  The Euclidean distribution applies just as well to any population of sources with a given luminosity distribution so long as this distribution does not evolve in time (and therefore space).  One consequence  of this type of distribution is that the integrated diffuse flux from such a distribution,
\begin{equation}
I=\frac{1}{4\pi}\int_{S_{max}}^{0}\frac{dN}{dS}SdS=k^\prime [S^{-1/2}]_{S_{max}}^0=\infty \;,
\end{equation}
diverges in the limit of infinitely large distributions.  Such distribution must be truncated at some minimum flux $S_{min}$ in order to avoid this problem.  This is commonly referred to as Olbers' paradox.

\begin{figure}[h]
\epsfxsize=\hsize
\centerline{\epsfbox{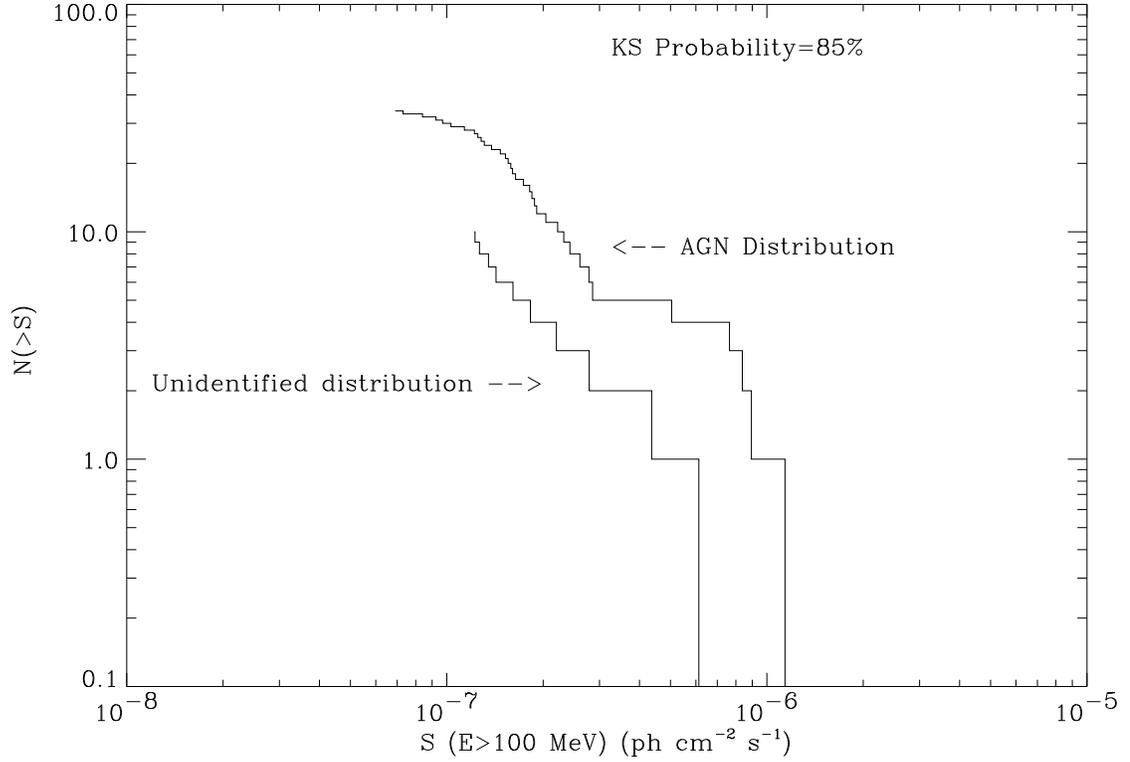}}
\caption{The measured $logN/logS$ function for the identified AGN high latitude sources and the unidentified high latitude sources. The Kolmogorov-Smirnoff (KS) statistic is calculated and shows that the distributions are consistent with each other. }
\label{ID_noID}
\end{figure}

	\fig{ID_noID} shows the measured $logN-logS$ for the detected blazars.  Also shown on this plot is the source flux distribution for the unidentified sources.  While the statistics are poor, there is no evidence for there being any detectable difference between these two groups of sources.  The Kolmogorov-Smirnov probability that these source distributions are drawn from the same underlying distribution is shown in the figure to be $85\%$.  Hereafter these distributions will be considered jointly.

\begin{figure}[h]
\epsfxsize=\hsize
\centerline{\epsfbox{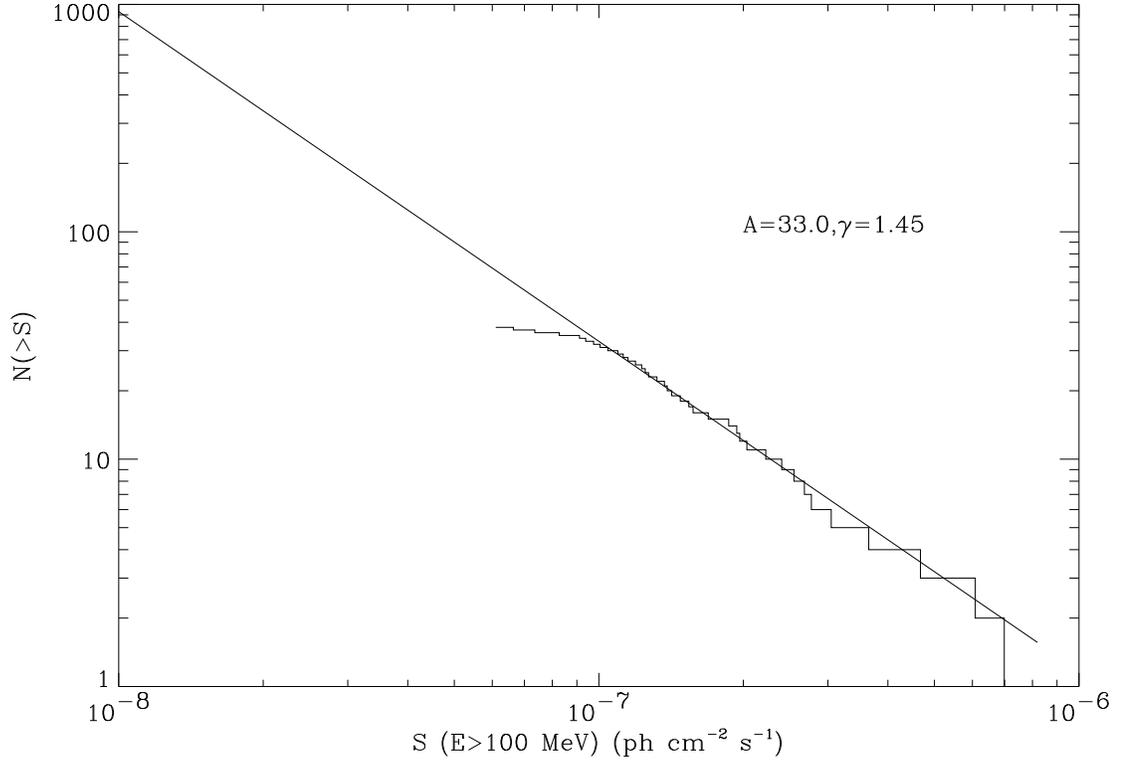}}
\caption{The measured $logN/logS$ function for the combined high latitude sources with $\mid b \mid >20^\circ$.  The best fit power law $logN-logS$ relationship is also shown.  This best fit function is slightly flatter than Euclidean.  The roll over below $1\times10^{-7}$ is most likely caused by the nonuniform detection threshold. }
\label{lognlogs.bestpow}
\end{figure}

	The simplest parameterization of the high latitude point source flux distribution is given by a truncated power law.
\begin{eqnarray}
	\frac{dN}{dS}=\left\{ \begin{array}{ll}
		A (S/S_0)^{-\gamma} & \mbox{$S>S_{min}$}\\
		0 & \mbox{$S<S_{min}$}
		\end{array}
	\right. 
\end{eqnarray}
and $A,\gamma,S_{min}$ are considered free parameters. \fig{lognlogs.bestpow} shows the measured high latitude $logN-logS$ function as well as the best fit power law parameterization.  The best fit parameters of this fit are $A=33.1,\gamma=1.45$.  This nearly Euclidean $logN-logS$ function leads to divergent values of the total diffuse flux if it is integrated down to arbitrarily small source fluxes.  It must break or roll over somewhere below the detection threshold.  The roll over visible at the smallest fluxes is most likely due to the non uniform detection threshold due to the variable exposure.

	A more realistic parameterization allows sources to be distributed not only in space but also over a range of intrinsic luminosities.  The number of sources per unit volume per unit luminosity,
\begin{equation}
	\rho(r,L)=\frac{dN}{dVdL} \;,
\end{equation}
is called the {\em luminosity function}.  Knowledge of the luminosity function allows one to calculate the observed $logN-logS$ function and can be integrated over all space and luminosities in order to calculate the total intensity contributed by a source class.
	
	In the case of blazars, the sources in question are located at cosmological distances (i.e. redshifts $z\sim1$).  In order to obtain a $logN-logS$ function from a luminosity function a cosmology must be assumed.  For the purposes of this analysis a flat universe will be assumed ($\Omega=1,\Lambda=0$ or alternatively $q_0=0.5$).  Furthermore the Hubble constant, $H_0$, which relates recession velocity (most commonly measured by redshift,$z$) to distance will be taken to be $ 75 \;{\rm km}\; {\rm s}^{-1} \;{\rm Mpc}^{-1}$.

	Using this cosmological model allows a source flux $S$ to be calculated given a luminosity $L$,
\begin{equation}
	S=(\frac{1}{\theta}) \frac{L}{d_l^2} \;,
\end{equation}
where $d_l$ is the {\em luminosity distance} which for our choice of cosmology is given by (\cite{Weedman86}),
\begin{equation}
	d_l=\frac{2c}{H_0 (1+z - \sqrt{(1+z)})} \;,
\end{equation}
and we have introduced the concept that the radiation emitted by a source may be beamed into a solid angle $\theta$.  One further complication is introduced if one is not dealing with bolometric flux but rather some discreet spectral window.  Because the frequency of a given photon depends on the redshift at which it is observed, one must adjust for the shifting of a spectral window with redshift.  This correction is called the K-correction.  Given power law energy spectrum with spectral index $\alpha$ the flux becomes,
\begin{equation}
	S=(\frac{1}{\theta}) \frac{L(1+z)^{1-\alpha}}{d_l^2} \;,
\end{equation}

	We can now compute an observed $logN-logS$ function given a luminosity function by substituting the relation $dL=dS(z)$ from equation 5.13 into the integral,
\begin{equation}
	\frac{dN}{dS}=\int_0^{z_{max}}dz \frac{dV}{dz} \frac{dN}{dVdL}\frac{(1+z)^{1-\alpha}}{\theta d_l^2} \;.
\end{equation}
where $dV/dz$ reflects the fact that the volume element per unit redshift changes as a function of redshift as the proper distance changes with redshift.  This relation can in turn be integrated in order to determine the contribution of the point sources described by a given luminosity function to the diffuse background.  
\begin{equation}
	I_{{\rm Sources}}=\int_{S_{min}}^\infty \frac{S}{\Omega} \frac{\theta}{4\pi} \frac{dN}{dS} dS \; ,
\end{equation}
where $\Omega$ is the total surveyed solid angle and the factor $\frac{\theta}{4\pi}$ accounts for the fact that as sources are more tightly beamed there is a higher probability that the emission will be beamed away from the observer.  Note that the beaming angle cancels out of this equation making the total contribution to the background insensitive to the beaming of the sources.  $S_{min}$ is determined by the minimum luminosity as well as by the maximum redshift.

\subsection{Direct Luminosity Function Constraints}

	The identified \egret\ blazars constitute a set of objects for which we now have measured \gammaray\ fluxes as well as measured redshifts from optical measurements.  This population can be used to constrain the luminosity function.  This has been done by Chiang \etal\ (\cite{Chiang95}).  In this work,  the luminosity evolution of the \egret\ blazars was deduced from the distribution of detected blazars using the $V/V_{max}$ test.  For a given detection threshold and a model cosmology, an accessible volume of space can be defined as the region of the universe within which a source of given luminosity is able to be detected by an instrument at earth.  If a given source remains at constant luminosity throughout its lifetime, it is equally likely to be found in the nearby half of this volume as the distant half.  As a result, the distribution of the statistic $V/V_{max}$ should have a mean of 0.5 in the limit of zero source evolution.  If a source tends to get brighter with time it will more likely be found in the near half of the total volume and as a result the mean $V/V_{max}$ would be smaller than 0.5.  Chiang \etal\ measure $V/V_{max}$ to be 0.7 indicating that \gammaray\ blazars were more luminous in the past.  This hypothesis is known as {\em pure luminosity evolution}.  The alternative hypothesis is that the blazars were more numerous in the past which is known as {\em pure density evolution}.

	The evolution was extracted by assuming functional form for the evolution,
\begin{equation}
	L(z)=L_0f(z)=L_0(1+z)^\beta \;,
\end{equation}
and adjusting the free parameter $\beta$ until $V/V_{max}$ was restored to 0.5.  This was achieved for $\beta=2.6\pm0.3$.

	Having measured the evolution, the sources were all de-evolved into the same epoch ($z=0$) and the luminosity function was determined by fitting a broken power law model to the luminosity function,
\begin{eqnarray}
	\frac{dN}{dV dL}&=& N_0(\frac{L}{L_B})^{-\gamma_1} , L<L_B \nonumber \\ 
			&=& N_0(\frac{L}{L_B})^{-\gamma_2} , L>L_B	
\end{eqnarray}
where $L_B$ is the break luminosity determined to be $10^{46} {\rm erg}\;{\rm s}^{-1}$ and the power law indices were constrained by the data to be,
\begin{eqnarray}
	\gamma_1&=&2.9\pm 2.0 \nonumber \\
	\gamma_2&=&1.6\pm 0.4 \nonumber \\
\end{eqnarray}
The large uncertainty in $\gamma_1$ reflects the small number of sources used to constraint the low luminosity end of this luminosity function. 

	Integrating over this luminosity function allows one to estimate the contribution of the blazars to the diffuse background.  In order to do this it is necessary to adopt a cut-off luminosity below which no sources exist.  This is required in order to prevent divergence of the total flux from these sources.  Because no such cut-off is directly observable in the \gammaray\ source distributions, the best that can be done is to adopt the lowest de-evolved luminosity as this cut-off.  This choice results in a lower limit for the total diffuse contribution because the true cut-off must be smaller than this value.  The resulting constraints are quite weak,
\begin{equation}
	I_{AGN}=\{0.5 -14.0\} \times 10^{-5} \;.
\end{equation}
All that can be said is that these sources contribute significantly to the diffuse background.

\subsection{Scaled Radio Luminosity Function Constraints}

	The large numbers of high latitude point sources that have been associated with flat spectrum radio quasars (FSRQ's) suggests that an appropriate test luminosity function to use would be the luminosity function for these sources which has been inferred from the large population of detected FSRQ's.  Dunlop \& Peacock (\cite{Dunlop90}) have analyzed several hundred flat spectrum quasars for which the redshifts are either directly measured or inferred from the relation between $K$-band flux and redshift for faint galaxies.  The data constrain the luminosity under a variety of parameterizations quite well to redshifts of about 2.  Beyond this there is some indication of a redshift cutoff at $z\simeq 5$.  If one again adopts the assumption of {\em pure luminosity evolution}, the luminosity function can be parameterized as,
\begin{equation}
\rho_r(L_r,z)=10^{-8.15}\{(\frac{L_r}{L_c(z)})^{0.83} +(\frac{L_r}{L_c(z)})^{1.96}\}^{-1} \;,
\end{equation}
where,
\begin{equation}
\log_{10} L_c(z)=25.26 + 1.18z -0.28z^2 \;,
\end{equation}
and the units of radio luminosity $L_r$ and co-moving density $\rho_r$ are, respectively, ${\rm W}\;{\rm Hz}^{-1}{\rm sr}^{-1}$ and ${\rm Mpc}^{-3}({\rm unit\;interval\;of}\;\log_{10}L_r)^{-1}$.

	In order to understand whether this luminosity applies to the \gammaray\ luminosity of AGN's, it must be ascertained whether the radio and \gammaray\ fluxes correlate.  Salamon \& Stecker (\cite{Salamon94}) report a strong correlation between radio luminosity and \gammaray\ luminosity.  However, the correlation they observe is between luminosities is strongly biased by their choice to correlate luminosities rather than fluxes.  Mattox (\cite{MattoxAPS}) has correlated fluxes using a larger sample of marginal \egret\ detected blazars using a rank order statistic and finds a significant correlation.  However, recent work by (\cite{Muecke96}) finds no evidence for a correlation between concurrently measured \gammaray\ and radio flux. A possible explanation for the weakness in the correlation is due to the fact that radio emission is believed to originate both in the relativistic jet as well in radio lobes in which decelerating electrons can create large radio fluxes.  While the core radio flux may well be correlated with the \gammaray\ emission, the radio lobe emission should not be and depending on the relative strengths of these two radio components the correlation could be made weaker or stronger.  Furthermore the large time variability of these sources makes the choice of what fluxes to correlate somewhat ambiguous.  While the evidence of this correlation is weak at best, it will be taken as an ansatz which will allow us to proceed.  More correlated VLBI radio observations (e.g. \cite{Grandi96}, \cite{Maraschi94}) will help to resolve this issue.

	The second challenge to the direct extension of the RLF to the \gammaray\ blazars is the relatively few radio loud AGN's that have been observed to emit in the \gammaray\ spectral region.  In order to produce such intense sources of high energy \gammarays\ the AGN emission must be beamed in order to avoid $\gamma\gamma$ attenuation which would prevent the \gammarays\ from escaping the source.  Salamon \& Stecker thus propose that the \gammaray\ emission is beamed into a smaller solid angle than is the radio emission.  This leads to the consequence that the earth can lie within the radio cone and outside the \gammaray\ cone to explain the discrepancy in the counting statistics.  Dermer (\cite{Dermer95}) has provided a theoretical justification for this assumption by demonstrating that synchrotron self Compton radiation will indeed be more tightly beamed in a relativistic jet than would synchrotron radiation.

\begin{figure}[h]
\epsfysize=4in
\centerline{\epsfbox{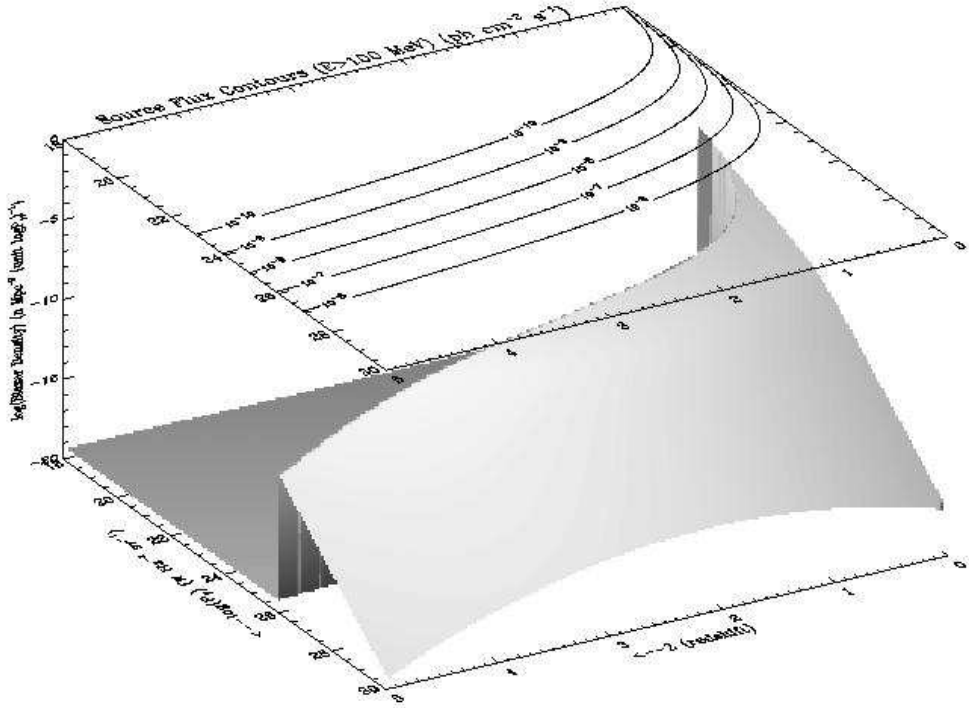}}
\caption{The \gammaray\ luminosity function for blazars modelled as discussed in this chapter.  The surface plot shows the density of blazars $(\rho)$, as a function of \gammaray\ luminosity ($L_\gamma$) and redshift ($z$).  The superimposed contour plot shows the contours of source flux as seen by the \egret\ telescope.  The luminosity function is truncated at the \egret\ detection threshold to demonstrate the region of $(L_\gamma,z)$ space which is inaccessible to \egret.}
\label{lumin.egret}
\end{figure}

	Taken together, these two assumptions provide a two parameter prescription for defining a \gammaray\ luminosity function (GLF),
\begin{equation}
\rho_\gamma(L_\gamma,z)=(\theta_\gamma/\theta_r)^2\rho_r((L_r/10^\xi),z)\;,
\end{equation}
This luminosity can be used as outlined above to generate an observed luminosity function.  \fig{lumin.egret} shows a surface plot of this luminosity function to help visualize the various distributions.  The surface has been truncated at the approximate \egret\ detection threshold for reasonable values of $\xi$ in order to illustrate the visible region of the universe.

\begin{figure}[h]
\epsfxsize=\hsize
\centerline{\epsfbox{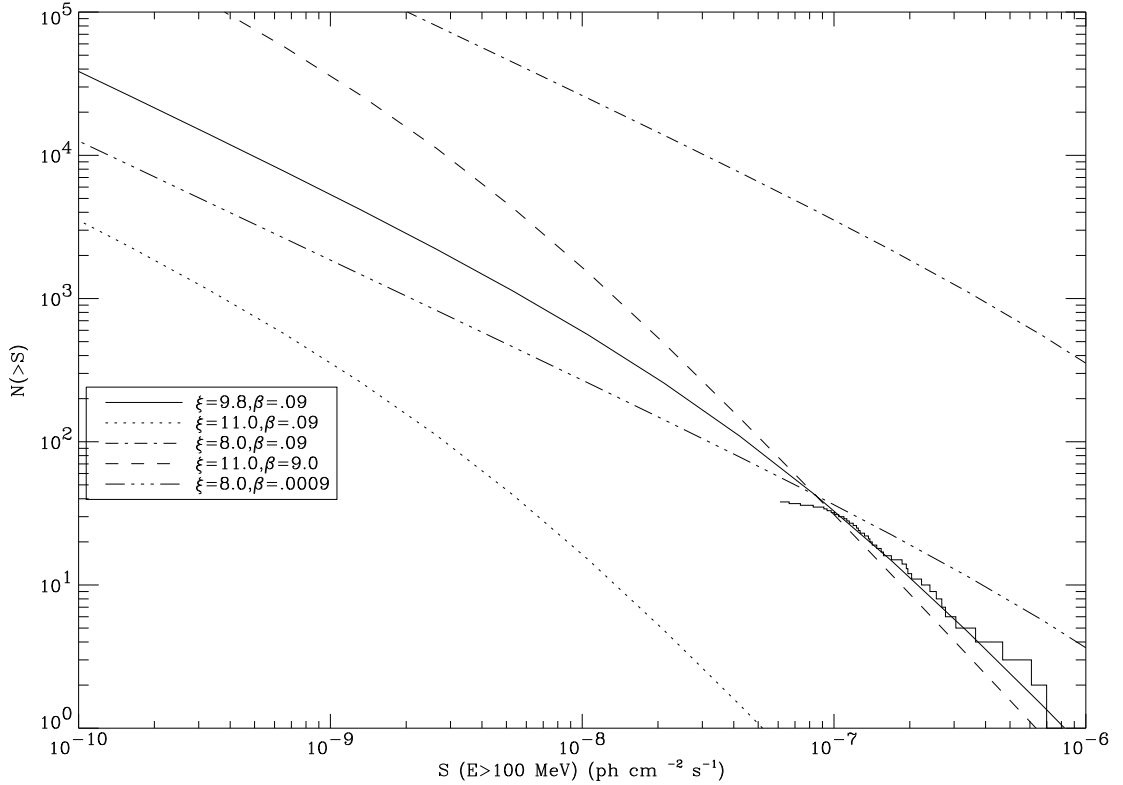}}
\caption{A family of $logN-logS$ functions generated using the scaled FSRQ luminosity function and several choices of the parameters $\xi,\beta$.  Also shown are the detected point sources with $\mid b\mid>30^\circ$.}
\label{steck.eg}
\end{figure}

	\fig{steck.eg} shows a family of $logN/logS$ functions  generated using this scaled FSRQ luminosity function for several choices of the parameters $\xi,\beta$.  The effect of decreasing $\xi$ is to increase the \gammaray\ intensity for a given radio flux.  As a result, a greater fraction of the universe becomes detectable and the $logN-logS$ function shifts to higher values.  There is a characteristic knee in the source distributions in this parameterization.  The other result of changing the $\xi$ parameter is to shift the position of this knee.  The second parameter $\beta$ merely shifts the $logN-logS$ distribution vertically.  As a result it is possible to adjust both parameters so as to keep the number of sources above some detection threshold constant as the slope at this threshold is adjusted.  In this way, the detected sources restrict the acceptable parameter space to a one dimensional band in $\xi,\beta$ space. The shape of the detected $logN-logS$ function above the detection threshold in principal provides an additional constraint but there is not much leverage in this range and the statistics are poor.

\begin{figure}[h]
\epsfxsize=4.0in
\centerline{\epsfbox{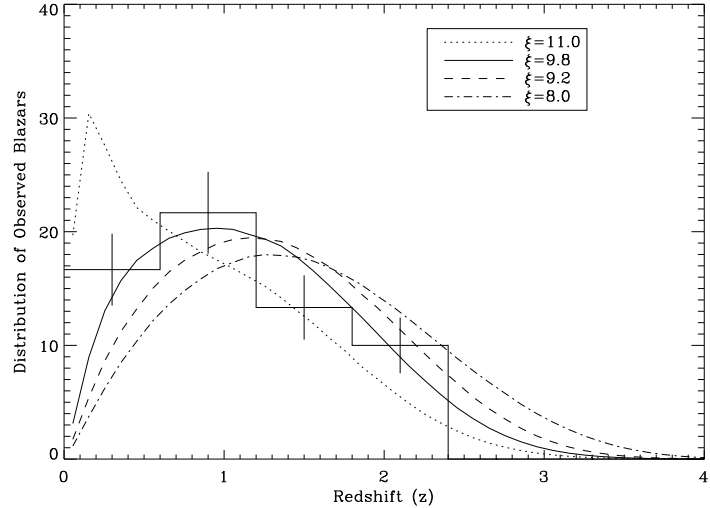}}
\caption{The measured redshift distribution for the detected blazars.  Also shown are the redshift distributions calculated using the scaled FSRQ luminosity function. The mean redshift of the distribution constrains the acceptable range of $\xi$.}
\label{zdist}
\end{figure}

	A more useful constraint is provided by the observed redshift distribution of the detected blazars, shown in \fig{zdist}.  The redshifts were compiled from the NASA/IPAC Extra-galactic Database (NED).  Shown superimposed are the redshift distributions calculated using the scaled FSRQ luminosity function and assuming a detection threshold of $1\times10^{-7}$.  Such distributions were calculated for a range of values of $\xi$.  The parameter $\beta$ does not effect the redshift distribution.  As mentioned above, low values of $\xi$ mean that more of the universe becomes visible to \egret\ leading to a larger mean redshift.  

	The combination of these two constraints limits the parameter space available to $9.0<\xi<11.0$.  The second parameter $\beta$ is then adjusted correspondingly in order to conserve the number of detected sources.  The resultant range is $.0001<\beta<9.0$.  These limits will be revisited in detail next chapter.

	Again it is possible to translate these constraints into contributions to the IDGRB.  Integrating over the luminosity functions using the scaled radio cut-off luminosity as the \gammaray\ luminosity cut-off yields,
\begin{equation}
	I_{AGN}=\{0.3 - 2.6\} \times 10^{-5} \;.
\end{equation}
This is slightly more restrictive than the constraints due solely to the \gammaray\ data but not sufficient to determine whether the bulk of the IDGRB is due to unresolved point sources.

\section{Discussion}

	The summed \egret\ data from the first $3\frac{1}{2}$ years of the mission have been reanalyzed using the more stringent albedo cuts motivated by Chapter 3. 55 sources were detected with $\sigma\geq4$.  Of these 37 of these sources have been identified with FSRQ's.  Another 17 are unidentified and the last is the Large Magellanic Cloud.  At least one of these sources is revealed by the elimination of the albedo background which had obscured it in the P12 analysis.

	A series of tests have attempted to ascertain whether the unidentified should be grouped with the identified blazars.  Results of the analysis of their spatial distribution, time variability, and source distribution are inconclusive.  There may be an indication that there is a galactic, time invariant component to the unidentified sources.  This has already been suggested by (\cite{Kanbach96}).  If these sources are indeed blazars the question of why no radio counterparts are detected must be answered.  The time variability of these sources may provide the answer. It may be that these sources were observed in a relatively radio quiet state.  There is, however, evidence that these sources are less variable in the radio than in the \gammaray\ band.  This makes such an explanation seem less likely.  Without strong evidence to the contrary, the unidentified sources will be treated as being due to the same blazar population for the rest of this discussion.

	The observed sources place constraints on the luminosity function which models the distribution of these sources in space and luminosity.  The work of (\cite{Chiang95}) has shown that while this is possible, the resulting luminosity function constraints are not very sensitive to low luminosity sources due to the small number of detections.

	This situation can be compared to the situation in the X-ray band in which the {\em Einstein Observatory} discovered many hundred high latitude sources. A uniform sample of 427 of these sources were  associated with spectroscopically identified AGN and used to constrain a luminosity function (\cite{Maccacaro91}).  These sources were sufficient to not only extract the evolution parameters but also to constrain the luminosity function over four orders of magnitude of luminosity.  This data was sufficient to resolve not only a break in the luminosity function power law but also to show that pure density evolution was not consistent with the data.  As a result the X-ray Seyferts have now been determined to be the dominant source of the hard X-ray background (\cite{Zdziarski95}).

	Adopting a scaled radio luminosity is problematic as well.  This resulting correlation between radio and \gammaray\ flux is not seen very strongly.  While it is possible to fit this luminosity function to the data assuming a smaller beaming angle for the \gammaray\ sources, the total contribution to the IDGRB is still not well determined.

	Because of the poor statistics of the detected blazars it is necessary to attempt to constrain the luminosity function by other means.  In the following chapter, indirect constraints on the blazar luminosity function will be evaluated using the technique of fluctuation analysis.

\chapter{Fluctuation Analysis}

	The observations described in the preceding chapters have established two results: an isotropic diffuse \gammaray\ flux of $1.33\times 10^{-5}$ is present, and a distribution of high latitude, isotropically distributed sources has been measured.  The logical question that follows is , ``Do unresolved members of this source population constitute the the IDGRB or is there some other truly diffuse source of this radiation?".  As shown in the preceding chapter, the detections alone are not sufficient to answer this question directly.

	Similar problems confronted scientists working on the origin of the X-ray background.  While it is now generally accepted that this background is due to unresolved Seyfert galaxies, this was not clear from such missions as HEAO-1.  One technique that was derived in order to deal with this problem in the X-ray band was the use of fluctuation analysis.

	Fluctuation analysis is a technique to search for the  signature of excess variance in an ensemble of observations of a quantity: in this case the intensity of the IDGRB.  This excess variance is used as a tracer for the inevitable random clumping of discreet sources.  The method was first developed for radio data (\cite{Scheuer57}) and has been used extensively in the analysis of the X-ray background (\cite{Shafer83}, \cite{Hamilton87}).  The approach presented here is directly adapted from the analysis used by Shafer.

\subsection{Distribution of Surface Brightness}

	The goal of this section is to derive the distribution of surface brightness, $P(I)$ in the presence of a population of point sources as seen by an instrument with a known point spread function.  Given a test $logN/logS$ relationship ($\frac{dN}{dS}$), we first determine the probability of picking a source at random from the sky and having it contribute an intensity $I$ to a pixel subtending a solid angle $\Omega$. ( Note that $I$ is a measure of surface brightness whereas $S$ is a measure of source flux.)
\begin{equation}
	P_1(I)=\frac{n(I)}{\mu} \;,
\end{equation}
where $\mu=\int n(I) dI$ represents the total number of sources which contribute to the intensity in the pixel and $n(I)$ represents the number which contribute and intensity between $I$ and $I+dI$.  
	
	Knowledge of the point spread function allows us to obtain $n(I)$ in the general case in which sources are not perfectly localized.
\begin{equation}
n(I)=\int_{\alpha,\delta} \frac{dN}{dS}(S(I,\alpha,\delta)) d\Omega\;,
\end{equation}
where $S(I,\alpha,\delta)$ is the flux of a source located at $\alpha,\delta$ that contributes an intensity $I$ to a pixel at $\alpha_0,\delta_0$,
\begin{equation}
S(I,\alpha,\delta)=\frac{I \Omega}{\int_{\alpha_0,\delta_0}PSF(\alpha,\delta,\alpha_0,\delta_0)d\Omega_0 }\;,
\end{equation}
and $PSF(\alpha,\delta,\alpha_0,\delta_0)$ is the appropriate EGRET point spread function.

	The general effect of an extended point spread function is to allow more sources to each contribute a smaller intensity to a given pixel.  As a result wider functions tend to wash out fluctuations.

	We now have a method for analytically determining the probability of obtaining an intensity measurement of $I$ if one source is near enough to a pixel to contribute to its intensity.  In order to find the probability of making such a measurement with an arbitrary number of sources allowed to contribute, we proceed by adding one source at a time. The probability of two sources jointly contributing an intensity $I$ to a given pixel is, 
\begin{equation}
	P_2(I)=\int_0^{I-I^\prime} P_1(I-I^\prime) P_1(I^\prime) dI^\prime \;.
\end{equation}
This is simply a convolution integral which is more easily evaluated in Fourier space.
\begin{equation}
{\cal F}(P_2(I))={\cal F}(P_1(I)){\cal F}(P_1(I)) = {\cal F}(P_1(I))^2 \;,
\end{equation}
where ${\cal F}(P)$ is the Fourier transform of the distribution $P$.  It follows that the probability of measuring an intensity $I$ given $n$ contributing sources is given by,
\begin{equation}
	 P_n(I)={\cal F}^{-1}({\cal F}(P_1(I))^n) \;.
\end{equation}

	The probability of $n$ sources being near enough to contribute to the intensity of a given pixel is determined by the Poisson probability with mean $\mu$.  Putting all this together we can now calculate the distribution of intensities in a given pixel with an arbitrary number of sources allowed to contribute.
\begin{equation}
	 P(I)=\sum_0^{\infty} \frac{e^{-\mu}\mu^n}{n!}{\cal F}^{-1}(({\cal F}(n(I)/\mu))^n) \;.
\end{equation}
Exploiting the linearity of the Fourier transform allows us to simplify this sum,
\begin{eqnarray}
	 P(I)&=&e^{-\mu}{\cal F}^{-1}\{\sum_0^{\infty} \frac{\mu^n {\cal F}(n(I)/\mu))^n}{n!}\} \nonumber \\
	&=& e^{-\mu}{\cal F}^{-1}(e^{{\cal F}(n(I))})\nonumber \\
	&=&{\cal F}^{-1}(e^{( {\cal F}(n(I)) -\mu)})\;. 
\end{eqnarray}

\begin{figure}[h]
\epsfysize=3.0in
\centerline{\epsfbox{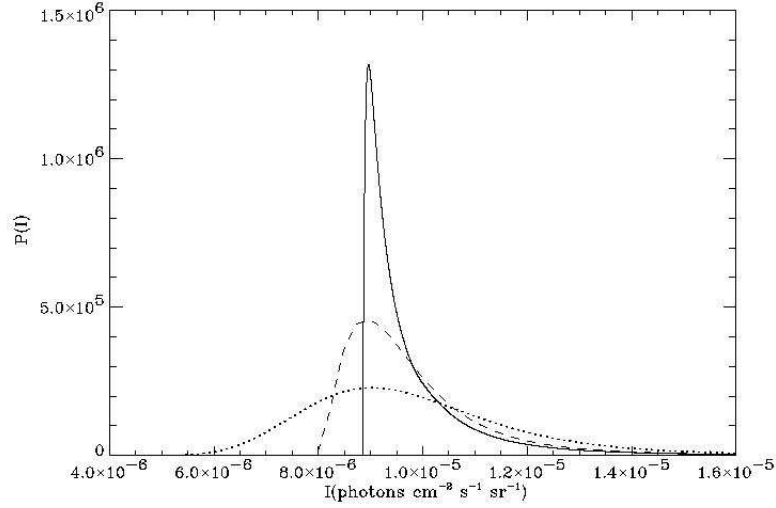}}
\caption{$P(I)$ for three different $logN/logS$ models with increasing point source contributions and decreasing true diffuse intensity.  Point source contributions of 10\% (solid line), 50\% (dashed line), and 100\% (dotted line) are shown.}
\label{fluc.fig.1}
\end{figure}

	$P(I)$ describes the variation in surface brightness across the sky due to the set of point sources.  \fig{fluc.fig.1} shows a series of $P(I)$ distributions for various $logN/logS$ relations.  In each case the total diffuse intensity is constant but the fraction of that total which is due to unresolved point sources increases from 10\% to 100\%.  The increasing variance in these distributions is due to the increasing numbers of point sources that contribute to these distributions.  It is this signal that indicates the presence of unresolved point sources.

	Because what is actually measured in a given pixel on the sky is a number of observed \gammarays\ and not a continuous variable $I$, the next step is to calculate the distribution of observed counts, $n_{ij}$, in a given pixel in the sky for which the galactic background intensity is modelled to be $G_{ij}$ and to which the there is exposure ${\cal E}_{ij}$.  This is simply the distribution of surface brightness $P(I)$ convolved with the Poisson probability of measuring $n_{ij}$ photons in that pixel.
\begin{equation}
P_C(n_{ij})_{ij}=\int_{I=0}^{\infty}\frac{e^{(-{\cal E}_{ij}(I+G_{ij}))}({\cal E}_{ij}(I+G_{ij}))^{n_{ij}}}{n_{ij}!}P(I) dI \;.
\end{equation}
It should be noted that the distribution $P_C(n_{ij})_{ij}$ bears an extra set of subscripts to denote the fact that this distribution only applies to a particular pixel.  This distribution must be calculated for each point in the sky to account for the varying exposure and galactic foreground emission.

\subsection{Fluctuation Constraints}

	The above discussion presents a methodology for obtaining a distribution of counts in a set of pixels given a test $logN/logS$ relationship.  We now turn our attention to solving the inversion problem that is presented by the actual data.  We wish to constrain model $logN/logS$ relationships given a set of measured counts in each pixel, $\{n_{ij}^o\}$.  We use the technique of maximum likelihood for this purpose.  If the $logN/logS$ relationship is parameterized by a set of parameters $\{\alpha_k\}$ then the likelihood is defined as,
\begin{equation}
	L(\{n_{ij}^o\};\{\alpha_k\})=\prod_{ij} P_C(n_{ij}^o;\{\alpha_k\})_{ij}\;.
\end{equation}
Again it is more useful to deal with the test statistic $T_S$,
\begin{equation}
	T_S=2 ln L(\{n_{ij}^o\};\{\alpha_k\})=\sum_{ij} 2ln( P_C(n_{ij}^o;\{\alpha_k\})_{ij}) \;.
\end{equation}
This quantity can then be maximized with respect to the parameters of the $logN/logS$ relation in order to find the best fit.

	The advantage of this method is that the behavior of $T_S$ in the null hypothesis is known. As discussed in Appendix A, $T_S$ is distributed as $\chi_\nu^2$ about the best fit value, where $\nu$ is the number of  free parameters (\cite{Eadie71}).  This allows one to calculate confidence regions for the parameters $\alpha_k$. 

	Maximizing $T_S$ reveals the most likely $logN/logS$ model given the data but it does not ensure that the absolute fit to the data is an acceptable one.  This is only the case if the parameterization includes the true distribution.  Because each data point is selected from a different distribution, it is not possible to simply compare the collection of data points to any one distribution to test the absolute fit.  Instead it is necessary to construct the distribution of the following quantity,
\begin{equation}
	y_{ij}=\int_0^{x_{ij}} P(x^\prime)_{ij} dx^\prime \;.
\end{equation}
This new statistic has the property that it is distributed uniformly on the interval $y=\{0,1\}$ if the values $x_{ij}$ are distributed as $P(x)_{ij}$. This can be seen easily by changing variables,
\begin{eqnarray}
	P^\prime(y_{ij})&=&\frac{dx}{dy}P(x_{ij})_{ij} \\ \nonumber
			&=& \frac{1}{P(x_{ij})_{ij}}P(x_{ij})_{ij}\\ \nonumber
			&=& 1
\end{eqnarray}
In our case the independent integral is over counts and so the integral becomes a sum,
\begin{equation}
	y_{ij}=\sum_0^{n_{ij}^o-1} {P_C(n;\{\alpha_k\})_{ij}} +  \frac{1}{2} P_C(n_{ij}^o;\{\alpha_k\})_{ij} \;,
\end{equation}
where the last term is divided by two in order to deal with the discontinuity at the end point of the integral.  Comparing the distribution of $\{y_{ij}\}$ with a uniform distribution gives a direct measurement of whether the absolute fit to the model is an acceptable one and as a result whether the parameterization is valid.

\subsection{Additional Constraints}

	The analysis that has been described thus far does not include the constraint that any choice of $logN/logS$ relationship that is used must match the observed distribution above the detection threshold.  Because this constraint should be independent of the fluctuation constraint, the joint likelihood is multiplicative.  If we detect $N_{obs}$ sources above the detection threshold,
\begin{equation}
	L(\{n_{ij}^o\},N_{obs};\alpha_k)=L(\{n_{ij}^o\};\alpha_k)\; \times \; P_{\rm detections}(N_{obs};\alpha_k)\;.
\end{equation}

	The probability of observing $N_{obs}$ sources above the detection threshold given a $logN/logsS$ relationship which predicts $N_{\alpha_k}$ sources above the detection threshold, is

\begin{equation}
	P_{detections}(N_{\alpha_k})=\frac{e^{-N_{\alpha_k}}N_{\alpha_k}^{N_{obs}}}{N_{obs}!} \;,
\end{equation}
So the likelihood function that describes both constraints is
\begin{equation}
	T_S=2 ln L(n,N;\alpha_k) = 2 (\sum_{ij}ln({P_C}(n_{ij};\alpha_k)_{ij}) ) -2N_{\alpha_k} + 2N_{obs}ln(N_{\alpha_k})\;,
\end{equation}
where we have dropped model independent terms.  This function is optimized with respect to the free parameters $\alpha_k$.

\section{Implementation}

	The implementation of the above technique will now be outlined.  The \egret\ photons are binned in celestial coordinate bins such that each bin has equal solid area.  This is done by increasing the latitude interval with  increasing latitude by a factor $1/cos(\delta)$.  For a given choice of input $logN-logS$ parameters, the distribution $n(I)$ is calculated by numerical integration using equation 6.3.  In general the tail of the point spread function causes $n(I)$ to get very large at small values of $I$.  In other words, a great many distant sources contribute small intensities to a pixel.  Computationally it is impossible to account for all of these sources explicitly, however, their effect is indistinguishable from a true diffuse flux because of the large numbers of these sources.  Consequently, all sources that contribute an intensity smaller than $I_0$ are converted into a set of sources with intensity equal $I_0$ such that the total intensity these sources contributes remains unchanged.  If $I_0$ is chosen to be sufficiently small,  the fluctuation analysis should remain unaffected by this computational approximation.  In the following analysis $I_0$ is chosen to be $1\times10^{-9}$ which is four orders of magnitude smaller than the diffuse intensity.  The analysis is completely insensitive to the form of $n(I)$ at this point.

	The distribution $n(I)$ is then used to calculate $P(I)$ using equation 6.9.  The Numerical Recipes subroutine {\tt dfour1} is used to perform the necessary Fourier transforms (\cite{Press92}).  A discreet imaginary array $2^{16}$ elements long was used to define $n(I)$ and $P(I)$. The step size used was $1\times10^{-9}$. The accuracy of the transform and inverse transform is gauged by checking that the quantity,
\begin{equation}
	\overline{I}=\sum_{i=0}^n P(I_i) I_i \;,
\end{equation}
is indeed consistent with the input value of the diffuse intensity due to point sources.
	
	For each pixel, the distribution $P_C(n)_{ij}$ was constructed.  The quantities ${\cal{E}}_{ij},G_{ij}$ were read from the exposure and background maps respectively.  The mean counts in a given pixel is given by,
\begin{equation}
	\overline{n}=(I_{{\rm IDGRB}}+G_{ij}) {\cal{E}}_{ij} \;.
\end{equation}
The integral over $I$ in equation 6.10 is carried out up to a maximum value of 
\begin{equation}
	I_{{\rm max}}=\frac{(\overline{n} + 10 \sqrt{\overline{n}})} {{\cal{E}}_{ij}} \;.
\end{equation}
The severity of this approximation is gauged again by check the integral,
\begin{equation}
	k=\sum_{n=0}^{n_{{\rm max}}}{P_C(n)}
\end{equation}
If this value deviates from 1.0 by more than a few percent then $n_{max}$ must be increased because there is a significant probability that $n$ exceeds $n_{{\rm max}}$.

	Finally the $T_S$ is calculated using equation 6.19.  The whole process is repeated for a different set of $logN/logS$ parameters and the best fit set of parameters is selected.  Confidence regions are ascertained using the contours of $T_S -(T_S)_{{\rm max}}$.  95\% confidence regions imply $T_S$ drops of 5.8 for 2 free parameter models and 7.7 for 3 free parameter models.

	A consistency check is then performed to ensure that the best fit model is a absolute good fit.  This is done by constructing the distribution of $y_{ij}$ and looking at the $\chi^2/\nu$ probability.  If it is not close to unity, the parameterization is inappropriate to the data.

\subsection{Spatial Scales}

	The spatial scale at which one searches for fluctuations is of critical importance to this analysis.  It is important to optimize the sensitivity of the analysis to changes in the point source distribution while minimizing any contaminating signal from other sources of emission.

	It is self evident that the fluctuation signal is washed out on large spatial scales because too many sources are allowed to contribute to the intensity in a given pixel.  However, the spatial scale of the test bins cannot be arbitrarily reduced as the analysis starts to become dominated by the Poisson noise associated with counting a small number of photons.  The most sensitive scale is therefore a compromise between these considerations. 

	This issue was investigated for the \egret\ data using simulated data.  Simulated \egret\ counts maps were generated using the galactic diffuse model and the combined exposure maps (see Appendix A).  A model of the sky is constructed including diffuse galactic emission, diffuse isotropic emission, and a list of point sources.  This set of point sources was sampled randomly from an input Euclidean point source distribution, distributed randomly across the sky, and convolved with the \egret\ point spread function.  The resulting maps were Poisson sampled in order to generate a typical \egret\ observation of this model sky consistent with the \egret\ exposure map.  This data can then be treated in the same manner as the flight data.

	For the purposes of this simulation a Euclidean $logN/logS$ relationship was used.  This $logN/logS$ relation was then extended below the detection threshold and truncated at a specific value.
\begin{eqnarray}
	\frac{dN}{dS}=\left\{ \begin{array}{ll}
		A (S/S_0)^{-1.5} {\rm srcs}\;{\rm sr}^{-1}& \mbox{$S>S_{min}$}\\
		0 & \mbox{$S<S_{min}$}
		\end{array}
	\right. 
\end{eqnarray}
Another way of parameterizing the same set of functions is using the integrated contribution to the diffuse flux, 
\begin{equation}
	\epsilon=\frac{\int_{S_{min}}^{S_{max}}S\frac{dN}{dS} dS}{I_{{\rm IDGRB}}} \;.
\end{equation}
In the simulation $S_{min}$ was chosen to be $2\times10^{-8}$.  This choice gives $\epsilon= 10\%$.  The data were then analyzed using a set of different values for $\epsilon$ and using a variety of pixel sizes.  The goal of this simulation is twofold: to provide a consistency check of the fluctuation calculation, and to find the spatial scale which shows the maximum sensitivity.

\begin{figure}[h]
\epsfysize=3.0in
\centerline{\epsfbox{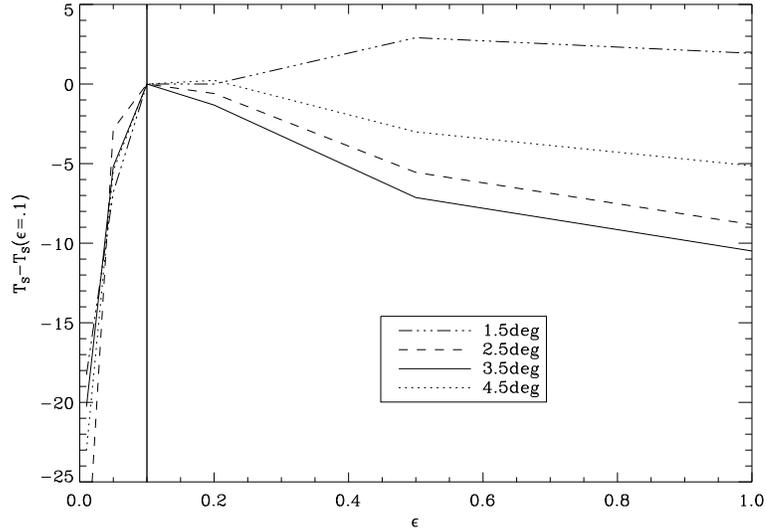}}
\caption{Likelihood curves as a function of $\epsilon$ on several spatial scales.  The input value of $\epsilon$ was $0.1$ which is consistent with the maximum likelihood value for all spatial scales except for the smallest.  The fluctuation analysis is shown to be most sensitive at $3.5^{\circ}$.}
\label{simscale}
\end{figure}

	\fig{simscale} shows the results of this simulation.  The first important result is that the likelihood ($T_S$) is maximized at the input $\epsilon$.  This verifies that the calculation of $P(I)$ and $P_C(n)_{ij}$ is accurate.  While this is not true in the case of the smallest spatial scale ($1.5^{\circ}\times1.5^{\circ}$ pixels), the likelihood function at this scale is so flat for $\epsilon > 0.1$ that this does not constitute a significant discrepancy. 

	The sensitivity of the analysis is measured by the steepness of the likelihood function around the best fit value of $\epsilon$.  This steepness reaches a maximum in the case of $3.5^{\circ}\times 3.5^{\circ}$ pixels.  At this scale, the average bin contains $\sim 50 \; {\rm photons}$.  Smaller pixels start to become dominated by Poisson fluctuations and larger pixels contain too many sources.

\begin{figure}[h]
\epsfysize=3.0in
\centerline{\epsfbox{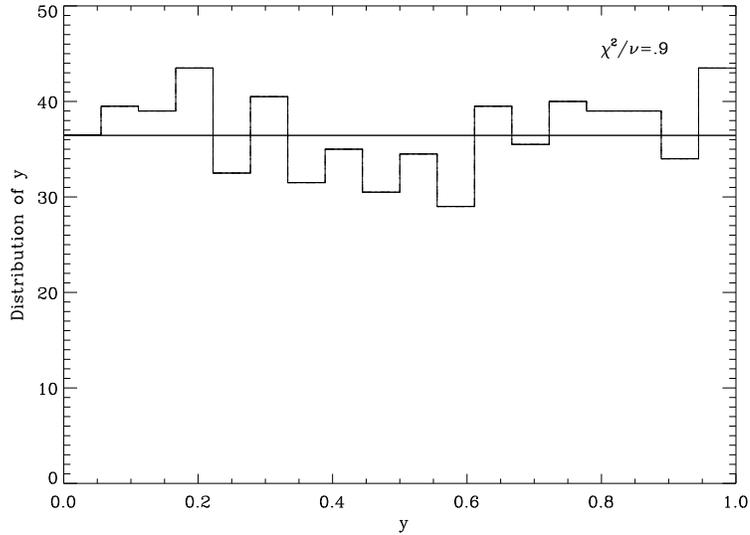}}
\caption{The distribution of $y$ as defined in equation 6.16 for the best fit model to the simulated data.  The value of $\chi^2/\nu$ shows that the distribution is entirely consistent with a uniform distribution.}
\label{sim_ydist}
\end{figure}

	The distribution of $y$ provides the final consistency check of this method.  \fig{sim_ydist} shows the distribution of $y$ for $S_{min}=2\times10^{-8}$ as well as the $\chi^2/\nu$ assuming a this data is drawn from a uniform distribution.  The result that $\chi^2/\nu$ is nearly unity is evidence that the calculated distributions $P_C(n)_{ij}$ are accurate.  This also implies that higher order effects that are not being taken into account are not seriously impairing the analysis.  The principal complication that is not treated in this analysis is the effect of the non-uniform exposure on the sky on $P(I)$.  In 
calculating $P(I)$ exposure is assumed to be uniform and the detection threshold is not allowed to vary across the sky.  Because the point spread function reduces the effects of distant sources this effect is not critical as evidenced by the relatively good agreement between simulations using the actual \egret\ exposure maps and the calculated distributions.

\subsection{Foreground Emission Uncertainty}

	The above simulations are carried out using a perfect model for the the foreground galactic diffuse \gammaray\ emission.  When analyzing the flight data, the potential effects of inaccuracies in the galactic emission model must be addressed.

\begin{figure}[h]
\epsfysize=3.0in
\centerline{\epsfbox{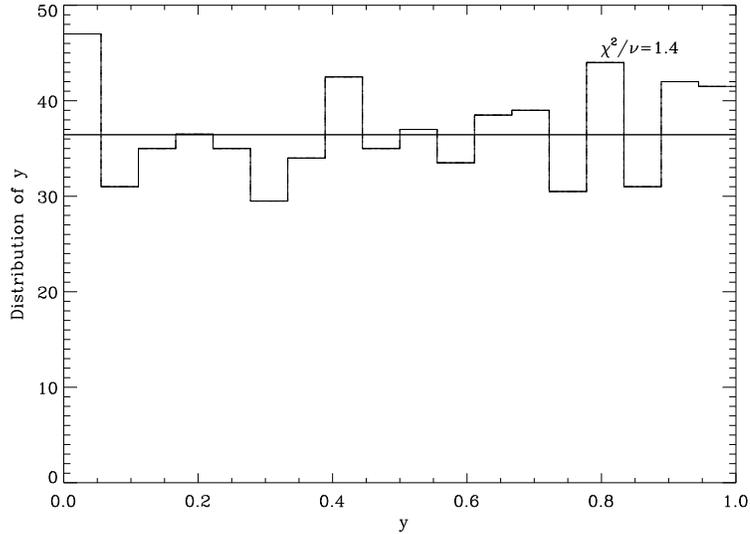}}
\caption{The distribution of $y$ as defined in equation 6.16 for the best fit model to the flight data using the galactic diffuse model derived from radio data.  The value of $\chi^2/\nu$ shows that the distribution is inconsistent with a uniform distribution and thus that the diffuse model is inconsistent with the data.}
\label{ydist_bad}
\end{figure}

	The diffuse gas model was discussed in the previous chapter.  The large  scale excesses at high latitude cannot be explained by point source fluctuations.  This point is emphasized in \fig{ydist_bad}.  This figure shows the distribution of $y$ for an input $logN-logS$ in which the point sources make up the entire IDGRB.  This model represents the best fit to the data and yet it is a demonstrably poor fit to the data as evidenced by the bowing of the $y$ distribution.  The data shows more variance than can be accounted for using point sources alone. In order to perform meaningful fluctuation analysis, it was necessary to use a diffuse model that gives isotropic residuals on large spatial scales.  Such a model is generated by the maximum likelihood fit to the diffuse background described in the previous chapter.

\begin{figure}[ht]
\epsfysize=3.0in
\centerline{\epsfbox{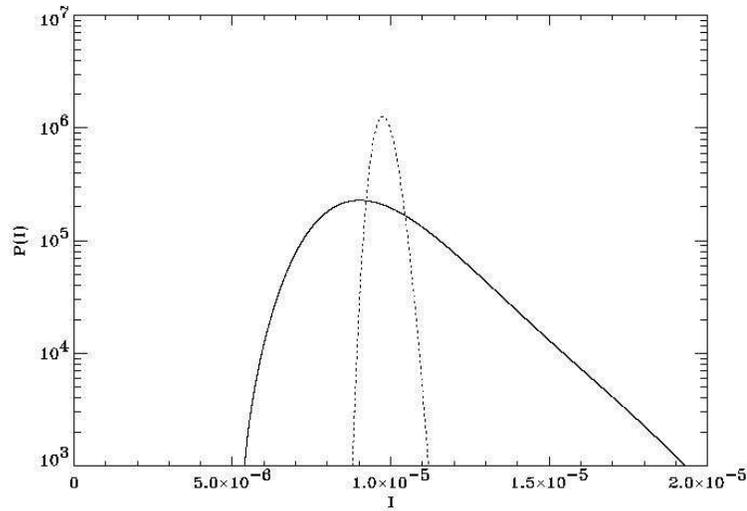}}
\caption{$P(I)$ distributions calculated using $(15^\circ)^2$ pixels (dashed curve) and for $(2.5^\circ)^2$ pixels (solid curve).  A Euclidean extrapolation of the detected source distribution was used to calculate these curves. The sharpness of the dashed curve demonstrates the absence of significant fluctuations on large spatial scales.}
\label{fluc.fig.4}
\end{figure}

	Using the data to constrain the diffuse model introduces potential biases to the analysis that must be guarded against.  The process of fitting the diffuse model to the data can erase real fluctuations on the spatial scale of the likelihood fit.  Thus if the diffuse parameters were fit on the scale of a few square degrees, the fluctuations would be absorbed into the foreground diffuse model and the fluctuation analysis would be biased toward smaller point source contributions.  \fig{fluc.fig.4} shows the relative fluctuations using $15^\circ$ radius bins compared to the fluctuations on a $3.5^\circ$ scale.  The relative absence of a signature due to point sources on the larger scale ensures that no significant point source signal is being absorbed into the galactic foreground model.

	A more important bias is caused by small scale fluctuations in the galactic foreground diffuse model that are unaccounted for when the likelihood scaling is done.  If the unmodelled source of emission that is removed through likelihood analysis exhibits spatial structure on the scale of the fluctuations, this structure could bias the analysis toward models containing higher point source contributions. 

	If it is assumed that the additional non-isotropic diffuse radiation beyond what is directly modelled is produced by some interaction between cosmic rays and photons or gas, then its spatial distribution should be similar to that of the modelled gas.  Under this assumption the bias caused by small scale fluctuations in this radiation can be directly estimated.  The integral in equation 6.10 becomes a double integral,
\begin{equation}
P_C(n_{ij})_{ij}=\int_{I=0}^{\infty}\int_{G_{ij}=0}^{\infty}\frac{e^{(-{\cal E}_{ij}(I+G_{ij}))}({\cal E}_{ij}(I+G_{ij}))^{n_{ij}}}{n_{ij}!}P(I)P^{\prime}(G_{ij}) dG_{ij} dI \;,
\end{equation}
where $P^{\prime}(G_{ij})$ expresses the uncertainty in the foreground emission, $G_{ij}$, on the spatial scales under investigation.  The galactic diffuse model can be used to directly measure $P^{\prime}(G_{ij})$.  This is done by calculating the residual intensity between a a pixel of solid angle $3.5^\circ \times 3.5^\circ$ and the mean intensity within $15^{\circ}$ of this pixel for a set of independent pixels across the sky.  This distribution is well approximated by a Gaussian with a relative error of 14\%.  This distribution is used in calculating $P_C(n_{ij})_{ij}$ in order to account for this bias.

\section{Results}

\begin{figure}[ht]
\epsfysize=3.5in
\centerline{\epsfbox{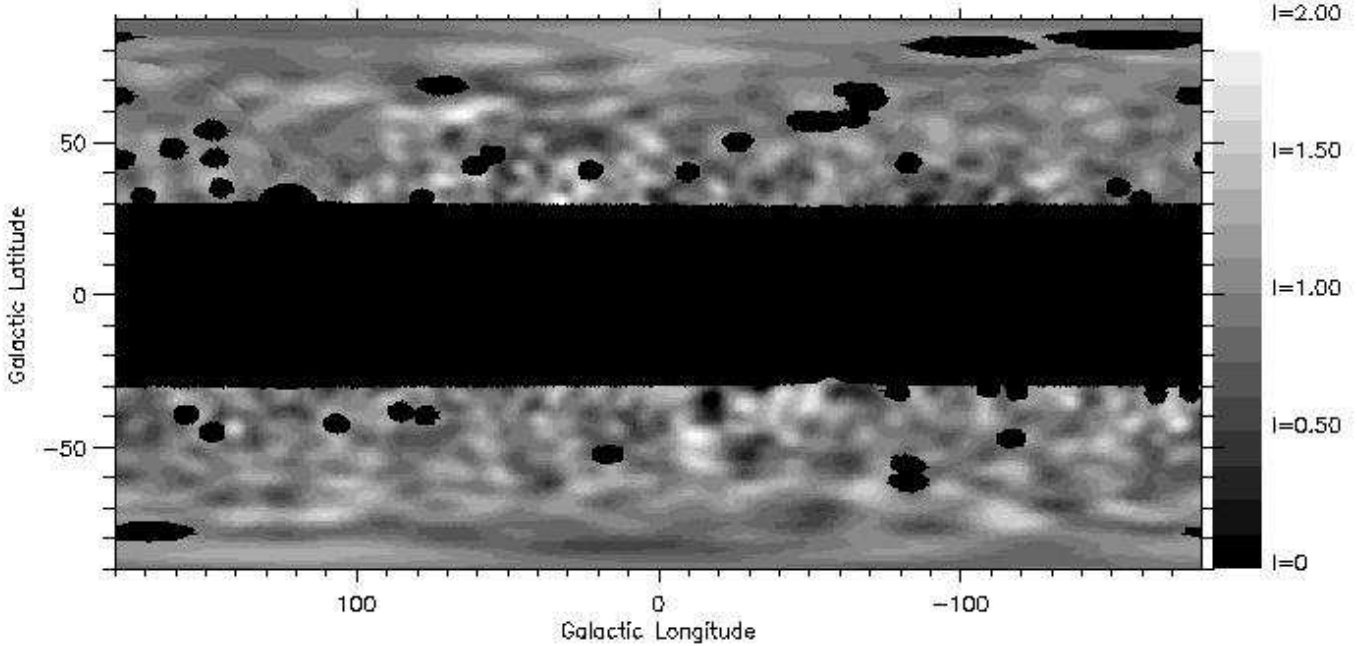}}
\caption{The residuals from the likelihood fit to the diffuse model.  Regions that are excluded from the analysis are shown in black.  These are regions near strong point sources and the galactic plane.}
\label{region}
\end{figure}

	The data analyzed using these techniques is shown in \fig{region}.  This plot shows the residual intensity from the likelihood fit to the diffuse model.  Regions that are excluded due to their proximity to detected points sources or the galactic plane are shown in black.

\begin{figure}[ht]
\epsfysize=4.0in
\centerline{\epsfbox{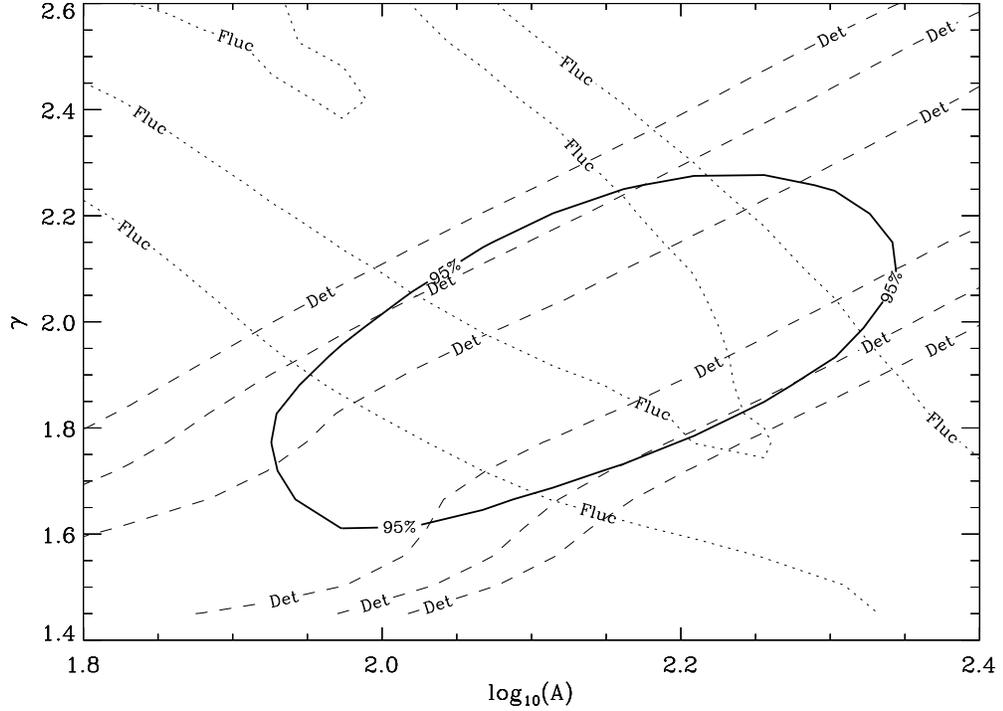}}
\caption{The likelihood contours in the space of power law source distributions.  The dashed lines indicate the constraints imposed by the directly detected sources.  Not surprisingly these constraints are only restrictive in one dimension as they are insensitive to the slope of the power law below the detection threshold.  The dotted lines indicate the fluctuation constraints.  These constraints are sensitive to both parameters but are more sensitive to the slope parameter.  The joint 95\% confidence interval is bounded by the solid line.}
\label{contour.power}
\end{figure}

	In order to understand how the various contributions to the likelihood function affect the constraints \fig{contour.power} shows the likelihood contours of the various components of the likelihood function separately as well as jointly.
The plot shows likelihood contours in the space of power law $logN-logS$ relations parameterized by $A,\gamma$ as defined by equation 6.24.  For the purpose of this plot, $S_{min}$ was fixed at $1\times 10^{-10}$.

	The contributions of the fluctuations to the overall constraints is clearly visible in this plot.  The constraint imposed by the detected point sources is manifestly insensitive to the slope of the $logN-logS$ function below the detection threshold.  As a result this constraint is linear in the function parameter space.  The fluctuation analysis is sensitive to both parameters but is more sensitive to the overall slope of the $logN-logS$ function.  The joint constraint, shown as a solid line in the contour plot, is thus bounded in both dimensions.

	The minimum point source contribution which is allowed by the data is determined by exploring the three dimensional parameter space in which the added dimension represents the position of the break in the $logN-logS$ function.  The 95\% confidence region in this parameter space achieves its minimum point source contribution at the point given by $(A,\gamma,S_{min})=(20,.6,<1\times10^{-8})$.  The analysis is not sensitive to values of $S_{min}<1\times10^{-8}$. This is because there are a sufficiently large number of sources below this flux to wash out their fluctuations.

	The constraints on steeper $logN-logS$ functions are more interesting.  It should be noted that Euclidean $logN-logS$ functions are strongly excluded as the maximum allowed value of $\gamma$ is 1.27.   It is clear that that the fluctuation analysis indicates a flattening of the source distribution below the detection threshold.

\begin{figure}[ht]
\epsfysize=4.0in
\centerline{\epsfbox{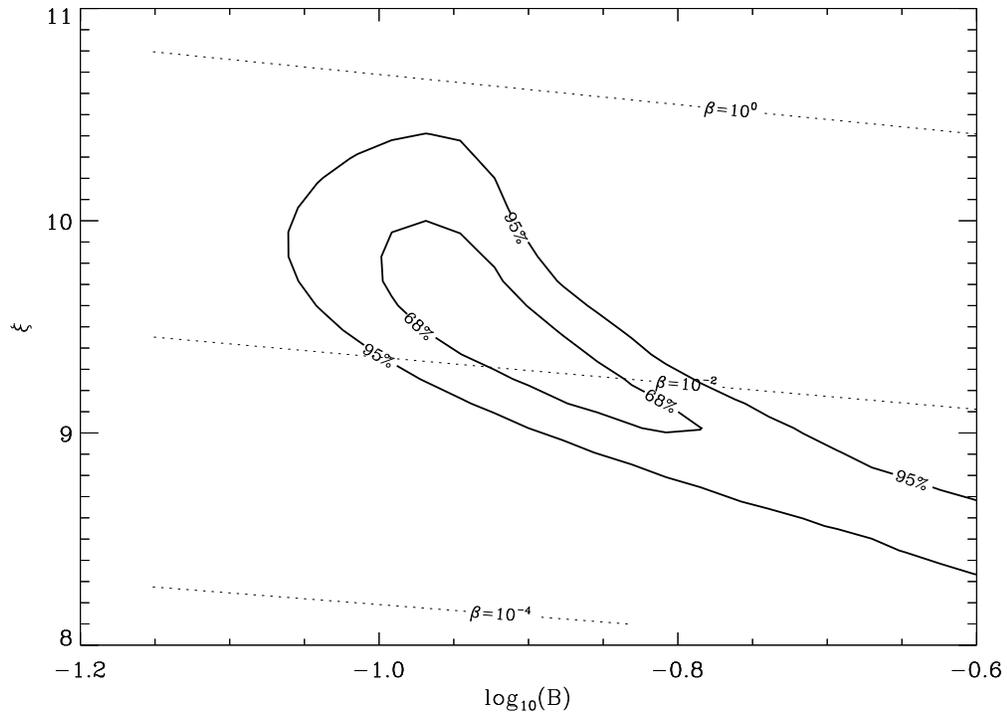}}
\caption{The likelihood contours in the space of scaled radio luminosity function source distributions.  The 68\% and 95\% contours are shown as solid lines.  Because the two parameters $\xi,\beta$ are highly correlated the parameter space is displayed as a function of a third variable $B$ which minimizes this correlation.  The contours of $\beta$ are overlayed in order to illustrate the range of this parameter allowed by the data.  The data are unbounded for small values of $\xi$ because the minimum slope of the $logN-logS$ relationship is determined by the luminosity function. }
\label{condraw_steck}
\end{figure}

	Any source distribution that approximates a power law is constrained according to the analysis presented above.  If the scaled radio luminosity function for flat spectrum radio quasars is used to generate a $logN-logS$ function as described last chapter, the result is not well approximated by a power law.  This is because there is a characteristic knee in the distribution.  For this reason this parameterization was treated separately.  As described last chapter, this method leads to a two parameter model of the $logN-logS$ function, $(\xi,\beta)$, where,
\begin{eqnarray}
	L_r&=& 10^{\xi}L_\gamma \\
	\beta&=&(\theta_\gamma/\theta_r)^2 \\
\end{eqnarray}
The constraints on these two parameters are highly correlated so the likelihood contours in this parameterization are shown in \fig{condraw_steck} in the space of $\xi,B(\xi,\beta)$ where $B$ is chosen to minimize the correlation between the uncertainty in $\xi$ and the uncertainty in $B$.  Contours of $\beta$ are shown overlayed on this plot.  It is important to note that if the knee in the $logN-logS$ function occurs at fluxes higher than the detected sources, the slope of the $logN-logS$ function does not vary as one changes the parameters.  At this point the parameterization becomes essentially a one dimensional power law model in which the normalization of the $logN-logS$ relationship is the only free parameter. For this reason the constraints are unbounded at the low end.  Physically what this represents is the fact that the scaling proportional constant between the \gammaray\ and radio luminosities can be made arbitrarily large as long as the \gammaray\ beaming factor is made  sufficiently small to retain the correct number of observed sources.

\begin{figure}[h]
\epsfxsize=\hsize
\centerline{\epsfbox{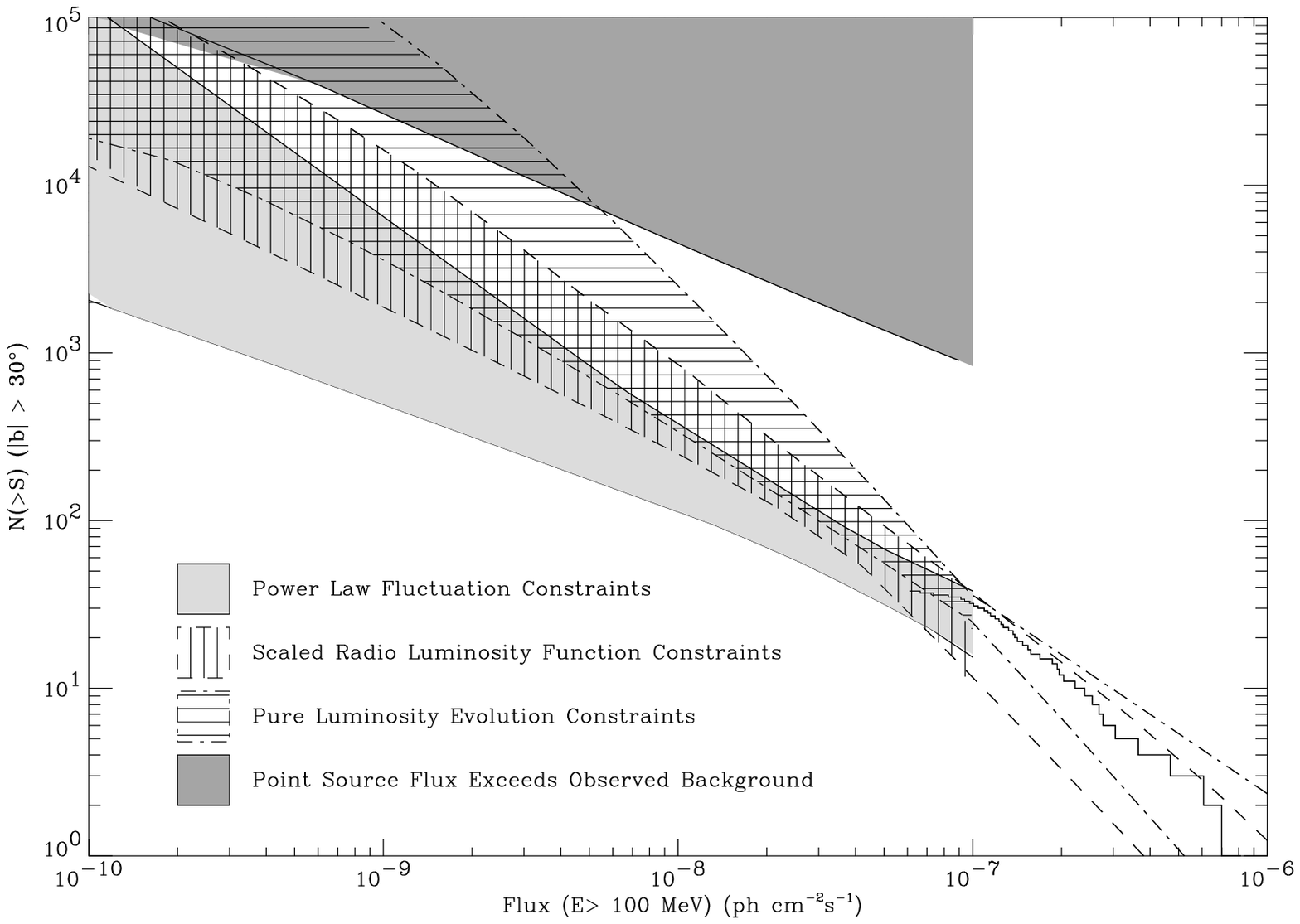}}
\caption{$logN/logS$ constraints.  The jagged line represents the flux distribution of the detected sources with $\sigma\geq4$.  The light gray region shows the region within which the extrapolated power law $logN/logS$ function must lie.  The horizontally hatched region displays the constraints on this function if it obeys a scaled version of the radio luminosity function.  The directly deconvolved \gammaray\ luminosity function under the assumption of pure luminosity evolution is constrained by the vertically hatched region.  The black region indicates the areas in this space in which the integrated intensity from the point sources exceeds the measured IDGB. }
\label{constraints.fluc}
\end{figure}

	 It is more instructive to display both these constraints by projecting the confidence interval back into the observed distribution space rather than the parameter space.  \fig{constraints.fluc} shows the fluctuation constraints as a gray shaded area on a $logN-logS$ plot.  All the source distribution functions allowed by the fluctuation analysis presented above lie within this region.  It should be noted that these boundaries are curved because they represent the locus of tangent lines to the bounding $logN-logS$ functions as a function of source flux.  The constituent $logN-logS$ functions are nonetheless straight power law lines.  The fluctuation analysis allows one to constrain the $logN-logS$ distribution an order of magnitude below the detection threshold.  In this representation the flattening of the $logN-logS$ distribution suggested by the fluctuation analysis is manifest.

	The constraints on the source distributions generated using a scaled radio luminosity are shown as a vertically hatched region.  The flattest allowed source distribution is comparatively steeper than the corresponding power law constraint.  This is a result of the fact that the parameterization does not allow flatter $logN-logS$ functions.  The luminosity function itself determines the lower bound on flat distributions.  The steeper models are, however, constrained by the fluctuations.  Here the upper bound is higher than for the power law constraints.  The steepening of the source distribution when parameterized in this way allows models with comparatively large numbers of sources below the detection threshold to remain consistent with the observed sources at the detection threshold.

	Also shown in \fig{constraints.fluc} are the family of source distribution models generated using the pure luminosity evolution model of (\cite{Chiang95}).  The region spanned by these models is shown as the horizontally hatched region in \fig{constraints.fluc}.  These models are steeper than the corresponding functions generated using a scaled radio luminosity function.  Since the source distributions generated by this model are essentially power laws, the gray areas on the plot represent the region of this space that is allowed for this model by the fluctuation analysis.  There is however a zone of consistency between the two regions. 

	The final constraint illustrated in \fig{constraints.fluc} is the limit imposed by the direct measurement of the IDGRB.  The summed contribution of the unresolved point sources must not exceed the total flux of the observed sky.  Because this constraint is on the integral flux, 
\begin{equation}
I_{{\rm Sources}}=\int_{S_{{\rm Threshold}}}^{S_{min}} S \frac{dN}{dS} dS \;,
\end{equation}
this region is not uniquely determined in this plot.  The region shown in dark gray represents the region within which a power law extending directly from the detected sources at the detection threshold yields more than the observed IDGRB. Because most functions discussed here are near power laws it is a good approximation to the region excluded by direct observation of the diffuse. It should be noted that most of the family of curves predicted by the pure luminosity evolution model are excluded by direct observation of the diffuse background.

\section{Discussion}

	The fluctuation analysis described here allows the next decade of point source flux to be probed without having to resolve the sources individually.  The limitations of this technique stem principally from the difficulty encountered in trying to separate variations in sky intensity caused by foreground emission from point source fluctuations.  Furthermore, there is 
no way to treat the identified and unidentified sources independently because both contribute to the fluctuation analysis in an indistinguishable way.

	Despite these difficulties, this analysis is able to place interesting constraints on the sources below the detection threshold.  Table 6.1 summarizes the constraints that result from this analysis.

\begin{table} [h]
\centering
\caption
{Summary of Fluctuation Constraints}
\bigskip
\small
\begin{tabular}{l|cr}\hline\hline
 Parameterization & Parameter Constraints & Diffuse Contribution \\ 
\hline
Power Law	& $\gamma=\{0.61,1.27\}$	&  $S_{min}=0: \{6\%,100\%\} $\\
		& $A=\{69.2,177.8\}    $	& $S_{min}=10^{-8}:\{4\%,15\%\}$ \\
\hline
Radio Luminosity Function	& $\xi=\{9.1,10.4\}$	&  $\{21\%,120\%\}$ \\
				& $\beta=\{.005,.4\}$&  \\ \hline
Pure Luminosity Evolution	& $\gamma_1=\{1.0,1.9\}$&  $\{5\%,14\%\}$ \\
				& $\gamma_2=\{2.0,2.6\}$&  \\

\hline \hline

 \end{tabular}
\end{table}

	The weakest constraints are those provided by using only the direct limits on the $logN-logS$ relationship without using the measured redshifts to constrain a luminosity function  The fluctuation analysis has revealed the existence of at least an additional 100 point sources  in the decade of flux below the \egret\ detection threshold.  These sources make up $4\%$ of the measured IDGRB.  Furthermore, the analysis described above has limited the number of sources in this decade to be less than $\sim300$ making up $15\%$ of the IDGRB.  In the absence of a means of determining the cut-off in the source distribution, the fluctuation analysis permits a large range of models in which the entire IDGRB is composed of unresolved point sources.  By itself it thus provides only comparatively weak constraints of the point source contribution to the IDGRB.

	If these source distributions are put in the context of a cosmological model through the introduction of a luminosity function, more restrictive constraints are possible.  Two luminosity functions have been explored in the last two chapters.  In each case, the directly observed sources provide the most restrictive lower limits on the point source contribution while the fluctuation analysis provides the more restrictive upper limit.

	The first model described uses the inferred luminosity function from the FSRQ's measured at radio wavelengths.  Under the ansatz that there is some scaling law between the radio flux and the \gammaray\ flux this luminosity function can be used to describe the \gammaray\ blazars provided the \gammaray\ emission is assumed to be beamed into a smaller solid angle than the radio emission.  

	There are three independent constraints on such a luminosity function:the source flux distribution of the detected sources, the redshift distribution of the detected sources and the fluctuation constraint on the sources below the detection threshold. As shown in the previous chapter, the redshift distribution constraint restricts the scaling parameter $\xi$ to the range $9.0<\xi<11.0$.  Models with smaller values of $\xi$ tend to predict sources at a higher mean redshift the is observed in the detected blazars and conversely models with larger values of $\xi$ tend to have distributions with smaller than detected mean redshift.

	The detected source distribution imposes a a relationship between the two model parameters $\xi,\beta$.  In order to predict the correct number of sources above the detection threshold of $\sim 1\times 10^{-7}$ as $\xi$ decreases, the beaming angle for blazar \gammaray\ emission relative to the radio emission must be similarly reduced.  The result is that the luminosity functions tend to predict larger numbers of sources with high \gammaray\ flux.  In order to make the number of sources above the detection threshold remain constant, these additional sources must be assumed to be beamed away from the earth which places a tighter constraint on their beaming angle.

	The fluctuation analysis is insensitive to the lower bound on $\xi$.  This is because the shape of the $logN-logS$ relationship is invariant to changes in $\xi$ once the knee in the distribution has been shifted to fluxes larger than the detection threshold.  On the other hand, large values of $\xi$ are efficiently constrained using this analysis because as the knee shifts through the detection threshold, the resulting number of sources predicted by the luminosity function below the detection threshold varies rapidly with increasing $\xi$.  The analysis presented above shows that the upper limit on $\xi$ can set at $\xi<10.4$ using this analysis.

	The real power of this approach lies in the fact that the constraints on the high flux tail of the luminosity function can be used to determine the two scaling parameters allowing the comparatively well constrained radio luminosity function to be used.  This luminosity function can then be integrated over the entire distribution which extends many orders of magnitude below the detection threshold in order to estimate the total blazar contribution to the IDGRB.  The results of this integration are shown in Table 6.1 for consolidated constraints provided by three independent measurements.  These joint constraints cannot exclude models in which unresolved blazars contribute $100\%$ of the IDGRB, however, this contribution can be as low as $20\%$.  Significant amounts of true diffuse emission cannot be ruled out on the basis of this analysis.

	The second luminosity function investigated was determined through the direct measurement of the \gammaray\ blazar luminosity evolution.  This model assumed pure luminosity evolution was responsible for the departure of the value of $V/V_{max}$ from 0.5.  Under this assumption a broken power law luminosity function was constrained using the de-evolved source luminosities of the detected blazars with measured redshifts.  This luminosity is also parameterized by two free parameters: the power law index above and below a break luminosity.  As shown in the preceding chapter, the power law index below the break luminosity is poorly constrained by the detected sources because very few low luminosity sources are detected by \egret.  As reported previously this index is restricted to the range $1.0<\gamma_1<4.9$.  The point source contribution to the diffuse background is, however, dominated by low luminosity sources and as a result reflects this uncertainty.

	The fluctuation analysis has again added an upper limit to the steepness of the $logN-logS$ distributions that result from this model.  In general the family of $logN-logS$ distributions which result from this model tend to be steeper than those resulting from the luminosity function previously discussed.  Most of the parameter space allowed by the detected sources is excluded when one considers the fluctuation constraints.  The fluctuation constraints limit the parameter $\gamma_1$ to the range $\{1.0,1.9\}$.  This reflects the fact that the fluctuations indicate a flattening of the $logN-logS$ distribution below the detection threshold.  

	As mentioned in the preceding chapter, a minimum luminosity must be assumed in order to evaluate the detection threshold.  Because this model does not assume a correlation between the radio and \gammaray\ flux, the cut-off luminosity is determined by the minimum observed luminosity.  As a result the integrated intensity due to unresolved sources in this model is really an upper limit to the point source contribution as the cut-off luminosity may be far smaller than the minimum detected luminosity.  As shown in Table 6.1, the integrated point source contribution for this luminosity function after applying the fluctuation constraints is $\epsilon=\{5\%,14\%\}$.  These are much smaller than the estimates made using the detections alone which predicted that the entire IDGRB would be made up of sources in the next 3 decades of flux below the detection threshold.  These model were biased by the steep $logN-logS$ distribution above the detection threshold.  This may be the result of the biasing of sources near the detection threshold toward higher detected sources.  This is caused by sources just below the detection threshold contributing to the flux of detected sources (\cite{Schmidt86}).  This effect is demonstrated in simulations described in Chapter 7.

	There remain many uncertainties associated with the point source contribution to the IDGRB.  The preceding analysis suggests that \gammaray\ blazars contribute significantly to the diffuse extra-galactic \gammaray\ background, however, it is not possible to say that they constitute the entire IDGRB.  Several outstanding questions remain.  These include:  the nature of the unidentified sources, the correlation or non-correlation between radio and \gammaray\ fluxes, the validity of pure luminosity evolution, and the cutoff luminosity of the luminosity functions.  Many of these outstanding issues will be addressed in the following chapter which describes what might be possible with a next generation \gammaray\ telescope.

\chapter{The Gamma-ray Large Area Space Telescope}

	\egret\ has provided an unprecedented view of the \gammaray\ universe.  It has brought \gammaray\ astronomy into the mainstream of astrophysical research.  This discussion of the diffuse background as measured by \egret\ highlights some of the limitations of this instrument.  While some of these limitations reflect the inherent limitations of satellite based \gammaray\ astronomy, others can be overcome in the future using the knowledge gained from the \egret\ mission. 

	This chapter will first outline some of the scientific issues pertaining the study of the  study of the diffuse \gammaray\ background that are beyond the reach of the \egret\ instrument and the requirements these issues make on a future \gammaray\ telescope.  The rest of the chapter will be dedicated to the description of one such possible future instrument: the Gamma-ray Large Area Space Telescope (GLAST).  The design of this telescope has been developed by a large collaboration of physicists (\cite{Michelson95}) led by groups at Stanford University and the Stanford Linear Accelerator Center (SLAC).  The design of this instrument will be summarized here as a prelude to a discussion of the scientific capabilities of such an instrument.  A virtual calibration of this instrument has been carried out through Monte Carlo simulation by W.B.Atwood at SLAC.  This work provides the basis for the evaluation of the instrument's scientific performance.

\section{Unresolved Scientific Issues}

\subsection{The Nature of the Isotropic Diffuse Gamma-ray Background}

	\egret\ has confirmed the existence of a high energy isotropic diffuse \gammaray\ background.  The intensity of this emission has been measured with unprecedented accuracy. Considerable uncertainty about its intensity remains due to the remaining uncertainty associated with our limited understanding of the foreground diffuse emission.

	As explained in the preceding chapters, the physical explanation of this background now centers on the radiation produced in relativistic jets emanating from massive black holes in active galaxies.  Such sources have been positively detected by \egret\ and unresolved members of this population undoubtedly contribute to the IDGRB.  Because only the brightest examples of these sources are observed, there remains considerable uncertainty as to their total contribution to the total IDGRB.  Fluctuation analysis has placed an upper limit to this contribution as being $>5\%$ of the total but there remains the possibility that much of the observed intensity is the result of a truly diffuse source.  In order to further constrain this contribution it will be necessary to achieve an increase in point source sensitivity in order to place more meaningful constraints on the \gammaray\ blazar luminosity function.

	In light of the difficulty of exactly accounting for the integrated blazar contribution to the IDGRB, a definitive discovery of a truly diffuse \gammaray\ background will most likely be discovered through spectral signatures.  As discussed in Chapter 1, the most intriguing possibility for such a discovery would result from the annihilation of a WIMP with some characteristic energy.  This sort of particle would result in a spectral bump (in the case of an unstable WIMP) or a spectral line (in the case of a stable WIMP) in the high energy diffuse spectrum.  Interesting constraints on such theories could be obtained with improved counting statistics at high energies.

\subsection{Understanding the Active Galactic Nucleus Acceleration Mechanism}

	In addition to measuring the contribution of active galaxies to the IDGRB, a detailed understanding of the physics of the acceleration mechanism in these sources remains beyond the scope of \egret. These sources are distributed over a wide range of redshifts ($.003<z<2.28$) and have can broadly be classified as blazars (a term encompassing BL Lacs, OVV and HPQ quasars).  Many are associated with superluminal radio sources.  The huge flux emitted in the \gammaray\ energy band as well as the time variability observed on time scales as short as a few days, all seem to point to massive black holes as the central acceleration engine.

	While there is general agreement that these extraordinary sources that the energy source is accretion onto a massive black hole, there is little agreement on the specific models used to explain the multiwavelength emission as well as the time variability.  The latter points strongly to a beamed highly relativistic jet being involved but disagreement persists as to the composition of such a jet (electon-positron jet versus electron-proton jet) (\cite{Mannheim92}, \cite{Blandford95}).

	The most powerful constraints on the physics of the acceleration mechanism arise from correlated multiwavelength studies that span the huge
spectral space in which these sources emit (\cite{vonMontigny95}).  There is currently some controversy about the correlation of \gammaray\ flares with optical and radio flares (\cite{Wagner95a}, \cite{Wagner95b}).  While the \egret\ team has led multiwavelength campaigns in order to address particular sources with simultaneous observations, it is impossible to predict the occurrence of such events due to the fairly small and uncertain duty cycle of blazars.  A large solid angle \gammaray\ monitor would allow much more powerful insight into the emission mechanisms as ground based observers in other spectral bands could be alerted to the occurrence of a blazar flare allowing correlated multiwavelength studies to be performed on a large population of flares.

\subsection{What is the Nature of Unidentified Gamma-ray sources?}

	Perhaps the most intriguing question to emerge from the \egret\ mission is the question of the nature of the many unidentified point sources that have been identified based on their \gammaray\ emission.  There are 80 sources reported in the second \egret\ catalog with no counterparts in other wavelength bands. Of these 32 lie within $10^\circ$ of the galactic plane and the remaining 48 are at higher latitudes. 

	The identified sources fall into two categories: nearby pulsars on the galactic plane, and blazars and BL Lac's off the galactic plane.  The only identified sources that do not fall into this category is the Large Magellanic Cloud (LMC).  It remains an open question whether the unidentified sources represent further examples of these two classes for which the corresponding radio and optical counterparts have yet to be discovered or whether they constitute an entirely new class of object.  

	The Geminga pulsar provides one clue as to the nature of the unidentified galactic plane sources.  While this source is the second brightest point source in the \gammaray\ sky, it shows no evidence of a radio counterpart.  This puzzle was resolved when \egret\ detected pulsations consistent with the period of the X-ray pulsations discovered using Einstein Observatory data (\cite{Halpern92}).  It is now believed that the radio and \gammaray\ emissions are beamed into different solid angles allowing the \gammarays\ from a source to be detected without a radio counterpart (\cite{Romani95}).  If this is the case then it is possible that some fraction of the unidentified sources on the galactic plane are simply pulsars whose radio emission is simply beamed away from the earth (\cite{Yadigaroglu95}).  A search for pulsations in these sources is underway (e.g. \cite{Brazier96}). Furthermore, it is possible that a faint radio counterpart has not yet been detected in the radio band.  Several searches of \egret\ error boxes have been performed (\cite{Nice94}).

	Another possible class of objects that may contribute to this population is that of Supernova Remnants (SNR). (\cite{Sturner95}) have found a statistical correlation between these objects and \egret\ sources.  An increase in resolution will greatly constrain this correlation.

	The high latitude unidentified sources have been discussed in some detail in Chapter 5.  There is little observational evidence with which to distinguish these sources from the blazars.  It is possible that they exhibit a smaller degree of time variability but this is most likely due to a bias caused by truncation at the detection threshold. As presented in Chapter 5 there is some suggestion of a slight excess of sources near the galactic plane.  This could be an indication that a fairly local population of galactic sources contributes to these detection but the evidence for this is statistically weak due to the small number of sources detected.  An increase in resolution is critical in narrowing the list of candidate counterparts to these sources.

\subsection{Closing the Gap with Ground Based Gamma-ray Astronomy}

	\gammarays\ with energies greater than $\sim 300\;{\rm GeV}$ cause sufficiently large electromagnetic showers when they interact with the earth's atmosphere, that the Cerenkov light from these showers can be imaged by ground based \gammaray\ telescopes (\cite{Weekes89}).  While the rejection of cosmic ray backgrounds is challenging, these telescopes have large effective areas which allow them to overcome the statistical challenge presented by the falling spectrum of sources in this region of the spectrum.

	To date 4 \gammaray\ sources have been detected by ground based sources.  The Crab nebula remains the brightest source in the sky at TeV energies.  The \egret\ pulsar 1706-44 has also been detected.  The spectra of these sources are well joined to the \egret\ data by synchrotron self-Compton models of these nebulae.  No evidence of pulsation has been detected at these high energies (\cite{Weekes89}).

	Perhaps more intriguing is the detection by the Whipple observatory of Mrk 421  (\cite{Punch92}) and subsequently Mrk 501.  These are both relatively nearby \egret\ detected BL Lacs with hard spectra.  Here again the spectra seem to match the extrapolation from the \egret\ data.  These detections suggest that the \gammaray\ production mechanism in blazars extends out to TeV energies.  The non-detections of the stronger but more distant blazars such as 3C273 and 3C279 (\cite{Kerrick95}) raises the possibility that these sources are extinguished at higher energies by $\gamma-\gamma$ interactions between these high energy \gammarays\ an the intergalactic IR radiation field (\cite{Stecker92}).
	
	  Unfortunately, the cut-off in these sources occurs below the region of the spectrum accessible to ground based telescopes and above the the maximum energy to which \egret\ is sensitive.  This hole in the explored spectrum extends from $\sim10 \;{\rm GeV}$ to $\sim300  \;{\rm GeV}$.  Not only is this an important region of the electromagnetic spectrum to explore for the reasons given above but the total lack of data in this band leaves open the door for unexpected new discoveries.

\section{Design Criteria}

	The above considerations can be used to inform the design of a future \gammaray\ telescope. The most obvious requirement of such a telescope is that it possess a greater active area than \egret.  This is of crucial importance in improving the counting statistics which remain a limiting constraint on \gammaray\ astronomy.  This in turn would increase the point source sensitivity as well as extending the accessible high energy spectral range, both of which are crucial performance criteria.  Because of the inherent limitations on absolute detector size imposed by the difficulty of putting a large instrument in orbit, the effective area goal should be to design an instrument of geometrical area comparable to \egret\ whose detection efficiency approaches unity.  The maximum \egret\ efficiency is of order 10\%.   

	A second generic requirement of a future \gammaray\ telescope is that the point source localization be improved. To some extent this will be a consequence of larger effective area and improved statistics, but it is more directly a function of the instrumental point spread function.  A reduction in the PSF width would allow more accurate source positions which will in turn aid the search for counterparts in other spectral bands.  Furthermore, improved resolution will allow a more detailed analysis of the structure of the galactic diffuse emission which will in turn allow more accurate models of this radiation to be constructed.  In order to achieve an improvement in resolution for high energy \gammarays\ in a pair conversion telescope, the tracker element spacing must be reduced or the tracker depth increased or both.  For low energy photons whose resolution is limited by multiple scattering it is necessary to attempt to minimize the thickness of the converter foils in order to preserve the directional information as the electron/positron pair travels through the tracker planes.

	In addition to these rather straightforward telescope improvements, the \egret\ data has added new constraints on future instruments.  The unexpectedly high degree of time variability of many of the detected sources places a premium on the ability to monitor many sources continuously.  This requirement argues strongly in favor of an instrument with an enlarged field of view in order to maximize the chances of observing flaring sources at all phases of their emission cycle.  In a pair conversion telescope, this capability requires a flattened aspect ratio in order to permit oblique photons to travel through a tracker to a calorimeter.

	Furthermore, the intriguing detections and non-detections of \egret\ sources at TeV energies emphasize the importance of extending the high energy reach of a future telescope.  The self-veto problems along with the falling spectrum of sources at high energy combine to limit \egret's range to $E<10\;{\rm GeV}$.  If this problem could be overcome in a larger telescope it should be possible to close the gap with the ground based telescopes at around 100 GeV.  

\section{Gamma-ray Large Area Space Telescope (GLAST) Baseline Design}

\subsection{Introduction}

	In the more than twenty years since the design of \egret\ there have been major technological advances that allow the basic principals of the \egret\ design to be upgraded to obtain significantly increased performance.  Most important among these are the improvements in silicon strip technology driven by the application of this technology to high energy physics detectors, and the vast improvements in available microprocessor power.  A collaboration between scientists at Stanford University, the Stanford Linear Accelerator Center (SLAC), the Naval Reseach Laboratory (NRL), and other institutions (\cite{Michelson95}) has incorporated these technological advances into a prototype \gammaray\ telescope design.  This instrument has been given the name: Gamma-ray Large Area Space Telescope (GLAST).

\begin{figure}[p]
\epsfysize=4.0in
\centerline{\epsfbox{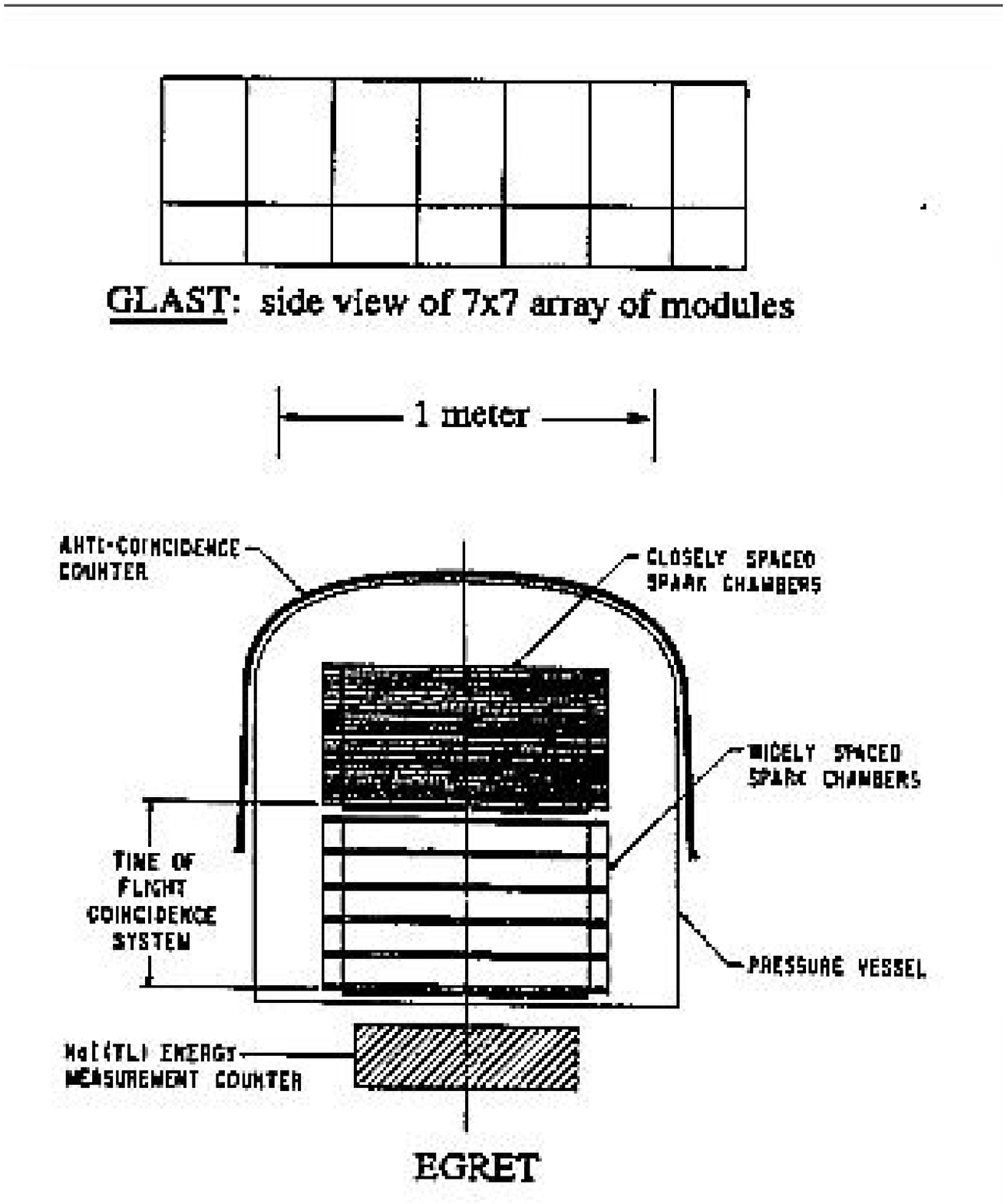}}
\centerline{GLAST and \egret\ cross-sections drawn to scale.}
\vspace{0.5in}
\epsfysize=3.0in
\centerline{\epsfbox{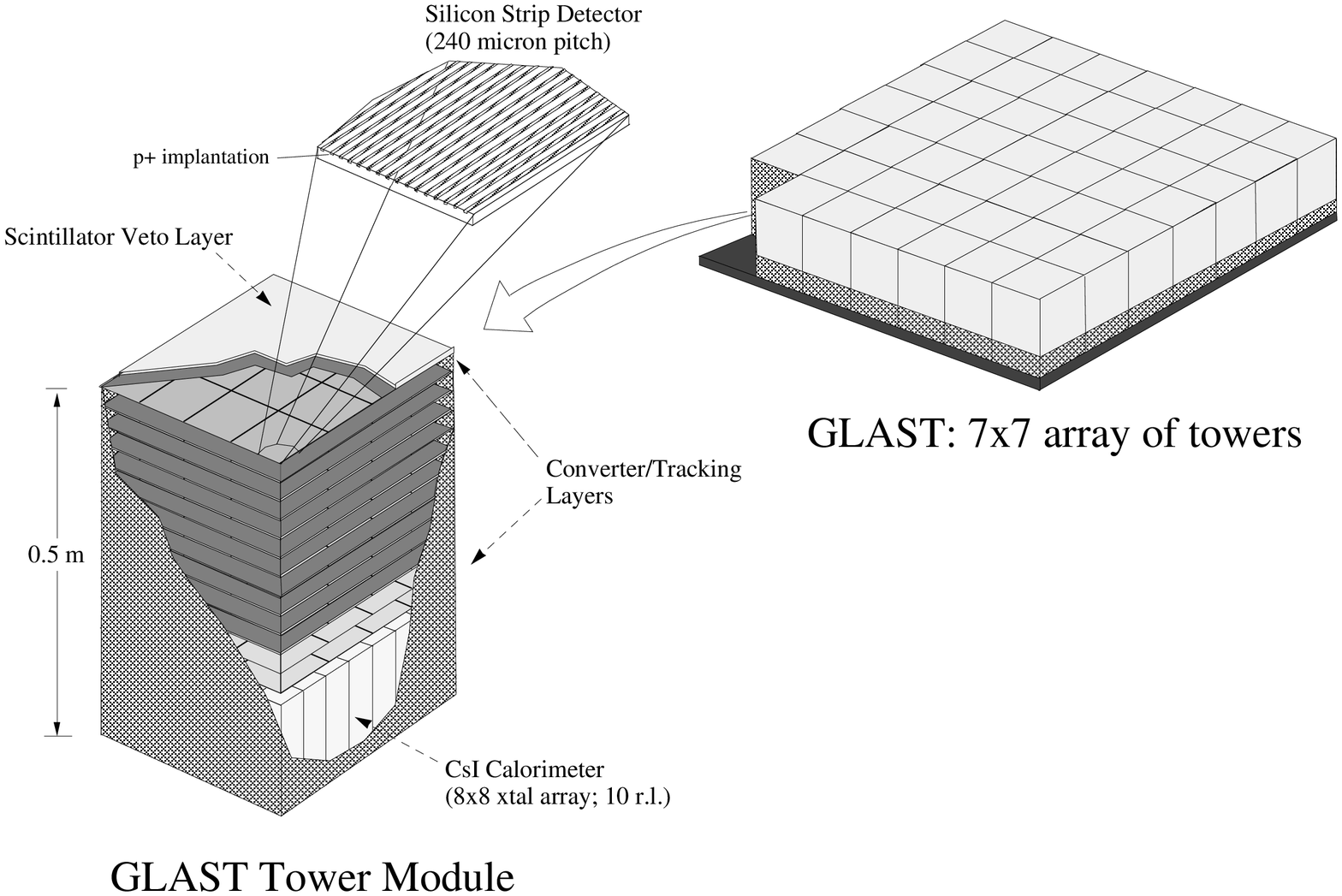}}
\caption{Schematic of the GLAST instrument showing one tower's internal structure.}
\label{GLAST1}
\end{figure}

	\fig{GLAST1} shows a schematic view of the proposed GLAST instrument as well as a similar view of \egret\ to scale.  The GLAST detector consists of a 7x7 set of towers.  Each tower consists of three familiar subsystems.  The surface of the tower is covered with an anti-coincidence scintillator layer.  The readout for each tower's anticoincidence layer is independent of the others.  Beneath this layer in the tracker.  This tracker consists of 12 planes of silicon strip arrays.  Each plane in the tracker has two sets of strip detectors that can accurately measure the charged particle tracks in two orthogonal dimensions.  The first ten planes are covered thin layer of lead to serve as converter material to initiate the electromagnetic showers.  Finally, the showers are collected in a calorimeter at the bottom of the tower.  The design of this calorimeter is still in progress but for the purposes of this discussion it will be assumed to be an array of CsI crystals each 10 radiation lengths long.

	The schematic views of \egret\ and GLAST illustrate several major advantages of GLAST over \egret.  First, the added size of GLAST results in increased effective area.  Furthermore the aspect ratio of GLAST is much flatter than \egret.  This results in a greatly increased field of view which as discussed above, is of critical importance for monitoring time variable sources.  This increase in field of view is made possible by the absence of a time of flight system that is used by \egret\ to reject upward moving cosmic-ray showers.  This innovation is made possible by the technological advances already mentioned and will be discussed in detail below.

\subsection{Veto Layer}

	The \egret\ veto layer functioned very well in rejecting cosmic ray backgrounds (see Chapter 3).  The drawback in the design of the \egret\  AC dome is the lack of segmentation.  All fluorescent light in the AC dome is collected by a set of phototubes whose signals are coadded.  As a result no positional information is available from the AC dome.  This is problematic because  high energy \gammaray\ events create large showers in the calorimeter.  It is not unlikely that a charged particle from such a shower should travel back through the spark chamber and trigger the AC dome.  The end result of this is that \egret\ tends to self-veto on high energy events.

	This effect, called the `backsplash effect', has been minimized in the GLAST design by segmentation of the veto layer.  Each individual tower retains its own layer of plastic scintillator. This plastic scintillator layer is read out with waveshifting optical fibers and Avalanche Photo Diodes (APD's).  If an event originates in a particular tower whose veto layer received a hit then the event is rejected.  However, if this veto hit occurs in any other tower, it is assumed to be backsplash and the event is processed.

\subsection{Tracker}

	As mentioned previously, developments in the use of silicon strip technology for particle detection provide the main technical stimulus for the GLAST tracker.  These solid state devices replace the obsolete spark chamber technology used aboard \egret.  The principal advantages of silicon strip technology over spark chambers are:
\begin{itemize}
\item Silicon strips can achieve $\sim10$ times better position resolution.
\item They exhibit no appreciable dead-time.
\item There are no consumables that limit mission lifetime.
\item They do not require high voltage.
\item The absence of a chamber gas and a replenishment system gives improved safety and reliability.
\end{itemize}
Taken together, these advantages amount to a very significant improvement in performance.

	The tower tracker planes are constructed from industry standard silicon wafers of 6 cm x 6 cm with a thickness of $300-500 \; \mu {\rm m}$.  A tower plane consists of two 3x3 chip arrays on top of each other; one layer each for the $x$ and $y$ position measurement.  Double sided devices could theoretically be used but the increase in cost does not justify the added performance.  Each chip has diode strips implanted on them to establish a readout pattern.  The strip separation (pitch) is chosen as a compromise between resolution and channel number which is limited by power requirements as will be discussed below.  The baseline design uses a strip pitch of $240 \; \mu {\rm m}$.  This pitch does not present a manufacturing challenge.  Strip pitches of $50 \; \mu {\rm m}$ are routinely used in particle physics experiments.  The chosen pitch will result in an rms position resolution of $\sim 69\;\mu{\rm m}$ in $x$ and $y$.

	All but the final two tracker planes are covered with a 0.05 radiation length lead foil.  The omission of converter material from the final two layers is a result of the requirement that at least three planes register hits in order to reconstruct a track.  Any conversion that would take place in the final two layers would be unable to produce a valid trigger.  The inclusion of converter material in these two layers simply serves to degrade the resolution of prior conversions through multiple scattering.

	One critical design feature of the tracker is that the converter material be very close to the first silicon plane.  It is apparent from the examination of a shower development that this is important to preserve low energy position resolution.  A conversion in a lead foil will immediately give a hit in the first silicon layer. This will give a very accurate measurement of the initial conversion point.  After traveling through the second converter, the secondary pair will undergo multiple scattering.  However, this scattering will be of minimal importance because the secondaries will have a high probability of interacting in the second silicon layer.  Because this second layer is very close to the converter, there is a very small lever arm for the scattering to degrade the position.  As a result the first two layers after conversion retain a large amount of directional information at low energies.

	At higher energies the directional resolution is no longer limited by multiple scattering but rather the geometric resolution resulting from finite pitch size.  In the current design in which the planes are separated by 3 cm the
rms angular resolution approaches $0.03^\circ$ in the limit of infinite energies.

\subsection{Calorimeter}

	The photon energy is collected and read out in the calorimeter.  The calorimeter must provide good energy resolution in the energy range from 20 MeV to 100 GeV.  As in the case of the veto layer, pixelation is introduced in order
to give a more complete image of a shower which will aid in background rejection.

	The current GLAST baseline design calls for the use of thallium doped cesium iodide crystals.  CsI(Tl) provides excellent energy resolution at modest cost, has a short radiation length (1.86 cm), provides a fast signal, and is radiation hard.  It is also a rugged material that has been used in space previously.  The calorimeter thickness is a compromise between weight constraints and high energy resolution.  The current design achieves this compromise by having 10 radiation length crystals.  Each tower site on top of a 64 element array of 3 cm x 3 cm crystals. 

	The calorimeter described above can resolve \gammaray\ energies out to $\sim 60 \;{\rm GeV}$.  At energies higher than this the bulk of the shower leaks out the back of the calorimeter.  Because GLAST sensitive areas are large enough to generate a significant number of such events, it would be desirable to be able to glean so information from these events.  This is particularly important in view of the fact that one of the major goals of a future satellite borne instrument will be to bridge the gap with air shower experiments.

	It is possible to retain some energy information from these massive showers if the shower can be imaged longitudinally within the calorimeter.  Several advanced calorimeter designs are under consideration which explore the feasibility of this idea.  These include longitudinal layering of the CsI crystals, and lead/scintillating fiber `spaghetti' calorimeters. 
\subsection{ Computing and Power Requirements}

	A recurring theme in the design outlined above is the segmentation of many of \egret's monolithic systems.  This segmentation increases the effective area by allowing more selective triggering.  The price that must be paid for this additional area is a non-trivial on-board computing requirement as well as a restrictive power requirement.

	In order to improve the tracking resolution, GLAST has increased the channel number dramatically from the numbers used on \egret.  A quick channel number count,
\begin{eqnarray}
	{\rm Channels} &=& (\frac{1}{240} \frac{{\rm Channels}}{\mu{\rm m}}) (10^4\frac{\mu{\rm m}}{{\rm cm}})( 6 \frac{{\rm cm}}{{\rm Chip}})( 9 \frac{{\rm Chip}}{{\rm Plane}})( 2 \frac{{\rm Plane}}{{\rm Level}})( 12 \frac{{\rm Level}}{{\rm Tower}})( 49 \frac{{\rm Tower}}{{\rm GLAST}})  \nonumber \\
       & \simeq & 3\times 10^{6}  \;
\end{eqnarray}
indicates that 3 million channels will be included.  If the strips are ganged together across an entire tower the number of channels is reduced to $\sim 10^{6}$.  Each of these channels requires power for a preamplifier and readout chip.  Current state of the art techniques in low power electronics indicate that a strip of this capacitance can be read out using optimized electronics requiring only $350 \; \mu{\rm W}$.  This places the power requirement for the entire tracker at $\sim 350\; W$.  This represents a significant increase over \egret\ which consumes only $\sim150$ W for the entire instrument, however, it does not pose a serious power generation challenge.

	Another issue raised by this power requirement is the ability to dissipate the heat generated by this power.  No high $Z$ materials such as copper are permitted inside the tracker because they would cause scattering.  Current schemes use aluminum to conduct away the heat.  Berrylium is also under investigation.

	Perhaps the most fundamental departure of the GLAST design from that of \egret\ is the trigger scheme.  The \egret\ instrument was designed to have a very selective trigger.  The requirements of a valid TOF measurement, a threshold energy in the calorimeter, and the absence of a veto from the AC dome combine to limit the trigger rate to 0.7 Hz.  This was done because the spark chamber gas degrades with each trigger.  Even with this selective trigger, the \egret\ instrumental sensitivity had dropped to 40\% of its pre-flight value before the latest spark chamber gas replenishment (\cite{Sreekumar96_inst}).  Furthermore, the spark chamber requires a 100 msec recharging time which implies that a much faster trigger rate would result in severe dead time complications.

	Because GLAST will not share these spark chamber limitations, the GLAST trigger can be much less selective.  This is reflected at a design level by the complete absence of a TOF system.  The philosophy of the GLAST design is to record as many events as possible while relying on pattern recognition in the hit patterns to reject background.

	As mentioned previously, the cosmic-ray background in low earth orbit is typically $10^4$ times greater than the corresponding \gammaray\ signal at a given energy threshold.  In a particle physics environment, it would be possible to record both background and signal events which would later be analyzed in order to distinguish the hadronic events from the electromagnetic ones.  In a satellite environment, however, the trigger rate is limited by the availability of ground communication bandwidth.  If a conservative upper limit to the communication rate is adopted to be $\sim100\; {\rm kbits}\; {\rm s}^{-1}$ then we can estimate the maximum allowable trigger rate based on the average event size.  The $\sim10^6$ channels are addressable with a 20-bit word.  If an average event contains $\sim100$ hits, then $250 \; {\rm bytes}$ are required just to describe the tracker pattern.  This should almost be doubled in order to account for calorimeter data as well as other signals (e.g. current pointing  knowledge, etc.) to give an average event length of $\sim 500 {\rm bytes}$.  The maximum tolerable trigger rate is then $\sim 25\; {\rm Hz}$.
	
	The challenge in implementing this sort of trigger scheme is that it requires non-negligible on-board computing power.  It is estimated that this trigger scheme could be implemented using a  15 MIP processor.  The technological advances in microprocessor have resulted in space qualified chips capable of delivering this sort of computing power.  Radiation hard versions of the commercial RISC processors ( e.g. IBM RAD6000) will most likely be available for a GLAST launch.  This is one of the main technology improvements that lead to the increased power of GLAST.

\section{Monte Carlo Simulations}

	In order to gain insight into the performance of the GLAST instrument, the powerful technique of Monte Carlo simulation has been employed.  Software developed for the simulation of particle physics detectors has been used to simulate both \gammaray\ and hadronic showers in a virtual GLAST and the results are then used to characterize the detector.  The simulations described below have been carried out at SLAC by W.B.Atwood and at University of Washington by T.Burnett.

	The simulations described below use the state of the art simulation package named Gismo (\cite{Atwood92}).  This program incorporates the QED interactions from the EGS4 code (pair production, bremsstrahlung, Compton scattering, Moller scattering, Bhabba scattering, pair annihilation, etc.)) and hadronic interactions from the Gheisha code ( proton-nucleus interactions, $\pi$-nucleus interactions, $\pi^0$ decay, charged $K_S$ decay, neutral $K_S$ decay, etc.).  Showers are calculated down to an energy of 200 KeV at which point the energy is assumed to be absorbed.  To date more than $10^6$ photons showers have been simulated by Gismo using the GLAST baseline geometry.

\begin{figure}[p]
\epsfysize=3.0in
\centerline{\epsfbox{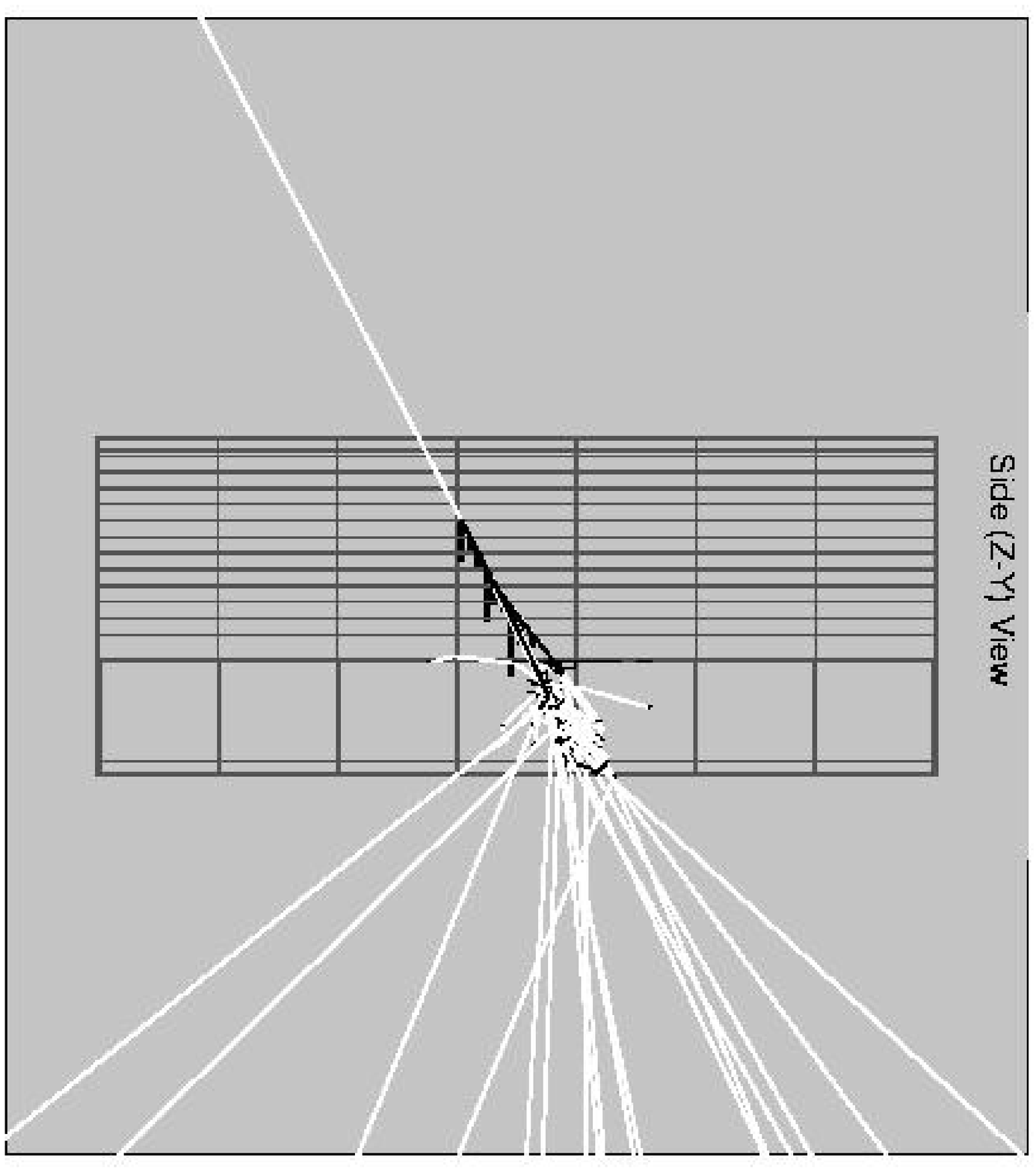}}
\vspace{0.2in}
\epsfysize=3.0in
\centerline{\epsfbox{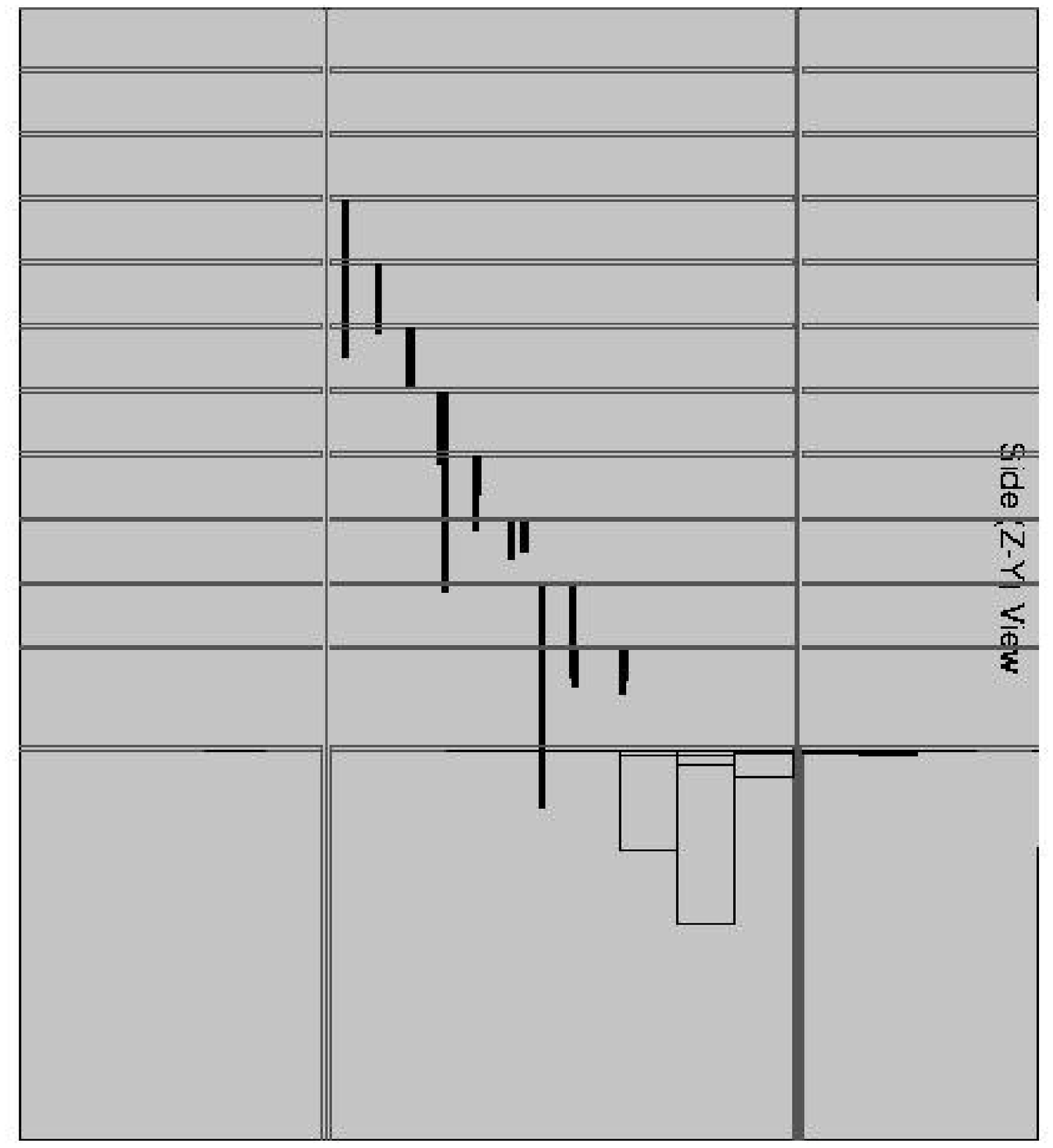}}
\caption{1 GeV photon simulation event.  The particle tracks are shown in white and the energy deposited in a given strip is shown as a vertical black line.  The energy deposition pattern in the calorimeter is also shown as a histogram over calorimeter elements.  The lower panel shows a close-up of the hit pattern alone.  Notice the collimation of the tracks.}
\label{G1000_all}
\end{figure}

\begin{figure}[p]
\epsfysize=3.0in
\centerline{\epsfbox{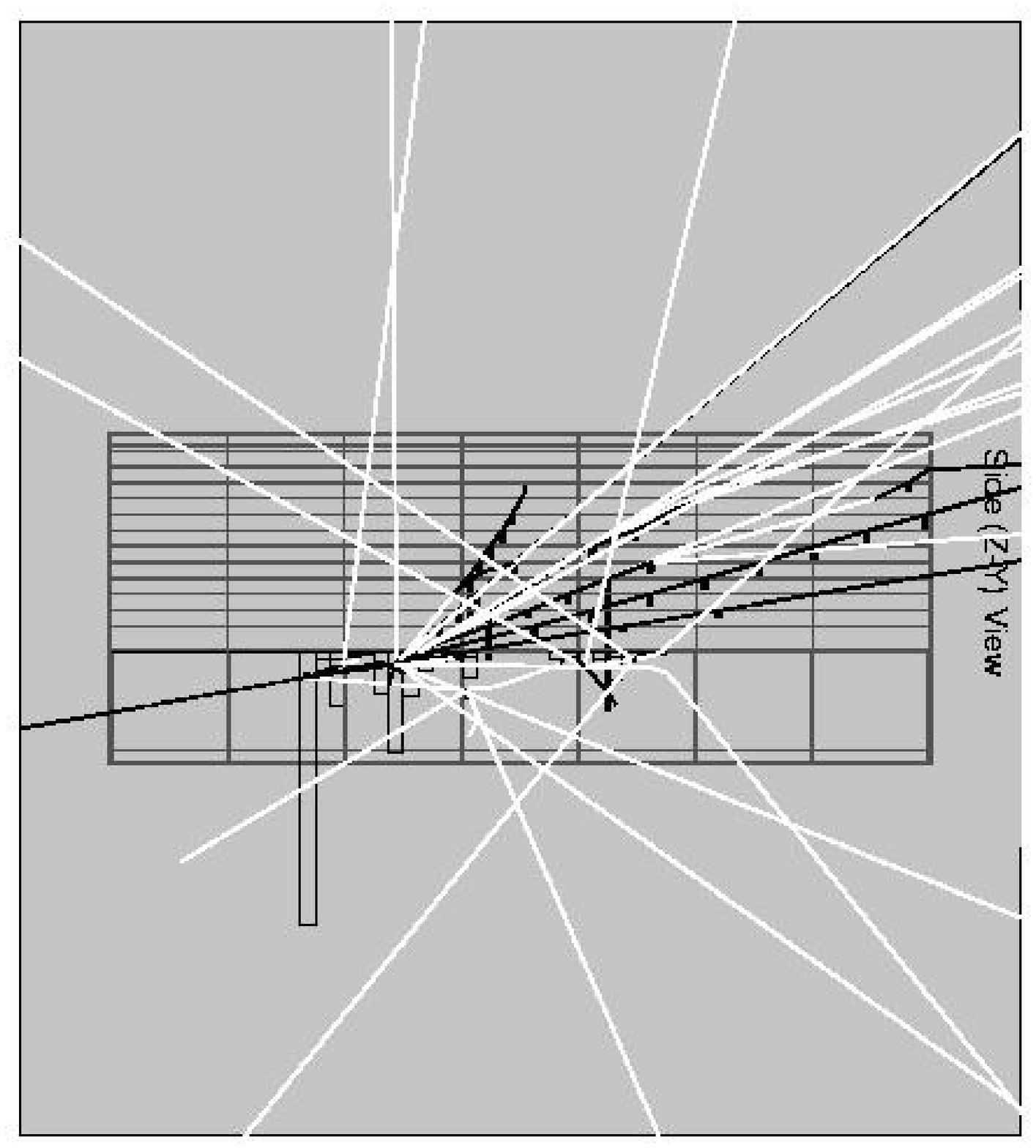}}
\vspace{0.2in}
\epsfysize=3.0in
\centerline{\epsfbox{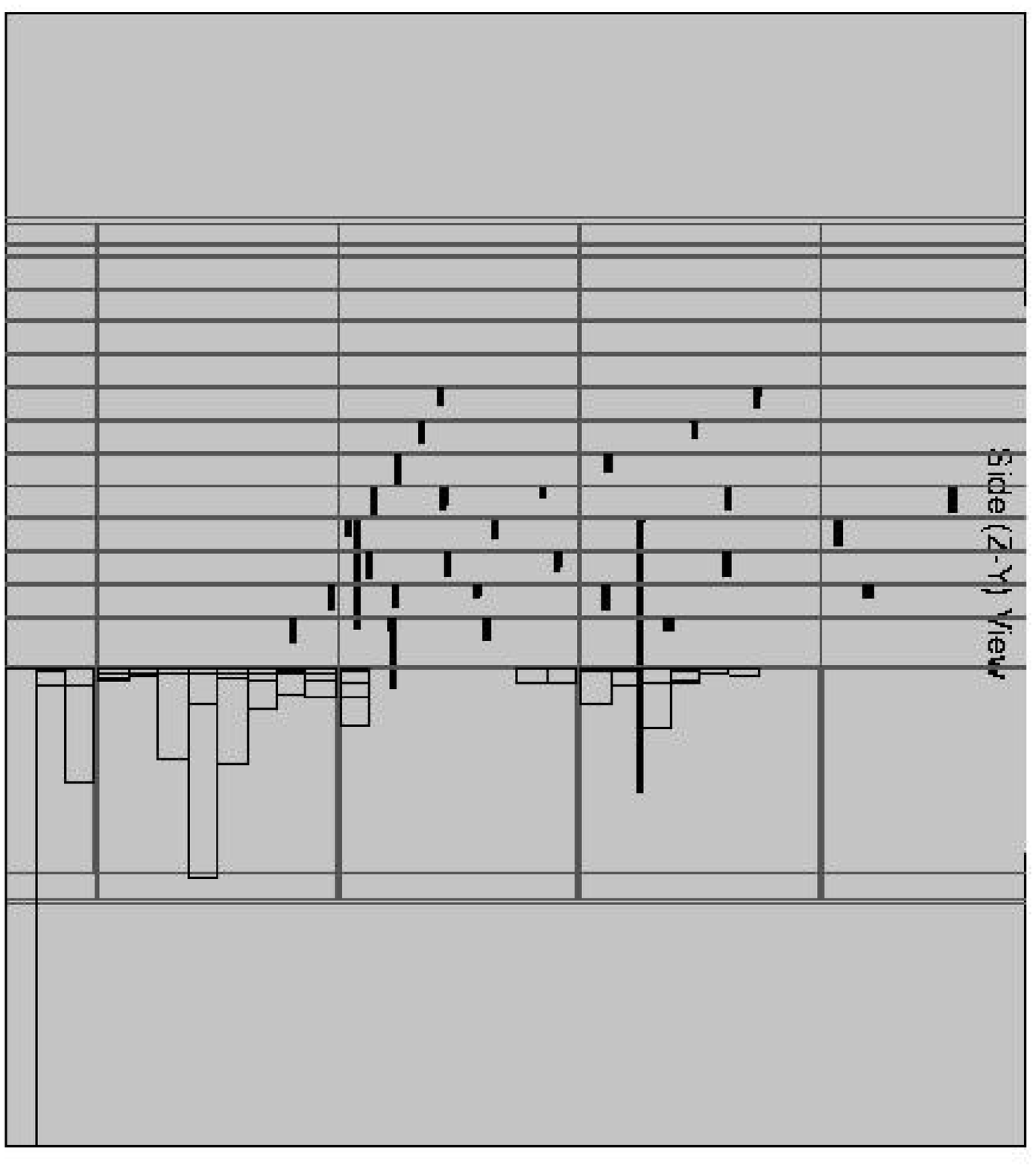}}
\caption{15 GeV proton simulation event.  Again both particle tracks and resultant hit pattern are shown.  Such a proton would miss the veto layer and thus would have to be rejected on the basis of the track pattern.  Notice the track scatter compared to the electromagnetic shower produced by the \gammaray\ event.}
\label{P15_all}
\end{figure}

	\fig{G1000_all} and \fig{P15_all} show a comparison between a simulated electromagnetic shower caused by a 1 GeV \gammaray\ and a hadronic shower caused by a 15 GeV proton which enters GLAST from such an angle as to miss the veto layer.  The particle tracks are shown in the first image and the resultant hit pattern in the detector is shown in the second frame.  The comparatively tighter, more collimated shower is produced by the \gammaray.  This provides one of the principal background rejection techniques as will be described below.

\subsection{Background Rejection}

	The GLAST trigger design is multi-level.  The level 1 trigger is simply the requirement that the silicon tracker show hits in three consecutive planes.  The Monte Carlo simulations described above using a realistic cosmic ray spectrum indicate that this level one trigger rate will be $\sim5\; {\rm kHz}$.  The level 2 trigger requires the best fit track to be reconstructed.  A clean \gammaray\ track will have no hits prior the conversion point whereas a hadronic shower is likely to have scattered hits above the best fit track conversion point.  The simulation results indicate that a cut on the distance of closest approach of a hit prior to the best fit conversion point to the best fit track, can give a background reduction ratio of 250:1 while only eliminating a few percent of the good \gammarays\ (\cite{Michelson95}).  The second level trigger reduces the trigger rate to $\sim20\; {\rm Hz}$ which is sufficient to allow the data to be funneled through the communications bottleneck.

	Down-linked data can then be subject to further pattern recognition cuts in much the same manner as \egret\ data.  A series of cuts on the Monte Carlo data have successfully demonstrated the ability to reject cosmic rays at the level of $5 \times 10^{-5}$ while preserving 80\% of the \gammaray\ induced triggers.  While this software can certainly be improved in order to give still further selectivity, the background rejection already demonstrated leads to an cosmic ray induced background an order of magnitude less than the extra-galactic diffuse radiation.

	It is worth highlighting the fact that such method of background rejection is already in use on the \egret\ data.  The discussion in Chapter 3 of the \egret\ instrumental background indicated that there is almost certainly a significant contribution of cosmic rays to the \egret\ hardware trigger rate.  Nevertheless,  these events appear to be very efficiently rejected by the \sage\  software package developed by the \egret\ team.

\subsection{Performance Parameters}

\begin{figure}[ht]
\epsfysize=6.0in
\centerline{\epsfbox{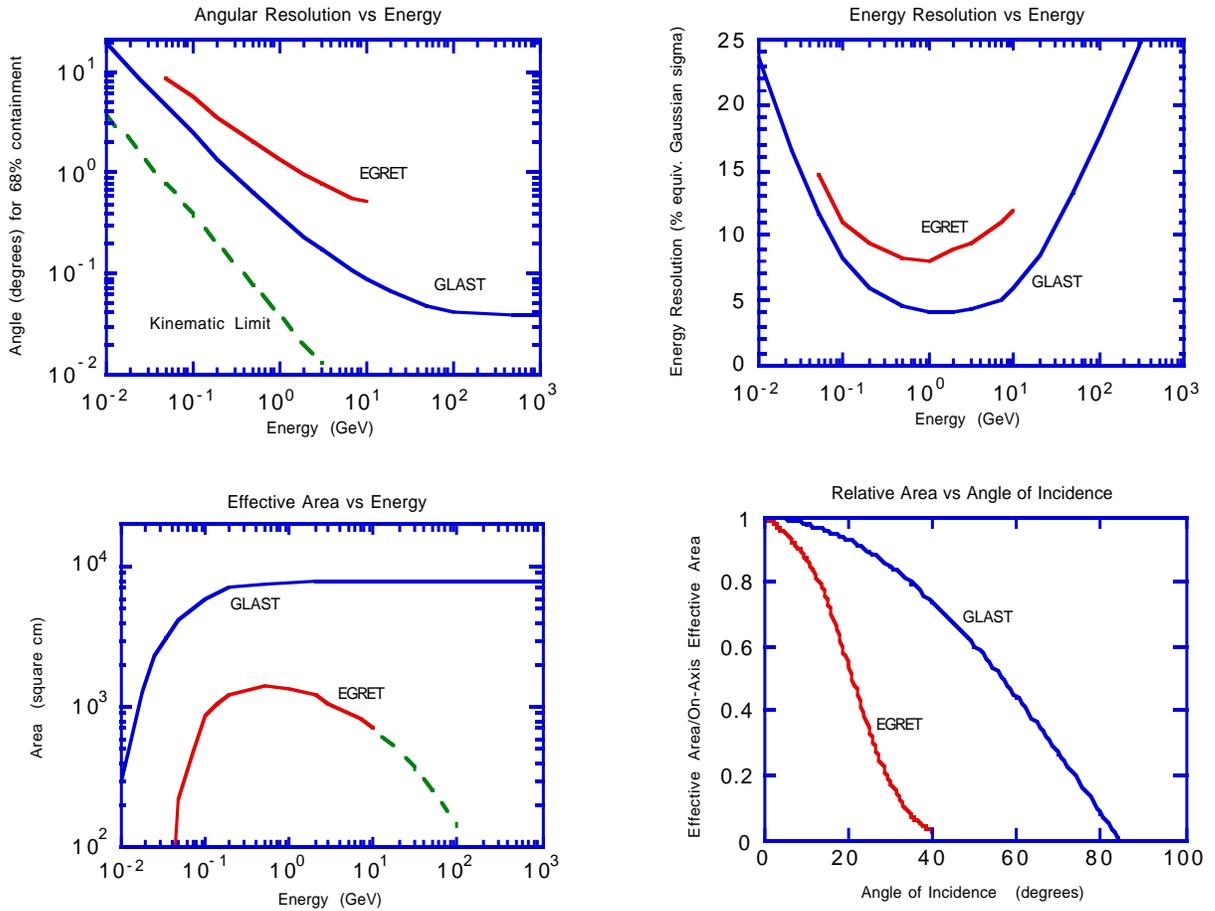}}
\caption{GLAST calibration as derived from the Gismo simulations.  Also shown for comparison are the \egret\ parameters.  Note in particular the dramatically improved sensitive area at high energy as well as the greatly expanded field of view.}
\label{GLASTPARAMS}
\end{figure}

	All the tools necessary for characterizing the performance of the GLAST instrument are now in place.  A point source of \gammarays\ is simulated using the Monte Carlo package.  The resultant events are subject to the flight trigger and passed through the track recognition software.  A virtual instrument calibration has been performed.  The results are shown in \fig{GLASTPARAMS}.

\section{Simulated Scientific Performance}

	 The techniques described above have provided a parametric description of the GLAST instrument.  Based on these parameters, the capability of GLAST to answer some of the relevant scientific issues will be examined in this section.

\subsection{Simulation Package}

	Several improvements in the GLAST performance lead to increased point source sensitivity.  The increased sensitive area improves the statistics.  The extended energy range allows one to make full use of the high energy photons which carry the most positional information, the sharper point spread function increases the signal to noise ratio, and the wider field of view increases the average coverage of a source which in turn leads to improved statistics.  The most direct way of quantifying the degree of improvement is again through Monte Carlo simulations.

	The first step in the construction of a simulation package is the calculation of the appropriate broad energy $PSF$ for GLAST.
\begin{equation}
	PSF(\theta)=\frac{1}{N} \int_{E_{min}}^{E_{max}}dE \; PSF(\theta,E) \int_{0}^{\infty}{E^\prime}^{-\alpha} d{E^\prime}\; EDF({E^\prime},E)\; SAR({E^\prime})\;,
\end{equation}
where $PSF(\theta),EDF({E^\prime},E),SAR({E^\prime})$ are the point spread function, energy dispersion function, and sensitive area function respectively.  In general, the spectral index $\alpha$ is chosen to be 2.0. 

\begin{figure}[p]
\epsfysize=3.0in
\centerline{\epsfbox{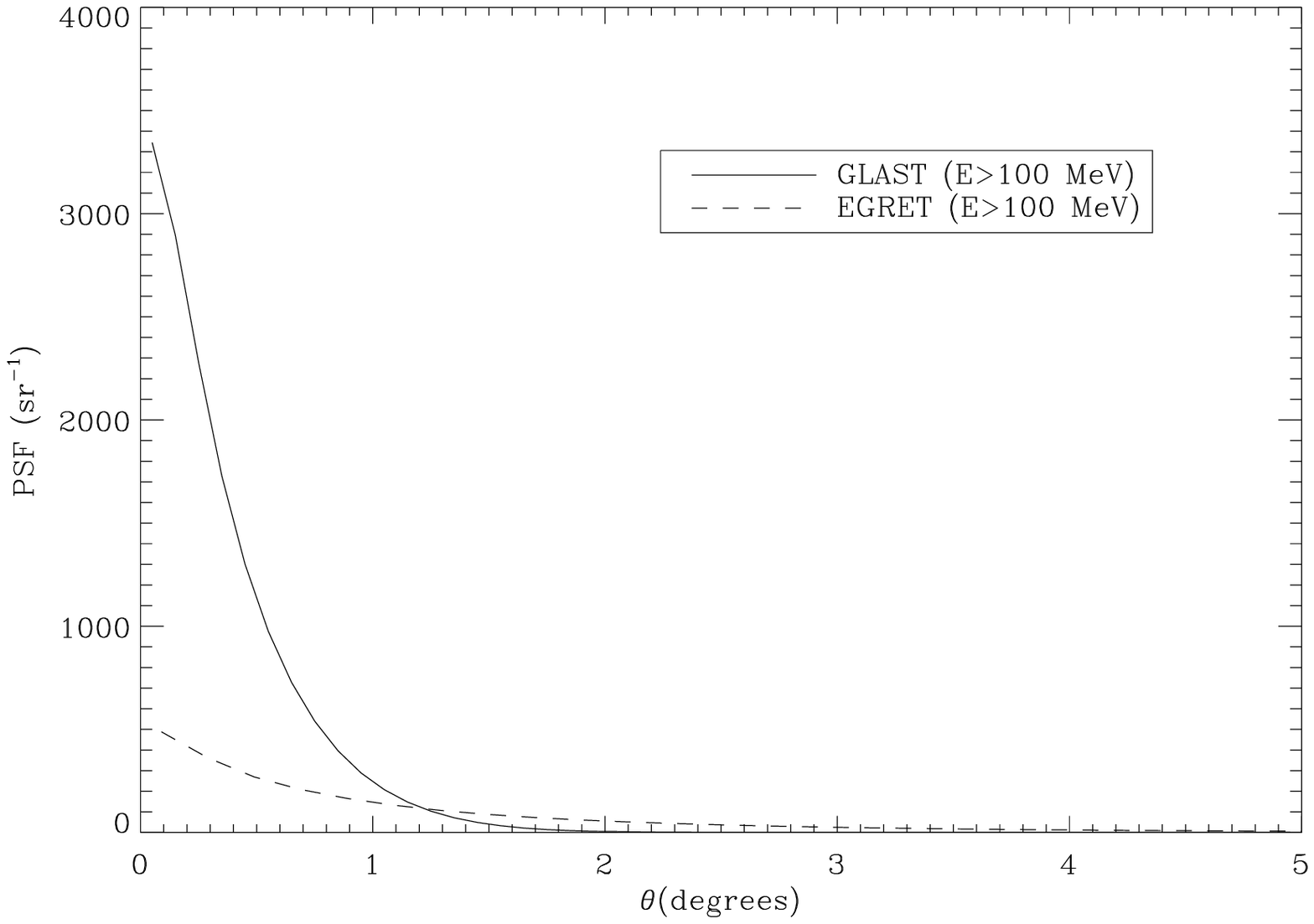}}
\vspace{0.2in}
\epsfysize=3.0in
\centerline{\epsfbox{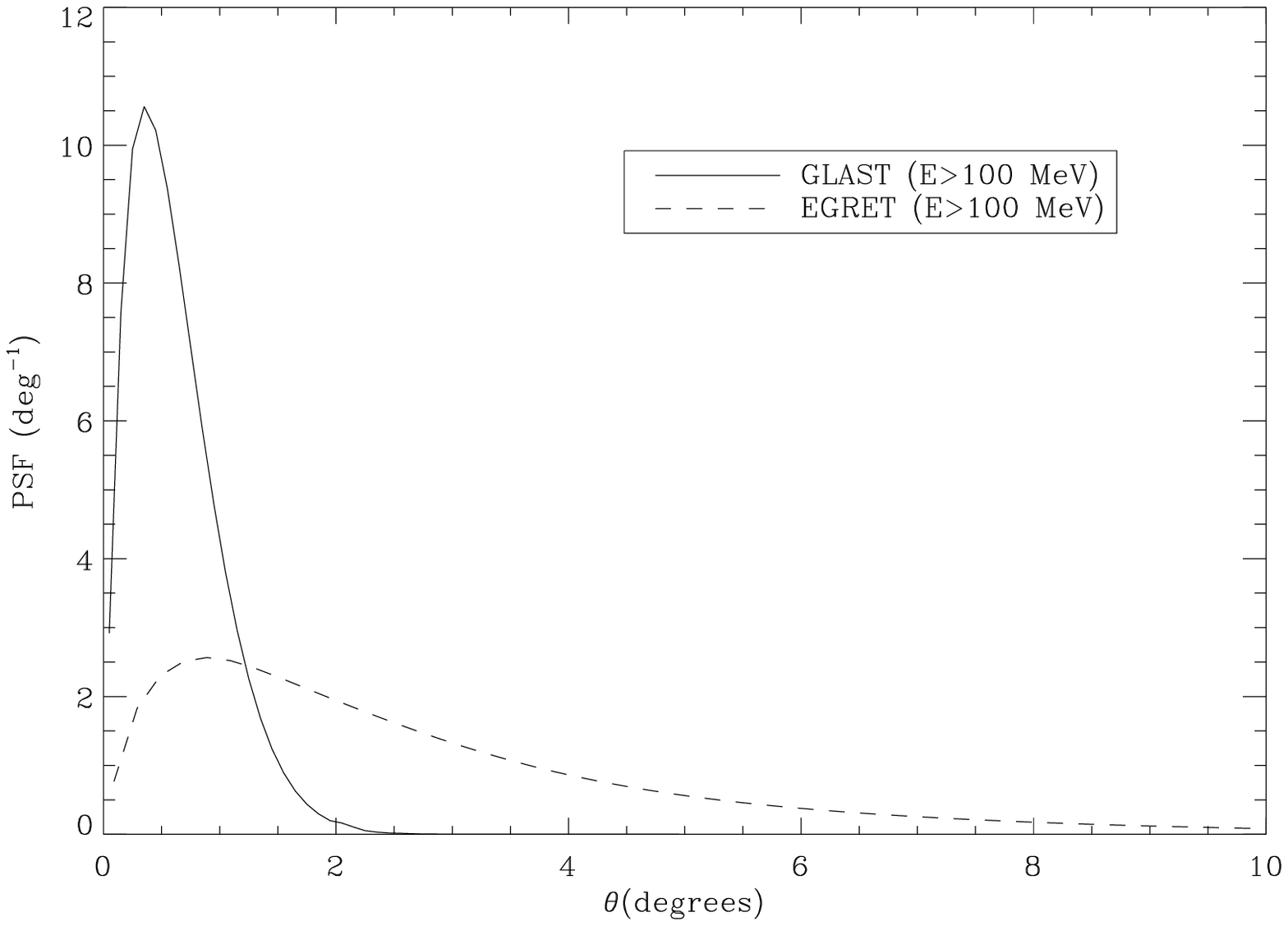}}
\caption{The simulated GLAST PSF compared to the measured \egret\ PSF.  The top panel shows the probability of detecting a photon per unit steradian whereas the bottom plot shows the probability per degree.  The central spike in the GLAST PSF is very pronounced.}
\label{PSFcomp}
\end{figure}

	\fig{PSFcomp} shows the GLAST PSF calculated using the parameters taken from the Monte Carlo results.  It should be noted that the increased importance of the high energy photons results in the greatly enhanced `spike' at the center of the $PSF$.

	This $PSF$ is then used to generate the GLAST background model.  For this the galactic diffuse model described in Chapter 4 was convolved with the GLAST $PSF$.  The results are shown below compared with the \egret\ background model.  The improvement in resolution is readily visible in this comparison.  This added resolution will provide powerful new data with which to constrain such models.  The construction of more accurate galactic models will be a crucial part of the future improvements in measurements of the IDGRB.  

\begin{figure}[ht]
\epsfysize=6.0in
\centerline{\epsfbox{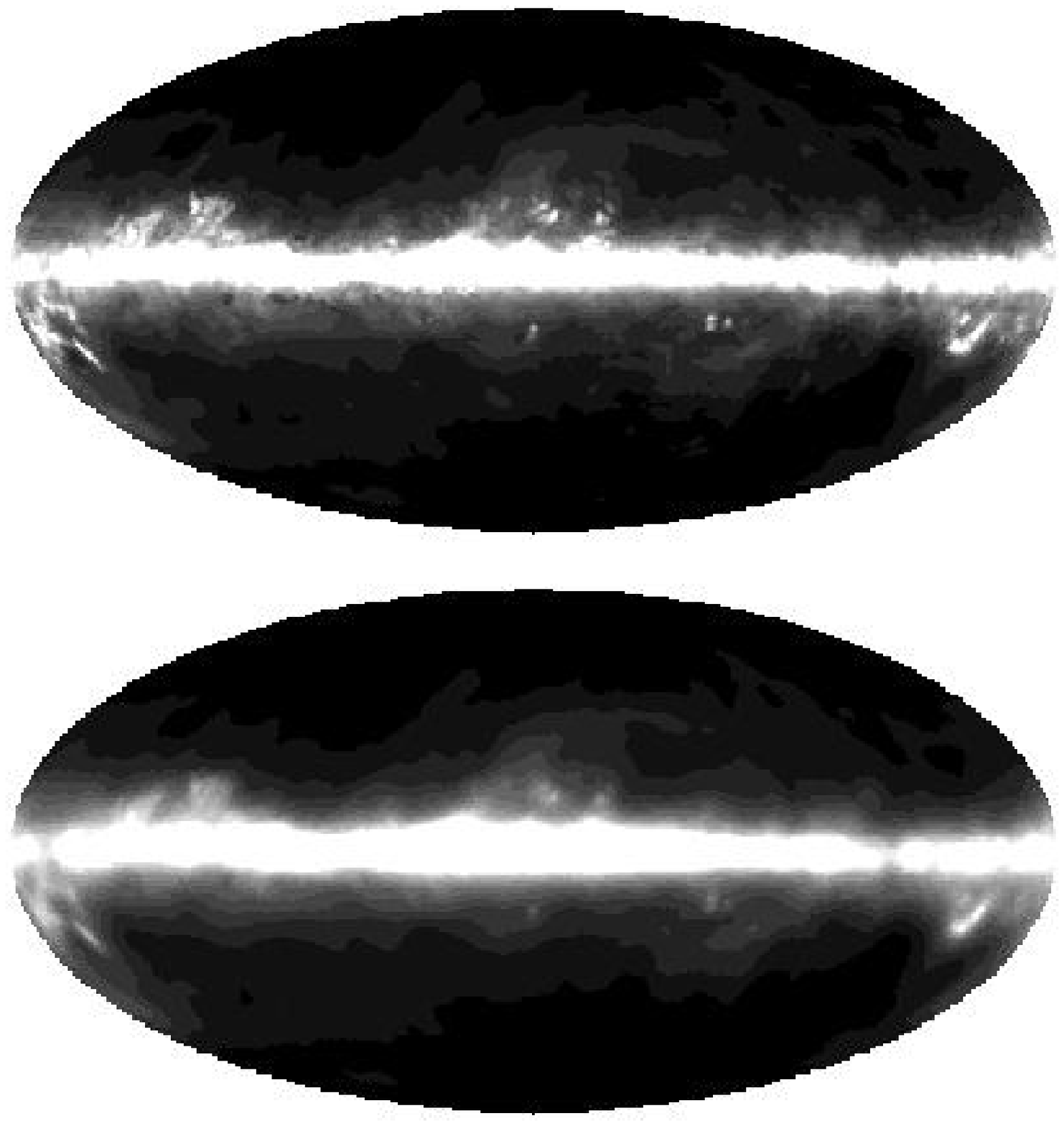}}
\caption{Comparison between the diffuse galactic emission with $E>100$ MeV as seen by GLAST (above) and \egret\ (below).}
\label{gascomp}
\end{figure}

	The exposure map can also be generated using the parameters that have been established.  The following assumptions are made about the viewing mode.  The telescope operates in a zenith pointed mode in which the pointing direction is swept across the sky in such a way as to always face away from the earth.  In this way the total exposure to the sky is maximized because dead time spent looking toward the earth is eliminated.  Furthermore, assuming a long enough integration, this exposure should be equally spread across the sky.  The total exposure to a source of spectral index $\alpha$ is then,
\begin{equation}
{\cal E}=\frac{\int_0^{\frac{\pi}{2}}(A_0 \cos\theta) 2\pi sin\theta d\theta}{4\pi} \times T_{live}=\frac{A_0T_{live}}{4} \;,
\end{equation}
where
\begin{equation}
A_0=\frac{\int_{E_{min}}^{E_{max}} SAR(E) E^{-\alpha} dE}{\int_{E_{min}}^{E_{max}}E^{-\alpha} dE} \;,
\end{equation}
is the spectrally weighted on-axis sensitive area.  Using the sensitive area results from above we find that for $E>100 \; {\rm MeV}$, $A_0=7300 {\rm cm}^2$ and consequently after a 1 year full sky survey the average exposure to any given point on the sky will be ${\cal E}=5.78\times10^{10} {\rm cm}^2 {\rm s}$.  This can be compared with the average exposure obtained by \egret\ ${\cal E}_{{\rm EGRET}}=4.3\times10^8 {\rm cm}^2 {\rm s}$. This amounts to a factor of 130 improvement over \egret.  If we consider where this improvement arises we find that there is about a factor 10 improvement in $A_0$, about a factor 4 improvement in field of view, and an additional factor 3 in live time improvement.  This last factor comes partly from the fact that GLAST can be operated in a zenith pointed mode because of its wide field of view whereas \egret\ must spend quite a bit of time looking at the earth.  An additional consideration is that the failure of the flight tape recorders increased the dead time by a factor of two above the nominal value.
 
	In order to analyse point sources in simulated GLAST data the \egret\ maximum likelihood analysis package was adapted to accept GLAST parameters.  The only substantive difference between this package and the \egret\ package was that in the case of GLAST a large amount of positional information is contained in the $E>1\;{\rm GeV}$ photons and it is important to preserve this information by not binning the data together in wide energy ranges as is done for \egret.  Instead analysis is done separately in the energy ranges $100<E<1000\;{\rm MeV}$ and $E>1\;{\rm GeV}$ and the two independent results are used in order to ascertain significances and point source locations.  Because the two data sets are independent the likelihoods are multiplicative and thus their logarithms are additive. As a result the likelihood values in the two energy ranges can be summed to yield a joint likelihood measurement.

\subsection{Point Source Sensitivity}

\begin{figure}[h]
\epsfysize=3.0in
\centerline{\epsfbox{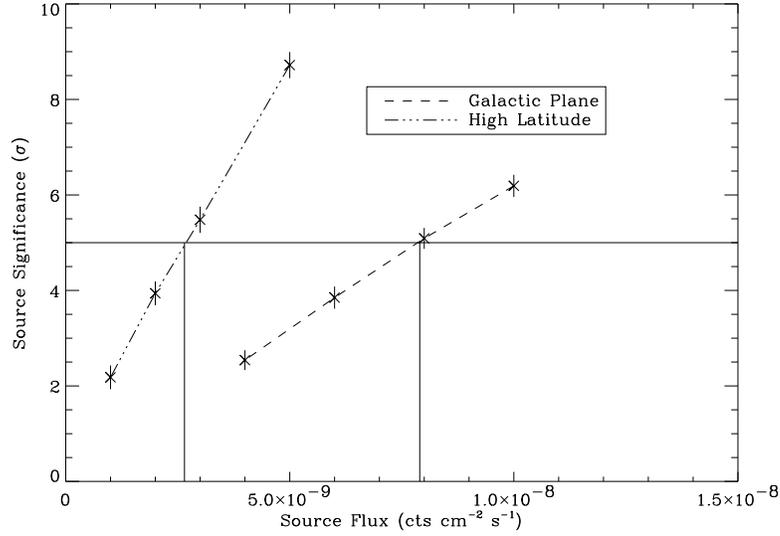}}
\caption{Point source significance versus point source flux.  The relationship is shown for typical points on the galactic plane and at high galactic latitude.}
\label{sens}
\end{figure}

	 The most critical measure of telescope performance is its point source sensitivity.  This is best expressed as a relation between point source strength and the significance of its detection.  As shown in Chapter 5, in the regime in which background counts dominate source counts,
\begin{equation}
\sigma=\frac{S\cal{E}}{\sqrt{{\cal{E}}(S+B)}} \simeq \sqrt{\frac{\cal{E}}{B}} S \;,
\end{equation}
the significance, $\sigma$, grows linearly with source flux $S$.  \fig{sens} shows the relation between average $\sigma$ and point source strength for two background conditions varying from high latitude to near the galactic center.  Fifty simulations were analyzed in the two energy bands for each point in order to ascertain the mean source detection confidence.

	The results indicate that the high latitude point source $5\sigma$ detection threshold is $\sim 2.6\times10^{-9}$ whereas near the galactic center it is $\sim8\times10^{-9}$.  This amounts to a factor 40 improvement over \egret\ performance.

\subsection{Point Source Location Accuracy}

	Correlated multiwavelength observations provide a powerful tool in the investigation of \gammaray\ sources.  The applicability of this method is severely limited by the relatively large size of the \egret\ error boxes.  A typical 95\% \egret\ error box is $\sim .5^\circ$ in diameter (\cite{Thompson96}).  The number of potential counterparts to these \gammaray\ sources that lie within these error boxes can be quite large.  A query of the NASA Extragalactic Database (NED) yields an average of $\sim10$ extra-galactic radio sources within the typical high latitude error box.  A critical performance criterion for GLAST is the improvement in this area.

	Distributions of the 95\% error box size were determined using the likelihood maps in the neighborhood of simulated sources.  Likelihood maps in each independent energy range were constructed and summed to yield a joint likelihood map from which the likelihood contours were measured.  The relationship between error box size and source strength was explored for sources at high latitudes and on the galactic plane.

\begin{figure}[h]
\epsfysize=3.0in
\centerline{\epsfbox{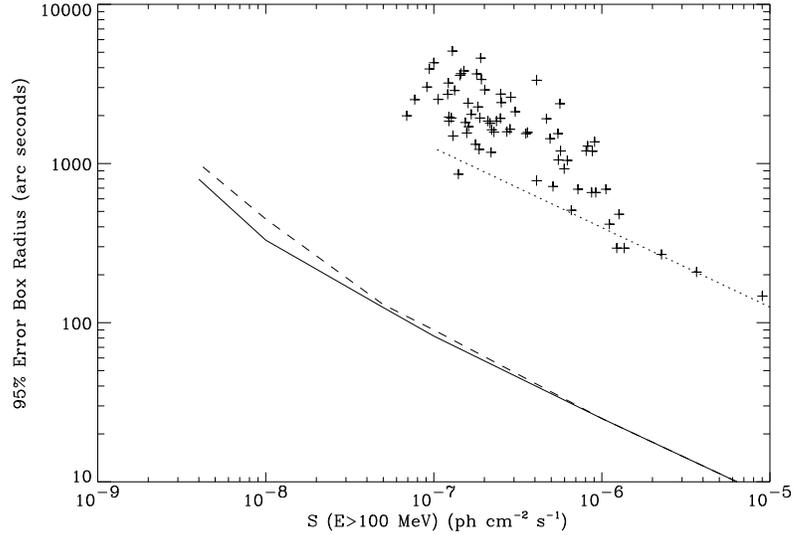}}
\caption{Scaling of the 95\% error box radius with source strength.  The crosses indicate the error box radius for the \egret\ catalog sources and the dotted line ins the fit to $S^{-1/2}$ at the high fluxes.  The solid line represents the results of a simulation of the GLAST performance at high latitude.  The dashed line shows the performance near the galactic center.}
\label{errorbox}
\end{figure}

	The interpretation of the results trend is straight forward.  In the limit of no background, the source location is determined by the width of the PSF, $\theta_{rms}$, and the source strength, $S$, which determines the total number of detected photons from the source, $N$.
\begin{equation}
\theta=\frac{\theta_{rms}}{\sqrt{N}} \;\; \alpha \;\; \theta_{rms} S^{-\frac{1}{2}} \;.
\end{equation}
As a result the confidence region should scale as $S^{-1/2}$ whenever the source dominates the background. This is clearly seen in the data for sources with source strengths greater than $\sim10^{-8}$.  Weaker sources on the plane start to show the effects of background fluctuations and their error boxes are correspondingly enlarged.  At high latitudes this effect begins at a lower flux.

\subsection{Point Source Confusion}

	All the above analyses have been completed using the most idealized case of an isolated point source on a perfectly modelled background.  On the actual sky this is never the case.  Assuming a simple Euclidean power law extrapolation to low source fluxes from the \egret\ data, we find that there will be $\sim 8660$ extra galactic sources with fluxes greater than the $5\sigma$ flux threshold at high latitudes of $3\times10^{-9}$.  This implies that the mean separation between sources is $\sim1.2^\circ$.  A certain fraction of these sources will inevitably be obscured by brighter objects.  Furthermore, sources
beneath the detection threshold will produce fluctuations which generate local
areas of discrepancy between the diffuse model and the data.  This will tend to obscure otherwise resolvable sources.

	This issue has been approached in two ways.  The simplest way to estimate the number of confused sources is to use the above calculation of GLAST error box sizes to estimate the number of sources which would overlap other sources.  For the purposes of this calculation a source is declared to be obscured if another source has an overlapping error box and is of equal or greater strength.  If a $logN-logS$ relationship is then assumed for the extra-galactic sources, an estimate can be made of the total number of obscured sources. 

	Source A is confused if the separation between it and a second source ($\theta$) is less than $\theta_{max}$,
\begin{equation}
	\theta_{max}=\theta_{95}(S_A) + \theta_{95}(S_B) \;,
\end{equation}
where $S_A$ and $S_B$ are the source fluxes of source A and B respectively.  The average number of sources that confuse source A is given by,
\begin{equation}
	N(S_A)=\int_{S_A}^\infty dS_B \int_0^{\theta_{max}(S_A,S_B)} d\theta \frac{d^2N}{dSd\Omega} 2\pi sin(\theta) \;,
\end{equation}
and thus the probability that a given source will be confused is,
\begin{equation}
	P=1-e^{-N(S_A)} \;.
\end{equation}
In the simplest case where we assume that the 95\% error box radius scales as $S^{-1/2}$ and that there is a Euclidean $logN-logS$ source distribution, this probability can be calculated analytically.
\begin{equation}
N(S_A)=\frac{31}{40} N_{0} \theta_{0}^2 (\frac{S_A}{S_0})^{-5/2} \;,
\end{equation}
where $N_0$ represents the number of sources in the sky above the reference flux $S_0$ and $\theta_0$ is the 95\% error box of a source with that strength.  Using the best fit Euclidean source distribution to the \egret\ data and the error box size determined above the number of confused sources at the detection threshold of $3\times10^{-9}$ is $\sim.03$.  Thus the probability of a source at the detection threshold being confused is $\sim3\%$.

	A second approach to this issue is to simply simulate an observation of a portion of the sky including a sample $logN-logS$.  A Poisson sample is then simulated and the standard likelihood techniques are then used to reconstruct the sources in the field.  The reconstructed $logN-logS$ can then be compared to the input distribution in order to ascertain the effects of obscuration by brighter sources as well as the effects of unresolved sources creating local inaccuracies in the background model.

\begin{figure}[p]
\epsfysize=2.5in
\centerline{\epsfbox{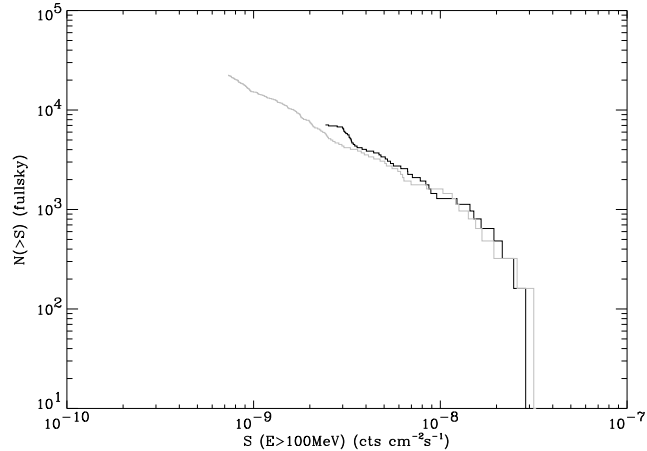}}
\caption{The results of a simulation of a GLAST observation of the high latitude sky.  The gray line shows the input distribution of sources.  This distribution was chosen at random within a $16^\circ \times 16^\circ$ field.  The black line shows the reconstructed source distribution after analyzing this field using the maximum likelihood techniques.  Sources are detected down to a threshold of $\sim 3\times10^{-9}$.  Near the detection threshold the source fluxes are biased toward slightly higher fluxes due to the contributions of unresolved sources.}
\label{reconstruct}
\end{figure}

	\fig{reconstruct} shows the comparison between the input $logN-logS$ curve and the corresponding reconstructed curve.  The detection threshold which is determined by the smallest detected source strength is $\sim 3\times 10^{-9}$.  This is slightly higher than the corresponding isolated source detection threshold because the fluctuations in the unresolved sources tend to lower the likelihood values.  Source confusion in this representation would appear as an inconsistency between the two distributions.  There appears to be little confusion until $S<5\times10^{-9}$ at which point the reconstructed source distribution deviates toward the slightly higher number of sources.  While this may at first seem counterintuitive, what is occurring is that unresolved sources in the vicinity of detected sources are contributing to the detected source's
intensity.  The result is to push that source to higher flux leading to an apparent over detection of sources at fluxes slightly higher than the detection threshold.  For a detailed discussion of this effect see (\cite{Schmidt86}).

\begin{figure}[p]
\epsfysize=2.5in
\centerline{\epsfbox{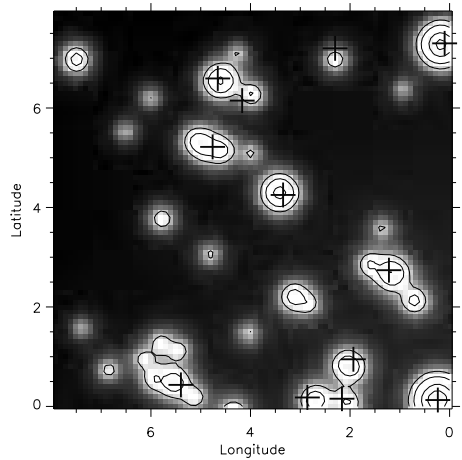}}
\caption{Simulated sky intensity for a high latitude region.  The map shows the modelled intensities of point sources in this region convolved with the GLAST PSF.  Also shown as cross hairs are the point source positions reconstructed using maximum likelihood from a Poisson sample of this map.  The correlation in positions shows that the reconstructed sources do indeed represent true sources.  In some cases multiple sources are shown to be reconstructed as a single source.}
\label{reconmap}
\end{figure}

	The detection accuracy can be examined in more detail by looking at the spatial maps of the simulated region.  \fig{reconmap} shows the intensity map of a small region in detail.  This map does not have Poisson statistical fluctuations added so as to present the actual model  used in the simulation.  The reconstructed source positions are shown as crosses.  One can see that in some cases, a strong source has been grouped together with some weaker sources leading to the detection of a source with a slightly increased flux.

	One way to quantify this effect is to integrate the additional flux which leaks in in the form of source confusion.  In the case of the $16^\circ\times16^\circ$ region simulated, the total integrated flux above $S=3\times10^{-9}$ was $3.0\times10^{-7}$ for the input sources and $3.3\times10^{-7}$ for the reconstructed sources.  As a result the total flux source confusion flux is $\sim 3\times10^{-8}$ which amounts to $\sim10\%$ of the total flux in the sky.  This number is of course dependent on the exact form of the $logN-logS$ relation chosen.  For the purposes of this simulation the steepest $logN-logS$ allowed by the fluctuation analysis of the \egret\ data was used so as to get as conservative an estimate of the source confusion problem as possible.

\subsection{Spectral Performance}

	The ability of \egret\ to explore the highest energy \gammarays\ was limited by two factors. Firstly, its relatively small area limited the number of high energy photons incident on the telescope.  Furthermore, those high energy photons that were incident on the telescope had a high probability of being self-vetoed which further limited the effective area.  In GLAST there has been an attempt to overcome these problems with increased detector size and by elimination of the backsplash problem through modularization of the veto layer.

\begin{figure}[h]
\epsfysize=3.0in
\centerline{\epsfbox{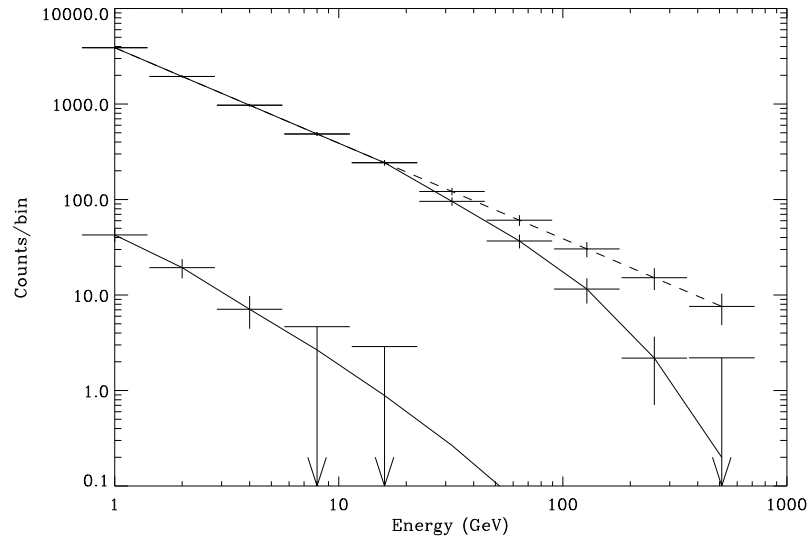}}
\caption{Measured spectrum of 3C279 after a one year GLAST exposure.  The dashed line shows the extrapolated spectrum from the \egret\ data.  The solid line assumes no intrinsic source cutoff but allows for extinction by the interstellar infrared radiation field as calculated by Stecker et al. (1993).  The lower graph shows the corresponding \egret\ spectra.  The GLAST spectra start to be limited by counting statistics at energies greater than $\sim 100 {\rm GeV}$.  It should nonetheless be possible to detect the spectral break which must occur at energies less than 1 TeV.}
\label{specglast}
\end{figure}

	The sensitive area of GLAST as a function of energy is shown in \fig{GLASTPARAMS}.  In contrast to the case of \egret, the sensitive area reaches a maximum around 1 GeV and remains at that level as energy increases.  What ultimately limits the sensitivity of GLAST is the falling spectra of the celestial sources of \gammarays.  \fig{specglast} demonstrates the capabilities of GLAST in the energy range $1 \;{\rm GeV} < E < 1 \;{\rm TeV}$.  The brightest extra-galactic source in the \egret\ sky is 3C279, a quasar that underwent a large flare during the first year of \egret\ observations (\cite{Kniffen93}).  This blazar has a spectral index of $\sim2.0$ and would be clearly visible to ground based air shower telescopes were its spectrum to continue unbroken to TeV energies.  Its non detection leads to the conclusion that either this source has an intrinsic spectral break in the region between 5 GeV and 500 GeV or that there is sufficient extinction of the high energy \gammarays\ through $\gamma-\gamma$ interactions with the interstellar infrared background to extinguish the source at high energies.

	\fig{specglast} shows the spectrum of 3C279 that would be produced by GLAST after a one year all-sky survey.  The source is assumed to have undergone a flare similar to that observed by \egret\ during the course of this year in order to produce an integrated time averaged flux above 100 MeV of $1.0 \times 10^{-6}$.  A spectral index of 2.0 is assumed consistent with the spectral index measured by \egret.  Two spectral models have been convolved with the GLAST response.  The first (shown by a dashed line) represents a simple extrapolation of the \egret\ spectrum to TeV energies.  This spectrum is inconsistent with the Whipple measurements at TeV energies and is shown for reference purposes.  The second spectral model assumes that there is no intrinsic source cut-off but estimates the infrared extinction as calculated by Stecker (\cite{Stecker93}).  In order to demonstrate the statistics of such a spectrum directly, the values plotted are the number of photons in each spectral bin.  It is worth pointing out that while the bin centered on 250 GeV contains only 2 photons, the chance that 2 background photons in this energy range arrive within the 140 arc second error box of these high energy photons is $5\times10^{-4}$.  The relative absence of background photons means that significant detections can be made with comparatively small number of photons if one has an a priori source position.  Also shown for comparison is the same model convolved with the \egret\ response.  Notice that the \egret\ self-vetoes on approximately half the converted photons at 8 GeV.

	The GLAST spectrum of 3C279 demonstrates the ability of GLAST to produce good quality spectra out to energies of $\sim100 \;{\rm GeV}$ for the brightest \egret\ sources.  Beyond this the spectra will be limited by statistics but this instrument should nonetheless be capable of resolving the spectral break that must occur below 1 TeV.  Of course an intrinsic spectral break in the source at lower energies would be easily resolvable.  Detections of such breaks in the brightest 10 sources would provide crucial information as to the nature of the acceleration mechanism in blazars as well as having the potential to address fundamental issues of cosmology by gleaning information about the infrared extinction of distant sources. 

\subsection{Diffuse Spectrum}

	The extended energy reach of GLAST will allow the IDGRB to be probed out to higher energies.  Furthermore, the detection of several thousand blazars will allow the blazar spectrum to be accurately measured.  These two measurements will allow any remaining IDGRB to be probed for distinctive spectral features that might signal the presence of a non-AGN component of the IDGRB.

\begin{figure}[h]
\epsfysize=3.0in
\centerline{\epsfbox{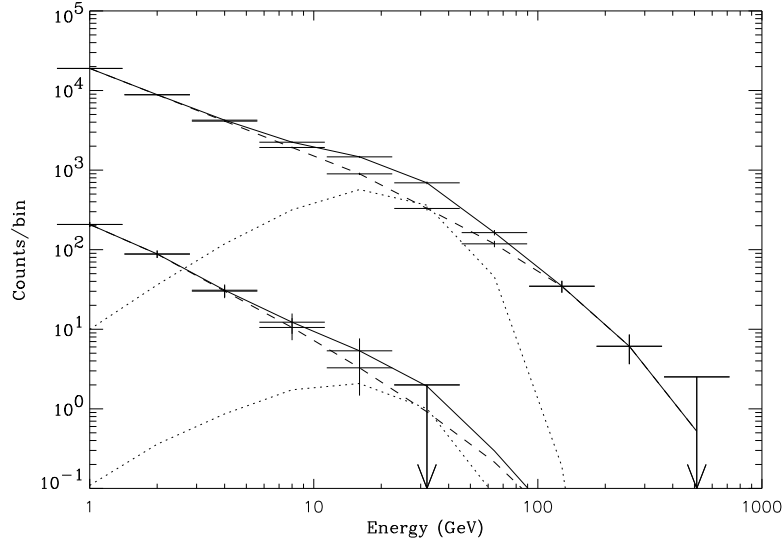}}
\caption{Diffuse extra-galactic spectra as measured by GLAST upper curves and \egret\ (lower curves).  The dashed lines indicate the IDGRB produced by unresolved AGNs.  The dotted line shows a possible WIMP decay spectrum.  The solid curve shows the summed spectrum.  GLAST is sensitive to this WIMP decay spectrum whereas \egret\ is not.}
\label{WIMP_z}
\end{figure}

	In Chapter 1, the possibility of WIMP decays or annihilations contributing to the IDGRB was discussed.  The sensitivity of GLAST to the spectral signals predicted by these models can now be assessed.  \fig{WIMP_z} shows IDGRB spectrum as observed by GLAST (upper set of curves) and \egret\ (lower set of curves).  We have assumed that a large fraction of the sky will be unsuitable for the measurement of the diffuse background because of the presence of strong point sources of strong foreground galactic emission.  As a result the data shown in the figure represents only 10\% of the sky.  Two assumptions are made in order to produce the various spectra shown. The dashed spectrum corresponds to an AGN-like spectrum scaled to account for $\sim50\%$ of the IDGRB. The remaining 50\% is assumed to be found in resolved AGNs.  The dotted line represents the predicted observed flux from a WIMP decay in the early universe.  Following the model described in chapter 1, this curve is calculated for a WIMP which decays at some characteristic redshift $z_D$ with mass $m_X$ and whose density is given by $\Omega_Xh^2$.  These parameters determine the spectrum shown using equation (1.3).  The simulation shown in figure x makes the following choice of parameters,
\begin{eqnarray}
	E_0 &=& \frac{m_X}{2(1+z_D)}= 10 {\rm GeV} \\ \nonumber
	\Omega_Xh^2&=& 10^{-14}
\end{eqnarray}

	The solid curve in \fig{WIMP_z} shows the sum of the AGN component and the WIMP decay component of the IDGRB.  For this choice of parameters a bump at $\sim 10 {\rm GeV}$ would be clearly visible.  Because, we have a shown that it is possible that $\sim50\%$ of the IDGRB will be resolved as point sources by GLAST, the dashed curve also represents the approximate summed spectrum over all detected AGNs.  In this way the IDGRB and the AGN component can be measured to the accuracy shown in the figure, allowing the WIMP decay component to be reconstructed if present.

	The corresponding \egret\ observation is also shown in \fig{WIMP_z}.  \egret\ is not sensitive to such spectral bumps.  This is due to its smaller collecting area and the reduced effective area caused by `backsplash'.

	Another spectral signal postulated in chapter 1 was the appearance of a \gammaray\ line caused by the annihilation of two WIMP's into \gammarays.  The sensitivity of GLAST and \egret\ to these lines has been calculated in the following way.  The \gammaray\ line is assumed to have negligible width compared to the instrumental energy resolution $\Delta E/E_0$. The signal counts from a \gammaray\ line of integrated intensity $I$ are then given by,
\begin{equation}
	N_S=I{\cal{E}}\int_{E_0-\Delta}^{E_0+\Delta} EDP(E,E_0) dE \;,
\end{equation}
where $EDP(E,E_0)$ is the instrumental energy dispersion function and $\cal{E}$ is the exposure.  If we assume this line is measured against a power law background then the background in this energy range is
\begin{equation}
	N_B=B_0{\cal {E}}\int_{E_0-\Delta}^{E_0+\Delta} E^{-\alpha} dE \;,
\end{equation}
where we have ignored the higher order correction caused by the convolution of the background spectrum with the instrumental dispersion functeon.  If we assume $EDP(E,E_0)$ to be a Gaussian with the relative width, $\sigma_E$, shown in \fig{GLASTPARAMS}, the signal to noise is maximized near $\Delta=\sigma_E$.  We can then evaluate the significance of a line detection as simply,
\begin{equation}
	\sigma_{{\rm Line}}=\frac{N_S}{\sqrt{N_S+N_B}}\;.
\end{equation}

\begin{figure}[h]
\epsfysize=3.0in
\centerline{\epsfbox{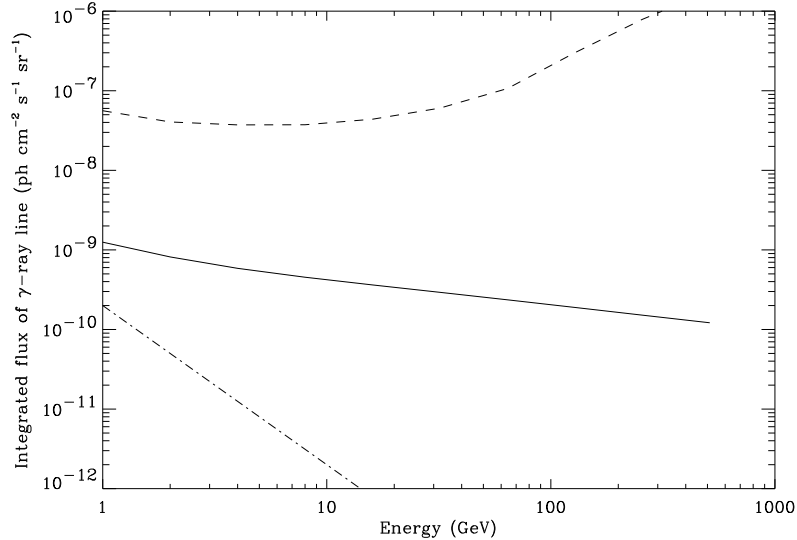}}
\caption{\Gammaray\ line sensitivities of GLAST and \egret.  The dashed (solid) curve shows the \gammaray\ line intensity that would result in a $2\sigma$ bump on the IDGRB when observed by \egret\ (GLAST).  The dot-dashed line indicates the estimated line intensity due to the annihilation of the lightest super-symmetric particle in the galactic halo.  Although cross-section used is highly speculative, it is likely that the detection of such WIMP annihilation lines will remain beyond the scope of GLAST.}
\label{wimpline}
\end{figure}

	\fig{wimpline} shows curves of constant significance as a function of $E_0=m_X$; the WIMP mass.  The dashed (solid) curve is shows the $2\sigma$ sensitivity of \egret\ (GLAST).  Also shown are the predictions by Kamionkowski (\cite{Kamionkowski95}) of the expected line flux from neutralino annihilation as a function of neutralino mass.  These predictions indicate that the sensitivity of GLAST will remain insufficient for the detection of these \gammaray\ lines.  However, the cross section for the coupling of WIMP's to \gammarays\ are highly speculative and contain large uncertainties.  GLAST could provide interesting constraints on these processes.

\section{Summary}

	\egret\ has provided a tantalizing look at some of the most violent processes in the universe while at the same time raised a number of questions which remain beyond its scope.  The motivation for probing the \gammaray\ universe in more detail is compelling.  The GLAST instrument concept described in this chapter represents an effort to develop a realistic instrument for the future of \gammaray\ astronomy.  The simulations presented here indicate that GLAST will become a powerful tool in the quest find answers for many of the questions raised by \egret.

\begin{table}[ht]
\centering
\tiny
\caption[Comparison of GLAST and \egret\ Performance]
{Comparison of GLAST and \egret\ Performance}
\bigskip
\begin{tabular}{lcc|cc}\hline\hline
& \multicolumn{2}{c}{\egret} & \multicolumn{2}{c}{GLAST} \\
\hline
				&High Latitude 	 &  Galactic Plane  	& High Latitude	&  Galactic Plane  \\
\hline
$5\sigma$ Detection Threshold	& 100     	 & 500			& 2.6		& 8      \\
($\times 10^{-9} ph cm^{-2} s{-1}$) & & & & \\[0.1in]
95\% Error box Radius (arcsec):  &    &  &  & \\
$S=10^{-6}$			&120 		&120 			& 25		& 25 \\
$S=10^{-7}$			&2500 		&2000 			& 80		& 90 \\
$S=10^{-8}$			&N/A 		&N/A 			& 330		& 430 \\[0.1in]
Number of Sources Detected      & 97      & 38   			& 400-14000	&  250-900   \\
\hline
\hline
\end{tabular}

\label{G_E.comp}
\end{table}

	Table 7.1 summarizes the improvements in point source detection achievable by GLAST over \egret.  The dramatic increase in the expected number of detectable sources represents a significant leap for the field of \gammaray\ astronomy.  Up to 40\% of the remaining isotropically distributed emission will be directly resolvable as point sources.  A further 25\% will be constrained by fluctuation analysis.  In addition, a population of several thousand blazars will be available.  If redshifts for a large fraction can be obtained, the luminosity function should be much better constrained.  As mentioned in chapter 5, the luminosity of the X-ray Seyferts has now been sufficiently well constrained with a comparable number of sources as to be able to identify these sources as the overwhelming source of the X-ray background. 

	In addition, the spectrum of the isotropic diffuse radiation should be very well measured, any discrepancy between the blazar spectrum and the diffuse spectrum could provide further evidence of a true diffuse component to the extra-galactic diffuse \gammaray\ background.  The ability of GLAST to detect spectral features of the sort that would be produced by WIMP decay has been established.

	The intriguing question of the nature of the unidentified \egret\ sources would be greatly elucidated by an instrument such as GLAST. Between the decrease in error box size which would reduce the number of counterparts in other wavelength bands, the ability to do long term monitoring of these sources which is afforded by the vast field of view of the telescope would allow multiwavelength campaigns to be easily mounted.  Such observations have proven to be critical to unraveling the mysteries of unidentified sources (e.g. Geminga).

	The ability of GLAST to extend the energy range of satellite borne \gammaray\ telescopes has also been demonstrated.  This development is particularly desirable in light of the fact that the ground based \gammaray\ telescopes are pushing to extend their effective spectral range to lower energies.  The ability to cross calibrate the two technologies will greatly enhance the usefulness of both telescopes.  Furthermore, the complementarity of a large FOV telescope with comparatively smaller effective area (GLAST) and a pointed telescope with larger effective area (future air shower telescope) should prove to be a very effective combination.

	Perhaps the single greatest possibility of an instrument such as GLAST is of course the truly unexpected and unpredictable discoveries that it will undoubtedly make.  In addition to opening a new spectral window onto the cosmos, the two decade increase in sensitivity will undoubtedly yield unexpected and exciting sources of \gammaray\ emission. As an example, AGN were detected by \egret\ in numbers that were totally unpredicted prior to launch.  It is the possibility of finding such staggering phenomena  in the future which should always be the primary motivation for pushing outward the frontiers of the detectable universe.

\appendix
\chapter{Maximum Likelihood Analysis}

\section{ Maximum Likelihood Analysis}

	Because of low gamma ray counting rates as well as the relatively intense and highly structured background, EGRET data analysis requires the use of statistical techniques.  An overview of the techniques used in the analysis of EGRET data is provided in (Fichtel et al. 1994).  The analysis undertaken by the EGRET team thus far includes time resolved studies of gamma ray pulsars (Nolan et al. 1993), diffuse studies (Bertsch et al. 1993), and point source studies (Mattox et al. 1994).  The principal tool used in the analysis of EGRET point sources is that of maximum likelihood.  The EGRET implementation of the maximum likelihood method is described extensively in Mattox et al. 1996 and will only be summarized here.

	Likelihood, introduced by Fischer (1925), expresses the joint probability that a data set are drawn from a hypothetical distribution.  Likelihood maximization is a useful technique for parameter estimation (Cash 1979) and has been used extensively in gamma ray astrophysics (Pollock et al. 1981).

	In a case like that of photon counting, where noise in a given pixel of solid angle is assumed to be Poissonian, likelihood for a set of pixels $ij$ with measured counts $n_{ij}$ is defined as:

\begin{equation}
	L_{0}=\prod_{ij}\frac{e^{-\theta_{ij}}\theta_{ij}^{n_{ij}}}{n_{ij}!} \;,
\end{equation}
where $\theta_{ij}=\theta(\alpha_k)_{ij}$ are the predictions of a model with parameters $\alpha_k$.  In the context of a given model, the most likely value of a parameter $\alpha_k$ is that value which maximizes the value of $L_{0}$. In practice $\ln{L_0}$ is more conveniently calculated:
\begin{equation}
	\ln{L_0} =\sum_{ij} n_{ij} \ln \theta_{ij} - \sum_{ij}\theta_{ij}-\sum_{ij}\ln n_{ij}!\;,
\end{equation}
This last term is model independent and can be dropped.

	The theory of maximum likelihood fitting also predicts that in the limit of large $N$ (where $N$ is the number of relevant photons), parameters will be normally distributed about the maximum value with a $\sigma$ of
\begin{equation}
	\sigma=-(H^{-1}_{\alpha \alpha})^{\frac{1}{2}} \;,
\end{equation}

where $\alpha$ is the parameter under estimation and $H^{-1}_{\alpha \beta}$ is the inverse of the Hessian matrix defined as,
\begin{equation}
	H_{\alpha \beta}={(\frac{\partial^2\ln L}{{\partial \alpha}{\partial \beta}})} \;,
\end{equation}
 This allows one to estimate not only expectation values but also uncertainties.

	 The likelihood ratio test (Neyman and Pearson 1928) extended the use of likelihood to hypothesis testing and significance evaluation.  This statistic is used for evaluating the extent to which one model better describes the data than a second such model.  It is defined as:
\begin{equation}
	TS=-2\ln\frac{{(L_i)}_{max}}{{(L_0)}_{max}}\;,
\end{equation}
where $L_i$ is the likelihood for a model with $m$ free parameters and $L_0$ is the likelihood for a model with $p<m$ free parameters.

The advantage of this test is that its distribution in the null hypothesis (model 0 is correct) is known analytically. Specifically this quantity was shown by Wilks (1938) to be distributed as ${\chi}^{2}_{n}$ where $n=m-p$ is the difference in the number of free parameters between the two models.

	In the case of EGRET data, likelihood is used to estimate the flux of a measured point source. The models used contain three free parameters: two parameters describe the diffuse emission described below and one describes the strength of a putative point source.
\begin{equation}
	\theta_{ij} = g_mG_{ij} + g_bE_{ij} + c_aPSF(\alpha,\delta)_{ij}\;,
\end{equation}
where $G_{ij}$ describes the structured galactic background, $E_{ij}$ the isotropic background, and $PSF(\alpha,\delta)_{ij}$ the point spread function for a point source at location $\alpha,\delta$. The free parameters in this model are the scaling factors $g_m,g_b,c_a$.  This model is used to find the best fit points source flux ($c_a$) and is compared to the model with no point source present ($c_a=0$) in order to determine $TS$ and thus the significance.

\section{ Monte Carlo Simulations}

		At the basis of this investigation is the creation of simulated data distributed according to a model of sources and background convolved with the EGRET telescope parameters.  This data can then be processed using EGRET data analysis techniques to test various responses. The creation of such simulated data requires two inputs: a spatially resolved background model and EGRET instrument parameters.   

\subsection{Instrumental Parameters}

	The EGRET instrument is characterized for the purposes of this study by the point spread function and the exposure map.  Comprehensive calibration of the EGRET instrument at SLAC indicate a Gaussian point spread function with energy dependent width($PSF(\theta,E)$).  All analysis described in this paper was done using photons with $E>100$ MeV with an averaged point spread function generated by weighting monoenergetic point spread functions with an assumed power law spectrum of the source and taking into account the energy dispersion of the EGRET instrument. 
\begin{equation}
	PSF(\theta)=\frac{1}{N} \int_{E_{min}}^{E_{max}}dE \; PSF(\theta,E) \int_{0}^{\infty}{E^\prime}^{-\alpha} d{E^\prime}\; EDF({E^\prime},E)\; SAR({E^\prime})\;,
\end{equation}

where $\alpha$ is the spectral index of the source, $EDF{E^\prime},E)$ is the energy dependent dispersion, $SAR({E^\prime})$ is the sensitive area.  The integration over ${E^\prime}$ represents an integral over true energies and the integration over $E$ represents an integral over measured energies and $N$ is the appropriate normalization.  In most cases a spectral index of 2.0 is used.  The background model is convolved with this $PSF$ in order to obtain a model of the observed background.  All added point sources are also convolved with this $PSF$.

\begin{figure}[h]
\epsfysize=3.0in
\centerline{\epsfbox{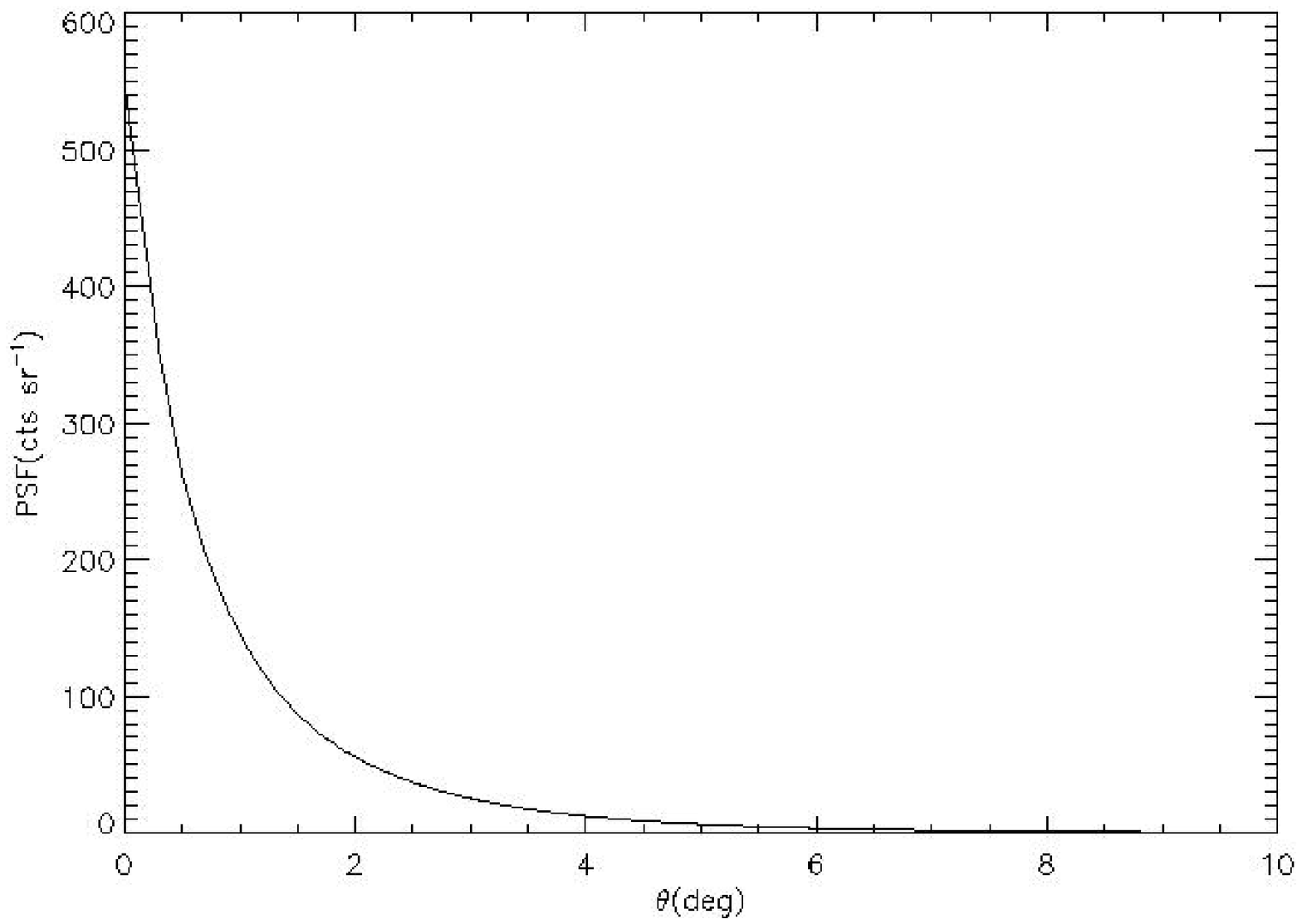}}
\caption{The \egret\ PSF for $E>100\;{\rm MeV}$.}
\label{sim.fig.1.epsi}
\end{figure}

	The exposure maps are generated in order to transform flux maps (e.g. galactic flux map) into predicted counts maps.  Sensitive area ($SAR$) as a function of photon energy, declination angle , and operational mode was calibrated at SLAC (Thompson et al. 1992).  Live time is recorded on board and these two quantities can be used to generate maps of exposure.

	Using the diffuse model with any arbitrary set of added point sources and exposure maps, one can calculate the rate of arrival of photons into any pixel of an observation.  These values form the smooth ``infinite statistics'' map.  The simulated observations are then generated from this smooth map by adding Poisson distributed deviates to each pixel.  The rejection method (Press et. al. 1992) is used for this purpose.  Output counts maps together with the exposure maps can then be analyzed exactly as flight data using the likelihood software developed by the EGRET team ( i.e. LIKE  (Mattox et al 1996)).  This program optimizes the likelihood with respect to the three free parameters: $g_m$ the diffuse galactic intensity; $g_b$ the diffuse isotropic intensity; $c_a$ the point source flux.

\section{ Point Source Analysis}

	Point source parameter fitting was investigated using statistical ensembles of observations containing test sources.  Sources of varying flux were placed in different points on the sky in different background conditions.  Both on and off axis sources were studied and the effect of lengthening the exposure was investigated. 

	Distributions of fitted flux, positions, and significance were generated.  Because likelihood analysis makes predictions not only about the value of these parameters but also their uncertainties, it was possible to check the first two moments of the distributions obtained against the likelihood results.

\subsection{ Flux Estimation}

\begin{figure}[h]
\epsfysize=3.0in
\centerline{\epsfbox{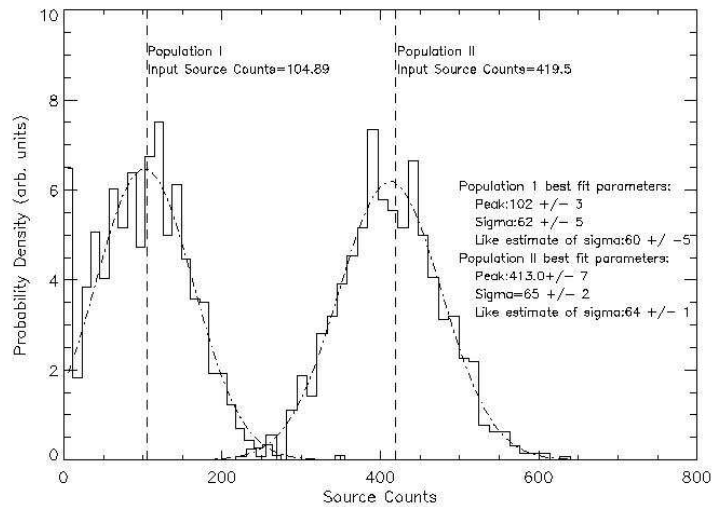}}
\caption{Two sets of simulations showing the distribution of measured counts given a simulated point source flux.  The distributions are peaked about the correct flux and the uncertainties in these measurements is consistent with the mean uncertainty derived from the likelihood function.}
\label{sim.fig.2}
\end{figure}

	\fig{sim.fig.2} shows the results obtained for sets of 1000 observations of the same source as one varies the flux.  The median value of the  fitted flux agrees well with the true value.  The effect of disallowing negative fluxes shows up in these distributions as a piling up of the distribution at zero flux. 

	The flux distributions are well approximated by Gaussian functions.  The distribution widths were calculated and compared to the results obtained by averaging the individual likelihood estimates of the flux uncertainty.  The agreement is good (\fig{sim.fig.2}).  One can see the transition between the background limited flux region (where the uncertainty is dominated by fluctuations in the background counts) to the source limited flux region (where the uncertainty is dominated by the fluctuations in the source counts).

\subsection{ Upper Limits}

	The most useful form of a point source upper limit is defined as follows: it is the point source flux above which higher true source fluxes can be ruled out with specified confidence based on the actual measurement of the flux.  In other words one would like to construct the distribution of true point  source flux given a measured flux $(P(S\mid m))$.  Such a distribution would typically be peaked about the measured flux and would have some characteristic width which would determine the $n\sigma$ upper limit.  Such distributions are not readily obtained from the data, however, and what one most often calculates is the point source flux that would result in a measured flux greater than or equal to the actual measured value with some specified confidence. 

	This latter upper limit is calculated in the following manner.  The measured flux ($m$)is obtained by maximizing the likelihood.  In the limit of large counts this flux is normally distributed with uncertainty obtained from the equation (3).  If this uncertainty is not strongly dependent on the point source strength, the upper limit of the source flux is then $S=m+n\sigma$.  That is one will have found the point source flux ($S$) 95\% of whose measured flux distribution lies above the actual measured flux if one chooses $n=2$.

	Monte Carlo simulations allows one to determine $P(m\mid S)$, the probability distribution of $m$ given $S$, and thus to test the various hypotheses used above to calculate upper limits.  The fact that measured flux is normally distributed has already been demonstrated (\fig{sim.fig.2}). Furthermore the likelihood estimate of the flux uncertainty has been confirmed within errors. The final assumption to be tested is whether the flux distribution changes as one increases the point source flux.  As is shown in fig x. the width of the measured flux distribution remains roughly fixed in the background dominated region.  Because upper limits are only calculated for the low significance sources, this assumption remains valid.

	With a Bayesian approach, one can use the knowledge of $P(m\mid S)$ to calculate $(P(S\mid m))$.  Using Bayes theorem:

\begin{equation}
	P(S\mid m)ds = P(m\mid S)*P(S)ds/\int_{0}^{\infty}P(m\mid S)*P(S)ds\;,
\end{equation}

where $P(S)$  is the prior distribution of $S$. The problem then becomes what to use as a prior distribution of S.  If one uses a uniform prior $P(S)=k$, one finds $P(S\mid m)=P(m\mid S)$ and the two definitions of an upper limit become equivalent.  If, however, one uses a prior based on a more realistic distribution of sources(e.g. one estimated from a measured $\log N ( \log S)$ relationship), one can significantly reduce the upper limits.  However, it is probably better in most cases to assume the most conservative prior which in this case is that of uniformity.

\subsection{Position Estimation}

	The method used to evaluate the position contours for a given source is outlined in Mattox et al. (1994).  If one tests the hypothesis that there is a source of some strength at position $l_0,b_0$ against the null hypothesis that there is no source present, one obtains a likelihood result.   One can now make this hypothesis the new null hypothesis and evaluate the likelihood of a model containing a source of the same strength located at a position $l_i,b_i$.  This new likelihood ratio can be optimized to determine the best fit position $l_{max},b_{max}$.

	One can determine position confidence intervals in much the same way as for flux. If the source were in fact located at a position $l_0,b_0$ (the null hypothesis), the position likelihood ratio would be distributed as $\chi_2^2$ (two additional degrees of freedom $l_i,b_i$).  We can define :
\begin{equation}
	TS_{position}(l_0,b_0)= lnL(l_{max},b_{max})-lnL(l_0,b_0) \; ,
\end{equation}
 Thus the locus of points $l_0,b_0$ giving a $TS_{position}$ value of $x$ determines the $y\%$ confidence contour for containing the source. Where $x$ and $y$ are related by the formula:
\begin{equation}
	y=\int_{0}^{x} \chi_{2}^{2}(TS) dTS \;,
\end{equation}

\begin{figure}[h]
\epsfysize=3.0in
\centerline{\epsfbox{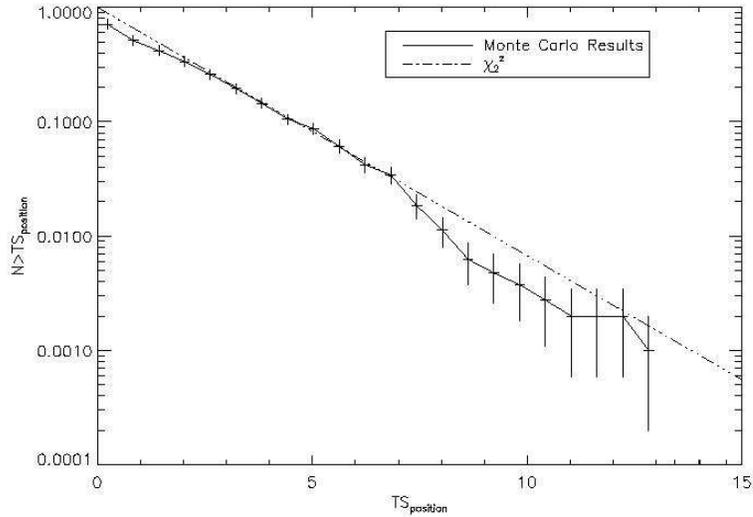}}
\caption{Cumulative distribution of the TS drop between the best fit position and the true simulated point source position. The distribution is consistent with $\chi_2^2$ except near zero where the finite bin size used in the analysis biased the distribution.}
\label{sim.fig.3}
\end{figure}

	The distribution of this test statistic in the null hypothesis was tested using Monte Carlo techniques. Maps of the position test statistic were generated for sample observations of a source, and the values of $TS_{position}(l_{true},b_{true})$ were recorded.  The resulting distribution was a good fit to $\chi_2^2$ (\fig{sim.fig.3}).  The slight underprediction near $TS_{position}=0$ results from the finite bin size used in calculating the likelihood maps. Differences in $TS_{position}$ less than some threshold were likely to fall in the same pixel and as a result be assigned a $TS_{position}$ of zero.
 
\subsection{ Inclination angle effects}
	Analysis of off axis sources have the added complication of map edges and poor exposure.  Source distributions for off axis observations were generated in order to test verify that this did not introduce systematic errors.

\begin{figure}[h]
\epsfysize=3.0in
\centerline{\epsfbox{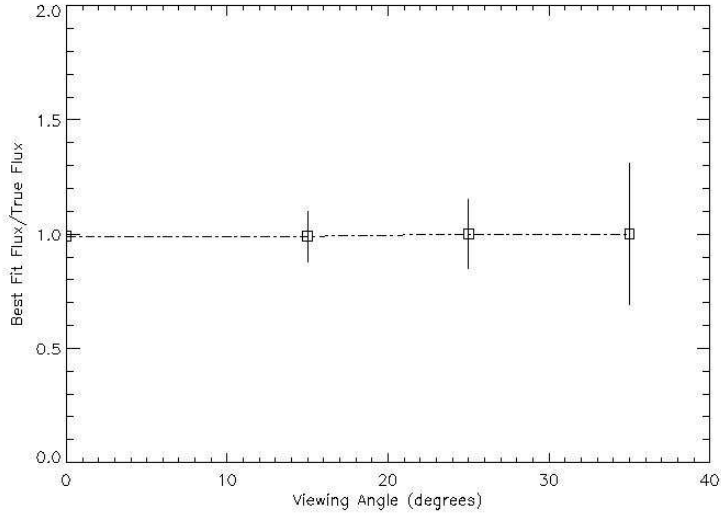}}
\caption{The distributions of best fit:true flux as one travels out toward the edge of the FOV. The software does not introduce a bias near the edge of the FOV.}
\label{sim.fig.4}
\end{figure}

	\fig{sim.fig.4} shows the fitted flux of a source as one moves  it away from the viewing axis.  Flux estimation is shown to be unaffected.

\section{ Source Significance and Spurious Detections}
\subsection{Significance}

	As mentioned previously, the likelihood ratio ($TS$) is used to determine point source significance. In the context of EGRET data, the significance of a point source can be characterized in several ways, each applicable in different scenarios.

	The most direct measure of significance is the probability that a likelihood ratio test at a given position with no source present would result in a $TS$ value greater than or equal to the measured value.  Significance defined in this way is determined by the distribution of $TS$ in the null hypothesis. Application of Wilks' theorem in this case predicts that $TS$ will be distributed as $\chi_{1} ^{2}$ ( free parameters in the null hypothesis=2 ($g_m,g_b$), free parameters in the alternate hypothesis = 3 ($g_m,g_b,c_a$)). For this particular distribution the significance is well approximated by
	$n_{\sigma}=\sqrt{TS}$ 
where $n_{\sigma}$ is the number of standard deviations. 

\begin{figure}[h]
\epsfysize=3.0in
\centerline{\epsfbox{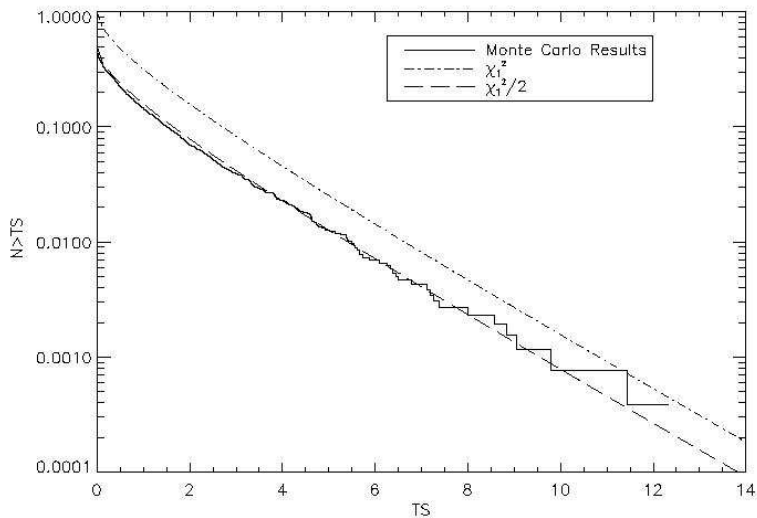}}
\caption{The distribution of TS in the null hypothesis.  No point sources were used in this simulation and the resulting distribution of TS is shown compared to $\chi^2_2$.  The discrepancy is exactly a factor 2 which results from the fact that negative fluxes are assigned TS=0.}
\label{sim.fig.5}
\end{figure}

	\fig{sim.fig.5} shows the result of several thousand independent likelihood ratio tests.  The discrepancy between the simulated distribution and $\chi_{1} ^{2}$ is a direct result of the physically motivated constraint that point source flux should be positive.  Statistical fluctuations are equally likely to result in point source like deficits as positive  excesses and such negative fluctuations result in positive $TS$ values if left unconstrained. These negative fluctuations do not correspond to physically meaningful point sources so they are assigned the value of zero flux and $TS$.  The result is that at every point except $TS$=0 the resulting distribution is a factor of two lower than the $\chi_{1} ^{2}$ prediction and all the remaining probability density is taken up at the origin.  The resulting single trial significance is correspondingly shifted to give slightly higher significance for a given value of $TS$.  This effect is small.  For example the $TS$ value corresponding to a significance of $3 \sigma \;$is 9.0 whereas the prediction was 10.3.  In general the simple expression $n_{\sigma}=\sqrt{TS}$ remains a reasonable approximation. 

	In searching for potential point sources, it is common to make many point source tests throughout the sky.  While the significance of any one source is determined as above, the significance of this set of trials is reduced through multiple trials.  Moreover, the extended $PSF$ couples nearby likelihood ratio trials making successive trials not completely independent.  Two trials separated by .5 degrees yielding elevated $TS$ values is not strong evidence for two separate sources. What is done is to generate likelihood maps in which likelihood ratio trials were made at the center of every .5deg x .5deg pixel in an observation. An "excess" is defined to be a set of contiguous pixels having $TS$ greater than or equal to some threshold.  The significance of such an excess is best defined as the chance probability of such an excess occurring per unit solid angle.

\begin{figure}[h]
\epsfysize=3.0in
\centerline{\epsfbox{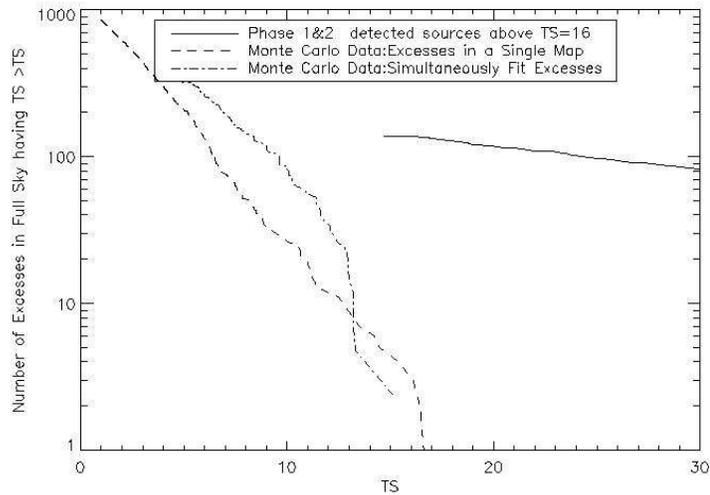}}
\caption{The distributions of spurious sources. The dashed line shows the distribution of TS within a given VP while the dot-dashed line shows the distribution that results from the simultaneous fitting of all excesses with $TS>5$.  Also shown are the phase 1+2 source distribution showing clear evidence for non-statistical fluctuations.}
\label{sim.fig.6}
\end{figure}

	\fig{sim.fig.6} shows the distribution of such excesses in several all sky likelihood maps generated with no point sources added.  The distribution of excesses represents the chance probability of finding a spurious source with specified TS or greater in the entire sky.  The distribution is no longer well approximated by $\chi_1^2$ but is significantly flatter.  The number of low significance excesses is less than what would be predicted from the simplistic argument about the number of trials because the significant excesses obscure less significant fluctuations.  The number of sources with $TS>16$ in the whole sky is $\sim 2$. The first EGRET source catalog (Fichtel et al. 1994) includes all excesses with $TS>16$ and as a result should be relatively free of spurious sources whose nature is purely statistical.

\begin{figure}[h]
\epsfysize=3.0in
\centerline{\epsfbox{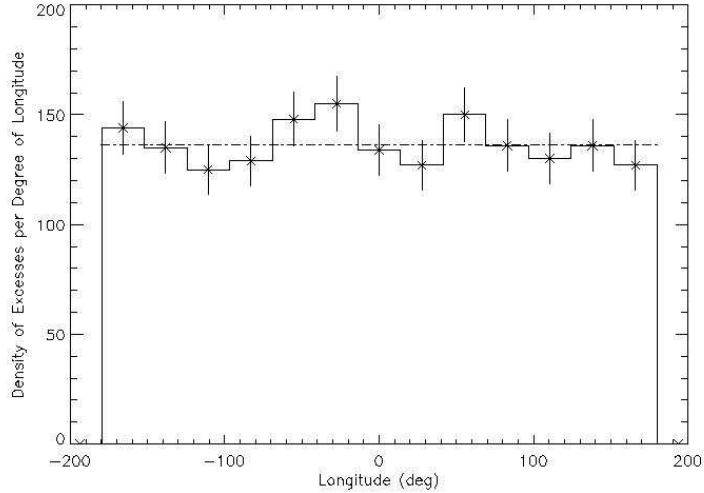}}
\caption{A distribution of TS in longitude for a simulation of the full sky with no input point sources.  There is no evidence for any nonuniformity in the distribution of TS.}
\label{sim.fig.7}
\end{figure}

	The independence of $TS$ on factors other than the number of free parameters can also be verified in this way.  The spatial distribution of excesses in the sky reveals a constant detection rate per solid angle across the sky with no correlation with exposure or background intensity (\fig{sim.fig.7}).

	Comparing these simulated $TS$ distributions with that obtained from actual flight data reveals the strong departure of the flight data from the null hypothesis.  \fig{sim.fig.6} shows a comparison between the differential distributions of $TS$ for simulated data with no point sources and for Phase I and II flight data.  It demonstrates the strong evidence for non-statistical excesses.  Some of these excesses are unambiguous such as the enormous significance of sources such as the Vela, Crab and Geminga, but at lower significance levels, the excesses present could be due to inaccuracies in the diffuse model as well as statistical fluctuations and weak point sources.

\section{ Computational Approximations: $R_{anal}$}

	The evaluation and maximization of the likelihood function is computationally intensive.  This computation time becomes prohibitive as one increases the number of pixels under analysis.  As a result, an expedient approximation is customarily made: only pixels within an angular distance $R_{anal}$ of a test point are considered when evaluating the likelihood function at that point.  An additional advantage of this truncation is that it makes the point source analysis somewhat independent of large scale inaccuracies in the background model.  The optimal value of $R_{anal}$ is one which minimizes computation time without introducing systematic analysis errors.
	
	A study of the effect of decreasing $R_{anal}$ on the best fit model parameters is shown in Table 2.  There is little sensitivity to changes in $R_{anal}$ down to values of $\sim 8$ degrees.  Beyond this point point source significance is reduced while flux increases.  This study provides justification for the conventional use of $R_{anal}=15$ degrees.

\begin{table} [h]
  \centering
  \caption{Effect of $R_{anal}$ on Parameter Estimation}
  \bigskip
  \begin{tabular}{|c|c|c|c|c|c|}
    \hline
    $R_{anal}$ & Input Source Counts & $c_a$   & $\sqrt{TS}$   & $g_m$    & $g_b$\\
    \hline
2	& 521.4	& 509.5	& 3.8	& 0.9	& 6.9\\
4	& 521.4	& 522.4	& 6.1	& 1.0	& 1.5\\
6	& 521.4	& 521.9	& 7.2	& 1.0	& 1.1\\
8	& 521.4	& 521.7	& 7.8	& 1.0	& 1.0\\
10	& 521.4	& 521.4	& 8.2	& 1.0	& 1.0\\
12	& 521.4	& 521.4	& 8.4	& 1.0	& 1.0\\
14	& 521.4	& 520.8	& 8.6	& 1.0	& 1.0\\
16	& 521.4	& 521.0	& 8.8	& 1.0	& 1.0\\
18	& 521.4	& 521.1	& 8.9	& 1.0	& 1.0\\
20	& 521.4	& 521.5	& 9.0	& 1.0	& 1.0\\
  \hline
  \end{tabular}
\end{table}

\section{ Multiple Source Analysis}
	The principal shortcoming of the likelihood technique as it has been thus far described is that it is only capable of fitting one point source and diffuse emission.  If point sources are sufficiently spaced ( $>R_{anal}$) this does not pose any problem but of course actual sources can be quite closely spaced.
	The solution to this problem has been to develop an iterative technique to effectively do multi-source fitting.  The likelihood model is then defined as,
\begin{equation}
	\theta_{ij} = g_mG_{ij} + g_bE_{ij} + c_aPSF(\alpha,\delta)_{ij} + \sum_k c_kPSF(\alpha_k,\delta_k)_{ij}\;,
\end{equation}

where $c_k,\alpha_k,\delta_k$ represent the best fit point source parameters of all sources other than the active source currently under optimization.  This technique amounts to altering the diffuse model to include features that account for the other sources present in an observation.  In practice what is done is to find the strongest point sources in an observation and add them to the diffuse model which is then used to search for further sources.  When these subsequent sources are found, the iterative technique is used to simultaneously fit all of them. This process is continued until no excesses with $TS > $ some threshold.

	This technique was tested to see if it accurately reproduced a set of sources inserted into the model.  Tables 3,4 show a comparison between the source parameters input into a simulated observation of the Virgo region and what was found upon analysis of this data. It would have been preferable to do an analysis of a statistical ensemble of such observations but the amount of computing time required made this approach impossible.  The analysis was carried down to a threshold of $TS>12$. No spurious sources appear in this observation . The average number of spurious sources with $TS > 12$ in this amount of solid angle is $\sim 2$ so this is not unlikely.  The fluxes and positions of the true sources are all within the respective error boxes of the measured data.  The analysis demonstrates the power of the likelihood technique even when the sources must be dealt with simultaneously.

\begin{table} [h]
  \centering
  \caption{Sources Input into Virgo Region (from Phase 1 Analysis)}
  \bigskip
  \begin{tabular}{|ll|cc|c|}
    \hline
EGRET Name & Identification & \multicolumn{2}{c|}{Position (J2000)}& Flux\\
  &  & $ra$ & $dec$ & ($10^{-8}cts\; cm^{-2}s^{-1}$)\\
\hline
GRO J1256-05 & 3C 279 &194.06 &-5.79 & 170\\
GRO J1232+02 & 3C 273& 188.25 & 2.82 & 20\\
GRO J1224+22 & & 186.08 & 22.75 & 13\\
GRO J1230-02 & & 187.69 & -2.96 & 10\\
\hline
  \end{tabular}
\end{table}

\begin{table} [h]
  \centering
  \caption{Results of Multisource Fit for Sources with $TS>12$}
  \bigskip
  \begin{tabular}{|l|cc|cc|cc|c|c|c|c|}
    \hline
 Name & \multicolumn{2}{c|}{Position} & \multicolumn{2}{c|}{Err.(arcmin)} & \multicolumn{2}{c|}{Flux} & $TS$ & $c_a$ & $g_m$ & $g_b$ \\ 
 & $ra$ & $dec$ & 68\% & 95\% & val. & err. & & & & \\
\hline
GRO J1256-05 & 194.12 & -5.76 & 5. & 7. & 167.6 & 5. & 2432.8 & 1301.3  & 1.37 &   0.64\\
GRO J1231+03 & 187.94  & 3.12 & 20. & 33. & 21.3 & 2.7 & 102.7 & 185.9 & 1.01 &  0.95\\
GRO J1224+22 & 186.08 & 22.80 & 21.& 48. &  11.1 & 2.8 & 24.2 & 66.1 & 0.42 &  1.46\\
GRO J1227-02 & 186.97 & -2.60 & 35.&  53. & 9.4 & 2.3 &  23.2 & 75.7 &  1.45 &  0.64\\
\hline
 \end{tabular}
\end{table}

	The distribution of spurious sources needs to be evaluated for this technique.  It is easy to see that the distribution of excesses will differ from the simple case of one source fitting.  Whereas the distribution of excesses for the first likelihood map is known in the null hypothesis, after altering the diffuse model to match this data and reevaluating the likelihood map, the new distribution will differ.  Monte Carlo techniques have allowed the direct measurement of the probability per unit solid angle of finding an excess with $TS$ greater than some threshold.

	What was done was to generate simulated observations with no point sources present and to analyze these data in exactly the same manner as flight data.  Likelihood maps are generated and strong sources identified and fit.  These sources are then added to the diffuse map and new likelihood maps are generated.  This process is repeated until no excesses above some threshold are present.  The results are shown in \fig{sim.fig.6}.  At higher values of $TS$, the results are consistent with what was found for single source analysis.  As one gets to lower values of $TS$ the distribution rises more steeply as one would expect given the fact that significant fluctuations are no longer allowed to obscure smaller ones.


\begin{thebibliography}{}


\bibitem[Alcock \etal\ 1993]{Alcock93}\reference Alcock, C., \etal\ 1993, \nature, 365, 621

\bibitem[Atwood \etal\ 1992]{Atwood92}\reference Atwood, W.B, \etal\ 1992, International Journal of Modern Physics C, 3, 459

\bibitem[Aubourg \etal\ 1993]{Aubourg93}\reference Aubourg, E., \etal\ 1993, \nature, 365, 623

\bibitem[Barcons  1990]{Barcons90}\reference Barcons, X., \& Fabian, A.C.  1990, \mnras, 243, 366

\bibitem[Bertsch \etal\ 1993]{Bertsch93}\reference Bertsch, D.L., \etal\ 1993, \apj, 416, 587

\bibitem[Biermann 1995]{Biermann95}\reference Biermann, P.L.,  1995, Space Sci. Rev., 74, 385

\bibitem[Bignami \etal\ 1981]{Bignami81}\reference Bignami, G.F., \etal\  1981, \aap, 93, 71

\bibitem[Bignami \etal\ 1975]{Bignami75}\reference Bignami, G. F., \etal\ 1975, 
Space Sci. Instr., 1, 245

\bibitem[Blandford \& Levinson 1975]{Blandford95}\reference Bignami, G. F., \etal\ 1975, 
Space Sci. Instr., 1, 245

\bibitem[Brazier \etal\ 1996]{Brazier96}\reference Brazier, K.T.S., \etal\  1996, Astron. \& Astrophys. Supp. Series, 116, 187

\bibitem[Breitschwerdt 1991]{Breitschwerdt91}\reference Blandford, R.D., \& Levinson, A.,  1995, \apj, 441, 79

\bibitem[Bunner 1989]{Bunner89}\reference Bunner, A. N. 1989, 
\GROworkshop, 1-11

\bibitem[Caravallo 1971]{Carvallo71}\reference Carvallo, G. \& Gould, R.J.  1971, Nuov. Cimento, 2B, 77

\bibitem[Cash 1977]{Cash77}\reference Cash, Webster, 1977, Ap. J., 228, 939

\bibitem[Chen  \etal\ 1995]{Chen95}\reference Chen, A., \etal\ 1995, \apjl, 445, L109

\bibitem[Chi \& Wolfendale 1991]{Chi91}\reference Chi, X. \& Wolfendale, A.W. 1991, Journal of Physics G, 17 , 987

\bibitem[Chiang \etal\ 1995]{Chiang95}\reference Chiang, J., \etal\ 1995, \apj, 452,156

\bibitem[Cisneros 1973]{Cisneros73}\reference Cisneros, A. 1973, \physrev, D7, 362

\bibitem[Cline 1990]{Cline1990}\reference Cline, David B,, \& Gao, Yi-Tian  1990, \apj, 348, 33

\bibitem[Combes \etal\ 1975]{Combes75}\reference Combes, F, \etal\  1975, Astrophys. Space Sci., 37, 151

\bibitem[Dar 1995]{Dar95}\reference Dar, Arnon, \& Shaviv, Nir J.  1995, Phys. Rev. Let., 75,17

\bibitem[DePaolis \etal\ 1995]{DePaolis95}\reference DePaolis, F., \etal\  1995, \aap, 295, 567

\bibitem[Dermer 1995]{Dermer95}\reference Dermer, Charles 1995, \apjl, 446,L63

\bibitem[Dermer 1986]{Dermer86}\reference Dermer, C.D.  1986, \aap, 157, 223

\bibitem[Dunlop \& Peacock 1990]{Dunlop90}\reference Dunlop, J.S., \& Peacock, J.A. 1990, \mnras, 247,19

\bibitem[Eadie \etal\ 1971]{Eadie71}\reference Eadie, W.T., et al., 1971, Statistical Methods in Experimental Physics, Noth-Holland Physics Pubishing Co. Inc.

\bibitem[Ellis \etal\ 1984]{Ellis84}\reference Ellis, J.,  \etal\ 1984, Nucl. Phys., B238, 453

\bibitem[Fierro 1994]{Fierro94}\reference Fierro, J. M. 1994, ``Sensitive Area
Calculation'', EGRET document: \\
EGRET/SU/JMF/94/AUG/02

\bibitem[Fichtel \etal\ 1972]{Fichtel72}\reference Fichtel, C.E., \etal\  1972, \apj, 171, 31

\bibitem[Fichtel \etal\ 1978]{Fichtel78}\reference Fichtel, C.E., \etal\  1978, \apj, 222, 833

\bibitem[Fichtel \etal\ 1991]{Fichtel91}\reference Fichtel, C.E., \etal\  1991, \apj, 374, 134

\bibitem[Fichtel \etal\ 1981]{Fichtel81}\reference Fichtel, C.E., \& Trombka, J.I. 1981, Gamma Ray Astrophysics (NASA SP-453).

\bibitem[Fichtel \etal\ 1975]{Fichtel75}\reference Fichtel, C. E., Hartman, 
R. C., Kniffen, D. A., Thompson, D. J., Bignami, G. F., \"Ogelman, H. B., 
\"Ozel, M. E., \& T\"umer, T. 1975, \apj, 198, 163

\bibitem[Fichtel \etal\ 1994]{Fichtel94}\reference Fichtel, C. E., \etal\ 1994, 
\apjs, 94, 551

\reference Fichtel, C.E., et al., 1994b, Proc. of the 2nd Compton Gamma-Ray Observatory Symposium, College Park, MD,Sept. 20-22, 1993, AIP, in press

\bibitem[Fishman \etal\ 1989]{Fishman89}\reference Fishman, G. E., \etal\ 1989,
\GROworkshop, 2-39

\bibitem[Gao \etal\ 1990]{Gao90}\reference Gao, Yi-Tian, \etal\ 1990, \apjl, 361, L37

\bibitem[Gao \etal\ 1991]{Gao91}\reference Gao, Yi-Tian, \etal\ 1991, \aap, 249, 1

\bibitem[Gao \etal\ 1990]{Gao290}\reference Gao, Yi-Tian, \etal\ 1990, \apj, 357, L1

\bibitem[Gehrels 1992]{Gehrels92}\reference Gehrels, Neil 1992, \nim, A313, 513

\bibitem[Grandi \etal\ 1996]{Grandi96}\reference Grandi, P. \etal\ 1996, \apj, 459, 73

\bibitem[Halpern \etal\  1992]{Halpern92}\reference Halpern, J.P., \etal\  1992, \nature, 357, 222

\bibitem[Hamilton  1987]{Hamilton87}\reference Hamilton, T.T., \& Helfand, D.J.  1987, \apj, 318, 93

\bibitem[Hunter \etal\ 1996]{Hunter96}\reference Hunter, S.D., \etal\ 1996, \apj, submitted

\bibitem[Hunter 1991]{Hunter91}\reference Hunter, S.D.,  1991, \nim, A307,520

\bibitem[Iwan \etal\ 1982]{Iwan82}\reference Iwan, DeAnn, \etal\ 1982, \apj, 260, 111

\bibitem[Johnson \etal\ 1993]{Johnson93}\reference Johnson, W. N., \etal\ 1993,
\apjs, 86, 693

\bibitem[Kamionkowski 1995]{Kamionkowski95}\reference Kamionkowski, Marc 1995, in The Gamma Ray Sky with CGRO and SIGMA,
ed. M. Signore, P. Salati, \& G. Vedrenne 
(Dordrecht: Kluwer Academic Publishers), 113

\bibitem[Kanbach \etal\ 1974]{Kanbach74}\reference Kanbach, G., 
\etal\ 1974, Journal of Geophysical Research, 79, 5159

\bibitem[Kanbach \etal\ 1974]{Kanbach89}\reference Kanbach, G., 
\etal\ 1989, Proc. of the Gamma-ray Science Workshop, held April 1989 Greenbelt MD, 2-1

\reference Kanbach, G. et al. 1988, Space Sci. Rev., 49, 69

\bibitem[Kanbach \etal\ 1996]{Kanbach96}\reference Kanbach, G., 
\etal\ 1996, Proc. of the Third Compton Symposium, Astron. \& Astrophys. Supp.

\bibitem[Kappadath \etal\ 1995]{Kappadath95}\reference Kappadath, S.C.., \etal\  1995, Proc. 24th Internatl. Cosmic Ray Conf. (XXIV ICRC: Rome 1995), Vol.2,230

\bibitem[Kerrick \etal\ 1995]{Kerrick95}\reference Kerrick, A.D., et al.  1995, \apj, 452, 588

\bibitem[Kinzer \etal\ 1996]{Kinzer96}\reference Kinzer, R.L., \etal\ 1996, 
\apj, submitted

\bibitem[Kniffen \etal\ 1995]{Kniffen95}\reference Kniffen, D., \etal\ 1995, 
Proc. of the Third Compton Symposium, Astron. \& Astrophys. Supp.

\bibitem[Kniffen \etal\ 1993]{Kniffen93}\reference Kniffen, D., \etal\ 1993, 
\apj, 411, 133

\bibitem[Kraushaar \etal\ 1972]{Kraushaar72}\reference Kraushaar, W, \etal\ 1972, \apj, 177, 341

\bibitem[Kutner \& Leung 1985]{Kutner85}\reference Kutner, M.L. \& Leung, C.M. 1985, \apj, 291, 188

\bibitem[Maloney \& Black 1988]{Maloney88}\reference Maloney, P. \& Black, J.H. 1988, \apj, 325, 389

\bibitem[Maccacaro \etal\ 1991]{Maccacaro91}\reference Maccacaro, Tommaso  \etal\ 1991, \apj, 374, 117

\bibitem[MacGibbon \etal\ 1993]{MacGibbon93}\reference MacGibbon, J.H.,  \etal\ 1993, \prd , 47, 2283

\bibitem[Mannheim 1992]{Mannheim92}\reference Mannheim, P.D.  1992, \apj , 391, 429

\bibitem[Maraschi \etal\ 1994]{Maraschi94}\reference Maraschi, L.,  \etal\ 1994, \apjl , 435, L91

\bibitem[Mattox \etal\ 1996]{Mattox96}\reference Mattox, J.R. , \etal\  1996, \apj, 461, 396

\bibitem[Mattox \etal\ 1987]{Mattox87}\reference Mattox, J. R., \etal\ 1987, 
Nucl. Instr. Meth., B24/25, 888

\bibitem[Mattox 1995]{MattoxAPS}\reference Mattox, J. R. 1995, 
American Physical Society Proceedings, Washington D.C.

\bibitem[Michelson \etal\  1995]{Michelson95}\reference Michelson, P.F. \etal\ 1995, NASA Instrument Concept Proposal

\reference Michelson, P.F., et al., 1994, Proc. of the 2nd Compton Gamma-Ray Observatory Symposium, College Park, MD,Sept. 20-22, 1993, AIP, in press

\bibitem[von Montigny \etal\ 1995]{vonMontigny95}\reference von Montigny, C.,  \etal\  1995, \apj, 440, 525

\bibitem[Morris  1984]{Morris84}\reference Morris, Daniel J.  1984, Journal of Geophysical Reseach, 89, 10685

\bibitem[Muecke \etal\ 1996]{Muecke96}\reference Muecke, A., \etal\  1996, submitted to \aap

\reference Neynman, J., Pearson, E.S., 1928, Biometrika, 20A, 175

\bibitem[Nice \etal\ 1994]{Nice94}\reference Nice, D.J., \etal\ 1994, Proc. of the Second Compton Symposium, Held: College Park MD, AIP Conference Proc., 304, 82

\bibitem[Nolan \etal\ 1993]{Nolan93}\reference Nolan, P.L., \etal\ 1993, \apj, 409, 697

\bibitem[Omnes 1972]{Omnes72}\reference Omnes, R. 1972, Physs Rep., 3 , 1

\bibitem[Osborne  1995]{Osborne95}\reference Osborne, J.L., Wolfendale, A.W., Zhang, L. 1995, J. Phys. G: Nucl. Part. Phys., 21, 429

\reference Pollock, A.M.T., et al.,1981, A\&A, 94,116

\bibitem[Press \etal\ 1992]{Press92}\reference Press, W. H., Teukolsky, S. A.,
Vetterling, W. T., \& Flannery, B. P. 1992, Numerical Recipes, 2nd edition
(New York: Cambridge University Press), 623

\bibitem[Punch \etal\ 1992]{Punch92}\reference Punch, M., et al.  1992, \nature, 358, 6386, 477

\bibitem[Ramani 1976]{Ramani76}\reference Ramani, A, \& Puget, J.L.  1976, \aap, 51, 411

\bibitem[Romani \etal\ 1995]{Romani95}\reference Romani, R.W, \etal\  1995, \apj, 438, 314

\bibitem[Salamon \& Stecker 1994]{Salamon94}\reference Salamon, M.H., \& Stecker, F.W.  1994, \apjl, 430, L21

\bibitem[Schmitt 1986]{Schmidt86}\reference Schmidt, Jurgen H.M.M., \& Maccacaro, Tommaso  1986, \apj, 310, 334

\bibitem[Scheuer 1957]{Scheuer57}\reference Scheuer, P.A.G., 1957, Proc. Cambridge Phil. Soc., 53, 764

\bibitem[Sch\"onfelder \etal\ 1978]{Schoenfelder78}\reference Sch\"onfelder, V., \etal\  1978, \nature, 274, 344

\bibitem[Sch\"onfelder \etal\ 1993]{Schonfelder93}\reference Sch\"onfelder, V., 
\etal\ 1993, \apjs, 86, 657

\bibitem[Shafer  1983]{Shafer83}\reference Shafer, Richard A.  1983, NASA Technical Memorandum 85029

\bibitem[Silk 1987]{Silk87}\reference Silk, Joseph, \& Bloemen, Hans 1987, \apjl, 313, L47

\bibitem[Silk 1984]{Silk84}\reference Silk, J.,  \& Srednicki, M. 1984, \prl , 53, 624

\bibitem[Sodroski \etal\ 1994]{Sodroski94}\reference Sodroski, T.J., \etal\  1994, \apj, 428, 638

\bibitem[Srednicki 1986]{Srednicki86}\reference  Srednicki, M. \& Silk, J.,  1986, \prl , 56, 263

\bibitem[Sreekumar \etal\ 1996]{Sreekumar96}\reference Sreekumar, P., \etal\ 1996, in preparation

\bibitem[Sreekumar \etal\ 1992]{Sreekumar92}\reference Sreekumar, P., \etal\ 1992, \apjl, 400, L67

\bibitem[Stecker 1970]{Stecker70}\reference Stecker, F.W. 1970, Ap\&SS, 6, 377

\bibitem[Stecker 1978]{Stecker78}\reference Stecker, F.W. 1978, \nature, 273, 493

\bibitem[Stecker 1992]{Stecker92}\reference Stecker, F.W., De Jager O.C., \& Salamon M.H.  1992, \apj, 390, L49

\bibitem[Stecker \etal\ 1993]{Stecker93}\reference Stecker, F.W., \etal\.  1992, \apjl, 415, L71

\bibitem[Stecker 1972]{Stecker72}\reference Stecker, F.W. \& Puget, J.L.  1972, \apj, 178, 57

\bibitem[Sturner \etal\ 1995]{Sturner95}\reference Sterner, S.J., \& Dermer, C.D.  1995, \aap, 293, L17

\bibitem[Strong \etal\ 1988]{Strong88}\reference Strong, A.W., \etal\  1991, \aap, 207, 1

\bibitem[Strong \& Youssefi 1995]{Strong95}\reference Strong, A.W. \& Youssefi, G. 1995, Proc. 24th Internatl. Cosmic Ray Conf. (XXIV ICRC: Rome 1995), Vol.2

\bibitem[Strong \etal\ 1993]{Strong93}\reference Strong, A. W., \etal\ 1993, 
\aaps, 97, 133

\bibitem[Sreekumar 1996]{Sreekumar96_inst}\reference Sreekumar P., et al.  1996, in preparation

\bibitem[Taylor \& Cordes 1993]{Taylor93}\reference Taylor, J.H. \& Cordes, J.M.  1993, \apj, 411, 674

\bibitem[Thompson \etal\ 1995]{Thompson96}\reference Thompson, D.J., \etal\ 1995, \apjs, 101, 259

\bibitem[Thompson \etal\ 1993]{Thompson93}\reference Thompson, D. J., 
\etal\ 1993, \apjs, 86, 629

\bibitem[Thompson \etal\ 1981]{Thompson81}\reference Thompson, D. J., 
\etal\ 1981, J. of Geophys. Res., 86, 1265

\bibitem[Trombka \etal\ 1977]{Trombka77}\reference Trombka, J.L., \etal\ 1977, \apj, 212,925

\bibitem[Wagner \etal\ 1995a]{Wagner95a}\reference Wagner, S.J.,  \etal\  1995, \apjl, 454, L97

\bibitem[Wagner \etal\ 1995b]{Wagner95b}\reference Wagner, S.J.,  \etal\  1995, \aap, 298, 688

\bibitem[Wagoner 1974]{Wagoner74}\reference Wagoner, R.V.,  Fowler, W.A., \& Hoyle, F.  1974, \apj, 148, 3

\bibitem[Weedman 1986]{Weedman86}\reference Weedman, Daniel W.  1986, Quasar Astronomy, (Cambridge University Press: Cambridge)

\bibitem[Weekes \etal\ 1989]{Weekes89}\reference Weekes, T.C., et al.  1989, \apj, 342, 379

\bibitem[Wilks 1938]{Wilks38}\reference Wilks, S.S., 1938, Ann. Math. Stat., 9, 60

\bibitem[Wright \etal\ 1991]{Wright91}\reference Wright, E.L., \etal\  1991, \apj, 381, 200

\bibitem[Yadigaroglu \& Romani 1995]{Yadigaroglu95}\reference Yadigaroglu, I.A. \& Romani, R.W. 1995, \apj, 449, 543

\bibitem[Youssefi \& Strong 1995]{Youssefi95}\reference Youssefi, G. \& Strong, A.W., 1995, Proc. 24th Internatl. Cosmic Ray Conf. (XXIV ICRC: Rome 1995), Vol.2

\bibitem[Zombeck 1990]{Zombeck90}\reference Zombeck, Martin V.,  1990, Handbook of Space Astronomy and Astrophysics, 2nd Edition, (Cambridge: Cambridge University Press)

\bibitem[Zdziarski \etal\ 1995]{Zdziarski95}\reference Zdziarski, A.A., \etal\  1995, \apjl, 438, L63


\end{thebibliography}
\end{document}